\documentclass[a4paper, 11pt]{article}
\pdfoutput=1 % if your are submitting a pdflatex (i.e., if you have
\usepackage{jheppub} % for details on the use of the package, please
\usepackage[T1]{fontenc} % if needed    
                     
\usepackage{hyperref}  
\usepackage{amsmath}  
\usepackage{amssymb}  
\usepackage{latexsym}
\usepackage{graphicx}
\usepackage{subfig} 
\usepackage{empheq}
\usepackage{braket}
\usepackage{amssymb}
\usepackage{amsmath}
\usepackage{amsthm}
\usepackage{mathrsfs}
\usepackage{framed}
\usepackage{hyperref}
\usepackage{slashed}
\usepackage{array}

\hypersetup{colorlinks=true}
\hypersetup{linkcolor=blue}
\hypersetup{citecolor=blue}
\hypersetup{urlcolor=blue}
\numberwithin{equation}{section}

\usepackage
{datetime}
\usepackage{fancyhdr}
\usepackage{tikz}
\usetikzlibrary{arrows,shapes}
\usetikzlibrary{trees}
\usetikzlibrary{matrix,arrows} 				% For commutative diagram
\usetikzlibrary{positioning}				% For "above of=" commands
\usetikzlibrary{calc,through}				% For coordinates
\usetikzlibrary{decorations.pathreplacing}  % For curly braces
\usepackage{pgffor}							% For repeating patterns

 \usetikzlibrary{fit}

\usetikzlibrary{decorations.pathmorphing}	% For Feynman Diagrams
\usetikzlibrary{decorations.markings}
\tikzset{
    vector/.style={decorate, decoration={snake}, draw},
	provector/.style={decorate, decoration={snake,amplitude=2.5pt}, draw},
	antivector/.style={decorate, decoration={snake,amplitude=-2.5pt}, draw},
    fermion/.style={draw=black, postaction={decorate},
        decoration={markings,mark=at position .55 with {\arrow[draw=black]{>}}}},
    fermionbar/.style={draw=black, postaction={decorate},
        decoration={markings,mark=at position .55 with {\arrow[draw=black]{<}}}},
    fermionnoarrow/.style={draw=black},
    gluon/.style={decorate, draw=black,
        decoration={coil,amplitude=4pt, segment length=5pt}},
    scalar/.style={dashed,draw=black, postaction={decorate},
        decoration={markings,mark=at position .55 with {\arrow[draw=black]{>}}}},
    scalarbar/.style={dashed,draw=black, postaction={decorate},
        decoration={markings,mark=at position .55 with {\arrow[draw=black]{<}}}},
    realscalar/.style={draw=black}, 
    electron/.style={draw=black, postaction={decorate},
        decoration={markings,mark=at position .55 with {\arrow[draw=black]{>}}}},
	bigvector/.style={decorate, decoration={snake,amplitude=4pt}, draw},
    phir/.style={draw=blue, postaction={decorate},},
   phil/.style={dashed,draw=blue,},
     phiav/.style={draw=cyan, postaction={decorate},},
    phidif/.style={ draw=cyan,},  
       chir/.style={draw=red, postaction={decorate},},
   chil/.style={dashed,draw=red,},
}

\usepackage[american,cuteinductors,smartlabels]{circuitikz}

\usetikzlibrary{calc}
\ctikzset{bipoles/thickness=1}
\ctikzset{bipoles/length=0.8cm}
\ctikzset{bipoles/diode/height=.375}
\ctikzset{bipoles/diode/width=.3}
\ctikzset{tripoles/thyristor/height=.8}
\ctikzset{tripoles/thyristor/width=1}
\ctikzset{bipoles/vsourceam/height/.initial=.7}
\ctikzset{bipoles/vsourceam/width/.initial=.7}
\tikzstyle{every node}=[font=\small]
\tikzstyle{every path}=[line width=0.8pt,line cap=round,line join=round]
\usetikzlibrary{positioning}

\newcommand{\phipropagator}[7]{% x-coordinate, y-coordinate,
\draw [#6,ultra thick] (#1,#2) -- (#1/2+#3/2,#2/2+#4/2);
\draw [#7,ultra thick] ((#1/2+#3/2,#2/2+#4/2) -- (#3,#4);	
\node at (#1/2+#3/2+.5,#2/2+#4/2+.5) {$#5$};	
} 

\newcommand{\phipropagatorr}[5]{% x_1-coordinate, y_1-coordinate,x_2-coordinate, y_2-coordinate 
\phipropagator{#1}{#2}{#3}{#4}{#5}{phir}{phir}
}

\newcommand{\phipropagatorp}[5]{% x_1-coordinate, y_1-coordinate,x_2-coordinate, y_2-coordinate 
\phipropagator{#1}{#2}{#3}{#4}{#5}{phir}{phil}
} 

\newcommand{\phipropagatorm}[5]{% x_1-coordinate, y_1-coordinate,x_2-coordinate, y_2-coordinate 
\phipropagator{#1}{#2}{#3}{#4}{#5}{phil}{phir}
}

\newcommand{\phipropagatorl }[5]{% x_1-coordinate, y_1-coordinate,x_2-coordinate, y_2-coordinate 
\phipropagator{#1}{#2}{#3}{#4}{#5}{phil}{phil}
}

\newcommand{\wtchipropagatorp}[4]{% x-coordinate, y-coordinate,
\draw [chir,ultra thick](#1,#2) -- (#1/2+#3/2,#2/2+#4/2);
\draw [chil,ultra thick] (#1,#2) -- (#3,#4);	
}

\newcommand{\phipropagatora}[5]{% x-coordinate, y-coordinate,
\begin{scope}[shift={(#1,#2)},rotate=#3]
\begin{scope}[scale=#4]
\draw [phiav, ultra thick] (0,0) -- (1,0); 
\draw [phiav, ultra thick] (1,0) -- (2,0);

\begin{scope}[shift={(1,0)} ]  
\draw [phiav, ultra thick] (.1,.2) -- (.1,-.2);	
\draw [phiav, ultra thick] (-.1,.2) -- (-.1,-.2);	
\end{scope}	 
\end{scope}    
\end{scope}    

}

\newcommand{\phipropagatorf}[5]{% x-coordinate, y-coordinate,
\begin{scope}[shift={(#1,#2)},rotate=#3]
\begin{scope}[scale=#4]
\draw [phiav, ultra thick] (0,0) -- (1,0); 
\draw [phiav, ultra thick][->](2,0) -- (1,0);

\begin{scope}[shift={(1,0)} ]  	
\draw [phiav, ultra thick] (.05,.2) -- (.05,-.2);	
\end{scope}	 
\end{scope}    
\end{scope}    

}

\newcommand{\phipropagatorb}[5]{% x-coordinate, y-coordinate,
\begin{scope}[shift={(#1,#2)},rotate=#3]
\begin{scope}[scale=#4]
\draw [phiav, ultra thick][->] (0,0) -- (1,0); 
\draw [phiav, ultra thick] (2,0) -- (1,0);

\begin{scope}[shift={(1,0)} ]  	
\draw [phiav, ultra thick] (-.05,.2) -- (-.05,-.2);	
\end{scope}	 
\end{scope}    
\end{scope}    

}

\newcommand{\phipropagatord}[5]{% x-coordinate, y-coordinate,
\begin{scope}[shift={(#1,#2)},rotate=#3]
\begin{scope}[scale=#4]
\draw [phiav, ultra thick][->] (0,0) -- (1,0); 
\draw [phiav, ultra thick][->] (2,0) -- (1,0);
\draw [phiav, ultra thick] (2,0) -- (0,0);

\end{scope}    
\end{scope}    

}

\newcommand{\phihalfpropagatora}[5]{% x-coordinate, y-coordinate,
\begin{scope}[shift={(#1,#2)},rotate=#3]
\begin{scope}[scale=#4]
\draw [phiav, ultra thick] (0,0) -- (1,0); 
\begin{scope}[shift={(1,0)} ]  
\draw [phiav, ultra thick] (-.1,.2) -- (-.1,-.2);	
\end{scope}	 
\end{scope}    
\end{scope}    

}

\newcommand{\phihalfpropagatord}[5]{% x-coordinate, y-coordinate,
\begin{scope}[shift={(#1,#2)},rotate=#3]
\begin{scope}[scale=#4]
\draw [phiav, ultra thick][->] (0,0) -- (1,0); 
\end{scope}    
\end{scope}    

}

\newcommand{\drawphiavdifdiagaa}[3]{% x-coordinate, y-coordinate % angle/orientation,
\begin{scope}[shift={(#1,#2)},rotate=#3]
\draw [phiav, ultra thick, domain=180:360] plot ({1*cos(\x)}, {1*sin(\x)});
\draw [phiav, ultra thick, domain=0:360] plot ({1*cos(\x)}, {1*sin(\x)});
\begin{scope}[shift={(1,0)}] 
\draw [phiav, ultra thick] (.2,.1) -- (-.2,.1);	
\draw [phiav, ultra thick] (.2,-.1) -- (-.2,-.1);

\end{scope}	 
\node at (-1,0) {$\times $};	 
\end{scope}    
}

\newcommand{\drawphiavdifdiagaf}[3]{% x-coordinate, y-coordinate % angle/orientation,
\begin{scope}[shift={(#1,#2)},rotate=#3]
\draw [phidif, ultra thick, domain=180:360][->] plot ({1*cos(\x)}, {1*sin(\x)}); 
\draw [phiav, ultra thick, domain=0:180] plot ({1*cos(\x)}, {1*sin(\x)});

\begin{scope}[shift={(1,0)} ]  	
\draw [phiav, ultra thick] (-.2,-.05) -- (.2,-.05);	 
\end{scope}	  
\node at (-1,0) {$\times $};	 
\end{scope}     
}

\newcommand{\drawphiavdifdiagbaa}[3]{% x-coordinate, y-coordinate,% angle/orientation
\begin{scope}[shift={(#1,#2)}]
\begin{scope}[rotate=#3]
\draw [phiav, ultra thick, domain=260:360] plot ({1*cos(\x)}, {1*sin(\x)});
\draw [phiav, ultra thick, domain=180:270] plot ({1*cos(\x)}, {1*sin(\x)});
\draw [phiav, ultra thick, domain=80:180][] plot ({1*cos(\x)}, {1*sin(\x)});
\draw [phiav, ultra thick, domain=0:90] plot ({1*cos(\x)}, {1*sin(\x)});
\node at (1,0) {$\times $};	
\node at (-1,0) {$\times $};		

\begin{scope}[shift={(0,-1)}] 
\draw [phiav, ultra thick] (.1,.2) -- (.1,-.2);	
\draw [phiav, ultra thick] (-.1,.2) -- (-.1,-.2);
\end{scope}

\begin{scope}[shift={(0,1)}] 
\draw [phiav, ultra thick] (.1,.2) -- (.1,-.2);	
\draw [phiav, ultra thick] (-.1,.2) -- (-.1,-.2);
\end{scope}  
\end{scope}    
\end{scope}     
}

\newcommand{\drawphiavdifdiagbaf}[3]{% x-coordinate, y-coordinate,% angle/orientation
\begin{scope}[shift={(#1,#2)}]
\begin{scope}[rotate=#3]
\draw [phidif, ultra thick, domain=360:270][->] plot ({1*cos(\x)}, {1*sin(\x)}); 
\draw [phiav, ultra thick, domain=180:280] plot ({1*cos(\x)}, {1*sin(\x)});
\draw [phiav, ultra thick, domain=80:180][] plot ({1*cos(\x)}, {1*sin(\x)});
\draw [phiav, ultra thick, domain=0:100] plot ({1*cos(\x)}, {1*sin(\x)});
\node at (1,0) {$\times $};	
\node at (-1,0) {$\times $};	 
\begin{scope}[shift={(0,1)}] 
\draw [phiav, ultra thick] (.1,.2) -- (.1,-.2);	
\draw [phiav, ultra thick] (-.1,.2) -- (-.1,-.2);
\end{scope}	  

\begin{scope}[shift={(0,-1)}] 
\draw [phiav, ultra thick] (.1,.2) -- (.1,-.2);	
\end{scope}

\end{scope}    
\end{scope}    
} 
 
\newcommand{\drawphiavdifdiagbab}[3]{% x-coordinate, y-coordinate,% angle/orientation
\begin{scope}[shift={(#1,#2)}]
\begin{scope}[rotate=#3]
\draw [phiav, ultra thick, domain=260:360] plot ({1*cos(\x)}, {1*sin(\x)});
\draw [phidif, ultra thick, domain=180:270][->] plot ({1*cos(\x)}, {1*sin(\x)});
\draw [phiav, ultra thick, domain=80:180] plot ({1*cos(\x)}, {1*sin(\x)});
\draw [phiav, ultra thick, domain=0:100] plot ({1*cos(\x)}, {1*sin(\x)});
\node at (1,0) {$\times $};	
\node at (-1,0) {$\times $};		

\begin{scope}[shift={(0,1)}] 
\draw [phiav, ultra thick] (.1,.2) -- (.1,-.2);	
\draw [phiav, ultra thick] (-.1,.2) -- (-.1,-.2);
\end{scope}	 

\begin{scope}[shift={(0,-1)}] 
\draw [phiav, ultra thick] (-.1,.2) -- (-.1,-.2);	
\end{scope}
  
\end{scope}    
\end{scope}    
}

\newcommand{\drawphiavdifdiagbfb}[3]{% x-coordinate, y-coordinate,% angle/orientation
\begin{scope}[shift={(#1,#2)}]
\begin{scope}[rotate=#3]
\draw [phidif, ultra thick, domain=260:360][<-] plot ({1*cos(\x)}, {1*sin(\x)});
\draw [phiav, ultra thick, domain=180:280] plot ({1*cos(\x)}, {1*sin(\x)});
\draw [phiav, ultra thick, domain=80:180]plot ({1*cos(\x)}, {1*sin(\x)});
\draw [phidif, ultra thick, domain=0:100][->] plot ({1*cos(\x)}, {1*sin(\x)});
\node at (1,0) {$\times $};	
\node at (-1,0) {$\times $};	 

\begin{scope}[shift={(0,1)}] 
\draw [phiav, ultra thick] (-.1,.2) -- (-.1,-.2);
\end{scope}	 

\begin{scope}[shift={(0,-1)}] 
\draw [phiav, ultra thick] (-.1,.2) -- (-.1,-.2);	
\end{scope}

\end{scope}    
\end{scope}    
}  

\newcommand{\drawphiavdifdiagbff}[3]{% x-coordinate, y-coordinate,% angle/orientation
\begin{scope}[shift={(#1,#2)}]
\begin{scope}[rotate=#3]
\draw [phidif, ultra thick, domain=260:360][<-] plot ({1*cos(\x)}, {1*sin(\x)});
\draw [phiav, ultra thick, domain=180:280] plot ({1*cos(\x)}, {1*sin(\x)});
\draw [phidif, ultra thick, domain=80:180][<-] plot ({1*cos(\x)}, {1*sin(\x)});
\draw [phiav, ultra thick, domain=0:100][] plot ({1*cos(\x)}, {1*sin(\x)});
\node at (1,0) {$\times $};	
\node at (-1,0) {$\times $};	

\begin{scope}[shift={(0,1)}] 
\draw [phiav, ultra thick] (.1,.2) -- (.1,-.2);	
\end{scope}	 

\begin{scope}[shift={(0,-1)}] 
\draw [phiav, ultra thick] (-.1,.2) -- (-.1,-.2);	
\end{scope}	
  
\end{scope}    
\end{scope}    
} 

\newcommand{\counterterm}[3]{% x-coordinate, y-coordinate,% angle/orientation
\begin{scope}[shift={(#1,#2)}]
\filldraw (0,0) circle (#3);  
\end{scope}    
}

\tikzstyle{block} = [draw, rectangle, 
    minimum height=3em, minimum width=6em]
\def\phidif{\phi_{d}} 
\def\phiav{\phi_{a}} 
\def\phir{\phi_{\textrm{R}}} 
\def\phil{\phi_{\textrm{L}}}

\def\im{\textrm{Im}} 
\def\re{\textrm{Re}} 
\def\ln{\textrm{ln}} 

\def\tr{\textrm{tr}}
\def\msbar{ {\overline {\textrm{MS}}} }

\def\diver{\textrm{div}}

\hbadness=\maxdimen
\vbadness=\maxdimen

\allowdisplaybreaks[1]

\title{\boldmath Renormalization in Open Quantum Field theory I: Scalar field theory}
\author[a]{Avinash}
\author[b]{, Chandan Jana}
\author[b]{, R. Loganayagam}
\author[c]{, Arnab Rudra}
\affiliation[a]{Indian Institute of Science,\\
C.V. Raman Avenue, Bangalore 560012, India.}
\affiliation[b]{International Centre for Theoretical Sciences (ICTS-TIFR)\\ 
Shivakote, Hesaraghatta Hobli, Bengaluru 560089, India.}
\affiliation[c]{Center for Quantum Mathematics and Physics (QMAP)\\
Department of Physics, University of California, Davis, CA 95616 USA}
\emailAdd{baidyaavinash@gmail.com} 
\emailAdd{chandan.jana@icts.res.in} 
\emailAdd{nayagam@gmail.com} 
\emailAdd{rudra.arnab@gmail.com}  

\abstract{While the notion of open quantum systems is itself old, most of the existing studies deal with quantum mechanical systems rather than quantum field theories. 
 After  a brief review of  field theoretical/path integral tools currently available to deal with open quantum field theories, 
 we  go on to apply these tools to an open version of  $\phi^3+\phi^4$ theory in four spacetime dimensions and demonstrate its one loop renormalizability (including the renormalizability of the  Lindblad structure).}

\begin{document}
\maketitle
\raggedbottom

%\newpage
 
\section{Introduction and Motivation}
\label{sec:intro}

Effective field theories are one of the great success stories of theoretical physics. From our understanding of elementary particles of the standard model to current cosmological models of evolution of the universe, from the theory of critical phenomena to polymer physics, the range and success of effective field theories is wide and diverse. The concept and the techniques of renormalisation, in particular have become textbook material and essential tools in the toolkit of many a theoretical physicist. Over the past few decades, String theory has further enriched this structure with its system of dualities, including the shocking suggestion that many theories of quantum gravity are really large $N$ quantum field theories in disguise.  

Despite all these successes, there are a variety of phenomena which still resist a clear understanding from the standard effective field theory viewpoint. A large class of them involve dissipation and information loss in evolution. It may be because the systems are open quantum systems in contact with an environment. Or the system might effectively behave like an open system because coarse-graining has traced out some  degrees of freedom into which the system dissipates. To tackle these systems, one needs to develop a quantum field theory of \emph{mixed states} where we can trace out degrees of freedom, run on a renormalisation flow and study dualities. 

This is not a new question. Two of the founders of quantum field theory - Schwinger and Feynman addressed these questions early on and made seminal contributions to the quantum field theories of density matrices. These are the notions of a Schwinger-Keldysh path integral \cite{Schwinger:1960qe, Keldysh:1964ud} and the Feynman-Vernon influence functionals \cite{Feynman:1963fq, Vernon:1959pla} - the first addressing how to set up the path-integral for unitary evolution of density matrices by doubling the fields and the second addressing how coarse-graining in a free theory leads to a density matrix path-integral with non-unitary evolution.  

The third classic result in this direction is by Veltman who, in the quest to give diagrammatic proofs of Cutkosky's cutting rules \cite{Cutkosky:1960sp}, effectively reinvented the Schwinger-Keldysh path integral and proved that the corresponding correlators obey the largest time equation \cite{Veltman:1963th, tHooft:1973pz}. The fourth important advance towards the effective theory of mixed states is the  discovery of the quantum master equation by Gorini-Kossakowski-Sudarshan \cite{Gorini:1975nb} and Lindblad \cite{Lindblad:1975ef}. The quantum master equation prescribes a specific form for the Feynman-Vernon influence functional \cite{Feynman:1963fq, Vernon:1959pla} using the constraints that  evolution should preserve the trace of the density matrix (trace-preserving) as well as keep the eigenvalues of the density matrix stably non-negative (complete positivity). We will review these ideas and their inter-relations in due turn. Our goal here is to construct a simple relativistic field theory which 
elucidates these ideas.

Before we move on to the subject of the paper, let us remind the reader of the broader motivations which drive this work. First of all, the theory of open quantum systems is a field with many recent advancements and is of experimental relevance to fields  like quantum optics, cold atom physics, non-equilibrium driven systems and quantum information. (See \cite{Breuer:2002pc, weiss2008quantum, carmichael2009open, rivas2011open, schaller2014open} for  textbook treatments of the subject.) It makes logical sense to test these ideas against relativistic QFTs and how they change under Wilsonian renormalisation.\footnote{We should mention that in the non-relativistic context, various interacting models and their 1-loop renormalisation have already been studied. We will refer the reader to chapter 8 of \cite{Kamenev} for textbook examples of 1-loop renormalisation in non-relativistic non-unitary QFTs. The examples include Hohenberg-Halperin classification of dynamics near classical critical 
points, reaction diffusion models, their critical behavior/scaling and surface growth models including the famous KPZ equations. A more detailed exposition is available in \cite{täuber2014critical}.} Second, open relativistic QFTs are very relevant by themselves in heavy ion physics and cosmology \cite{Jordan:1986ug, Calzetta:1986ey, Weinberg:2005vy, Calzetta:2008iqa} . Third motivation is to better understand the apparently non-unitary evolution engendered by black holes and to give a quantitative characterization of the information loss. In particular, AdS/CFT suggests that exterior of black holes is naturally dual to open conformal field theories. Hence, it is reasonable to expect that developing  the theory of  open conformal field theories would tell us how to think about horizons in quantum gravity. 
 
In this work, we take a modest step towards answering these questions by setting up 
and studying the simplest looking open quantum field theory : the open version of scalar $ \phi^3+\phi^4 $ in $d=4$ space-time dimensions. One can characterise the effective theory of density matrix of $\phi^3+\phi^4$ theory by a Schwinger-Keldysh (SK) effective action. This action involves the ket field $\phir$ as well as the bra field $\phil$ describing the two side evolution of the density matrix.
It takes the form
\begin{equation}
\begin{split} 
S_{\phi} &= -\int d^d x \Bigl[ \frac{1}{2} z\ (\partial \phir)^2 + \frac{1}{2} m^2  \phir^2+ \frac{\lambda_3}{3!}\phir^3+\frac{\lambda_4}{4!} \phir^4+\frac{\sigma_3}{2!}\phir^2\phil + \frac{\sigma_4}{3!} \phir^3 \phil \Bigr]\\
&\qquad+\int d^d x \Bigl[ \frac{1}{2} z^\star (\partial \phil)^2 + \frac{1}{2} {m^2}^\star \phil^2 +\frac{\lambda_3^\star }{3!}\phil^3+ \frac{\lambda_4^\star}{4!} \phil^4+ \frac{\sigma_3^\star }{2!}\phil^2\phir+\frac{\sigma_4^\star}{3!} \phil^3 \phir\Bigr] \\
&\qquad + i \int d^d x \Bigl[ z_\Delta\ (\partial \phir).(\partial \phil)  + m^2_\Delta \phir \phil+\frac{{\lambda_\Delta}}{2!2!} \phir^2 \phil^2 \Bigr]
\label{Smacro}
\end{split}
\end{equation}
This is the most general local, power-counting renormalisable, Lorentz invariant and
CPT invariant action that could be written down involving $\phir$ and $\phil$. Note that
CPT acts as an anti-linear, anti-unitary symmetry exchanging $\phir$ and $\phil$ and taking $i\mapsto (-i)$. It can be easily checked that, under this anti-linear, anti-unitary flip $e^{iS}$ remains invariant
provided the couplings appearing in the last line of action $\{ z_\Delta,m^2_\Delta, \lambda_\Delta\}$ are real. This action along with a future boundary condition identifying $\phir$ and $\phil$ at future infinity defines the SK effective theory which we will study in this paper.

There are two features of the above action which makes it distinct from the SK effective action of the unitary $\phi^3+\phi^4$ theory. First, there are interaction terms which couple the ket field $\phir$ with the bra field $\phil$. Such cross couplings necessarily violate unitarity and indicate the breakdown of the usual Cutkosky cutting rules .
They are  also necessarily a part of `influence functionals' as defined by Feynman and Vernon and are generated only when a part of the system is traced out \cite{Feynman:1963fq, Vernon:1959pla}. A more obvious way the above action violates unitarity is due to the fact that $S$ is not purely real. If we turn off all cross couplings between $\phir$ and $\phil$ and set to zero
all imaginary couplings in $S$, we recover  the SK effective action of the unitary $\phi^4$ theory :
\begin{equation}
\begin{split} 
S_{\phi,\text{Unitary}} &= -\int d^d x \Bigl[ \frac{1}{2} z\ (\partial \phir)^2 + \frac{1}{2} m^2  \phir^2+ \frac{\lambda_3}{3!}\phir^3+\frac{\lambda_4}{4!} \phir^4\Bigr]\\
&\qquad+\int d^d x \Bigl[ \frac{1}{2} z (\partial \phil)^2 + \frac{1}{2} {m}^2 \phil^2 +\frac{\lambda_3 }{3!}\phil^3+ \frac{\lambda_4}{4!} \phil^4\Bigr]
\label{SUmacro}
\end{split}
\end{equation}
where all couplings are taken to be real. Our aim is to deform $\phi^4$ theory away from this familiar unitary limit and study the theory defined in \eqref{Smacro} 
via perturbation theory.

The first question one could ask is whether this theory is renormalisable in perturbation theory, i.e., whether, away from unitary limit, the one-loop divergences in this theory can be absorbed into counter
terms of the same form. We answer this in affirmative in this work and compute the 1-loop  beta functions  to be
\begin{equation}\label{eq:betamassintro}
\begin{split}
\frac{d m^2}{d\ \ln\, \mu}
&=\frac{1}{(4\pi)^2} (\lambda_3+\sigma_3^\star)  (\lambda_3+2\sigma_3-\sigma_3^\star)+ \frac{m^2}{(4\pi)^2} \Bigl[ \lambda_4+2\sigma_4 -i\lambda_\Delta\ \Bigr]\\
\frac{d m_\Delta^2}{d\ \ln\, \mu}
&=-\frac{4}{(4\pi)^2} \im  \ \sigma_3\ (\re \ \lambda_3+ \ \re \ \sigma_3)+ \frac{2}{(4\pi)^2} \re \Bigl[ m^2 (\lambda_\Delta+i\sigma_4) \Bigr]
\end{split}
\end{equation}
for the mass terms,
\begin{equation}\label{eq:betacubicintro}
\begin{split}
\frac{d \lambda_3}{d\ \ln\, \mu}
&=\frac{3}{(4\pi)^2} \Bigl[ \lambda_4(\lambda_3+\sigma_3)+ \sigma_4(\lambda_3+\sigma_3^\star)+i \lambda_\Delta(\sigma_3-\sigma_3^\star) \Bigr] \\
\frac{d \sigma_3}{d\ \ln\, \mu}
&=\frac{1}{(4\pi)^2} \Bigl[ (\lambda_4+2\sigma_4^\star)(\sigma_3-\sigma_3^\star) +\sigma_4 (\lambda_3^\star+2\lambda_3+3\sigma_3)-i\lambda_\Delta(\lambda_3^\star+2\lambda_3+3\sigma_3^\star)
 \Bigr] \\
\end{split}
\end{equation}
for the cubic couplings, and
\begin{equation}\label{eq:betaquarticintro}
\begin{split}
\frac{d \lambda_4}{d\ \ln\, \mu}
&= \frac{3}{(4\pi)^2}  (\lambda_4+2\ \sigma_4-i\lambda_\Delta)(\lambda_4+i \lambda_\Delta)\\
\frac{d \sigma_4}{d\ \ln\, \mu}
&=\frac{3}{(4\pi)^2} (\lambda_4+\sigma_4+\sigma_4^\star+i\lambda_\Delta)(\sigma_4 - i \lambda_\Delta)\\
\frac{d \lambda_\Delta}{d\ \ln\, \mu}
&=\frac{1}{(4\pi)^2i} \Bigl[ (\lambda_4+2\sigma_4^\star)(\sigma_4^\star+ i \lambda_\Delta)+3i \sigma_4 \lambda_\Delta -c.c. \Bigr]\\
\end{split}
\end{equation}
for the quartic couplings. Note that at $1$-loop level we can set $z=1$ and $z_\Delta=0$ since there is no field renormalisation. These equations constitute the central result of this paper.

The above set of $\beta$ functions have a remarkable property which is made evident
by deriving the $1$-loop renormalisation running of certain combinations of couplings :
\begin{equation}\label{eq:betalindlbladintro}
\begin{split}
\frac{d}{d\ \ln\, \mu}{(\im\,m^2 - m_\Delta^2)}&=\frac{2}{(4\pi)^2}\Bigg[(\im\, \lambda_3+3\, \im\,  \sigma_3)(\re\, \lambda_3+\re\,  \sigma_3) \\
&\qquad\quad
+  (\im\lambda_4\ +4\im\,\sigma_4  -3\lambda_\Delta )(\re\,m^2)\Bigg]\\
\frac{d}{d\ \ln\, \mu}{(\im\,  \lambda_3+3\im\,  \sigma_3)}
&=\frac{3}{(4\pi)^2}\Bigg[(\re\, \lambda_4+2\re\, \sigma_4) (\im\, \lambda_3+3\im\, \sigma_3)
\\
&\qquad\quad
+(\re\, \lambda_3+\re\, \sigma_3)(\im\, \lambda_4 +4\im\, \sigma_4 -3\lambda_\Delta)\Bigg]
\\
  \frac{d}{d\ \ln\, \mu}{(\im\,  \lambda_4+4\im\,  \sigma_4-3\lambda_\Delta)}
&=
\frac{6}{(4\pi)^2} (\im  \ \lambda_4+ 4\ \im  \ \sigma_4 -3 \lambda_\Delta)  (\re \ \lambda_4+ 2 \re \ \sigma_4) 
\end{split}
\end{equation}
These equations show that the conditions 
\begin{equation}
\label{eq:LindbladCondn}
\im\, z- z_\Delta=0\ ,\quad \im\,m^2 - m_\Delta^2=0\ ,\quad \im\,  \lambda_3+3\ \im\,  \sigma_3 =0\ ,\quad
\im  \ \lambda_4+ 4\ \im  \ \sigma_4 -3 \lambda_\Delta=0 \ ,
\end{equation}
are preserved under renormalisation! We will prove a non-renormalisation theorem at all orders in perturbation theory to prove that the above conditions are never corrected at any order in loops. One can think of this as  violating Gell-Mann's totalitarian principle \cite{Gell-Mann:1956iqa} that ``Everything not forbidden is compulsory" (or as there being new principles in open quantum field theory which forbid some combinations from appearing in perturbation theory).  This kind of fine-tuning of couplings which are still protected under renormalisation is a hallmark of open quantum field theories and is a signature of microscopic unitarity \cite{Baidya:2017ab}. 
 
We will now move to briefly describe the significance of the above conditions. We will give three related derivations of the conditions above in this work:
\begin{enumerate}
\item In the Schwinger-Keldysh formalism, the microscopic unitarity demands that difference operators (i.e., operators of the form $O_R-O_L$ ) have trivial correlators. This, as a statement about correlation functions, should hold even in the coarse-grained open effective field theory. The decoupling of difference operators then naturally lead to the conditions above.

\item Relatedly, while the open EFT is non-unitary, one can demand that a certain weaker version of Veltman's largest time equation be obeyed. This then  leads to the conditions above.

\item The trace preserving and the complete positivity of the evolution demands that the Feynman-Vernon influence functional be of the Lindblad form.
Insisting that the dynamics of the open EFT be of the Lindblad form naturally leads to the conditions above.
\end{enumerate}
Thus, a certain weak form of unitarity still holds in the open EFT and is explicitly realized by the conditions above. And once these conditions are satisfied,
the structure is robust against perturbative renormalisation.

%\begin{equation}
%\begin{split} 
%\delta \phir =\delta \phil =\bar{\epsilon} \psi + \epsilon \bar{\psi}\ , \qquad  \delta \psi =  \epsilon (\phir-\phil)\ , \qquad  \delta \bar{\psi} =  -\bar{\epsilon} (\phir-\phil)
%\end{split}
%\end{equation}
%
%\begin{equation}
%\begin{split} 
%S_\psi &= -\int d^d x \Bigl[  z_\psi\ (\partial \bar{\psi}).(\partial \psi)  +  \Bigr(m_\psi^2+y_3 \phir + y_3^\star \phil+ \frac{1}{2!} (y_4 \phir^2+y_\Delta \phir \phil+y_4^\star \phil^2) \Bigl) \bar{\psi} \psi \Bigr]
%\label{SPsi1}
%\end{split}
%\end{equation}
%\begin{equation}
%\begin{split} 
%z_\psi &= \re\ z\ ,\qquad  m_\psi^2= \re\ m^2\ ,\qquad \\
%y_3&=\frac{1}{3} (\re\ \lambda_3 + \re\ \sigma_3) + \frac{i}{4} (\im\ \lambda_3-\im \sigma_3) \ ,\qquad\\
%y_4&=\frac{1}{3} (\re\ \lambda_4 + \re\ \sigma_4) + \frac{i}{4} (\im\ \lambda_4+\lambda_\Delta) \ ,\qquad\\
%y_\Delta &= \frac{1}{3} (\re\ \lambda_4 + 4\ \re\ \sigma_4) 
%\end{split}
%\end{equation}

There is a fourth way of deriving the same conditions, whose deeper significance we will leave for 
our future work. Say one adds to the above action for the open EFT two Grassmann odd ghost fields $g$ and $\bar{g}$ and demand that the following
Grassmann odd symmetry hold for the entire theory :
\begin{equation}
\begin{split} 
\delta \phir =\delta \phil =\bar{\epsilon} g + \epsilon \bar{g}\ , \qquad  \delta g =  \epsilon (\phir-\phil)\ , \qquad  \delta \bar{g} =  -\bar{\epsilon} (\phir-\phil)\ .
\end{split}
\end{equation}
This symmetry then fixes the $\phi$ self-couplings to obey equation \eqref{eq:LindbladCondn}. Further, the ghost action is completely fixed to be
\begin{equation}
\begin{split} 
S_g &= -\int d^d x \Bigl[  z_g\ (\partial \bar{g}).(\partial g)  +  \Bigr(m_g^2+\mathscr{Y}_3 \phir + \mathscr{Y}_3^\star \phil+ \frac{1}{2!} (\mathscr{Y}_4 \phir^2+\mathscr{Y}_\Delta \phir \phil+\mathscr{Y}_4^\star \phil^2) \Bigl) \bar{g} g \Bigr]
\label{SPsi2}
\end{split}
\end{equation}
where
\begin{equation}
\begin{split} 
z_g &= \re\ z\ ,\qquad  m_g^2= \re\ m^2\ ,\qquad \\
\mathscr{Y}_3&=\frac{1}{3} (\re\ \lambda_3 + \re\ \sigma_3) + \frac{i}{4} (\im\ \lambda_3-\im \sigma_3) \ ,\qquad\\
\mathscr{Y}_4&=\frac{1}{3} (\re\ \lambda_4 + \re\ \sigma_4) + \frac{i}{4} (\im\ \lambda_4+\lambda_\Delta) \ ,\qquad\\
\mathscr{Y}_\Delta &= \frac{1}{3} (\re\ \lambda_4 + 4\ \re\ \sigma_4) 
\end{split}
\end{equation}
If the boundary conditions/initial states are chosen such that the ghosts do not propagate, our computations of the beta functions still hold. We will leave a detailed examination
of these issues to the future work. We will also not address in this work various other crucial questions on the derivation of a open EFT : first is the problem of infrared divergences
in the unitary theory which need to be tackled correctly to yield a sensible open EFT. Second is the related question of the appropriate initial states and dealing with various
transient effects.The third question we will comment on but leave out a detailed discussion of, is the modification of the cutting rules in the open EFT. We hope to return to these questions in the future.

\paragraph{Organization of the paper}
The rest of the paper is organized as follows. In the rest of the introduction, we will very briefly review the relevant background for our work. This includes the concepts of Schwinger-Keldysh
path integrals, their relation to Veltman's cutting rules,  Feynman-Vernon influence functionals for open EFTs and the Lindblad form for the evolution. The readers who are familiar with these 
concepts are encouraged to skim through these subsections in order to familiarize themselves with our notation.

In section \ref{sec:macrointro} we will write down the action for the open EFT  and set up the propagators and Feynman rules. We will also discuss the conditions under which the evolution density matrix of the theory is of Lindblad form. In section \ref{sec:oneloopmass} we compute the one loop beta function for various coupling constants. The result of the section is summarized in \ref{subsec:oneloopsummary}. In section \ref{sec:avgdifbasis}, we rewrite the theory in average-difference basis and we illustrate the great simplification that happens in this basis. The details of the computation in this basis can be found in appendix \ref{sec:appendixavgdiff}. In section \ref{subsec:avgdifallorder}, we present a proof that the Lindblad condition is never violated under perturbative corrections. Section \ref{sec:conclusion} consists of the conclusion of our analysis and various future directions.  Appendix \ref{sec:conventions} describes some of our notations and conventions. Computation of the various one loop Passarino-Veltman integrals required for open EFT can be found in  appendix \ref{sec:onelooppvintegrals} and in appendix \ref{sec:onelooppvavgdiff}.

\subsection{Basics of Schwinger-Keldysh theory}
\label{sec:sktheorybasics}

The Schwinger-Keldysh(SK) path integrals have been reviewed in \cite{Chou:1984es, Haehl:2016pec, Calzetta:2008iqa, Kamenev, Kamenev:2009jj, Sieberer:2015svu}. Here we will mention some key features : given a unitary QFT and a
initial density matrix $\rho(t=t_i)= \rho_i$, we define the SK path integral via 
\begin{equation}\label{eq:SKdef}
\begin{split} 
\mathcal{Z}_{SK}[J_R,J_L] \equiv \text{Tr}\Bigl\{\ U[J_R]\  \rho_i\ (U[J_L])^\dag\  \Bigr\}
\end{split}
\end{equation} 
Here, $U[J]$ is the unitary evolution operator of the quantum field theory deformed by sources $J$ for some operators of the theory. This path integral is a generator of all correlation functions
with at most one time-ordering violation. This should be contrasted with the Feynman path-integral which can compute only completely time-ordered correlators. \\

One could in principle consider the generating functions for correlators with arbitrary number of time-ordering violations \cite{Haehl:2017qfl} (for example, the correlator used to obtain the Lyapunov exponent involves two time-ordering violations \cite{Larkin:1969aa})
but, in this work, we will limit ourselves to just the usual SK path-integral. The Schwinger-Keldysh path integral gives a convenient way to access the evolution of the most general mixed state
in quantum field theory including the real time dynamics at finite temperature. It is an essential tool in the non-equilibrium description of QFTs which is directly defined in Lorentzian signature 
without any need for analytic continuation from the Euclidean description.

Given an action $S[\phi, J]$ of the unitary QFT, we can give a path-integral representation of $\mathcal{Z}_{SK}[J_R,J_L] $ by introducing a ket field $\phir$ and a bra field $\phil$ :
\begin{equation}\label{eq:SKdefPI}
\begin{split} 
\mathcal{Z}_{SK}[J_R,J_L] \equiv \int^{\phir(t=\infty)=\phil(t=\infty)}_{\rho_i(\phir,\phil)} [d\phir][d\phil]\  e^{ iS[\phir, J_R]- i S[\phil, J_L] }
\end{split}
\end{equation}
The lower limit is the statement that near $t=t_i$ the boundary condition for the path-integral is weighed by the initial density matrix $\rho_i$. The upper limit is 
the statement that the bra and the ket fields should be set equal at far future and summed over in order to correctly reproduce the trace. The factors $e^{ iS[\phir, J_R]}$
and $e^{- i S[\phil, J_L] }$ correctly reproduce the evolution operators $U[J_R]$ and $(U[J_L])^\dag$ respectively. 

If the unitary QFT is in a perturbative regime, the above path integral can be used to set up the Feynman rules \cite{Kamenev, Chou:1984es}.

\begin{enumerate}
\item In a unitary QFT, there are no vertices coupling the bra and the ket fields. The bra vertices are complex conjugates of ket vertices. 
\item The ket propagator is time-ordered while the bra propagator is anti-time-ordered. In addition to these, SK boundary conditions also induce a  bra-ket propagator
which is the on-shell propagator (obtained by putting the exchanged particle on-shell). We will term these propagators as \emph{cut propagators}. The terminology here is borrowed from 
the discussion of Cutkosky cutting rules where one thinks of the dividing lines between the bra and ket parts of the diagram as a `cut' of the diagram where particles go on-shell.
\end{enumerate}
We will call these rules as Veltman rules after Veltman who re-derived these rules in his study of unitarity \cite{Veltman:1963th, tHooft:1973pz} .  To reiterate, a fundamental feature of Veltman  rules is the fact that 
in a unitary theory, bra and ket fields talk only via cut propagators but not via cut vertices. As we will see in the following, this ceases to be true in an open QFT where, as 
Feynman and Vernon \cite{Feynman:1963fq, Vernon:1959pla} showed, there are novel cut vertices which signal non-unitarity. 

One of the fundamental features of the Veltman rules is a statement called the largest time equation which is fundamental to Veltman's approach to proving perturbative unitarity and cutting rules. The largest time equation is  a direct consequence of the definition  of SK path integral in equation \eqref{eq:SKdef} as reviewed in \cite{Haehl:2016pec}. We will briefly summarise below the argument for the largest time equation 
and its relation to SK formalism. We will refer the reader to \cite{Veltman:1994wz} or \cite{Haehl:2016pec} for more details.

In the SK path integral, consider the case where the sources obey $J_R=J_L=J(x)$ beyond a particular point of time $t=t_f$. One can then argue that the path-integral is in fact independent of the  source $J(x)$ in the future of $t_f$. This follows from  unitarity : the  contributions of $U[J_R]$ and $U[J_L]^\dag$ have to cancel each other in $\mathcal{Z}_{SK}$ if $J_R=J_L$ by unitarity.

To convert the above observation into a statement about correlators, we begin by  noting that the source $J(x)$ couples to difference operators $O_R-O_L$ in the SK path integral.
If we  differentiate the path-integral \eqref{eq:SKdefPI} with respect to the common source $J(x)$, it follows that one is basically computing a correlator with the difference operators $O_R-O_L$ placed in the future of $t_f$. The independence of $\mathcal{Z}_{SK}$ on  $J(x)$ then implies the vanishing of the correlators with the future-most (or the largest time) operators as difference operators $O_R-O_L$.

Microscopic unitarity thus requires that correlators of purely difference operators are trivial and any macroscopic open EFT should faithfully reproduce this condition. One of the main motivations of this work is to understand how these conditions get renormalized and the relation of these conditions to the Lindbladian form studied in open quantum system context.

\subsection{Basics of Lindblad theory and Effective theory}
\label{sec:Lindbladtheorybasics}
\begin{figure}
\begin{center}
\begin{tikzpicture}[scale=0.5] 
\phipropagatorp{0}{0}{3}{0}{}
\node at (1.5,0) {$\times$};
\node at (4,0) {$=$};
\node at (5,0) {$\Big($};
%\phipropagatorp{6}{0}{9}{0}{}
%\node at (10,0) {$+$};
\phipropagatorr{6}{0}{9}{0}{}
\node at (9,0) {$\times$};
%\node at (10.5,0.5) {$\text{Env.}$};
\wtchipropagatorp{9}{0}{12}{0}{} 
\node at (12,0) {$\times$};
\phipropagatorl{12}{0}{15}{0}{}
\node at (18,0) {$\Big)_{\text{Unitary QFT}}$};	

\begin{scope}[shift={(9.6,0.5)}]
\draw [ultra thick] [->] (0,2) -- (.9,.15);
\node at (0,2.5) {cut propagator of Env. field};
\end{scope}

\end{tikzpicture}
\end{center}

\caption{Feynman-Vernon vertex of an open QFT}
\label{fig:vertFeynVer} 
\end{figure}
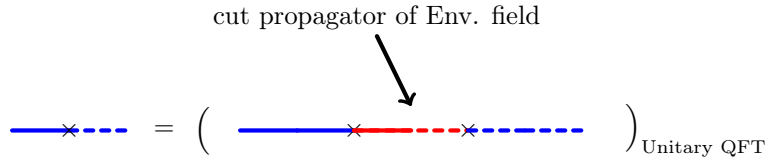

Following Feynman-Vernon \cite{Feynman:1963fq, Vernon:1959pla}, we can integrate out the `environment' fields in the Schwinger-Keldysh path integral and obtain an  effective path integral for the quantum system under question. This inevitably induces a  coupling between the bra and ket fields (called Feynman-Vernon(FV) coupling in the following) as shown schematically in the figure~\ref{fig:vertFeynVer}. 
Here the red-line represents the  `environment' fields of Feynman-Vernon which couples to the system field via a linear coupling. These `environment' fields  when traced/integrated out 
induce the unitarity violating FV coupling for the fields describing the open quantum field theory. \\
Note that  the propagator that induces FV coupling is necessarily a cut propagator of the environment which means that the FV coupling is only induced in the regime where the `environment' fields go on-shell. This also explains why, in usual QFT where we integrate out heavy fields that can never go on-shell in vacuum, no FV coupling or effective non-unitarity is induced by Wilsonian
RG. \footnote{Note that this is true about dilute states which are near vacuum state. A counterexample is at finite temperature where thermal fluctuations of the environment can and do contribute to the influence functional.} We will assume that the open QFT that we are studying in this paper arises from some hitherto unspecified microscopic theory \`a la Feynman-Vernon.

The FV couplings induced by integrating out environment fields need not always be local. A local description for the resultant open QFT is often accomplished by working with a limit 
where the time scales in the environment are assumed to be very fast compared to the rate at which the information flows from the system to the environment. In this approximation
(often termed Born-Markov approximation), one expects a nice local non-unitary EFT and our intent here is to study renormalisation in such an EFT. In the context of open quantum 
mechanical systems, under a clear separation of timescales, one can derive the Lindblad  equation (or the quantum master equation) \cite{Gorini:1975nb,Lindblad:1975ef,Breuer:2002pc}
for the reduced density matrix of the form 
\begin{equation}
\begin{split} 
i\hbar \frac{d\rho}{dt} = [H,\rho] +i \sum_{\alpha,\beta} \Gamma_{\alpha\beta} \Bigl(
L_{_\beta} \rho L_\alpha^\dag   - \frac{1}{2} L_\alpha^\dag L_{_\beta} \rho -\frac{1}{2}\rho\,  L_\alpha^\dag L_\beta\Bigr) \ .
\end{split}
\end{equation}
Here, $H$ is the Hamiltonian of the system  leading to the unitary part of the evolution, whereas the non-unitary (Feynman-Vernon) part of the evolution comes from rest of the terms in RHS. The non-unitarity is captured by a set of operators $L_\alpha$ and a set of couplings $\Gamma_{\alpha\beta}$ of the system. It is easily checked that the form above implies 
\begin{eqnarray}
\frac{d}{dt}\, \tr\,  \rho =0 \qquad ,
\nonumber
\end{eqnarray}
i.e., it is trace-preserving. Further, if $ \Gamma_{\alpha\beta}$ is a positive matrix, one can show that the above equation describes a dissipative system which keeps the eigenvalues of $\rho$ non-negative. These two properties (along with linearity in $\rho$) qualify Lindblad form of evolution as a physically sensible
dynamics describing an open quantum system. The above equation in Schr\"{o}dinger picture has an equivalent Heisenberg picture description via an evolution equation for operators :
\begin{equation}
\begin{split}
i\hbar \frac{d\mathcal{A}}{dt} = [\mathcal{A},H] +i \sum_{\alpha,\beta} \Gamma_{\alpha\beta} \Bigl(
L_\alpha^\dag  \mathcal{A}  L_\beta  - \frac{1}{2} L_\alpha^\dag L_\beta \mathcal{A} -\frac{1}{2}\mathcal{A} L_\alpha^\dag L_\beta\Bigr) \ .
\end{split}
\end{equation}
Equivalently, one can obtain a path-integral description by adding to the Schwinger-Keldysh action of the system, an influence functional term of the form \cite{Sieberer:2015svu}
\begin{equation}
S_{FV}= i\int  \sum_{\alpha,\beta} \Gamma_{\alpha\beta} \Bigl(
L_\alpha^\dag[\phir]  L_\beta[\phil]  - \frac{1}{2} L_\alpha^\dag[\phir] L_\beta[\phir] -\frac{1}{2} L_\alpha^\dag[\phil] L_\beta[\phil] \Bigr)
\end{equation}
where we have indicated the way the action should be written in terms of the bra and ket fields in order to correctly reproduce Lindblad dynamics. We note that the Lindblad form of the influence functional 
has a particular structure which relates the $\phir$-$\phil$ cross-terms with the imaginary parts of both the $\phir$ action and $\phil$ action.

Let us note some important features of the above expression. If we set $\phir=\phil$ in the action above, it vanishes. It is clear that this is exactly the calculation done few lines above in the Schr\"{o}dinger picture to show that Lindblad evolution is trace-preserving. This is also related to the difference operator decoupling mentioned in the last subsection in the context of Schwinger-Keldysh path 
integrals. Thus, trace preserving property in the Schr\"{o}dinger picture becomes difference operator decoupling at the level of SK path integral for the EFT.

We also note that if we take one of the Lindblad operators say $L_\beta$ to be an identity operator, the Lindblad form then becomes a difference operator, i.e., it can be written as a difference 
between an operator made of ket fields and the same  operator evaluated over the bra fields. This is the form of SK action for a unitary QFT (c.f. equation \eqref{eq:SKdefPI}) and it merely shifts the 
system action. But when both Lindblad operators are not identity, one gets various cross terms and associated imaginary contributions to the pure $\phir$ and the pure $\phil$ action. Thus, once the cross
couplings are determined, one can use the Lindblad form to determine all imaginary couplings. This is the route we will take to write down the Lindblad conditions like the ones in equation \eqref{eq:LindbladCondn}. \\
Having finished this brief review of the necessary ideas, let us turn to the open $\phi^4$ theory whose renormalisation we want to study. We will begin by describing in detail the effective action and the associated Feynman rules in the next section.

\section{Introduction to Open effective theory}
\label{sec:macrointro}
 
Let us begin by writing down the action for the most general open quantum field theory,
consisting of a real scalar which can interact via cubic and quartic interactions, given in \eqref{eq:macroaction3}. The most general action, taking into 
account CPT symmetry(See for example, \cite{Haehl:2016pec}) and SK boundary conditions, is given by 
\begin{equation}
\begin{split} 
S &= -\int d^d x \Bigl[ \frac{1}{2} z\ (\partial \phir)^2 + \frac{1}{2} m^2\  \phir^2+ \frac{\lambda_4}{4!} \phir^4 + \frac{\sigma_4}{3!} \phir^3 \phil \Bigr]\\
&\qquad+\int d^d x \Bigl[ \frac{1}{2} z^\star (\partial \phil)^2 + \frac{1}{2} {m^2}^\star \ \phil^2 + \frac{\lambda_4^\star}{4!} \phil^4+ \frac{\sigma_4^\star}{3!} \phil^3 \phir\Bigr] \\
&\qquad + i \int d^d x \Bigl[ z_\Delta\ (\partial \phir).(\partial \phil)  + m^2_\Delta \ \phir \phil+\frac{{\lambda_\Delta}}{2!2!} \phir^2 \phil^2 \Bigr]
\\
&
\qquad 
-\int d^dx\left[\frac{\lambda_3}{3!}\phir^3-\frac{\lambda_3^\ast }{3!}\phil^3
	+\frac{\sigma_3}{2!}\phir^2\phil-\frac{\sigma_3^\star }{2!}\phil^2\phir
	\right]
\label{eq:macroaction3}
\end{split}
\end{equation}

\subsection{Lindblad condition}
\label{subsec:Lindbladcond}

Imposing CPT and  demanding that the action  \eqref{eq:macroaction3} should be of the Lindblad form, we get four constraints among the coupling constants - one for field renormalisation, one for the mass, one for the cubic coupling and one for quartic coupling terms. We begin by tabulating all the power counting renormalisable Lindblad terms in the $\phi^3+\phi^4$ theory in Table. \ref{tab:LindbladOps}. Also tabulated are the conditions resulting from insisting that our action be of Lindblad form (we call these  the Lindblad conditions). We will  now consider various parts of the action in turn 
and rewrite them in a way that the Lindblad conditions become manifest.

\begin{table}
\begin{center}
\begin{tabular}{|c|c|c|c|c|} 
\hline
\hline
Lindblad couplings & $L_\alpha^\dag[\phi]$ &$L_\beta[\phi]$& Imaginary coupling of  &Lindblad condition \\
 $\Gamma_{\alpha\beta}$ &  & &   $L_\alpha^\dag L_\beta$& \\
\hline
\hline
$z_\Delta$ & $\partial_\mu \phi$ &  $\partial_\mu \phi$ &  $ \im\, z$ &  $ \im\, z=z_\Delta$\\ [0.1cm]
$m_\Delta^2$ & $\phi$ & $\phi$ & $\im\,m^2$ & $\im\,m^2 = m_\Delta^2$ \\ [0.1cm]
$\frac{{\lambda_\Delta}}{2!2!} $ & $\phi^2$ & $\phi^2$ & $\im  \ \lambda_4$ & $\im  \ \lambda_4=3 \lambda_\Delta-4\ \im  \ \sigma_4 $ \\ [0.1cm]
$i\frac{\sigma_4}{3!} $ & $\phi^3$ & $\phi$ & $\im  \ \lambda_4$& \\ [0.1cm]
$-i\frac{\sigma_4^\star}{3!} $ & $\phi$ & $\phi^3$ & $\im  \ \lambda_4$& \\ [0.1cm]
$i\frac{\sigma_3}{2!} $ & $\phi^2$ & $\phi$ & $ \im\,  \lambda_3$ & $ \im\,  \lambda_3=-3\ \im\,  \sigma_3 $\\ [0.1cm]
$-i\frac{\sigma_3^\star}{2!} $ & $\phi$ & $\phi^2$ & $ \im\,  \lambda_3$& \\ [0.1cm]
\hline
\hline
\end{tabular}
\caption{\label{tab:LindbladOps} Renormalisable Lindblad operators for $\phi^3+\phi^4$ theory} 
\end{center}
\end{table}
 
\subsubsection*{Real terms of the action}
\label{subsec:macroreal} 
The real part of the action is given by
 \begin{equation}
\begin{split} 
\re\, [S] &= -\int d^d x \Bigl[ \frac{1}{2} \re[z]\ [(\partial \phir)^2-(\partial \phil)^2] + \frac{1}{2}\re[m^2] (\phir^2-\phil^2)\Bigr]\\
&\qquad -\int d^d x \Bigl[ \frac{\re\ \lambda_4}{4!} (\phir^4- \phil^4) + \frac{\re \sigma_4}{3!}\phir \phil (\phir^2-\phil^2) \Bigr]\\
&\qquad 
-\int d^dx\Bigl[\frac{\re\  \lambda_3}{3!}(\phir^3-\phil^3)
	+\frac{\re\ \sigma_3}{2!}\phir\phil(\phir-\phil)
	\Bigr]
\label{eq:macroactionRe}
\end{split}
\end{equation}
We note that CPT constrains this action to vanish when $\phir=\phil$. As a result, there are no conditions on these real couplings from the Lindblad structure. 

\subsubsection*{Imaginary Quadratic terms of the action}
\label{subsec:macroquadratic}

The imaginary part of the quadratic terms is given by 
\begin{equation}
\begin{split}
\im\, [S_2] &= -\int d^d x \Bigl[ \frac{1}{2} \im[z]\ ((\partial \phir)^2 +(\partial \phil)^2)-z_\Delta\ (\partial \phir).(\partial \phil) \Bigr]\\
&\qquad -\int d^d x \Bigl[  \frac{1}{2} \im[m^2] (\phir^2+\phil^2)- m^2_\Delta \phir \phil  \Bigr]\\
 &= -\int d^d x \Bigl[ \frac{1}{2} \im[z]\ (\partial \phir-\partial \phil)^2 +\frac{1}{2} \im[m^2] (\phir-\phil)^2 \Bigr]
 \\
&\qquad +\int d^d x \Bigl[  (z_\Delta-\im[z])\ (\partial \phir).(\partial \phil)+ (m^2_\Delta-\im[m^2]) \phir \phil  \Bigr]
\label{macroaction4}
\end{split}
\end{equation}
The Lindblad condition is given by  
\begin{equation}
\begin{split}
z_\Delta &= \im\, [z]\ ,\qquad m^2_\Delta=\im\, [m^2]\ , \qquad 
%\frac{{\lambda_\Delta}}{2!2!}=2\im\Big(\frac{\lambda_4}{4!} +\frac{\sigma_4}{3!} \Big)\ . 
\label{macroaction5}
\end{split}
\end{equation}

\subsubsection*{Imaginary Cubic coupling}
\label{subsec:macrocubic}
Now we compute the imaginary part of the cubic terms in the action 
\begin{eqnarray}
-\im\, [S_3]&=& 	\int d^dx\left[\frac{\im\, \lambda_3}{3!}\phir^3+\frac{\im\, \lambda_3 }{3!}\phil^3
	+\frac{\im\, \sigma_3}{2!}\phir^2\phil +\frac{\im\, \sigma_3}{2!}\phil^2\phir
	\right]
	\nonumber\\
	&=&\int d^dx\left[\frac{\im\, \lambda_3}{3!}(\phir-\phil)(\phir^2-\phil^2)+\left(\frac{\im\, \lambda_3 }{3!}
	+\frac{\im\, \sigma_3}{2!}\right)\left(\phir^2\phil +\phil^2\phir\right)
	\right] 
\label{macroaction7}
\end{eqnarray}
The Lindblad condition is given by   
\begin{eqnarray}
\begin{split}
&\frac{\im\, \lambda_3 }{3!}
	+\frac{\im\,  \sigma_3}{2!}= 0\\
\Rightarrow\ &\im\ \lambda_3 + 3 \im\ \sigma_3 =0	
\label{macroaction8}
\end{split}
\end{eqnarray}

\subsubsection*{Imaginary Quartic coupling}
\label{subsec:macroquartic}

The imaginary part of action at the level quartic coupling is given by 
\begin{equation}
\begin{split}
\im[S_4] 
&=-\int d^d x \Bigl[\frac{1}{4!} \im[\lambda_4] (\phir^4+\phil^4) + \frac{1}{3!} \im[\sigma_4] (\phir^3\phil+\phir\phil^3)-\frac{{\lambda_\Delta}}{2!2!} \phir^2 \phil^2\Bigr] \\ 
 &= -\int d^d x \Bigl[ 
 \Big(\frac{1}{4!} \im[\lambda_4] +\frac{1}{3!} \im[\sigma_4] \Big)(\phir^2-\phil^2)^2 + 
% \Bigr]\\
% &\qquad+\int d^d x \Bigl[ 
 \frac{1}{3!} \im[\sigma_4] (\phir-\phil)(\phir^3-\phil^3)\Bigr] \\
&\quad+\int d^d x  \Bigl[\frac{{\lambda_\Delta}}{2!2!}-2\, \im\Big(\frac{\lambda_4}{4!} +\frac{\sigma_4}{3!} \Big) \Bigr]\phir^2 \phil^2
\label{macroaction9}
\end{split}
\end{equation}
The Lindblad condition at for the quartic couplings is given by 
\begin{equation}
\begin{split}
&\frac{{\lambda_\Delta}}{2!2!}=2\, \im\, \left(\frac{\lambda_4}{4!} +\frac{\sigma_4}{3!} \right) \\
\Rightarrow\ & \im\ \lambda_4 + 4 \im\ \sigma_4 -3 \lambda_\Delta =0
\label{macroaction10}
\end{split}
\end{equation}

\subsection{Exact propagators}
\label{subsec:exactprop}

The  ket field $\phir$ and the bra field $\phil$ in SK path-integral satisfy the following boundary condition \eqref{eq:SKdefPI}
\begin{equation}
\phir(t=\infty)=\phil(t=\infty)
\end{equation}

Owing to this boundary condition and the mixing term between $\phir$ and $\phil$ fields, the kinetic matrix derived from the action 
\eqref{eq:macroaction3} is given by
\begin{equation}
\begin{split}
\mathcal{K}
=  
\left(
\begin{array}{cc}
i (z\ k^2+m^2-i\varepsilon) & z_\Delta k^2 + m_\Delta^2-2\ \varepsilon\ \Theta(-k^0)\\
z_\Delta k^2 + m_\Delta^2 -2\ \varepsilon\ \Theta(k^0)& -i (z^\star\ k^2+(m^2)^\star+i\varepsilon)  \\
\end{array}
\right)
\end{split}
\end{equation}
where the $\varepsilon$ prescription implements Schwinger-Keldysh boundary conditions. We define the kinetic matrix $\mathcal{K}$ by
\begin{equation}
\begin{split}
 iS \ni - \frac{1}{2} \Bigr( \phir(-k) \phil(-k) \Bigl) \mathcal{K} \Bigr(\begin{array}{c}  \phir(k)\\ \phil(k) \end{array} \Bigl)
\end{split}
\end{equation}

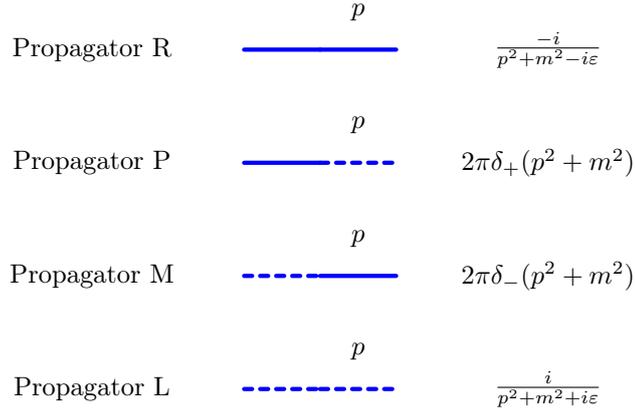
\begin{figure}[ht]
\begin{center}
	
\begin{tikzpicture} 
\phipropagatorr{0}{0}{2}{0}{p}
\node at (-2,0) { Propagator R};
\node at (4,0) {$\frac{-i}{p^2+m^2-i\varepsilon}$};

\phipropagatorp{0}{-1.5}{2}{-1.5}{p}
\node at (-2,-1.5) { Propagator P};
\node at (4,-1.5) {$2\pi \delta_{+}(p^2+m^2)
$};

\phipropagatorm{0}{-3}{2}{-3}{p}
\node at (-2,-3) { Propagator M};
\node at (4,-3) {$2\pi \delta_{-}(p^2+m^2)
$};

\phipropagatorl{0}{-4.5}{2}{-4.5}{p}
\node at (-2,-4.5) { Propagator L};
\node at (4,-4.5) {$\frac{i}{p^2+m^2+i\varepsilon}$};
\end{tikzpicture}

\end{center}

\caption{SK propagator for $\phi$ fields}
\label{fig:microskphipropagator} 
%fig:micro-sk-phi-propagator 
\end{figure}

Its inverse (viz., the propagator) can be written as
\begin{equation}\label{eq:invKin}
\begin{split}
\mathcal{K}^{-1} &\equiv \Biggl(\begin{array}{cc}  \langle \phir( -k )  \phir( k ) \rangle & \langle \phir( -k )  \phil( k ) \rangle \\ \langle \phil( -k )  \phir( k ) \rangle & \langle \phil( -k )  \phil( k ) \rangle \end{array} \Biggr) 
\\
&= \mathfrak{z}^{-1}
\Biggl(
\begin{array}{cc}
\frac{-i}{\re[zk^2+m^2]-i\varepsilon}& 2\pi \delta(\re[zk^2+m^2]) \Theta(-k^0)\\
2\pi \delta(\re[zk^2+m^2]) \ \Theta(k^0)&\frac{i}{\re[zk^2+m^2]+i\varepsilon}  \\
\end{array}
\Biggr) \\  
&\qquad + \mathfrak{z}^{-1} \frac{(-i)}{\re[zk^2+m^2]-i\varepsilon} \times \frac{i}{\re[zk^2+m^2]+i\varepsilon} \times
\Biggl(
\begin{array}{cc}
\im[zk^2+m^2]&  z_\Delta k^2 + m_\Delta^2\\
z_\Delta k^2 + m_\Delta^2 &\im[zk^2+m^2] \\
\end{array}
\Biggr)
\end{split}
\end{equation}
where,
\begin{equation}
\begin{split}
 \mathfrak{z} &\equiv 1+ \frac{(-i)}{\re[zk^2+m^2]-i\varepsilon} \times \frac{i}{\re[zk^2+m^2]+i\varepsilon} \times \Biggl( (\im[zk^2+m^2]-\varepsilon)^2 - (z_\Delta k^2 + m_\Delta^2-\varepsilon)^2 \Bigr)\\
 \end{split}
\end{equation}
Please note that when the Lindblad conditions \eqref{macroaction5} are satisfied, we have
\begin{eqnarray}
	\mathfrak{z}=1\ .
\end{eqnarray} 
Further, it can be easily checked that in this limit, the sum of diagonal entries in the propagator matrix is equal to the sum of off-diagonal entries, i.e., 
\begin{eqnarray}  
(\mathcal{K}^{-1})_{RR}+ (\mathcal{K}^{-1})_{LL} = (\mathcal{K}^{-1})_{RL}+(\mathcal{K}^{-1})_{LR} 
\end{eqnarray} 
The corresponding property in the unitary quantum field theory is the well-known relation between the various correlators in the Keldysh formalism \cite{Kamenev}. This can equivalently be reformulated as the vanishing of two point function of two difference correlators : 
\begin{eqnarray}
(\mathcal{K}^{-1})_{R-L,R-L}=0.
\end{eqnarray}
In this work, we will work in the limit where the non-unitary couplings $\im [m^2]$ and $m_\Delta^2$ are considered as  perturbations to $\re [m^2]$, and similarly, $\im[z^2]$ and 
$z_\Delta^2$ are considered small compared   to $\re [z^2]$. Further, since 1-loop correction to the propagators do not generate field renormalisation we can also set $z=1$. In this limit,  the propagators in equation \eqref{eq:invKin} reduced to those given by figure \ref{fig:microskphipropagator}.

%the inverse kinetic matrix is given by
%\begin{equation}
%\begin{split}
%\mathcal{K}^{-1}
%&= 
%\left(
%\begin{array}{cc}
%\frac{-i}{\re[zk^2+m^2]-i\varepsilon} & 2\pi \delta(\re[zk^2+m^2]) \Theta(-k^0)\\
%2\pi \delta(\re[zk^2+m^2]) \ \Theta(k^0)& \frac{i}{\re[zk^2+m^2]+i\varepsilon}  \\
%\end{array}
%\right)
%\end{split}
%\label{eq:invKin}
%\end{equation}
%
% In that limit,
%it's easy to see, 
%
%Notice also that our propagators here satisfy Veltman's largest time equation which is a salient feature of the theory and,  in general, for Schwinger-Keldysh theory. This comes about as a result of $\mathcal{Z}[0]=1$  (\avnote{to be referred to the relevant equation}) or the fact that the
%correlators of difference operators goes to zero (\avnote{It would be nice to show in our introduction that these two statements are equivalent}). 

\subsection{Feynman rules} 
\label{subsec:feynrules}

In this paper henceforth, we will set $z=z_\Delta=1$ (which is not renormalised at one-loop in d=4 dimensions). We will treat all other parameters in our action except the real part of $m^2$(i.e., $Re(m^2)$) perturbatively. This includes $ \lambda_3$,  $\sigma_3$, $ \lambda_4$, $ \sigma_4$
and $\lambda_\Delta$, as well as $\im\, m^2$ and $m_\Delta^2$.

The propagators of $\phi$ fields are given below.  We have used solid blue and dotted blue lines for $\phir$(ket fields) and $\phil$(bra fields) fields respectively. Note that in the cut propagators P and M the energy is restricted to flow from the ket field to the bra field.

We will now set up the Veltman rules for the vertices to compute SK correlators in the open $\phi^3+\phi^4$ theory: 
 
\begin{center}
\begin{tabular}{ | m{5em} | m{6cm}|  } 
\hline
Vertex & Factor  \\ 
\hline
\hline
$\phir^3$ & $(-i\lambda_3) (2\pi)^d \delta\left(\sum p\right)$ \\ 
\hline
$\phil^3$ & $(i\lambda_3^\star) (2\pi)^d \delta\left(\sum p\right)$  \\ 
\hline
$\phir^2\phil$  & $(-i\sigma_3) (2\pi)^d \delta\left(\sum p\right)$ \\

\hline
$\phir\phil^2$  & $(i\sigma_3^\star) (2\pi)^d \delta\left(\sum p\right)$ \\

\hline
$\phir^4$  & $(-i\lambda_4) (2\pi)^d \delta\left(\sum p\right)$ \\

\hline
$\phil^4$  & $(i\lambda_4^\star) (2\pi)^d \delta\left(\sum p\right)$ \\
\hline
$\phir^3\phil$  & $(-i\sigma_4) (2\pi)^d \delta\left(\sum p\right)$ \\
\hline
$\phir\phil^3$  & $(i\sigma_4^\star) (2\pi)^d \delta\left(\sum p\right)$ \\
\hline
$\phir^2\phil^2$  & $(-\lambda_\Delta)\ (2\pi)^d \delta\left(\sum p\right)$ \\
\hline
\hline
\end{tabular} 
\end{center}

\begin{figure}[ht] 
\begin{center}
\begin{tikzpicture}[line width=1 pt, scale=1]
%\begin{scope}[rotate=-80]

\begin{scope}[shift={(0,5)}]
\draw [phir, ultra thick]  (0,0) -- (1,0);
\draw [phir, ultra thick]  (0,0) -- (-1,1);
\draw [phir, ultra thick]  (0,0) -- (-1,-1);
\node at (0,-2) {\Large $-i\lambda_3$};	
\end{scope} 

\begin{scope}[shift={(5,5)}]
\draw [phil, ultra thick]  (0,0) -- (1,0);
\draw [phil, ultra thick]  (0,0) -- (-1,1);
\draw [phil, ultra thick]  (0,0) -- (-1,-1);
\node at (0,-2) {\Large $i\lambda_3^\star $};	
\end{scope} 

\begin{scope}[shift={(10,5)}]
\draw [phir, ultra thick]  (0,0) -- (1,0);
\draw [phir, ultra thick]  (0,0) -- (-1,1);
\draw [phil, ultra thick]  (0,0) -- (-1,-1);
\node at (0,-2) {\Large $-i\sigma_3$};	
\end{scope}

\begin{scope}[shift={(0,0)}]
\draw [phil, ultra thick]  (0,0) -- (1,0);
\draw [phil, ultra thick]  (0,0) -- (-1,1);
\draw [phir, ultra thick]  (0,0) -- (-1,-1);
\node at (0,-2) {\Large $i\sigma_3^\star$};	
\end{scope} 

\begin{scope}[shift={(5,0)}]
\draw [phir, ultra thick]  (0,0) -- (1,1);
\draw [phir, ultra thick]  (0,0) -- (1,-1);
\draw [phir, ultra thick]  (0,0) -- (-1,1);
\draw [phir, ultra thick]  (0,0) -- (-1,-1);
\node at (0,-2) {\Large $-i\lambda_4$};	
\end{scope} 

\begin{scope}[shift={(10,0)}]
\draw [phil, ultra thick]  (0,0) -- (1,1);
\draw [phil, ultra thick]  (0,0) -- (1,-1);
\draw [phil, ultra thick]  (0,0) -- (-1,1);
\draw [phil, ultra thick]  (0,0) -- (-1,-1);
\node at (0,-2) {\Large $i\lambda_4^\star $};	
\end{scope}

\begin{scope}[shift={(0,-5)}]
\draw [phir, ultra thick]  (0,0) -- (1,1);
\draw [phir, ultra thick]  (0,0) -- (1,-1);
\draw [phir, ultra thick]  (0,0) -- (-1,1);
\draw [phil, ultra thick]  (0,0) -- (-1,-1);
\node at (0,-2) {\Large $-i\sigma_4$};	
\end{scope}

\begin{scope}[shift={(5,-5)}] 
\draw [phil, ultra thick]  (0,0) -- (1,1);
\draw [phil, ultra thick]  (0,0) -- (1,-1);
\draw [phil, ultra thick]  (0,0) -- (-1,1);
\draw [phir, ultra thick]  (0,0) -- (-1,-1);
\node at (0,-2) {\Large $i\sigma_4^\star $};	
\end{scope}

\begin{scope}[shift={(10,-5)}]
\draw [phir, ultra thick]  (0,0) -- (1,1);
\draw [phir, ultra thick]  (0,0) -- (1,-1);
\draw [phil, ultra thick]  (0,0) -- (-1,1);
\draw [phil, ultra thick]  (0,0) -- (-1,-1);
\node at (0,-2) {\Large $-\lambda_\Delta$};	
\end{scope}

\end{tikzpicture}
\end{center}
\caption{Diagrammatic Representation of all the Tree level processes}
\label{fig:macrotree1} 
\end{figure}
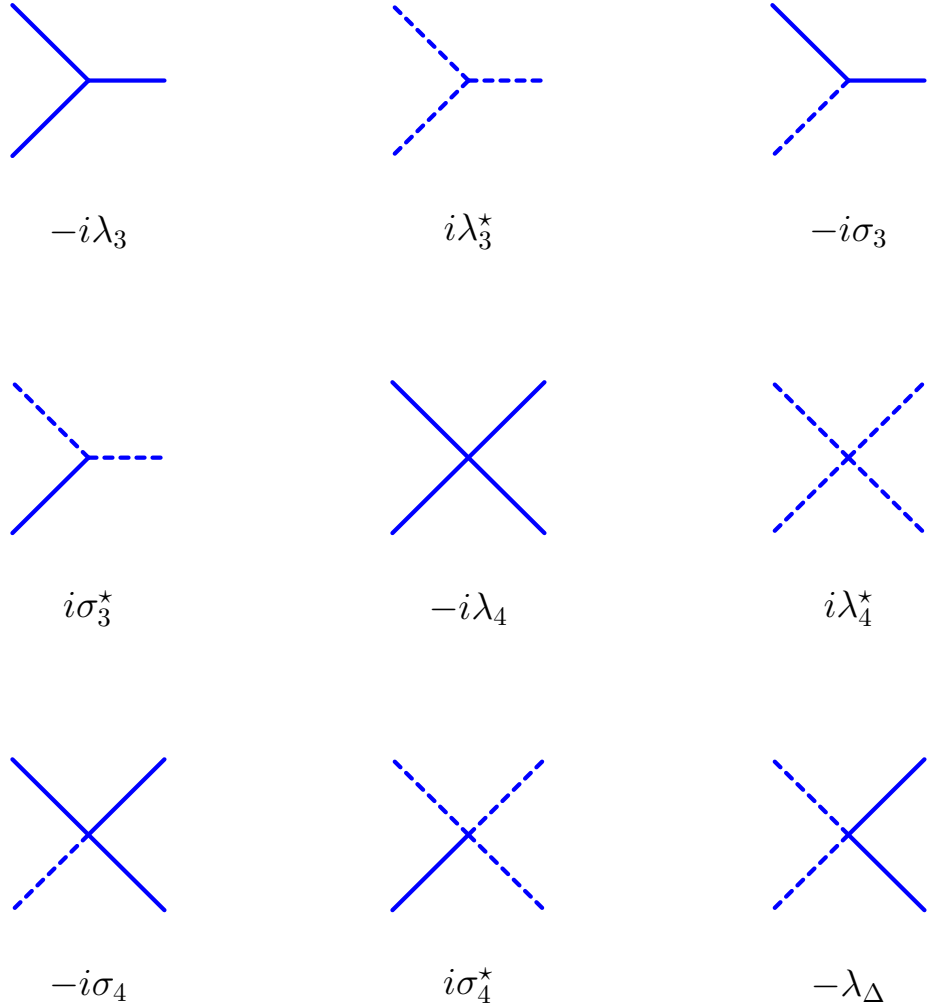

\subsection{Lindblad condition from tree level correlators } 
\label{subsec:treeunitarity}

In a unitary Schwinger-Keldysh theory, the correlator of difference operators vanishes to all order in perturbation theory. This is equivalent to Veltman's largest time equation (see for example \cite{Haehl:2016pec}).  One could ask  whether this statement continues to hold true in the non-unitary theory.
We have already remarked during our discussion  of propagators around equation \eqref{eq:invKin} that the quadratic Lindblad conditions are equivalent to the vanishing of difference operator two point functions. 
We can extend this to higher point functions simply. Consider the tree level correlator of three difference operators 
\begin{equation}\label{eq:diff3}
\begin{split}
	&\left\langle(\phir(p_1)-\phil(p_1))(\phir(p_2)-\phil(p_2))(\phir(p_3)-\phil(p_3))
	\right\rangle\\
&=-i(\lambda_3-\lambda_3^\ast)-4i	(\sigma_3-\sigma_3^\ast)
=2(\im\, \lambda_3+3\,\im\, \sigma_3) 
\end{split}
\end{equation}
the correlator of four difference operators is given by 
\begin{equation} 
\label{eq:diff4}
\begin{split}
&\left\langle(\phir(p_1)-\phil(p_1))(\phir(p_2)-\phil(p_2))(\phir(p_3)-\phil(p_3))(\phir(p_4)-\phil(p_4))
	\right\rangle \\
&=-i(\lambda_4-\lambda_4^\ast)-4i	(\sigma_4-\sigma_4^\ast)-6\lambda_\Delta
=2(\im\, \lambda_4+4\,\im\, \sigma_4-3 \lambda_\Delta) 
\end{split}
\end{equation}
The correlators of the three and the four difference operators are precisely given by the Lindblad violating couplings.  This implies that at tree level, the Lindblad conditions are the same as the  vanishing of correlators of difference operators. 

One can, in fact, show the following statement \cite{Baidya:2017ab}:  consider an open EFT, which is  obtained by tracing out  some subset of fields in an underlying unitary theory. Then, the unitarity of the underlying theory implies  that the open EFT satisfies the Lindblad condition. 

%\chapter{One loop mass renormalization in the macroscopic theory}

\section{One loop beta function }
\label{sec:oneloopmass}

In this section, we compute the beta function for all the mass terms and coupling constants that appear in the action of the open $\phi^3+\phi^4$ theory. The main aim in this section will be to demonstrate the following three claims :
\begin{enumerate}
\item Despite the novel UV divergences that occur in the open $\phi^3+\phi^4$ theory, one can use a simple extension of the standard counter-term method to deal with the divergences. Thus, the open $\phi^3+\phi^4$ theory is one-loop renormalisable.
\item Once these UV divergences are countered, the standard derivation of beta functions and RG running also goes through, except for the fact that one has to now also renormalise the non-unitary couplings.
\item  We will also demonstrate that the running of a certain combinations of the couplings, the ones which given by the Lindblad conditions (equation \eqref{macroaction5}, equation \eqref{macroaction8} and equation \eqref{macroaction10} respectively), under one-loop renormalisation are proportional to the Lindblad conditions.
\end{enumerate}

%It suggests the fact that the Lindblad conditions are not violated under one-loop renormalisation and hence, the trace remains preserved. Notice if that weren't the case, 
% the violation under renormalisation would lead to a violation of the trace-preserving criteria of the theory, which is unphysical. 
 We shall provide an all-order proof in the next section that the Lindblad conditions are never violated under perturbative corrections. Here we shall use the notations and results presented in appendix \ref{sec:onelooppvintegrals}.

%\paragraph{Note for Advanced reader: } 
%The content of this section are standard quantum field theory computations. suitably adopted to SK formalism, 
%that all of us learn in graduate school. Advanced readers can skip this part and can directly go to the results in section \eqref{}

\subsection{One loop beta function for \texorpdfstring{$m^2$}{m squared}}
\label{subsec:oneloopmass}
We will now begin a discussion of various loop diagrams. The simplest is perhaps the tadpole diagrams which can be cancelled by a counter-term linear in $\phir$ and $\phil$. It is easily demonstrated that the necessary counter-terms do not violate the Lindblad condition (See appendix\ref{appendix:tadpole}).

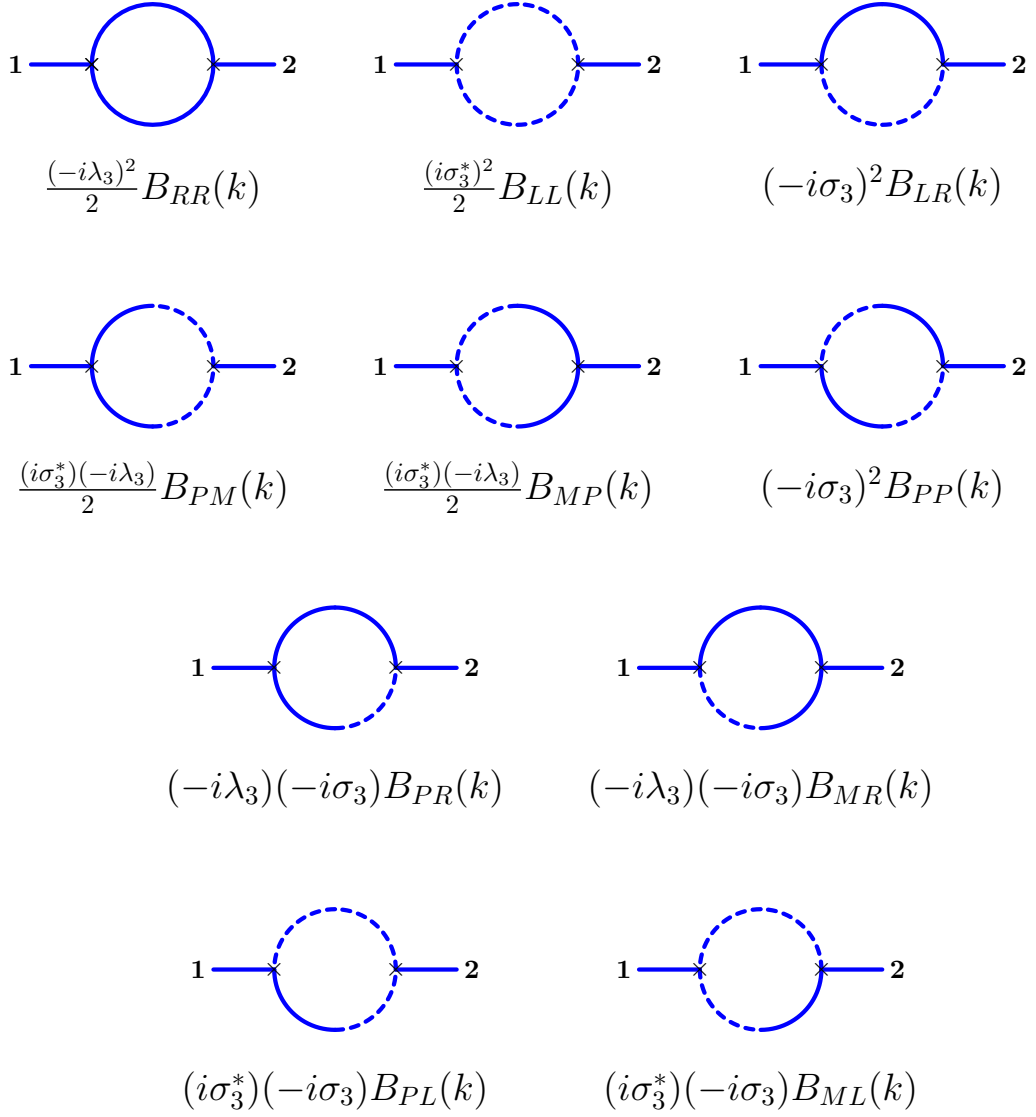
\begin{figure}
\begin{center}
\begin{tikzpicture}[line width=1 pt, scale=0.8]
%\begin{scope}[rotate=-80]
 
\begin{scope}[shift={(0,0)}]
\draw [phir, ultra thick] (1,0) -- (2,0);
\draw [phir, ultra thick] (-1,0) -- (-2,0);
\draw [phir, ultra thick, domain=0:180] plot ({1*cos(\x)}, {1*sin(\x)});
\draw [phir, ultra thick, domain=180:360] plot ({1*cos(\x)}, {1*sin(\x)}); 
\node at (-1,0) {$\times $};	
\node at (1,0) {$\times $};	
\node at (0,-2) {\Large $\frac{(-i\lambda_3)^2}{2} B_{RR}(k)$};	
\node at (-1,0) {$\times $};	

\node at (-2.25,0) {$\bf 1$};	
\node at (2.25,0) {$\bf 2$};

\end{scope}

\begin{scope}[shift={(6,0)}]
\draw [phir, ultra thick] (1,0) -- (2,0);
\draw [phir, ultra thick] (-1,0) -- (-2,0);
\draw [phil, ultra thick, domain=0:90] plot ({1*cos(\x)}, {1*sin(\x)});
\draw [phil, ultra thick, domain=90:180] plot ({1*cos(\x)}, {1*sin(\x)});
\draw [phil, ultra thick, domain=180:270] plot ({1*cos(\x)}, {1*sin(\x)});
\draw [phil, ultra thick, domain=270:360] plot ({1*cos(\x)}, {1*sin(\x)}); 
\node at (-1,0) {$\times $};	
\node at (1,0) {$\times $};	
\node at (0,-2) {\Large $\frac{(i\sigma_3^\ast)^2}{2} B_{LL}(k)$};

\node at (-2.25,0) {$\bf 1$};	
\node at (2.25,0) {$\bf 2$};		
	
\end{scope}

\begin{scope}[shift={(12,0)}]
\draw [phir, ultra thick] (1,0) -- (2,0);
\draw [phir, ultra thick] (-1,0) -- (-2,0);
\draw [phir, ultra thick, domain=0:90] plot ({1*cos(\x)}, {1*sin(\x)});
\draw [phir, ultra thick, domain=90:180] plot ({1*cos(\x)}, {1*sin(\x)});
\draw [phil, ultra thick, domain=180:270] plot ({1*cos(\x)}, {1*sin(\x)});
\draw [phil, ultra thick, domain=270:360] plot ({1*cos(\x)}, {1*sin(\x)}); 
\node at (-1,0) {$\times $};	
\node at (1,0) {$\times $};	
\node at (0,-2) {\Large $(-i\sigma_3)^2 B_{LR}(k)$};	

\node at (-2.25,0) {$\bf 1$};	
\node at (2.25,0) {$\bf 2$};		
	
\end{scope}

\begin{scope}[shift={(0,-5)}]
\draw [phir, ultra thick] (1,0) -- (2,0);
\draw [phir, ultra thick] (-1,0) -- (-2,0);
\draw [phil, ultra thick, domain=0:90] plot ({1*cos(\x)}, {1*sin(\x)});
\draw [phir, ultra thick, domain=90:180] plot ({1*cos(\x)}, {1*sin(\x)});
\draw [phir, ultra thick, domain=180:270] plot ({1*cos(\x)}, {1*sin(\x)});
\draw [phil, ultra thick, domain=270:360] plot ({1*cos(\x)}, {1*sin(\x)});  
\node at (-1,0) {$\times $};	
\node at (1,0) {$\times $};	
\node at (0,-2) {\Large $\frac{(i\sigma_3^\ast)(-i\lambda_3)}{2} B_{PM}(k)$};	

\node at (-2.25,0) {$\bf 1$};	
\node at (2.25,0) {$\bf 2$};		
	
\end{scope}

\begin{scope}[shift={(6,-5)}]
\draw [phir, ultra thick] (1,0) -- (2,0);
\draw [phir, ultra thick] (-1,0) -- (-2,0);
\draw [phir, ultra thick, domain=0:90] plot ({1*cos(\x)}, {1*sin(\x)});
\draw [phil, ultra thick, domain=90:180] plot ({1*cos(\x)}, {1*sin(\x)});
\draw [phil, ultra thick, domain=180:270] plot ({1*cos(\x)}, {1*sin(\x)});
\draw [phir, ultra thick, domain=270:360] plot ({1*cos(\x)}, {1*sin(\x)}); 
\node at (-1,0) {$\times $};	
\node at (1,0) {$\times $};	
\node at (0,-2) {\Large $\frac{(i\sigma_3^\ast)(-i\lambda_3)}{2} B_{MP}(k)$};	

\node at (-2.25,0) {$\bf 1$};	
\node at (2.25,0) {$\bf 2$};		
	
\end{scope}

\begin{scope}[shift={(12,-5)}]
\draw [phir, ultra thick] (1,0) -- (2,0);
\draw [phir, ultra thick] (-1,0) -- (-2,0);
\draw [phir, ultra thick, domain=0:90] plot ({1*cos(\x)}, {1*sin(\x)});
\draw [phil, ultra thick, domain=90:180] plot ({1*cos(\x)}, {1*sin(\x)});
\draw [phir, ultra thick, domain=180:270] plot ({1*cos(\x)}, {1*sin(\x)});
\draw [phil, ultra thick, domain=270:360] plot ({1*cos(\x)}, {1*sin(\x)}); 
\node at (-1,0) {$\times $};	
\node at (1,0) {$\times $};	
\node at (0,-2) {\Large $(-i\sigma_3)^2 B_{PP}(k)$};	
	
\node at (-2.25,0) {$\bf 1$};	
\node at (2.25,0) {$\bf 2$};		

\end{scope}

\begin{scope}[shift={(3,-10)}]
\draw [phir, ultra thick] (1,0) -- (2,0);
\draw [phir, ultra thick] (-1,0) -- (-2,0);
\draw [phir, ultra thick, domain=0:90] plot ({1*cos(\x)}, {1*sin(\x)});
\draw [phir, ultra thick, domain=90:180] plot ({1*cos(\x)}, {1*sin(\x)});
\draw [phir, ultra thick, domain=180:270] plot ({1*cos(\x)}, {1*sin(\x)});
\draw [phil, ultra thick, domain=270:360] plot ({1*cos(\x)}, {1*sin(\x)}); 
\node at (-1,0) {$\times $};	 
\node at (1,0) {$\times $};	
\node at (0,-2) {\Large $(-i\lambda_3)(-i\sigma_3) B_{PR}(k)$};	

\node at (-2.25,0) {$\bf 1$};	
\node at (2.25,0) {$\bf 2$};		
	
\end{scope}

\begin{scope}[shift={(10,-10)}]
\draw [phir, ultra thick] (1,0) -- (2,0);
\draw [phir, ultra thick] (-1,0) -- (-2,0);
\draw [phir, ultra thick, domain=0:90] plot ({1*cos(\x)}, {1*sin(\x)});
\draw [phir, ultra thick, domain=90:180] plot ({1*cos(\x)}, {1*sin(\x)});
\draw [phil, ultra thick, domain=180:270] plot ({1*cos(\x)}, {1*sin(\x)});
\draw [phir, ultra thick, domain=270:360] plot ({1*cos(\x)}, {1*sin(\x)}); 
\node at (-1,0) {$\times $};	
\node at (1,0) {$\times $};	
\node at (0,-2) {\Large $(-i\lambda_3)(-i\sigma_3) B_{MR}(k)$};

\node at (-2.25,0) {$\bf 1$};	
\node at (2.25,0) {$\bf 2$};	
	
\end{scope}

\begin{scope}[shift={(3,-15)}]
\draw [phir, ultra thick] (1,0) -- (2,0);
\draw [phir, ultra thick] (-1,0) -- (-2,0);
\draw [phil, ultra thick, domain=0:90] plot ({1*cos(\x)}, {1*sin(\x)});
\draw [phil, ultra thick, domain=90:180] plot ({1*cos(\x)}, {1*sin(\x)});
\draw [phir, ultra thick, domain=180:270] plot ({1*cos(\x)}, {1*sin(\x)});
\draw [phil, ultra thick, domain=270:360] plot ({1*cos(\x)}, {1*sin(\x)}); 
\node at (-1,0) {$\times $};	
\node at (1,0) {$\times $};	
\node at (0,-2) {\Large $(i\sigma_3^\ast)(-i\sigma_3) B_{PL} (k)$};	

\node at (-2.25,0) {$\bf 1$};	
\node at (2.25,0) {$\bf 2$};	
	
\end{scope}

\begin{scope}[shift={(10,-15)}]
\draw [phir, ultra thick] (1,0) -- (2,0);
\draw [phir, ultra thick] (-1,0) -- (-2,0);
\draw [phil, ultra thick, domain=0:90] plot ({1*cos(\x)}, {1*sin(\x)});
\draw [phil, ultra thick, domain=90:180] plot ({1*cos(\x)}, {1*sin(\x)});
\draw [phil, ultra thick, domain=180:270] plot ({1*cos(\x)}, {1*sin(\x)});
\draw [phir, ultra thick, domain=270:360] plot ({1*cos(\x)}, {1*sin(\x)});  
\node at (-1,0) {$\times $};	
\node at (1,0) {$\times $};	
\node at (0,-2) {\Large $(i\sigma_3^\ast)(-i\sigma_3) B_{ML}(k)$};	

\node at (-2.25,0) {$\bf 1$};	
\node at (2.25,0) {$\bf 2$};	
	
\end{scope}	  
 
\end{tikzpicture}
\end{center}
\caption{One Loop corrections to $ m^2$ due to cubic couplings}
\label{diag:macromassrenorm3}
\end{figure}

\begin{figure}[ht]
\begin{center}
\begin{tikzpicture}[line width=1 pt, scale=1]
%\begin{scope}[rotate=90]

\begin{scope}[shift={(0,0)}, rotate=-90]
\draw [phir, ultra thick, domain=0:360] plot ({1*cos(\x)}, {1*sin(\x)});
\draw [phir, ultra thick] (1,0) -- (1,2);
\draw [phir, ultra thick] (1,0) -- (1,-2);
\node at (1,0) {$\times $};
\node at (2,0) {\Large $ \frac{(-i\lambda_4)}{2} A_{R}$};
\end{scope}

\begin{scope}[shift={(5,0)}, rotate=-90]
\draw [phil, ultra thick, domain=0:360] plot ({1*cos(\x)}, {1*sin(\x)});
\draw [phir, ultra thick] (1,0) -- (1,2);
\draw [phir, ultra thick] (1,0) -- (1,-2);
\node at (1,0) {$\times $};
\node at (2,0) {\Large $\frac{(-\lambda_\Delta)}{2}A_{L}$};
\end{scope}

\begin{scope}[shift={(10,0)}, rotate=-90] 
%\begin{scope}[shift={(2.5,-4)}, rotate=-90]
\draw [phil, ultra thick, domain=0:180] plot ({1*cos(\x)}, {1*sin(\x)});
\draw [phir, ultra thick, domain=180:360] plot ({1*cos(\x)}, {1*sin(\x)});
\draw [phir, ultra thick] (1,0) -- (1,2);
\draw [phir, ultra thick] (1,0) -- (1,-2);
\node at (1,0) {$\times $};
\node at (2,0) {\Large $(-i\sigma_4) A_{M}$};
\end{scope}

\end{tikzpicture}
\end{center}
\caption{One loop correction to $m^2$ due to quartic couplings}
\label{fig:macromassrenorm1}
\end{figure}

\begin{figure}[ht]
\begin{center}
\begin{tikzpicture}[line width=1 pt, scale=1]
%\begin{scope}[rotate=-80]

\begin{scope}[shift={(0,0)}]
\draw [phir, ultra thick] (0,0) -- (1,0);
\draw [phir, ultra thick] (0,0) -- (-1,0);
\node at (0,0) {$\times $};	
\node at (0,-1) {\Large $-i\delta m^{2}$};	 

\node at (-1.25,0) {$\bf 1$};	
\node at (1.25,0) {$\bf 2$};	 
\filldraw (0,0) circle (.1);  

\end{scope}

\end{tikzpicture}
\end{center}
\caption{Diagrammatic representation of the one loop counter-term for  $m^2 $ }
\label{diag:masscounterterm3}
\end{figure}

Let us compute the one loop beta function for $m^2 $. We shall consider all the one loop Feynman diagrams that contributes to the process  $\phir  \rightarrow \phir $. One can verify that there are mainly two types of diagrams - one class of diagrams due to the cubic couplings, as depicted in figure \ref{diag:macromassrenorm3}, and the other class of diagrams due to quartic couplings, depicted in figure \ref{fig:macromassrenorm1}.  

The sum of the contribution from all the Feynman diagrams is given by 
\begin{eqnarray}
&&-im^2 
\nonumber\\&& 
+\frac{(-i\lambda_3)^2}{2} B_{RR}(k)+\frac{(i\sigma_3^\star)^2}{2}B_{LL}(k)	+(-i\sigma_3)^2 B_{LR}(k)
\nonumber\\&&
+\frac{(i\sigma_3^\ast) (-i\lambda_3)}{2}B_{PM}(k)+\frac{(i\sigma_3^\ast) (-i\lambda_3)}{2}B_{MP}(k)+(-i\sigma_3)^2B_{PP}(k)
\nonumber\\&&
+(-i\lambda_3)(-i\sigma_3)B_{PR}(k)+(-i\lambda_3)(-i\sigma_3)B_{MR}(k)
\nonumber\\&&
+(i\sigma_3^\ast)(-i\sigma_3)B_{PL}(k)+(i\sigma_3^\ast)(-i\sigma_3)B_{ML}(k) 
\nonumber\\&&
+\frac{(-i\lambda_4)}{2}A_R+\frac{(-\lambda_\Delta)}{2}A_L
+(-i\sigma_4)A_M 
\label{macromass1}  
\end{eqnarray}
 Using the results in \eqref{skpvsummary1a}-\eqref{skpvsummary1d}, we can see that the contribution is divergent and one needs to add one 
loop counter-terms $\delta m^{2}$, in the $\msbar$ scheme, to absorb the divergences. 
\begin{eqnarray}
\delta m^{2}\Big|_{\msbar} &	= 
&-\frac{1}{(4\pi)^2} \Bigl[ (\lambda_3)^2 -(\sigma_3^\star)^2+
2\left\{\lambda_3\sigma_3+|\sigma_3|^2 \right\}\ \Bigr] \Bigl[ \frac{1}{d-4} +\frac{1}{2}(\gamma_E-\ln\, 4\pi) \Bigr]
\nonumber\\
&&-\frac{1}{(4\pi)^2} \left[\lambda_4  -i \lambda_\Delta +2\sigma_4
	\right]\Bigl[ \frac{1}{d-4} +\frac{1}{2}(\gamma_E-\ln\, 4\pi) \Bigr]\left(\re\,m^2\right) 
\label{macromass2}  
\end{eqnarray} 
Using the standard methods of quantum field theory, one can then compute the one loop beta function as
\begin{eqnarray} 
\beta_{m^2}= \frac{1}{(4\pi)^2} \Bigl[ (\lambda_3)^2 -(\sigma_3^\star)^2+
2\left\{\lambda_3\sigma_3+|\sigma_3|^2 \right\}\ 
%\nonumber\\
%&&
+(\lambda_4-i\lambda_\Delta+2\sigma_4)\left(\re\,m^2\right) 
\Bigr]
\label{macromass3}  
\end{eqnarray} 
If set $\sigma_3=\sigma_4=\lambda_\Delta=0$, then we get back the standard results of $\phi^3+\phi^4$ theory in $d=4$ space-time dimensions.  

\subsection{One loop beta function for \texorpdfstring{$m_\Delta^2$}{mdelta}}
\label{subsec:oneloopmdelta}

\begin{figure}
\begin{center}
\begin{tikzpicture}[line width=1 pt, scale=0.8]

\begin{scope}[shift={(0,0)}]
\draw [phil, ultra thick] (1,0) -- (2,0);
\draw [phir, ultra thick] (-1,0) -- (-2,0);
\draw [phir, ultra thick, domain=0:180] plot ({1*cos(\x)}, {1*sin(\x)});
\draw [phir, ultra thick, domain=180:360] plot ({1*cos(\x)}, {1*sin(\x)}); 
\node at (-1,0) {$\times $};	
\node at (1,0) {$\times $};	
\node at (0,-2) {\Large $\frac{(-i\lambda_3)(-i\sigma_3)}{2}B_{RR}(k)$};	

\node at (-2.25,0) {$\bf 1$};	
\node at (2.25,0) {$\bf 2$};	
	
\end{scope}

\begin{scope}[shift={(6,0)}]
\draw [phil, ultra thick] (1,0) -- (2,0);
\draw [phir, ultra thick] (-1,0) -- (-2,0);
\draw [phil, ultra thick, domain=0:90] plot ({1*cos(\x)}, {1*sin(\x)});
\draw [phil, ultra thick, domain=90:180] plot ({1*cos(\x)}, {1*sin(\x)});
\draw [phil, ultra thick, domain=180:270] plot ({1*cos(\x)}, {1*sin(\x)});
\draw [phil, ultra thick, domain=270:360] plot ({1*cos(\x)}, {1*sin(\x)}); 
\node at (-1,0) {$\times $};	
\node at (1,0) {$\times $};	
\node at (0,-2) {\Large $\frac{(i\lambda^\star_3)(i\sigma_3^\ast)}{2}B_{LL}(k)$};	

\node at (-2.25,0) {$\bf 1$};	
\node at (2.25,0) {$\bf 2$};	
	
\end{scope}

\begin{scope}[shift={(12,0)}]
\draw [phil, ultra thick] (1,0) -- (2,0);
\draw [phir, ultra thick] (-1,0) -- (-2,0);
\draw [phir, ultra thick, domain=0:90] plot ({1*cos(\x)}, {1*sin(\x)});
\draw [phir, ultra thick, domain=90:180] plot ({1*cos(\x)}, {1*sin(\x)});
\draw [phil, ultra thick, domain=180:270] plot ({1*cos(\x)}, {1*sin(\x)});
\draw [phil, ultra thick, domain=270:360] plot ({1*cos(\x)}, {1*sin(\x)}); 
\node at (-1,0) {$\times $};	
\node at (1,0) {$\times $};	
\node at (0,-2) {\Large $(-i\sigma_3)(i\sigma^\star_3)B_{LR}(k)$};	

\node at (-2.25,0) {$\bf 1$};	
\node at (2.25,0) {$\bf 2$};	
	
\end{scope}

\begin{scope}[shift={(0,-5)}]
\draw [phil, ultra thick] (1,0) -- (2,0);
\draw [phir, ultra thick] (-1,0) -- (-2,0);
\draw [phil, ultra thick, domain=0:90] plot ({1*cos(\x)}, {1*sin(\x)});
\draw [phir, ultra thick, domain=90:180] plot ({1*cos(\x)}, {1*sin(\x)});
\draw [phir, ultra thick, domain=180:270] plot ({1*cos(\x)}, {1*sin(\x)});
\draw [phil, ultra thick, domain=270:360] plot ({1*cos(\x)}, {1*sin(\x)});  
\node at (-1,0) {$\times $};	
\node at (1,0) {$\times $};	
\node at (0,-2) {\Large $\frac{(-i\lambda_3)(i\lambda^\star_3)}{2}B_{PM}(k)$};	

\node at (-2.25,0) {$\bf 1$};	
\node at (2.25,0) {$\bf 2$};		
	
\end{scope}

\begin{scope}[shift={(6,-5)}]
\draw [phil, ultra thick] (1,0) -- (2,0);
\draw [phir, ultra thick] (-1,0) -- (-2,0);
\draw [phir, ultra thick, domain=0:90] plot ({1*cos(\x)}, {1*sin(\x)});
\draw [phil, ultra thick, domain=90:180] plot ({1*cos(\x)}, {1*sin(\x)});
\draw [phil, ultra thick, domain=180:270] plot ({1*cos(\x)}, {1*sin(\x)});
\draw [phir, ultra thick, domain=270:360] plot ({1*cos(\x)}, {1*sin(\x)}); 
\node at (-1,0) {$\times $};	
\node at (1,0) {$\times $};	
\node at (0,-2) {\Large $\frac{(-i\sigma_3)(i\sigma^\ast_3)}{2}B_{MP}(k)$};	

\node at (-2.25,0) {$\bf 1$};	
\node at (2.25,0) {$\bf 2$};	
	
\end{scope}

\begin{scope}[shift={(12,-5)}]
\draw [phil, ultra thick] (1,0) -- (2,0);
\draw [phir, ultra thick] (-1,0) -- (-2,0);
\draw [phir, ultra thick, domain=0:90] plot ({1*cos(\x)}, {1*sin(\x)});
\draw [phil, ultra thick, domain=90:180] plot ({1*cos(\x)}, {1*sin(\x)});
\draw [phir, ultra thick, domain=180:270] plot ({1*cos(\x)}, {1*sin(\x)});
\draw [phil, ultra thick, domain=270:360] plot ({1*cos(\x)}, {1*sin(\x)}); 
\node at (-1,0) {$\times $};	
\node at (1,0) {$\times $};	
\node at (0,-2) {\Large $(-i\sigma_3)(i\sigma^\star_3)B_{PP}(k)$};	

\node at (-2.25,0) {$\bf 1$};	
\node at (2.25,0) {$\bf 2$};	
	
\end{scope}

\begin{scope}[shift={(3,-10)}]
\draw [phil, ultra thick] (1,0) -- (2,0);
\draw [phir, ultra thick] (-1,0) -- (-2,0);
\draw [phir, ultra thick, domain=0:90] plot ({1*cos(\x)}, {1*sin(\x)});
\draw [phir, ultra thick, domain=90:180] plot ({1*cos(\x)}, {1*sin(\x)});
\draw [phir, ultra thick, domain=180:270] plot ({1*cos(\x)}, {1*sin(\x)});
\draw [phil, ultra thick, domain=270:360] plot ({1*cos(\x)}, {1*sin(\x)}); 
\node at (-1,0) {$\times $};	
\node at (1,0) {$\times $};	
\node at (0,-2) {\Large $(i\sigma^\star_3)(-i\lambda_3)B_{PR}(k)$};	

\node at (-2.25,0) {$\bf 1$};	
\node at (2.25,0) {$\bf 2$};	
	
\end{scope}

\begin{scope}[shift={(10,-10)}]
\draw [phil, ultra thick] (1,0) -- (2,0);
\draw [phir, ultra thick] (-1,0) -- (-2,0);
\draw [phir, ultra thick, domain=0:90] plot ({1*cos(\x)}, {1*sin(\x)});
\draw [phir, ultra thick, domain=90:180] plot ({1*cos(\x)}, {1*sin(\x)});
\draw [phil, ultra thick, domain=180:270] plot ({1*cos(\x)}, {1*sin(\x)});
\draw [phir, ultra thick, domain=270:360] plot ({1*cos(\x)}, {1*sin(\x)}); 
\node at (-1,0) {$\times $};	
\node at (1,0) {$\times $};	
\node at (0,-2) {\Large $(-i\sigma_3)^2B_{MR}(k)$};	

\node at (-2.25,0) {$\bf 1$};	
\node at (2.25,0) {$\bf 2$};	
	
\end{scope}

\begin{scope}[shift={(3,-15)}]
\draw [phil, ultra thick] (1,0) -- (2,0);
\draw [phir, ultra thick] (-1,0) -- (-2,0);
\draw [phil, ultra thick, domain=0:90] plot ({1*cos(\x)}, {1*sin(\x)});
\draw [phil, ultra thick, domain=90:180] plot ({1*cos(\x)}, {1*sin(\x)});
\draw [phir, ultra thick, domain=180:270] plot ({1*cos(\x)}, {1*sin(\x)});
\draw [phil, ultra thick, domain=270:360] plot ({1*cos(\x)}, {1*sin(\x)}); 
\node at (-1,0) {$\times $};	
\node at (1,0) {$\times $};	
\node at (0,-2) {\Large $(-i\sigma_3)(i\lambda^\star_3)B_{PL}(k)$};	

\node at (-2.25,0) {$\bf 1$};	
\node at (2.25,0) {$\bf 2$};	
	
\end{scope}

\begin{scope}[shift={(10,-15)}]
\draw [phil, ultra thick] (1,0) -- (2,0);
\draw [phir, ultra thick] (-1,0) -- (-2,0);
\draw [phil, ultra thick, domain=0:90] plot ({1*cos(\x)}, {1*sin(\x)});
\draw [phil, ultra thick, domain=90:180] plot ({1*cos(\x)}, {1*sin(\x)});
\draw [phil, ultra thick, domain=180:270] plot ({1*cos(\x)}, {1*sin(\x)});
\draw [phir, ultra thick, domain=270:360] plot ({1*cos(\x)}, {1*sin(\x)});  
\node at (-1,0) {$\times $};	 
\node at (1,0) {$\times $};	
\node at (0,-2) {\Large $(i\sigma^\star_3)^2B_{ML}(k)$};	

\node at (-2.25,0) {$\bf 1$};	
\node at (2.25,0) {$\bf 2$};
	
\end{scope}	 
\end{tikzpicture}
\end{center}
\caption{One Loop corrections to $ m^2_\Delta$ due to cubic couplings} 
\label{diag:macromassrenorm4}
\end{figure}

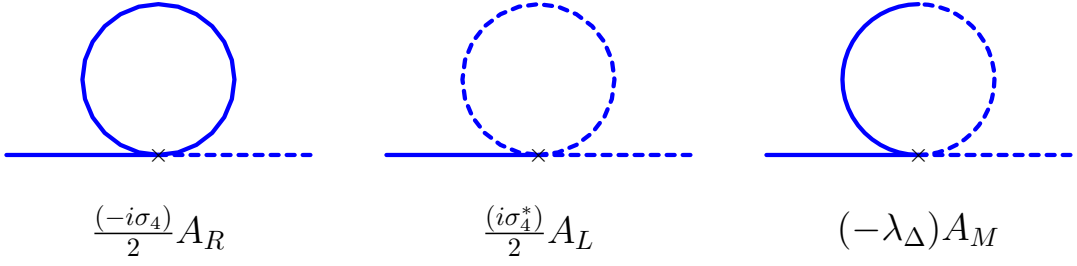
\begin{figure}[ht]
\begin{center}
\begin{tikzpicture}[line width=1 pt, scale=1]
%\begin{scope}[rotate=90]

\begin{scope}[shift={(0,0)}, rotate=-90]
\draw [phir, ultra thick, domain=0:360] plot ({1*cos(\x)}, {1*sin(\x)});
\draw [phil, ultra thick] (1,0) -- (1,2);
\draw [phir, ultra thick] (1,0) -- (1,-2);
\node at (1,0) {$\times $};
\node at (2,0) {\Large $\frac{(-i\sigma_4)}{2} A_{R}$};
\end{scope}

\begin{scope}[shift={(5,0)}, rotate=-90]
\draw [phil, ultra thick, domain=0:360] plot ({1*cos(\x)}, {1*sin(\x)});
\draw [phil, ultra thick] (1,0) -- (1,2);
\draw [phir, ultra thick] (1,0) -- (1,-2);
\node at (1,0) {$\times $};	
\node at (2,0) {\Large $\frac{(i\sigma_4^*)}{2} A_{L}$};
\end{scope}

\begin{scope}[shift={(10,0)}, rotate=-90] 
%\begin{scope}[shift={(2.5,-4)}, rotate=-90]
\draw [phil, ultra thick, domain=0:180] plot ({1*cos(\x)}, {1*sin(\x)});
\draw [phir, ultra thick, domain=180:360] plot ({1*cos(\x)}, {1*sin(\x)});
\draw [phil, ultra thick] (1,0) -- (1,2);
\draw [phir, ultra thick] (1,0) -- (1,-2); 
\node at (1,0) {$\times $};
\node at (2,0) {\Large $(-\lambda_\Delta) A_{M}$};
\end{scope}

\end{tikzpicture} 
\end{center}
\caption{One loop correction to $m^2_\Delta$ due to quartic couplings }
\label{fig:macromassrenorm2}
\end{figure}

Now, we will compute the one loop beta function for $m_\Delta^2$. As in the case of $m^2$, there will again be two classes of diagrams. The diagrams due to
cubic and quartic couplings are as shown in figure \ref{diag:macromassrenorm4} and in figure \ref{fig:macromassrenorm2} respectively. The sum over all the contributions is given by 
\begin{eqnarray}
&&-m_\Delta^2
\nonumber\\
&&
+	
\frac{(-i\lambda_3)(-i\sigma_3)}{2}B_{RR}(k)
+\frac{(i\lambda^\star_3)(i\sigma_3^\ast )}{2}B_{LL}(k)
+(-i\sigma_3)(i\sigma^\star_3)B_{LR}(k)
\nonumber\\
&&	
+
\frac{(-i\lambda_3)(i\lambda_3^\star)}{2}B_{PM}(k)
+
\frac{(-i\sigma_3)(i\sigma^\star_3)}{2}B_{MP}(k)
+
(-i\sigma_3)(i\sigma_3^\star)B_{PP}(k)
\label{macromass11}  
\\ 
&&	
+
(i\sigma^\star_3)(-i\lambda_3)B_{PR}(k)
+ 
(-i\sigma_3)^2B_{MR}(k)
+
(-i\sigma_3)(i\lambda^\star_3)B_{PL}(k)
+
(i\sigma^\star_3)^2 B_{ML}(k)
\nonumber\\&&
+\frac{(-i\sigma_4)}{2}A_R+\frac{(i\sigma^\star_4 )}{2}A_L
+(-\lambda_\Delta)A_M 
\nonumber
\end{eqnarray}
Some of these one loop contributions are divergent and one needs to add one loop counter-terms. The $ m_\Delta^2$ counter-term in $\msbar$ scheme is given by
\begin{eqnarray}
\delta m_\Delta^2\Big|_\msbar&=& -\frac{1}{(4\pi)^2}\Big[	-4(\re\, \lambda_3\,+\re \, \sigma_3)\im\, \sigma_3+(2\lambda_\Delta-2\im\,\sigma_4)(\re\,m^2)\Big]
\nonumber\\
&&
\Bigl[ \frac{1}{d-4} +\frac{1}{2}(\gamma_E-\ln\, 4\pi) \Bigr]
%\nonumber\\
\label{macromass12}   
\end{eqnarray}
and the beta function for $m_\Delta^2$ is given by 
\begin{eqnarray}
\beta_{ m_\Delta^2} = \frac{1}{(4\pi)^2} \left[-4(\re\,  \lambda_3\,+\re\,   \sigma_3\,)\, \im\,  \sigma_3 
 +(2\lambda_\Delta - 2 \im\,\sigma_4)(\re\,m^2)
 \right]  
\label{macromass13}
\end{eqnarray}  

\subsection{Checking Lindblad condition for mass renormalization}
\label{subsec:oneloopLindbladquadratic}

From equation \eqref{macromass3}, we find that the beta function for $\im\,m^2$ is given by 
\begin{eqnarray}
	\frac{d\,(\im\,m^2)}{d\ln\, \mu}
&=&\frac{1}{(4\pi)^2}\Big[2(\re\, \lambda_3\ +\re\, \sigma_3) (\im\, \lambda_3+\im\, \sigma_3)
\nonumber\\
&&\qquad\quad		+(\im\lambda_4\, +2\,\im\,\sigma_4   -\lambda_\Delta)(\re\,m^2)	 \Big] 		 
\label{Lindbladbeta1}
\end{eqnarray} 
Now, using equation \eqref{Lindbladbeta1} and equation \eqref{macromass13}, one gets the beta function for $(\im\,m^2-m_\Delta^2)$ 
\begin{eqnarray}
 \beta_{(\im\,m^2 - m_\Delta^2)}&=&\frac{2}{(4\pi)^2}[(\im\, \lambda_3+3\, \im\,  \sigma_3)(\re\, \lambda_3+\re\,  \sigma_3)
% \nonumber\\
%  && +\frac{1}{(4\pi)^2}
+  (\im\lambda_4\ +4\im\,\sigma_4  -3\lambda_\Delta )(\re\,m^2)]
   \nonumber\\
\label{Lindbladbeta2}
\end{eqnarray}
equation \eqref{Lindbladbeta2} shows that the one loop beta function for Lindblad violating mass terms vanish in the absence of Lindblad violating cubic (equation \eqref{macroaction8}) and 
quartic coupling (equation \eqref{macroaction10}) at the tree level. 

%\chapter{One loop Three point function}

%\section{One loop beta function for cubic couplings}
%\label{sec:oneloopcubiccoupling}

\subsection{One loop beta function for \texorpdfstring{$\lambda_3$}{L3}}
\label{subsec:onelooplambdathree}
 
\begin{figure}
\begin{center}
\begin{tikzpicture}[line width=1 pt, scale=0.8]
%\begin{scope}[rotate=-80]

\begin{scope}[shift={(0,0)}]
\draw [phir, ultra thick] (1,0) -- (2,0);
\draw [phir, ultra thick] (-1,0) -- (-2,1.5);
\draw [phir, ultra thick] (-1,0) -- (-2,-1.5);
\draw [phir, ultra thick, domain=0:180] plot ({1*cos(\x)}, {1*sin(\x)});
\draw [phir, ultra thick, domain=180:360] plot ({1*cos(\x)}, {1*sin(\x)}); 
\node at (-1,0) {$\times $};	
\node at (1,0) {$\times $};	
\node at (0,-2) {\Large $\frac{(-i\lambda_4)(-i\lambda_3)}{2} B_{RR}(k_3)$};	
\node at (-1,0) {$\times $};	

\node at (-2.25,1.5) {$\bf 1$};	 
\node at (-2.25,-1.5) {$\bf 2$};	
\node at (2.25,0) {$\bf 3$};

\end{scope}

\begin{scope}[shift={(6,0)}]
\draw [phir, ultra thick] (1,0) -- (2,0);
\draw [phir, ultra thick] (-1,0) -- (-2,1.5);
\draw [phir, ultra thick] (-1,0) -- (-2,-1.5);
\draw [phil, ultra thick, domain=0:90] plot ({1*cos(\x)}, {1*sin(\x)});
\draw [phil, ultra thick, domain=90:180] plot ({1*cos(\x)}, {1*sin(\x)});
\draw [phil, ultra thick, domain=180:270] plot ({1*cos(\x)}, {1*sin(\x)});
\draw [phil, ultra thick, domain=270:360] plot ({1*cos(\x)}, {1*sin(\x)}); 
\node at (-1,0) {$\times $};	
\node at (1,0) {$\times $};	
\node at (0,-2) {\Large $\frac{(-\lambda_\Delta)(i\sigma_3^\star)}{2} B_{LL}(k_3)$};

\node at (-2.25,1.5) {$\bf 1$};	
\node at (-2.25,-1.5) {$\bf 2$};	
\node at (2.25,0) {$\bf 3$};

\end{scope}

\begin{scope}[shift={(12,0)}]
\draw [phir, ultra thick] (1,0) -- (2,0);
\draw [phir, ultra thick] (-1,0) -- (-2,1.5);
\draw [phir, ultra thick] (-1,0) -- (-2,-1.5);
\draw [phir, ultra thick, domain=0:90] plot ({1*cos(\x)}, {1*sin(\x)});
\draw [phir, ultra thick, domain=90:180] plot ({1*cos(\x)}, {1*sin(\x)});
\draw [phil, ultra thick, domain=180:270] plot ({1*cos(\x)}, {1*sin(\x)});
\draw [phil, ultra thick, domain=270:360] plot ({1*cos(\x)}, {1*sin(\x)}); 
\node at (-1,0) {$\times $};	
\node at (1,0) {$\times $};	
\node at (0,-2) {\Large $(-i\sigma_4)(-i\sigma_3) B_{LR}(k_3)$};	

\node at (-2.25,1.5) {$\bf 1$};	
\node at (-2.25,-1.5) {$\bf 2$};	
\node at (2.25,0) {$\bf 3$};

\end{scope}

\begin{scope}[shift={(0,-5)}]
\draw [phir, ultra thick] (1,0) -- (2,0);
\draw [phir, ultra thick] (-1,0) -- (-2,1.5);
\draw [phir, ultra thick] (-1,0) -- (-2,-1.5);
\draw [phil, ultra thick, domain=0:90] plot ({1*cos(\x)}, {1*sin(\x)});
\draw [phir, ultra thick, domain=90:180] plot ({1*cos(\x)}, {1*sin(\x)});
\draw [phir, ultra thick, domain=180:270] plot ({1*cos(\x)}, {1*sin(\x)});
\draw [phil, ultra thick, domain=270:360] plot ({1*cos(\x)}, {1*sin(\x)});  
\node at (-1,0) {$\times $};	
\node at (1,0) {$\times $};	
\node at (0,-2) {\Large $\frac{(-i\lambda_4)(i\sigma^\star_3)}{2} B_{PM}(k_3)$};	

\node at (-2.25,1.5) {$\bf 1$};	
\node at (-2.25,-1.5) {$\bf 2$};	
\node at (2.25,0) {$\bf 3$};

\end{scope}

\begin{scope}[shift={(6,-5)}]
\draw [phir, ultra thick] (1,0) -- (2,0);

\draw [phir, ultra thick] (-1,0) -- (-2,1.5);
\draw [phir, ultra thick] (-1,0) -- (-2,-1.5);
\draw [phir, ultra thick, domain=0:90] plot ({1*cos(\x)}, {1*sin(\x)});
\draw [phil, ultra thick, domain=90:180] plot ({1*cos(\x)}, {1*sin(\x)});
\draw [phil, ultra thick, domain=180:270] plot ({1*cos(\x)}, {1*sin(\x)});
\draw [phir, ultra thick, domain=270:360] plot ({1*cos(\x)}, {1*sin(\x)}); 
\node at (-1,0) {$\times $};	
\node at (1,0) {$\times $};	
\node at (0,-2) {\Large $\frac{(-\lambda_\Delta)(-i\lambda_3)}{2} B_{MP}(k_3)$};	

\node at (-2.25,1.5) {$\bf 1$};	
\node at (-2.25,-1.5) {$\bf 2$};	
\node at (2.25,0) {$\bf 3$};

\end{scope}

\begin{scope}[shift={(12,-5)}]
\draw [phir, ultra thick] (1,0) -- (2,0);

\draw [phir, ultra thick] (-1,0) -- (-2,1.5);
\draw [phir, ultra thick] (-1,0) -- (-2,-1.5);
\draw [phir, ultra thick, domain=0:90] plot ({1*cos(\x)}, {1*sin(\x)});
\draw [phil, ultra thick, domain=90:180] plot ({1*cos(\x)}, {1*sin(\x)});
\draw [phir, ultra thick, domain=180:270] plot ({1*cos(\x)}, {1*sin(\x)});
\draw [phil, ultra thick, domain=270:360] plot ({1*cos(\x)}, {1*sin(\x)}); 
\node at (-1,0) {$\times $};	
\node at (1,0) {$\times $};	
\node at (0,-2) {\Large $(-i\sigma_4)(-i\sigma_3) B_{PP}(k_3)$};	
	
\node at (-2.25,1.5) {$\bf 1$};	
\node at (-2.25,-1.5) {$\bf 2$};	
\node at (2.25,0) {$\bf 3$};

\end{scope}

\begin{scope}[shift={(3,-10)}]
\draw [phir, ultra thick] (1,0) -- (2,0);

\draw [phir, ultra thick] (-1,0) -- (-2,1.5);
\draw [phir, ultra thick] (-1,0) -- (-2,-1.5);
\draw [phir, ultra thick, domain=0:90] plot ({1*cos(\x)}, {1*sin(\x)});
\draw [phir, ultra thick, domain=90:180] plot ({1*cos(\x)}, {1*sin(\x)});
\draw [phir, ultra thick, domain=180:270] plot ({1*cos(\x)}, {1*sin(\x)});
\draw [phil, ultra thick, domain=270:360] plot ({1*cos(\x)}, {1*sin(\x)}); 
\node at (-1,0) {$\times $};	 
\node at (1,0) {$\times $};	
\node at (0,-2) {\Large $(-i\lambda_4)(-i\sigma_3) B_{PR}(k_3)$};	

\node at (-2.25,1.5) {$\bf 1$};	
\node at (-2.25,-1.5) {$\bf 2$};	
\node at (2.25,0) {$\bf 3$};

\end{scope}

\begin{scope}[shift={(10,-10)}]
\draw [phir, ultra thick] (1,0) -- (2,0);

\draw [phir, ultra thick] (-1,0) -- (-2,1.5);
\draw [phir, ultra thick] (-1,0) -- (-2,-1.5);
\draw [phir, ultra thick, domain=0:90] plot ({1*cos(\x)}, {1*sin(\x)});
\draw [phir, ultra thick, domain=90:180] plot ({1*cos(\x)}, {1*sin(\x)});
\draw [phil, ultra thick, domain=180:270] plot ({1*cos(\x)}, {1*sin(\x)});
\draw [phir, ultra thick, domain=270:360] plot ({1*cos(\x)}, {1*sin(\x)}); 
\node at (-1,0) {$\times $};	
\node at (1,0) {$\times $};	
\node at (0,-2) {\Large $(-i\sigma_4)(-i\lambda_3) B_{MR}(k_3)$};

\node at (-2.25,1.5) {$\bf 1$};	
\node at (-2.25,-1.5) {$\bf 2$};	
\node at (2.25,0) {$\bf 3$};

\end{scope}

\begin{scope}[shift={(3,-15)}]
\draw [phir, ultra thick] (1,0) -- (2,0);

\draw [phir, ultra thick] (-1,0) -- (-2,1.5);
\draw [phir, ultra thick] (-1,0) -- (-2,-1.5);
\draw [phil, ultra thick, domain=0:90] plot ({1*cos(\x)}, {1*sin(\x)});
\draw [phil, ultra thick, domain=90:180] plot ({1*cos(\x)}, {1*sin(\x)});
\draw [phir, ultra thick, domain=180:270] plot ({1*cos(\x)}, {1*sin(\x)});
\draw [phil, ultra thick, domain=270:360] plot ({1*cos(\x)}, {1*sin(\x)}); 
\node at (-1,0) {$\times $};	
\node at (1,0) {$\times $};	
\node at (0,-2) {\Large $(-i\sigma_4)(i\sigma_3^\star) B_{PL} (k_3)$};	

\node at (-2.25,1.5) {$\bf 1$};	
\node at (-2.25,-1.5) {$\bf 2$};	
\node at (2.25,0) {$\bf 3$};

\end{scope}

\begin{scope}[shift={(10,-15)}]
\draw [phir, ultra thick] (1,0) -- (2,0);

\draw [phir, ultra thick] (-1,0) -- (-2,1.5);
\draw [phir, ultra thick] (-1,0) -- (-2,-1.5);
\draw [phil, ultra thick, domain=0:90] plot ({1*cos(\x)}, {1*sin(\x)});
\draw [phil, ultra thick, domain=90:180] plot ({1*cos(\x)}, {1*sin(\x)});
\draw [phil, ultra thick, domain=180:270] plot ({1*cos(\x)}, {1*sin(\x)});
\draw [phir, ultra thick, domain=270:360] plot ({1*cos(\x)}, {1*sin(\x)});  
\node at (-1,0) {$\times $};	
\node at (1,0) {$\times $};	
\node at (0,-2) {\Large $(-\lambda_\Delta)(-i\sigma_3) B_{ML}(k_3)$};	

\node at (-2.25,1.5) {$\bf 1$};	
\node at (-2.25,-1.5) {$\bf 2$};	
\node at (2.25,0) {$\bf 3$};	
	
\end{scope}	  

\end{tikzpicture}
\end{center}
\caption{Diagrammatic representation of the Ten  $1$-Loop Integrals $ \phir\phir \rightarrow \phir$ }
\label{diag:lambdathreeoneloop}
\end{figure}
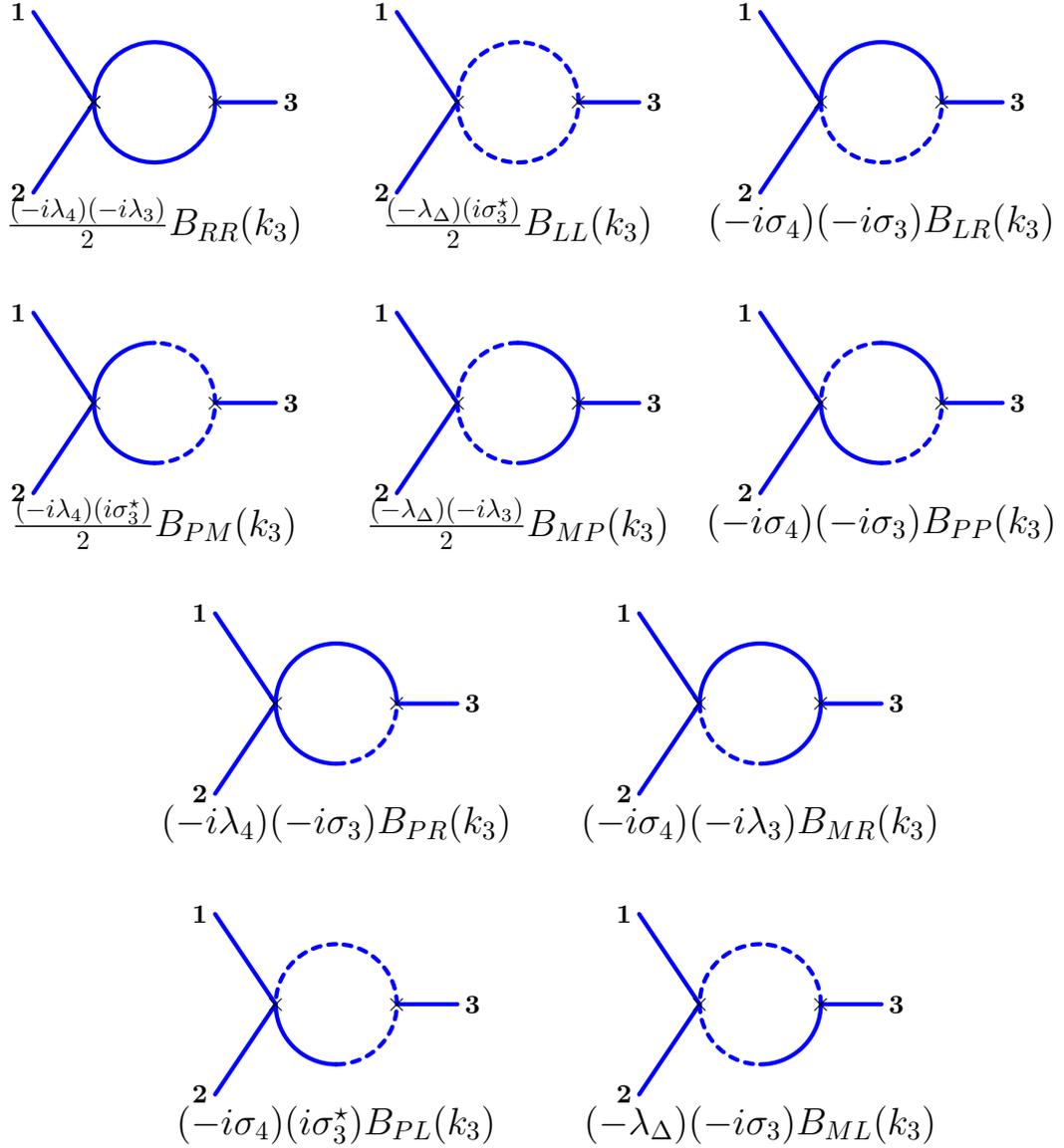

Now we will compute the one loop beta function for various cubic couplings.  The Passarino-Veltman $C$ and $D$ integrals will have no contribution to the one loop $\beta$ function for the cubic (and quartic) couplings, since they are UV finite \footnote{Note that the standard C and D integrals are  well-known to be UV finite in the Euclidean theory. Since the SK versions of these integrals are different analytic continuations of  these Euclidean integrals, they continue to be UV finite. We will leave the detailed computation including these finite contributions to future work.}. Hence, we shall not consider those Feynman diagrams in our analysis.  We begin with the beta function computation of $\lambda_3$. The diagrams for one of the channels are depicted in figure  \ref{diag:lambdathreeoneloop}. The other two channels are obtained by interchanging  $1\longleftrightarrow 3$ and $2\longleftrightarrow 3$.  The sum over the all the Feynman diagrams is given by 
\begin{equation}
\begin{split}
&-i\lambda_3 
\\
&+\frac{(-i\lambda_4)(-i\lambda_3)}{2} B_{RR}(k_3)+\frac{(-\lambda_\Delta)(i\sigma_3^\star)}{2}B_{LL}(k_3)	+(-i\sigma_4)(-i\sigma_3) B_{LR}(k_3)
\\
&+\frac{(-i\lambda_4)(i\sigma_3^\ast) }{2}B_{PM}(k_3)+\frac{(-\lambda_\Delta) (-i\lambda_3)}{2}B_{MP}(k_3)+(-i\sigma_4)(-i\sigma_3)B_{PP}(k_3)
\\
&+(-i\lambda_4)(-i\sigma_3)B_{PR}(k_3)+(-i\sigma_4)(-i\lambda_3)B_{MR}(k_3)
\\
&+(-i\sigma_4)(i\sigma_3^\ast)B_{PL}(k_3)+(-\lambda_\Delta)(-i\sigma_3)B_{ML}(k_3) 
\\
&+\textrm{Two more channels}
\end{split}
\label{lambdathree1}
\end{equation}
 Using the results in \eqref{skpvsummary1a}-\eqref{skpvsummary1d}, we see that the one loop contributions are divergent and we need to add one loop 
 counter-terms $\delta \lambda_3$  to cancel the divergences. 
\begin{eqnarray}
\delta \lambda_3\Big|_\msbar	&=&- \frac{3}{(4\pi)^2} \Bigl[\lambda_4\lambda_3 -2\lambda_\Delta\im\, \sigma_3
+\lambda_4\sigma_3+\sigma_4\lambda_3+\sigma_4\sigma_3^\ast \Bigr] 
\nonumber\\
&&\qquad
\Bigl[ \frac{1}{d-4} +\frac{1}{2}(\gamma_E-\ln\, 4\pi) \Bigr]
\label{lambdathree2}
\end{eqnarray}
Following the standard methods of quantum field theory, we compute the one loop beta function to be
\begin{eqnarray}
\beta_{\lambda_3}
&=\frac{3}{(4\pi)^2} \Bigl[\lambda_4\lambda_3 -2\lambda_\Delta\im\, \sigma_3
+\lambda_4\sigma_3+\sigma_4\lambda_3+\sigma_4\sigma_3^\ast \Bigr]
\label{lambdathree3}
\end{eqnarray}

\subsection{One loop beta function for \texorpdfstring{$\sigma_3$}{sigma3}}
\label{subsec:oneloopsigmathree}

\begin{figure}
\begin{center}
\begin{tikzpicture}[line width=1 pt, scale=0.8]

\begin{scope}[shift={(0,0)}]
\draw [phil, ultra thick] (1,0) -- (2,0);
\draw [phir, ultra thick] (-1,0) -- (-2,1.5);
\draw [phir, ultra thick] (-1,0) -- (-2,-1.5);
\draw [phir, ultra thick, domain=0:180] plot ({1*cos(\x)}, {1*sin(\x)});
\draw [phir, ultra thick, domain=180:360] plot ({1*cos(\x)}, {1*sin(\x)}); 
\node at (-1,0) {$\times $};	
\node at (1,0) {$\times $};	
\node at (0,-2) {\Large $\frac{(-i\lambda_4)(-i\sigma_3)}{2}B_{RR}(k_3)$};	

\node at (-2.25,1.5) {$\bf 1$};	
\node at (-2.25,-1.5) {$\bf 2$};	
\node at (2.25,0) {$\bf 3$};

\end{scope}
 
\begin{scope}[shift={(6,0)}]
\draw [phil, ultra thick] (1,0) -- (2,0);
\draw [phir, ultra thick] (-1,0) -- (-2,1.5);
\draw [phir, ultra thick] (-1,0) -- (-2,-1.5);
\draw [phil, ultra thick, domain=0:90] plot ({1*cos(\x)}, {1*sin(\x)});
\draw [phil, ultra thick, domain=90:180] plot ({1*cos(\x)}, {1*sin(\x)});
\draw [phil, ultra thick, domain=180:270] plot ({1*cos(\x)}, {1*sin(\x)});
\draw [phil, ultra thick, domain=270:360] plot ({1*cos(\x)}, {1*sin(\x)}); 
\node at (-1,0) {$\times $};	
\node at (1,0) {$\times $};	
\node at (0,-2) {\Large $\frac{(-\lambda_\Delta)(i\lambda^\star_3)}{2}B_{LL}(k_3)$};	

\node at (-2.25,1.5) {$\bf 1$};	
\node at (-2.25,-1.5) {$\bf 2$};	
\node at (2.25,0) {$\bf 3$};

\end{scope}

\begin{scope}[shift={(12,0)}]
\draw [phil, ultra thick] (1,0) -- (2,0);

\draw [phir, ultra thick] (-1,0) -- (-2,1.5);
\draw [phir, ultra thick] (-1,0) -- (-2,-1.5);
\draw [phir, ultra thick, domain=0:90] plot ({1*cos(\x)}, {1*sin(\x)});
\draw [phir, ultra thick, domain=90:180] plot ({1*cos(\x)}, {1*sin(\x)});
\draw [phil, ultra thick, domain=180:270] plot ({1*cos(\x)}, {1*sin(\x)});
\draw [phil, ultra thick, domain=270:360] plot ({1*cos(\x)}, {1*sin(\x)}); 
\node at (-1,0) {$\times $};	
\node at (1,0) {$\times $};	
\node at (0,-2) {\Large $(-i\sigma_4)(i\sigma_3^\ast)B_{LR}(k_3)$};	

\node at (-2.25,1.5) {$\bf 1$};	
\node at (-2.25,-1.5) {$\bf 2$};	
\node at (2.25,0) {$\bf 3$};

\end{scope}

\begin{scope}[shift={(0,-5)}]
\draw [phil, ultra thick] (1,0) -- (2,0);

\draw [phir, ultra thick] (-1,0) -- (-2,1.5);
\draw [phir, ultra thick] (-1,0) -- (-2,-1.5);
\draw [phil, ultra thick, domain=0:90] plot ({1*cos(\x)}, {1*sin(\x)});
\draw [phir, ultra thick, domain=90:180] plot ({1*cos(\x)}, {1*sin(\x)});
\draw [phir, ultra thick, domain=180:270] plot ({1*cos(\x)}, {1*sin(\x)});
\draw [phil, ultra thick, domain=270:360] plot ({1*cos(\x)}, {1*sin(\x)});  
\node at (-1,0) {$\times $};	
\node at (1,0) {$\times $};	
\node at (0,-2) {\Large $\frac{(-i\lambda_4)(i\lambda_3^\star)}{2}B_{PM }(k_3)$};	

\node at (-2.25,1.5) {$\bf 1$};	
\node at (-2.25,-1.5) {$\bf 2$};	
\node at (2.25,0) {$\bf 3$};

\end{scope}
 
\begin{scope}[shift={(6,-5)}]
\draw [phil, ultra thick] (1,0) -- (2,0);

\draw [phir, ultra thick] (-1,0) -- (-2,1.5);
\draw [phir, ultra thick] (-1,0) -- (-2,-1.5);
\draw [phir, ultra thick, domain=0:90] plot ({1*cos(\x)}, {1*sin(\x)});
\draw [phil, ultra thick, domain=90:180] plot ({1*cos(\x)}, {1*sin(\x)});
\draw [phil, ultra thick, domain=180:270] plot ({1*cos(\x)}, {1*sin(\x)});
\draw [phir, ultra thick, domain=270:360] plot ({1*cos(\x)}, {1*sin(\x)}); 
\node at (-1,0) {$\times $};	
\node at (1,0) {$\times $};	
\node at (0,-2) {\Large $\frac{(-\lambda_\Delta)(-i\sigma_3 )}{2}B_{MP}(k_3)$};	

\node at (-2.25,1.5) {$\bf 1$};	
\node at (-2.25,-1.5) {$\bf 2$};	
\node at (2.25,0) {$\bf 3$};

\end{scope}

\begin{scope}[shift={(12,-5)}]
\draw [phil, ultra thick] (1,0) -- (2,0);

\draw [phir, ultra thick] (-1,0) -- (-2,1.5);
\draw [phir, ultra thick] (-1,0) -- (-2,-1.5);
\draw [phir, ultra thick, domain=0:90] plot ({1*cos(\x)}, {1*sin(\x)});
\draw [phil, ultra thick, domain=90:180] plot ({1*cos(\x)}, {1*sin(\x)});
\draw [phir, ultra thick, domain=180:270] plot ({1*cos(\x)}, {1*sin(\x)});
\draw [phil, ultra thick, domain=270:360] plot ({1*cos(\x)}, {1*sin(\x)}); 
\node at (-1,0) {$\times $};	
\node at (1,0) {$\times $};	
\node at (0,-2) {\Large $(-i\sigma_4)(i\sigma_3^\ast  )B_{PP}(k_3)$};	

\node at (-2.25,1.5) {$\bf 1$};	
\node at (-2.25,-1.5) {$\bf 2$};	
\node at (2.25,0) {$\bf 3$};

\end{scope}

\begin{scope}[shift={(3,-10)}]
\draw [phil, ultra thick] (1,0) -- (2,0);

\draw [phir, ultra thick] (-1,0) -- (-2,1.5);
\draw [phir, ultra thick] (-1,0) -- (-2,-1.5);
\draw [phir, ultra thick, domain=0:90] plot ({1*cos(\x)}, {1*sin(\x)});
\draw [phir, ultra thick, domain=90:180] plot ({1*cos(\x)}, {1*sin(\x)});
\draw [phir, ultra thick, domain=180:270] plot ({1*cos(\x)}, {1*sin(\x)});
\draw [phil, ultra thick, domain=270:360] plot ({1*cos(\x)}, {1*sin(\x)}); 
\node at (-1,0) {$\times $};	
\node at (1,0) {$\times $};	
\node at (0,-2) {\Large $(-i\lambda_4)(i\sigma_3^\ast)B_{PR}(k_3)$};	

\node at (-2.25,1.5) {$\bf 1$};	
\node at (-2.25,-1.5) {$\bf 2$};	
\node at (2.25,0) {$\bf 3$};

\end{scope} 

\begin{scope}[shift={(10,-10)}]
\draw [phil, ultra thick] (1,0) -- (2,0);

\draw [phir, ultra thick] (-1,0) -- (-2,1.5);
\draw [phir, ultra thick] (-1,0) -- (-2,-1.5);
\draw [phir, ultra thick, domain=0:90] plot ({1*cos(\x)}, {1*sin(\x)});
\draw [phir, ultra thick, domain=90:180] plot ({1*cos(\x)}, {1*sin(\x)});
\draw [phil, ultra thick, domain=180:270] plot ({1*cos(\x)}, {1*sin(\x)});
\draw [phir, ultra thick, domain=270:360] plot ({1*cos(\x)}, {1*sin(\x)}); 
\node at (-1,0) {$\times $};	
\node at (1,0) {$\times $};	
\node at (0,-2) {\Large $(-i\sigma_4)(-i\sigma_3)B_{MR}(k_3)$};	

\node at (-2.25,1.5) {$\bf 1$};	
\node at (-2.25,-1.5) {$\bf 2$};	
\node at (2.25,0) {$\bf 3$};

\end{scope}

\begin{scope}[shift={(3,-15)}]
\draw [phil, ultra thick] (1,0) -- (2,0);

\draw [phir, ultra thick] (-1,0) -- (-2,1.5);
\draw [phir, ultra thick] (-1,0) -- (-2,-1.5);
\draw [phil, ultra thick, domain=0:90] plot ({1*cos(\x)}, {1*sin(\x)});
\draw [phil, ultra thick, domain=90:180] plot ({1*cos(\x)}, {1*sin(\x)});
\draw [phir, ultra thick, domain=180:270] plot ({1*cos(\x)}, {1*sin(\x)});
\draw [phil, ultra thick, domain=270:360] plot ({1*cos(\x)}, {1*sin(\x)}); 
\node at (-1,0) {$\times $};	
\node at (1,0) {$\times $};	
\node at (0,-2) {\Large $(-i\sigma_4)(i\lambda_3^\star)  B_{PL}(k_3)$};	

\node at (-2.25,1.5) {$\bf 1$};	
\node at (-2.25,-1.5) {$\bf 2$};	
\node at (2.25,0) {$\bf 3$};

\end{scope}

\begin{scope}[shift={(10,-15)}]
\draw [phil, ultra thick] (1,0) -- (2,0);

\draw [phir, ultra thick] (-1,0) -- (-2,1.5);
\draw [phir, ultra thick] (-1,0) -- (-2,-1.5);
\draw [phil, ultra thick, domain=0:90] plot ({1*cos(\x)}, {1*sin(\x)});
\draw [phil, ultra thick, domain=90:180] plot ({1*cos(\x)}, {1*sin(\x)});
\draw [phil, ultra thick, domain=180:270] plot ({1*cos(\x)}, {1*sin(\x)});
\draw [phir, ultra thick, domain=270:360] plot ({1*cos(\x)}, {1*sin(\x)});  
\node at (-1,0) {$\times $};	
\node at (1,0) {$\times $};	
\node at (0,-2) {\Large $(-\lambda_\Delta)(i\sigma_3^\ast)B_{ML}(k_3)$};	

\node at (-2.25,1.5) {$\bf 1$};	 
\node at (-2.25,-1.5) {$\bf 2$};	
\node at (2.25,0) {$\bf 3$};	
	
\end{scope}	
 
\end{tikzpicture}
\end{center}
\caption{Diagrammatic representation of the Ten  $1$-Loop Integrals $ \phir\phir  \rightarrow\phil  $}
\label{diag:sigmathreeoneloopI}
\end{figure}

As described in the previous subsection, we will only consider PV $B$ type diagrams for two of the channels are depicted in figure \ref{diag:sigmathreeoneloopI} and \ref{diag:sigmathreeoneloopII}. The remaining channel is obtained by  interchanging $1\longleftrightarrow 2$ in the diagrams in figure  \ref{diag:sigmathreeoneloopII}. The sum over all the contributions is given by  
\begin{eqnarray}
%i\mathcal{M}&=&
-i\sigma_3+i\mathcal{M}_1(k_3)+i\mathcal{M}_2(k_2)+i\mathcal{M}_2(k_1)
\label{sigmathree1}
\end{eqnarray} 
Here $i\sigma_3$ is the tree level contribution. The term  $i\mathcal{M}_1(k_3)$ denotes the sum over Feynman diagrams in figure  \ref{diag:sigmathreeoneloopI} whereas $i\mathcal{M}_2(k_2)$ denotes the sum over Feynman diagrams in figure  \ref{diag:sigmathreeoneloopII}. The contribution $i\mathcal{M}_2(k_1)$ is obtained by interchanging $1\leftrightarrow 2$ in figure  \ref{diag:sigmathreeoneloopII}. The expression for $i\mathcal{M}_1(k_3)$ is given by  
\begin{eqnarray}
i\mathcal{M}_1(k_3)&=& 
\frac{(-i\lambda_4)(-i\sigma_3)}{2} B_{RR}(k_3)+\frac{(-\lambda_\Delta)(i\lambda_3^\star)}{2}B_{LL}(k_3)	+(-i\sigma_4)(i\sigma_3^\ast) B_{LR}(k_3)
\nonumber\\&&
+\frac{(-i\lambda_4)(i\lambda_3^\ast) }{2}B_{PM}(k_3)+\frac{(-\lambda_\Delta) (-i\sigma_3)}{2}B_{MP}(k_3)+(-i\sigma_4)(i\sigma_3^\ast)B_{PP}(k_3)
\nonumber\\&&
+(-i\lambda_4)(i\sigma_3^\ast)B_{PR}(k_3)+(-i\sigma_4)(-i\sigma_3)B_{MR}(k_3)
\nonumber\\&&
+(-i\sigma_4)(i\lambda_3^\ast)B_{PL}(k_3)+(-\lambda_\Delta)(i\sigma_3^\ast)B_{ML}(k_3) 
\label{sigmathree2}
\end{eqnarray}
The divergent contributions from $i\mathcal{M}_1(k_3)$ is cancelled by the following counter-term   
\begin{eqnarray}
\delta \sigma_3\Big|_\msbar^{(1)}&=&-
\frac{1}{(4\pi)^2}
\Big[2i\lambda_4\,\im\,  \sigma_3 -i\lambda_\Delta(\sigma_3^\star+\lambda_3^\star) 
 +\sigma_4\lambda^\star_3
+\sigma_4\sigma_3\Big]
\nonumber\\
&&\qquad
\Bigl[ \frac{1}{d-4} +\frac{1}{2}(\gamma_E-\ln\, 4\pi) \Bigr]
\label{sigmathree3}
\end{eqnarray}
\begin{figure}
\begin{center}
\begin{tikzpicture}[line width=1 pt, scale=0.8]
%\begin{scope}[rotate=-80]

\begin{scope}[shift={(0,0)}]
\draw [phil, ultra thick] (0,1) -- (2,1.5);
\draw [phir, ultra thick] (0,-1) -- (-2,-1.5);
\draw [phir, ultra thick] (0,1) -- (-2,1.5);
\draw [phir, ultra thick, domain=-90:90] plot ({1*cos(\x)}, {1*sin(\x)});
\draw [phir, ultra thick, domain=90:270] plot ({1*cos(\x)}, {1*sin(\x)}); 
\node at (0,1) {$\times $};	
\node at (0,-1) {$\times $};	
\node at (0,-2.5) {\Large $\frac{(-i\sigma_4)(-i\lambda_3)}{2}B_{RR}(k_2)$};	

\node at (-2.25,1.5) {$\bf 1$};	 
\node at (-2.25,-1.5) {$\bf 2$};	
\node at (2.25,1.5) {$\bf 3$};
	
\end{scope}

\begin{scope}[shift={(6,0)}]
\draw [phil, ultra thick] (0,1) -- (2,1.5);
\draw [phir, ultra thick] (0,-1) -- (-2,-1.5);
\draw [phir, ultra thick] (0,1) -- (-2,1.5);
\draw [phil, ultra thick, domain=-90:90] plot ({1*cos(\x)}, {1*sin(\x)});
\draw [phil, ultra thick, domain=90:270] plot ({1*cos(\x)}, {1*sin(\x)}); 
\node at (0,1) {$\times $};	
\node at (0,-1) {$\times $};		
\node at (0,-2.5) {\Large $\frac{(i\sigma^\star_4)(i\sigma^\star_3)}{2}B_{LL}(k_2)$};	

\node at (-2.25,1.5) {$\bf 1$};	 
\node at (-2.25,-1.5) {$\bf 2$};	
\node at (2.25,1.5) {$\bf 3$};
	
\end{scope}

\begin{scope}[shift={(12,0)}]
\draw [phil, ultra thick] (0,1) -- (2,1.5);
\draw [phir, ultra thick] (0,-1) -- (-2,-1.5);
\draw [phir, ultra thick] (0,1) -- (-2,1.5);
\draw [phil, ultra thick, domain=-90:90] plot ({1*cos(\x)}, {1*sin(\x)});
\draw [phir , ultra thick, domain=90:270] plot ({1*cos(\x)}, {1*sin(\x)}); 
\node at (0,1) {$\times $};	
\node at (0,-1) {$\times $};	
\node at (0,-2.5) {\Large $(-\lambda_\Delta)(-i\sigma_3)B_{LR}(k_2)$};	
 
\node at (-2.25,1.5) {$\bf 1$};	 
\node at (-2.25,-1.5) {$\bf 2$};	
\node at (2.25,1.5) {$\bf 3$};
	
\end{scope}

\begin{scope}[shift={(0,-5)}]
\draw [phil, ultra thick] (0,1) -- (2,1.5);
\draw [phir, ultra thick] (0,-1) -- (-2,-1.5);
\draw [phir, ultra thick] (0,1) -- (-2,1.5);
\draw [phir, ultra thick, domain=0:90] plot ({1*cos(\x)}, {1*sin(\x)});
\draw [phir, ultra thick, domain=90:180] plot ({1*cos(\x)}, {1*sin(\x)});
\draw [phil, ultra thick, domain=180:270] plot ({1*cos(\x)}, {1*sin(\x)});
\draw [phil, ultra thick, domain=270:360] plot ({1*cos(\x)}, {1*sin(\x)});  
\node at (0,1) {$\times $};	
\node at (0,-1) {$\times $};
\node at (0,-2.5) {\Large $\frac{(-i\sigma_4)(i\sigma^\star_3)}{2}B_{PM}(k_2)$};	 

\node at (-2.25,1.5) {$\bf 1$};	 
\node at (-2.25,-1.5) {$\bf 2$};	
\node at (2.25,1.5) {$\bf 3$};
	
\end{scope}

\begin{scope}[shift={(6,-5)}]
\draw [phil, ultra thick] (0,1) -- (2,1.5);
\draw [phir, ultra thick] (0,-1) -- (-2,-1.5);
\draw [phir, ultra thick] (0,1) -- (-2,1.5);
\draw [phil, ultra thick, domain=0:90] plot ({1*cos(\x)}, {1*sin(\x)});
\draw [phil, ultra thick, domain=90:180] plot ({1*cos(\x)}, {1*sin(\x)});
\draw [phir, ultra thick, domain=180:270] plot ({1*cos(\x)}, {1*sin(\x)});
\draw [phir, ultra thick, domain=270:360] plot ({1*cos(\x)}, {1*sin(\x)}); 
\node at (0,1) {$\times $};	
\node at (0,-1) {$\times $};
\node at (0,-2.5) {\Large $\frac{(i\sigma^\star_4)(-i\lambda_3)}{2}B_{MP}(k_2)$};	

\node at (-2.25,1.5) {$\bf 1$};	 
\node at (-2.25,-1.5) {$\bf 2$};	
\node at (2.25,1.5) {$\bf 3$};
	
\end{scope}

\begin{scope}[shift={(12,-5)}]
\draw [phil, ultra thick] (0,1) -- (2,1.5);
\draw [phir, ultra thick] (0,-1) -- (-2,-1.5);
\draw [phir, ultra thick] (0,1) -- (-2,1.5);
\draw [phir, ultra thick, domain=0:90] plot ({1*cos(\x)}, {1*sin(\x)});
\draw [phil, ultra thick, domain=90:180] plot ({1*cos(\x)}, {1*sin(\x)});
\draw [phir, ultra thick, domain=180:270] plot ({1*cos(\x)}, {1*sin(\x)});
\draw [phil, ultra thick, domain=270:360] plot ({1*cos(\x)}, {1*sin(\x)}); 
\node at (0,1) {$\times $};	
\node at (0,-1) {$\times $};
\node at (0,-2.5) {\Large $(-\lambda_\Delta)(-i\sigma_3)B_{PP}(k_2)$};	

\node at (-2.25,1.5) {$\bf 1$};	 
\node at (-2.25,-1.5) {$\bf 2$};	
\node at (2.25,1.5) {$\bf 3$};
	
\end{scope}

\begin{scope}[shift={(3,-10)}]
\draw [phil, ultra thick] (0,1) -- (2,1.5);
\draw [phir, ultra thick] (0,-1) -- (-2,-1.5);
\draw [phir, ultra thick] (0,1) -- (-2,1.5);
\draw [phir, ultra thick, domain=0:90] plot ({1*cos(\x)}, {1*sin(\x)});
\draw [phir, ultra thick, domain=90:180] plot ({1*cos(\x)}, {1*sin(\x)});
\draw [phir, ultra thick, domain=180:270] plot ({1*cos(\x)}, {1*sin(\x)});
\draw [phil, ultra thick, domain=270:360] plot ({1*cos(\x)}, {1*sin(\x)}); 
\node at (0,1) {$\times $};	
\node at (0,-1) {$\times $};
\node at (0,-2.5) {\Large $(-i\sigma_4)(-i\sigma_3)B_{PR}(k_2)$};	

\node at (-2.25,1.5) {$\bf 1$};	 
\node at (-2.25,-1.5) {$\bf 2$};	
\node at (2.25,1.5) {$\bf 3$};
	
\end{scope}

\begin{scope}[shift={(10,-10)}]
\draw [phil, ultra thick] (0,1) -- (2,1.5);
\draw [phir, ultra thick] (0,-1) -- (-2,-1.5);
\draw [phir, ultra thick] (0,1) -- (-2,1.5);
\draw [phil, ultra thick, domain=0:90] plot ({1*cos(\x)}, {1*sin(\x)});
\draw [phir, ultra thick, domain=90:180] plot ({1*cos(\x)}, {1*sin(\x)});
\draw [phir, ultra thick, domain=180:270] plot ({1*cos(\x)}, {1*sin(\x)});
\draw [phir, ultra thick, domain=270:360] plot ({1*cos(\x)}, {1*sin(\x)}); 
\node at (0,1) {$\times $};	
\node at (0,-1) {$\times $};  
\node at (0,-2.5) {\Large $(-\lambda_\Delta)(-i\lambda_3)B_{MR}(k_2)$};	

\node at (-2.25,1.5) {$\bf 1$};	 
\node at (-2.25,-1.5) {$\bf 2$};	
\node at (2.25,1.5) {$\bf 3$};	
	
\end{scope} 

\begin{scope}[shift={(3,-15)}]
\draw [phil, ultra thick] (0,1) -- (2,1.5);
\draw [phir, ultra thick] (0,-1) -- (-2,-1.5);
\draw [phir, ultra thick] (0,1) -- (-2,1.5);
\draw [phil, ultra thick, domain=0:90] plot ({1*cos(\x)}, {1*sin(\x)});
\draw [phil, ultra thick, domain=90:180] plot ({1*cos(\x)}, {1*sin(\x)});
\draw [phir, ultra thick, domain=180:270] plot ({1*cos(\x)}, {1*sin(\x)});
\draw [phil, ultra thick, domain=270:360] plot ({1*cos(\x)}, {1*sin(\x)}); 
\node at (0,1) {$\times $};	
\node at (0,-1) {$\times $};
\node at (0,-2.5) {\Large $(i\sigma^\star_4)(-i\sigma_3)B_{ML}(k_2)$};	
	
\node at (-2.25,1.5) {$\bf 1$};	 
\node at (-2.25,-1.5) {$\bf 2$};	
\node at (2.25,1.5) {$\bf 3$};	
	
\end{scope}

\begin{scope}[shift={(10,-15)}]
\draw [phil, ultra thick] (0,1) -- (2,1.5);
\draw [phir, ultra thick] (0,-1) -- (-2,-1.5);
\draw [phir, ultra thick] (0,1) -- (-2,1.5);
  
\draw [phil, ultra thick, domain=0:90] plot ({1*cos(\x)}, {1*sin(\x)});
\draw [phir, ultra thick, domain=90:180] plot ({1*cos(\x)}, {1*sin(\x)});
\draw [phil, ultra thick, domain=180:270] plot ({1*cos(\x)}, {1*sin(\x)});
\draw [phil, ultra thick, domain=270:360] plot ({1*cos(\x)}, {1*sin(\x)});   
\node at (0,1) {$\times $};	
\node at (0,-1) {$\times $};
\node at (0,-2.5) {\Large $(-\lambda_\Delta) (i\sigma_3^\ast)B_{PL}(k_2)$};	

\node at (-2.25,1.5) {$\bf 1$};	  
\node at (-2.25,-1.5) {$\bf 2$};	
\node at (2.25,1.5) {$\bf 3$};
	
\end{scope}	  

\end{tikzpicture}
\end{center}
\caption{Diagrammatic representation of the Ten  $1$-Loop Integrals $ \phir\phir \rightarrow \phil  $}
\label{diag:sigmathreeoneloopII}
\end{figure} 

The expression for $i\mathcal{M}_2(k_2)$ is given by 
\begin{equation}
\begin{split}
i\mathcal{M}_2(k_2)&=
\frac{(-i\sigma_4)(-i\lambda_3)}{2} B_{RR}(k_2)+\frac{(i\sigma^\star_4)(i\sigma_3^\star)}{2}B_{LL}(k_2)	+(-\lambda_\Delta)(-i\sigma_3) B_{LR}(k_2)
\\
&+\frac{(-i\sigma_4)(i\sigma_3^\ast) }{2}B_{PM}(k_2)+\frac{(i\sigma^\star_4) (-i\lambda_3)}{2}B_{MP}(k_2)+(-\lambda_\Delta)(-i\sigma_3)B_{PP}(k_2)
\\
&+(-i\sigma_4)(-i\sigma_3)B_{PR}(k_2)+(-\lambda_\Delta)(-i\lambda_3)B_{MR}(k_2)+(i\sigma^\ast_4)(-i\sigma_3)B_{PL}(k_2)\\
&+(-\lambda_\Delta)(i\sigma_3^\ast)B_{ML}(k_2) 
\label{sigmathree4}
\end{split}
\end{equation}
The divergent contribution from $i\mathcal{M}_2(k_2)$ (and from  $i\mathcal{M}_2(k_1)$)  are cancelled by the following counter-term  
\begin{eqnarray}
\delta \sigma_3\Big|_\msbar^{(2)}&=&-
\frac{2}{(4\pi)^2}
 \Big[\sigma_4\lambda_3
+2i\im[\sigma_4\sigma_3]-i\lambda_\Delta(\lambda_3+\sigma_3^\ast)+ \sigma^\ast_4\sigma_3 \Big] 
\nonumber\\
&&
\qquad
\Big[ \frac{1}{d-4} +\frac{1}{2}(\gamma_E-\ln\, 4\pi) \Big]
\label{sigmathree5}
\end{eqnarray}  
Hence the total one loop beta function for ${\sigma_3}$ is given by
\begin{equation}
\begin{split}
\beta_{\sigma_3}\equiv	\frac{d \sigma_3}{d\ln\, \mu}
&=\frac{1}{(4\pi)^2}
\Bigg[2i\lambda_4\im\,  \sigma_3 -i\lambda_\Delta(\sigma_3^\star+\lambda_3^\star)
 +\sigma_4\lambda^\star_3+\sigma_4\sigma_3 \\  
&\qquad\qquad\,\,\,  
+2\sigma_4\lambda_3
+4i\im[\sigma_4\sigma_3]-2i\lambda_\Delta(\lambda_3+\sigma_3^\ast)+ 2\sigma^\ast_4\sigma_3 \Bigg]  
%\nonumber\\
\label{sigmathree6}
\end{split}
\end{equation}

\subsection{Checking Lindblad condition at the level of cubic couplings}
\label{subsec:oneloopLindbladcubic}

From equation \eqref{lambdathree3}, we obtain  the beta function for $\im\,\lambda_3$ as
\begin{eqnarray}
	\frac{d\,(\im\,\lambda_3)}{d\ln\, \mu}&=&\frac{3}{(4\pi)^2}\Bigg[
	\re\, \lambda_4\, (\im\, \lambda_3+\im\, \sigma_3)
+	\re\, \sigma_4\,(\im\, \lambda_3-\im\, \sigma_3)
\nonumber\\
&&\qquad\quad
+(\re\, \lambda_3+\re\, \sigma_3)(\im\, \lambda_4+\im\, \sigma_4 )
\Bigg]		
\end{eqnarray}
and the beta function of $\im\,\sigma_3$ can be computed from the imaginary part of equation \eqref{sigmathree6}. We obtain
\begin{eqnarray}	
\frac{d\,(\im\,\sigma_3)}{d\, \ln\, \mu}&=&\frac{1}{(4\pi)^2}	
\Bigg[
2\re\, \lambda_4\,\im\,  \sigma_3+\re\, \sigma_4 (7\, \im\, \sigma_3 +\im\, \lambda_3) +
\nonumber\\
&&\qquad\qquad
3(\re\, \lambda_3+\re\, \sigma_3)(\im\, \sigma_4-\lambda_\Delta)
%\nonumber\\
%&&
%\qquad\quad+
%2\re\, \sigma_4 (\im\, \lambda_3+3\im\, \sigma_3)+2(\re\, \lambda_3+\re\, \sigma_3)(\im\, \sigma_4-\lambda_\Delta)   
\Bigg] 
\end{eqnarray} 
Adding these two equations we get 
\begin{equation}
\begin{split}
\frac{d\,}{d\, \ln\, \mu}\left[\im\,  \lambda_3+3\im\,  \sigma_3\right]
&=\frac{3}{(4\pi)^2}\Bigg[(\re\, \lambda_4+2\re\, \sigma_4) (\im\, \lambda_3+3\im\, \sigma_3) \\
&\qquad\quad+(\re\, \lambda_3+\re\, \sigma_3)(\im\, \lambda_4 +4\im\, \sigma_4 -3\lambda_\Delta)\Bigg]
\end{split}\label{Lindbladbeta21}
\end{equation}  
Again, one can see that the one loop beta function for the Lindblad violating cubic coupling is zero when there is no Lindblad violating coupling in the  tree level Lagrangian.

%\chapter{One loop Four point function}

%\section{One loop beta function for the quartic couplings}
%\label{sec:oneloopquarticcoupling}

\subsection{One loop beta function for \texorpdfstring{$\lambda_4$}{lambda4}}
\label{subsec:onelooplambdafour}

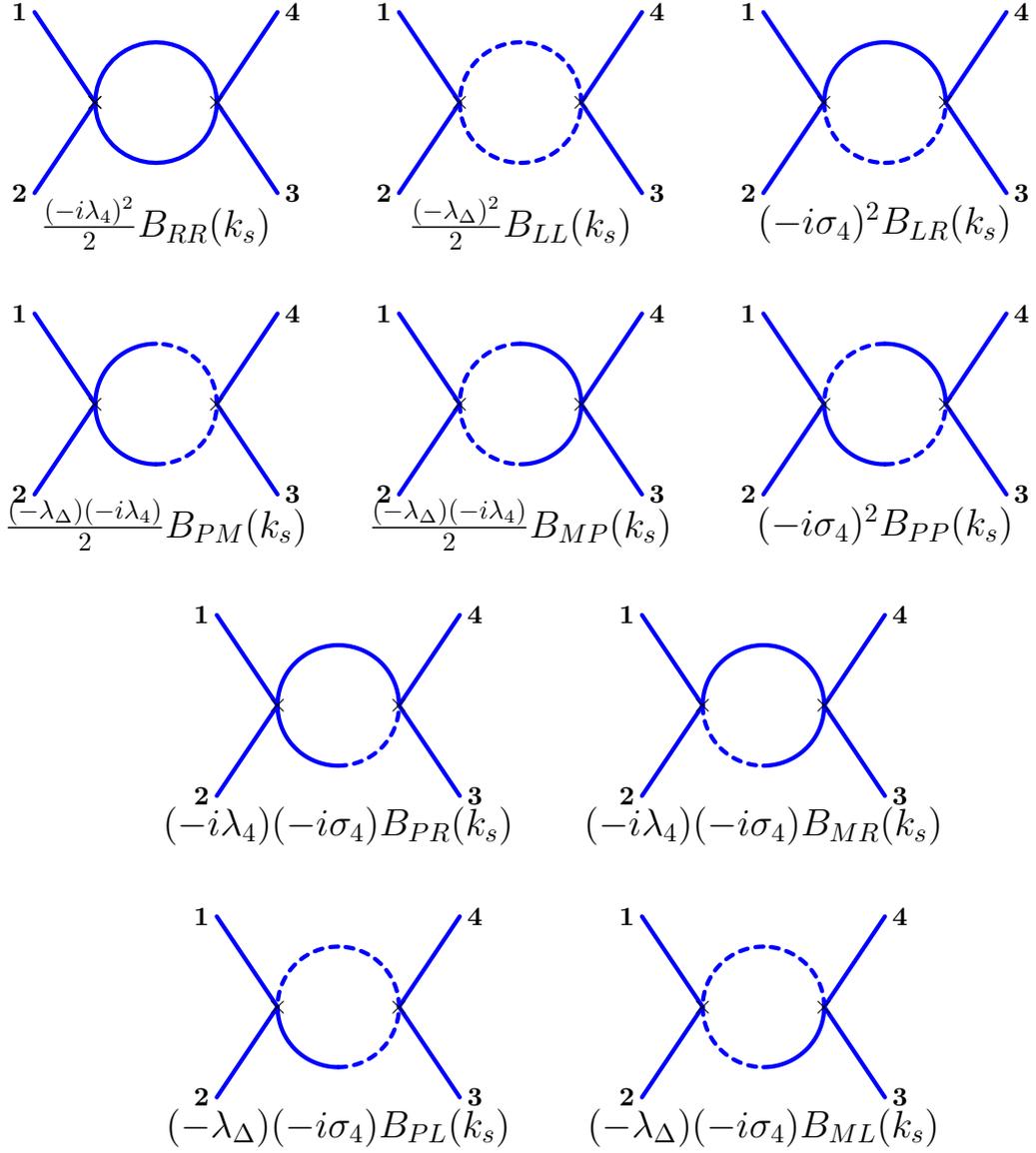
\begin{figure}
\begin{center}
\begin{tikzpicture}[line width=1 pt, scale=0.8]
%\begin{scope}[rotate=-80]

\begin{scope}[shift={(0,0)}]
\draw [phir, ultra thick] (1,0) -- (2,1.5);
\draw [phir, ultra thick] (1,0) -- (2,-1.5);
\draw [phir, ultra thick] (-1,0) -- (-2,1.5);
\draw [phir, ultra thick] (-1,0) -- (-2,-1.5);
\draw [phir, ultra thick, domain=0:180] plot ({1*cos(\x)}, {1*sin(\x)});
\draw [phir, ultra thick, domain=180:360] plot ({1*cos(\x)}, {1*sin(\x)}); 
\node at (-1,0) {$\times $};	
\node at (1,0) {$\times $};	
\node at (0,-2) {\Large $\frac{(-i\lambda_4)^2}{2} B_{RR}(k_s)$};	
\node at (-1,0) {$\times $};	

\node at (-2.25,1.5) {$\bf 1$};	
\node at (-2.25,-1.5) {$\bf 2$};	
\node at (2.25,-1.5) {$\bf 3$};	
\node at (2.25,1.5) {$\bf 4$};

\end{scope}

\begin{scope}[shift={(6,0)}]
\draw [phir, ultra thick] (1,0) -- (2,1.5);
\draw [phir, ultra thick] (1,0) -- (2,-1.5);
\draw [phir, ultra thick] (-1,0) -- (-2,1.5);
\draw [phir, ultra thick] (-1,0) -- (-2,-1.5);
\draw [phil, ultra thick, domain=0:90] plot ({1*cos(\x)}, {1*sin(\x)});
\draw [phil, ultra thick, domain=90:180] plot ({1*cos(\x)}, {1*sin(\x)});
\draw [phil, ultra thick, domain=180:270] plot ({1*cos(\x)}, {1*sin(\x)});
\draw [phil, ultra thick, domain=270:360] plot ({1*cos(\x)}, {1*sin(\x)}); 
\node at (-1,0) {$\times $};	
\node at (1,0) {$\times $};	
\node at (0,-2) {\Large $\frac{(-\lambda_\Delta)^2}{2} B_{LL}(k_s)$};

\node at (-2.25,1.5) {$\bf 1$};	
\node at (-2.25,-1.5) {$\bf 2$};	
\node at (2.25,-1.5) {$\bf 3$};	
\node at (2.25,1.5) {$\bf 4$};		
	
\end{scope}

\begin{scope}[shift={(12,0)}]
\draw [phir, ultra thick] (1,0) -- (2,1.5);
\draw [phir, ultra thick] (1,0) -- (2,-1.5);
\draw [phir, ultra thick] (-1,0) -- (-2,1.5);
\draw [phir, ultra thick] (-1,0) -- (-2,-1.5);
\draw [phir, ultra thick, domain=0:90] plot ({1*cos(\x)}, {1*sin(\x)});
\draw [phir, ultra thick, domain=90:180] plot ({1*cos(\x)}, {1*sin(\x)});
\draw [phil, ultra thick, domain=180:270] plot ({1*cos(\x)}, {1*sin(\x)});
\draw [phil, ultra thick, domain=270:360] plot ({1*cos(\x)}, {1*sin(\x)}); 
\node at (-1,0) {$\times $};	
\node at (1,0) {$\times $};	
\node at (0,-2) {\Large $(-i\sigma_4)^2 B_{LR}(k_s)$};	

\node at (-2.25,1.5) {$\bf 1$};	
\node at (-2.25,-1.5) {$\bf 2$};	
\node at (2.25,-1.5) {$\bf 3$};	
\node at (2.25,1.5) {$\bf 4$};	
	
\end{scope}

\begin{scope}[shift={(0,-5)}]
\draw [phir, ultra thick] (1,0) -- (2,1.5);
\draw [phir, ultra thick] (1,0) -- (2,-1.5);
\draw [phir, ultra thick] (-1,0) -- (-2,1.5);
\draw [phir, ultra thick] (-1,0) -- (-2,-1.5);
\draw [phil, ultra thick, domain=0:90] plot ({1*cos(\x)}, {1*sin(\x)});
\draw [phir, ultra thick, domain=90:180] plot ({1*cos(\x)}, {1*sin(\x)});
\draw [phir, ultra thick, domain=180:270] plot ({1*cos(\x)}, {1*sin(\x)});
\draw [phil, ultra thick, domain=270:360] plot ({1*cos(\x)}, {1*sin(\x)});  
\node at (-1,0) {$\times $};	
\node at (1,0) {$\times $};	
\node at (0,-2) {\Large $\frac{(-\lambda_\Delta)(-i\lambda_4)}{2} B_{PM}(k_s)$};	

\node at (-2.25,1.5) {$\bf 1$};	
\node at (-2.25,-1.5) {$\bf 2$};	
\node at (2.25,-1.5) {$\bf 3$};	
\node at (2.25,1.5) {$\bf 4$};	
	
\end{scope}

\begin{scope}[shift={(6,-5)}]
\draw [phir, ultra thick] (1,0) -- (2,1.5);
\draw [phir, ultra thick] (1,0) -- (2,-1.5);
\draw [phir, ultra thick] (-1,0) -- (-2,1.5);
\draw [phir, ultra thick] (-1,0) -- (-2,-1.5);
\draw [phir, ultra thick, domain=0:90] plot ({1*cos(\x)}, {1*sin(\x)});
\draw [phil, ultra thick, domain=90:180] plot ({1*cos(\x)}, {1*sin(\x)});
\draw [phil, ultra thick, domain=180:270] plot ({1*cos(\x)}, {1*sin(\x)});
\draw [phir, ultra thick, domain=270:360] plot ({1*cos(\x)}, {1*sin(\x)}); 
\node at (-1,0) {$\times $};	
\node at (1,0) {$\times $};	
\node at (0,-2) {\Large $\frac{(-\lambda_\Delta)(-i\lambda_4)}{2} B_{MP}(k_s)$};	

\node at (-2.25,1.5) {$\bf 1$};	
\node at (-2.25,-1.5) {$\bf 2$};	
\node at (2.25,-1.5) {$\bf 3$};	
\node at (2.25,1.5) {$\bf 4$};	
	
\end{scope}

\begin{scope}[shift={(12,-5)}]
\draw [phir, ultra thick] (1,0) -- (2,1.5);
\draw [phir, ultra thick] (1,0) -- (2,-1.5);
\draw [phir, ultra thick] (-1,0) -- (-2,1.5);
\draw [phir, ultra thick] (-1,0) -- (-2,-1.5);
\draw [phir, ultra thick, domain=0:90] plot ({1*cos(\x)}, {1*sin(\x)});
\draw [phil, ultra thick, domain=90:180] plot ({1*cos(\x)}, {1*sin(\x)});
\draw [phir, ultra thick, domain=180:270] plot ({1*cos(\x)}, {1*sin(\x)});
\draw [phil, ultra thick, domain=270:360] plot ({1*cos(\x)}, {1*sin(\x)}); 
\node at (-1,0) {$\times $};	
\node at (1,0) {$\times $};	
\node at (0,-2) {\Large $(-i\sigma_4)^2 B_{PP}(k_s)$};	
	
\node at (-2.25,1.5) {$\bf 1$};	
\node at (-2.25,-1.5) {$\bf 2$};	
\node at (2.25,-1.5) {$\bf 3$};	
\node at (2.25,1.5) {$\bf 4$};	

\end{scope}

\begin{scope}[shift={(3,-10)}]
\draw [phir, ultra thick] (1,0) -- (2,1.5);
\draw [phir, ultra thick] (1,0) -- (2,-1.5);
\draw [phir, ultra thick] (-1,0) -- (-2,1.5);
\draw [phir, ultra thick] (-1,0) -- (-2,-1.5);
\draw [phir, ultra thick, domain=0:90] plot ({1*cos(\x)}, {1*sin(\x)});
\draw [phir, ultra thick, domain=90:180] plot ({1*cos(\x)}, {1*sin(\x)});
\draw [phir, ultra thick, domain=180:270] plot ({1*cos(\x)}, {1*sin(\x)});
\draw [phil, ultra thick, domain=270:360] plot ({1*cos(\x)}, {1*sin(\x)}); 
\node at (-1,0) {$\times $};	 
\node at (1,0) {$\times $};	
\node at (0,-2) {\Large $(-i\lambda_4)(-i\sigma_4) B_{PR}(k_s)$};	

\node at (-2.25,1.5) {$\bf 1$};	
\node at (-2.25,-1.5) {$\bf 2$};	
\node at (2.25,-1.5) {$\bf 3$};	
\node at (2.25,1.5) {$\bf 4$};	
	
\end{scope}

\begin{scope}[shift={(10,-10)}]
\draw [phir, ultra thick] (1,0) -- (2,1.5);
\draw [phir, ultra thick] (1,0) -- (2,-1.5);
\draw [phir, ultra thick] (-1,0) -- (-2,1.5);
\draw [phir, ultra thick] (-1,0) -- (-2,-1.5);
\draw [phir, ultra thick, domain=0:90] plot ({1*cos(\x)}, {1*sin(\x)});
\draw [phir, ultra thick, domain=90:180] plot ({1*cos(\x)}, {1*sin(\x)});
\draw [phil, ultra thick, domain=180:270] plot ({1*cos(\x)}, {1*sin(\x)});
\draw [phir, ultra thick, domain=270:360] plot ({1*cos(\x)}, {1*sin(\x)}); 
\node at (-1,0) {$\times $};	
\node at (1,0) {$\times $};	
\node at (0,-2) {\Large $(-i\lambda_4)(-i\sigma_4) B_{MR}(k_s)$};

\node at (-2.25,1.5) {$\bf 1$};	
\node at (-2.25,-1.5) {$\bf 2$};	
\node at (2.25,-1.5) {$\bf 3$};	
\node at (2.25,1.5) {$\bf 4$};	
	
\end{scope}

\begin{scope}[shift={(3,-15)}]
\draw [phir, ultra thick] (1,0) -- (2,1.5);
\draw [phir, ultra thick] (1,0) -- (2,-1.5);
\draw [phir, ultra thick] (-1,0) -- (-2,1.5);
\draw [phir, ultra thick] (-1,0) -- (-2,-1.5);
\draw [phil, ultra thick, domain=0:90] plot ({1*cos(\x)}, {1*sin(\x)});
\draw [phil, ultra thick, domain=90:180] plot ({1*cos(\x)}, {1*sin(\x)});
\draw [phir, ultra thick, domain=180:270] plot ({1*cos(\x)}, {1*sin(\x)});
\draw [phil, ultra thick, domain=270:360] plot ({1*cos(\x)}, {1*sin(\x)}); 
\node at (-1,0) {$\times $};	
\node at (1,0) {$\times $};	
\node at (0,-2) {\Large $(-\lambda_\Delta)(-i\sigma_4) B_{PL} (k_s)$};	

\node at (-2.25,1.5) {$\bf 1$};	
\node at (-2.25,-1.5) {$\bf 2$};	
\node at (2.25,-1.5) {$\bf 3$};	
\node at (2.25,1.5) {$\bf 4$};	
	
\end{scope}

\begin{scope}[shift={(10,-15)}]
\draw [phir, ultra thick] (1,0) -- (2,1.5);
\draw [phir, ultra thick] (1,0) -- (2,-1.5);
\draw [phir, ultra thick] (-1,0) -- (-2,1.5);
\draw [phir, ultra thick] (-1,0) -- (-2,-1.5);
\draw [phil, ultra thick, domain=0:90] plot ({1*cos(\x)}, {1*sin(\x)});
\draw [phil, ultra thick, domain=90:180] plot ({1*cos(\x)}, {1*sin(\x)});
\draw [phil, ultra thick, domain=180:270] plot ({1*cos(\x)}, {1*sin(\x)});
\draw [phir, ultra thick, domain=270:360] plot ({1*cos(\x)}, {1*sin(\x)});  
\node at (-1,0) {$\times $};	
\node at (1,0) {$\times $};	
\node at (0,-2) {\Large $(-\lambda_\Delta)(-i\sigma_4) B_{ML}(k_s)$};	

\node at (-2.25,1.5) {$\bf 1$};	
\node at (-2.25,-1.5) {$\bf 2$};	
\node at (2.25,-1.5) {$\bf 3$};	
\node at (2.25,1.5) {$\bf 4$};	
	
\end{scope}	  

\end{tikzpicture}
\end{center}
\caption{Diagrammatic representation of the Ten  $1$-Loop Integrals $ \phir\phir \rightarrow \phir\phir $ (Here $k_s=k_1+k_2$)}
\label{diag:lambdaoneloop}
\end{figure}

\begin{figure}[ht]
\begin{center}
\begin{tikzpicture}[line width=1 pt, scale=1]
%\begin{scope}[rotate=-80]

\begin{scope}[shift={(0,0)}]
\draw [phir, ultra thick] (0,0) -- (1,1.5);
\draw [phir, ultra thick] (0,0) -- (1,-1.5);
\draw [phir, ultra thick] (0,0) -- (-1,1.5);
\draw [phir, ultra thick] (0,0) -- (-1,-1.5);
\node at (0,0) {$\times $};	
\node at (0,-2) {\Large $-i\delta\lambda_4$};	

\node at (-1.25,1.5) {$\bf 1$};	
\node at (-1.25,-1.5) {$\bf 2$};	 
\node at (1.25,-1.5) {$\bf 3$};	
\node at (1.25,1.5) {$\bf 4$};	
\filldraw (0,0) circle (.1);

\end{scope}

\end{tikzpicture}
\end{center}
\caption{Diagrammatic representation of the one loop counter-term for  $ \lambda_4 $ }
\label{diag:lambdaonecounterterm} 
\end{figure}
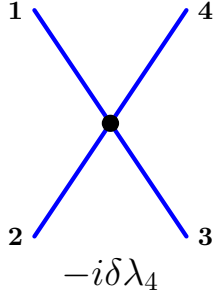 

Now we proceed to compute the one loop beta function for the quartic couplings. We will only consider the bubble diagrams since the triangle and box diagrams are finite. Let us consider all the one loop Feynman diagrams that contribute to the process  $\phir + \phir \rightarrow \phir + \phir$. All the diagrams are depicted in figure \ref{diag:lambdaoneloop}. The sum over all the Feynman diagrams is given by 
\begin{equation}
\label{lambdafour1} 
\begin{split}
&-i\lambda_4  \\
& +\frac{(-i\lambda_4)^2}{2} B_{RR}(k_s)+\frac{(-\lambda_\Delta)^2}{2}B_{LL}(k_s)	+(-i\sigma_4)^2 B_{LR}(k_s)\\
&+\frac{(-\lambda_\Delta) (-i\lambda_4)}{2}B_{PM}(k_s)+\frac{(-\lambda_\Delta) (-i\lambda_4)}{2}B_{MP}(k_s)+(-i\sigma_4)^2B_{PP}(k_s)\\
&+(-i\lambda_4)(-i\sigma_4)B_{PR}(k_s)+(-i\lambda_4)(-i\sigma_4)B_{MR}(k_s)\\
&+(-\lambda_\Delta)(-i\sigma_4)B_{PL}(k_s)+(-\lambda_\Delta)(-i\sigma_4)B_{ML}(k_s) \\
&+(k_s\longleftrightarrow k_t) +(k_s\longleftrightarrow k_u) 
\end{split}
\end{equation}
% Note that the one-loop corrections to $\lambda_4$ does not involve $\lambda_4^\star$ and $\sigma_4^\star$. 
Using the results in \eqref{skpvsummary1a}-\eqref{skpvsummary1d}, it's easy to see that the contribution is divergent and we need to add a one loop counter-term $\delta \lambda_4$
to cancel the divergences; 
\begin{equation}\label{eq:lambdafour2}
\delta \lambda_4\Big|_\msbar	= -\frac{3}{(4\pi)^2} \Bigl[ \lambda_4^2+2\  \sigma_4(\lambda_4+i \lambda_\Delta)+\lambda_\Delta^2\ \Bigr] \Bigl[ \frac{1}{d-4} +\frac{1}{2}(\gamma_E-\ln\, 4\pi) \Bigr]
\end{equation}
Using the standard methods of quantum field theory, we can compute the one loop beta function  to be
\begin{equation}
\label{eq:lambdafour3} 
\beta_{\lambda_4}
=\frac{3}{(4\pi)^2} \Bigl[ \lambda_4^2+2\  \sigma_4(\lambda_4+i \lambda_\Delta)+\lambda_\Delta^2\ \Bigr]= \frac{3}{(4\pi)^2}  (\lambda_4+2\ \sigma_4-i\lambda_\Delta)(\lambda_4+i \lambda_\Delta)
\end{equation}
By setting $\sigma_4=\lambda_\Delta=0$ we recover the standard result of unitary $\phi^4$ theory.

\newpage

\subsection{One loop beta function for \texorpdfstring{$\sigma_4$}{sigma4}}
\label{subsec:oneloopsigmafour}

\begin{figure}
\begin{center}
\begin{tikzpicture}[line width=1 pt, scale=0.8]

\begin{scope}[shift={(0,0)}]
\draw [phil, ultra thick] (1,0) -- (2,1.5);
\draw [phir, ultra thick] (1,0) -- (2,-1.5);
\draw [phir, ultra thick] (-1,0) -- (-2,1.5);
\draw [phir, ultra thick] (-1,0) -- (-2,-1.5);
\draw [phir, ultra thick, domain=0:180] plot ({1*cos(\x)}, {1*sin(\x)});
\draw [phir, ultra thick, domain=180:360] plot ({1*cos(\x)}, {1*sin(\x)}); 
\node at (-1,0) {$\times $};	
\node at (1,0) {$\times $};	
\node at (0,-2) {\Large $\frac{(-i\lambda_4)(-i\sigma_4)}{2}B_{RR}(k_s)$};	

\node at (-2.25,1.5) {$\bf 1$};	
\node at (-2.25,-1.5) {$\bf 2$};	
\node at (2.25,-1.5) {$\bf 3$};	
\node at (2.25,1.5) {$\bf 4$};	
	
\end{scope}
 
\begin{scope}[shift={(6,0)}]
\draw [phil, ultra thick] (1,0) -- (2,1.5);
\draw [phir, ultra thick] (1,0) -- (2,-1.5);
\draw [phir, ultra thick] (-1,0) -- (-2,1.5);
\draw [phir, ultra thick] (-1,0) -- (-2,-1.5);
\draw [phil, ultra thick, domain=0:90] plot ({1*cos(\x)}, {1*sin(\x)});
\draw [phil, ultra thick, domain=90:180] plot ({1*cos(\x)}, {1*sin(\x)});
\draw [phil, ultra thick, domain=180:270] plot ({1*cos(\x)}, {1*sin(\x)});
\draw [phil, ultra thick, domain=270:360] plot ({1*cos(\x)}, {1*sin(\x)}); 
\node at (-1,0) {$\times $};	
\node at (1,0) {$\times $};	
\node at (0,-2) {\Large $\frac{(i\sigma_4^\star)(-\lambda_\Delta)}{2}B_{LL}(k_s)$};	

\node at (-2.25,1.5) {$\bf 1$};	
\node at (-2.25,-1.5) {$\bf 2$};	
\node at (2.25,-1.5) {$\bf 3$};	
\node at (2.25,1.5) {$\bf 4$};	
	
\end{scope}

\begin{scope}[shift={(12,0)}]
\draw [phil, ultra thick] (1,0) -- (2,1.5);
\draw [phir, ultra thick] (1,0) -- (2,-1.5);
\draw [phir, ultra thick] (-1,0) -- (-2,1.5);
\draw [phir, ultra thick] (-1,0) -- (-2,-1.5);
\draw [phir, ultra thick, domain=0:90] plot ({1*cos(\x)}, {1*sin(\x)});
\draw [phir, ultra thick, domain=90:180] plot ({1*cos(\x)}, {1*sin(\x)});
\draw [phil, ultra thick, domain=180:270] plot ({1*cos(\x)}, {1*sin(\x)});
\draw [phil, ultra thick, domain=270:360] plot ({1*cos(\x)}, {1*sin(\x)}); 
\node at (-1,0) {$\times $};	
\node at (1,0) {$\times $};	
\node at (0,-2) {\Large $(-i\sigma_4)(-\lambda_\Delta)B_{LR}(k_s)$};	

\node at (-2.25,1.5) {$\bf 1$};	
\node at (-2.25,-1.5) {$\bf 2$};	
\node at (2.25,-1.5) {$\bf 3$};	
\node at (2.25,1.5) {$\bf 4$};	
	
\end{scope}

\begin{scope}[shift={(0,-5)}]
\draw [phil, ultra thick] (1,0) -- (2,1.5);
\draw [phir, ultra thick] (1,0) -- (2,-1.5);
\draw [phir, ultra thick] (-1,0) -- (-2,1.5);
\draw [phir, ultra thick] (-1,0) -- (-2,-1.5);
\draw [phil, ultra thick, domain=0:90] plot ({1*cos(\x)}, {1*sin(\x)});
\draw [phir, ultra thick, domain=90:180] plot ({1*cos(\x)}, {1*sin(\x)});
\draw [phir, ultra thick, domain=180:270] plot ({1*cos(\x)}, {1*sin(\x)});
\draw [phil, ultra thick, domain=270:360] plot ({1*cos(\x)}, {1*sin(\x)});  
\node at (-1,0) {$\times $};	
\node at (1,0) {$\times $};	
\node at (0,-2) {\Large $\frac{(-i\lambda_4)(i\sigma_4^\star)}{2}B_{PM }(k_s)$};	

\node at (-2.25,1.5) {$\bf 1$};	
\node at (-2.25,-1.5) {$\bf 2$};	
\node at (2.25,-1.5) {$\bf 3$};	
\node at (2.25,1.5) {$\bf 4$};	
	
\end{scope}
 
\begin{scope}[shift={(6,-5)}]
\draw [phil, ultra thick] (1,0) -- (2,1.5);
\draw [phir, ultra thick] (1,0) -- (2,-1.5);
\draw [phir, ultra thick] (-1,0) -- (-2,1.5);
\draw [phir, ultra thick] (-1,0) -- (-2,-1.5);
\draw [phir, ultra thick, domain=0:90] plot ({1*cos(\x)}, {1*sin(\x)});
\draw [phil, ultra thick, domain=90:180] plot ({1*cos(\x)}, {1*sin(\x)});
\draw [phil, ultra thick, domain=180:270] plot ({1*cos(\x)}, {1*sin(\x)});
\draw [phir, ultra thick, domain=270:360] plot ({1*cos(\x)}, {1*sin(\x)}); 
\node at (-1,0) {$\times $};	
\node at (1,0) {$\times $};	
\node at (0,-2) {\Large $\frac{(-\lambda_\Delta)(-i\sigma_4)}{2}B_{MP}(k_s)$};	

\node at (-2.25,1.5) {$\bf 1$};	
\node at (-2.25,-1.5) {$\bf 2$};	
\node at (2.25,-1.5) {$\bf 3$};	
\node at (2.25,1.5) {$\bf 4$};	
	
\end{scope}

\begin{scope}[shift={(12,-5)}]
\draw [phil, ultra thick] (1,0) -- (2,1.5);
\draw [phir, ultra thick] (1,0) -- (2,-1.5);
\draw [phir, ultra thick] (-1,0) -- (-2,1.5);
\draw [phir, ultra thick] (-1,0) -- (-2,-1.5);
\draw [phir, ultra thick, domain=0:90] plot ({1*cos(\x)}, {1*sin(\x)});
\draw [phil, ultra thick, domain=90:180] plot ({1*cos(\x)}, {1*sin(\x)});
\draw [phir, ultra thick, domain=180:270] plot ({1*cos(\x)}, {1*sin(\x)});
\draw [phil, ultra thick, domain=270:360] plot ({1*cos(\x)}, {1*sin(\x)}); 
\node at (-1,0) {$\times $};	
\node at (1,0) {$\times $};	
\node at (0,-2) {\Large $(-i\sigma_4)(-\lambda_\Delta)B_{PP}(k_s)$};	

\node at (-2.25,1.5) {$\bf 1$};	
\node at (-2.25,-1.5) {$\bf 2$};	
\node at (2.25,-1.5) {$\bf 3$};	
\node at (2.25,1.5) {$\bf 4$};	
	
\end{scope}

\begin{scope}[shift={(3,-10)}]
\draw [phil, ultra thick] (1,0) -- (2,1.5);
\draw [phir, ultra thick] (1,0) -- (2,-1.5);
\draw [phir, ultra thick] (-1,0) -- (-2,1.5);
\draw [phir, ultra thick] (-1,0) -- (-2,-1.5);
\draw [phir, ultra thick, domain=0:90] plot ({1*cos(\x)}, {1*sin(\x)});
\draw [phir, ultra thick, domain=90:180] plot ({1*cos(\x)}, {1*sin(\x)});
\draw [phir, ultra thick, domain=180:270] plot ({1*cos(\x)}, {1*sin(\x)});
\draw [phil, ultra thick, domain=270:360] plot ({1*cos(\x)}, {1*sin(\x)}); 
\node at (-1,0) {$\times $};	
\node at (1,0) {$\times $};	
\node at (0,-2) {\Large $(-i\lambda_4)(-\lambda_\Delta)B_{PR}(k_s)$};	

\node at (-2.25,1.5) {$\bf 1$};	
\node at (-2.25,-1.5) {$\bf 2$};	
\node at (2.25,-1.5) {$\bf 3$};	
\node at (2.25,1.5) {$\bf 4$};	
	
\end{scope} 

\begin{scope}[shift={(10,-10)}]
\draw [phil, ultra thick] (1,0) -- (2,1.5);
\draw [phir, ultra thick] (1,0) -- (2,-1.5);
\draw [phir, ultra thick] (-1,0) -- (-2,1.5);
\draw [phir, ultra thick] (-1,0) -- (-2,-1.5);
\draw [phir, ultra thick, domain=0:90] plot ({1*cos(\x)}, {1*sin(\x)});
\draw [phir, ultra thick, domain=90:180] plot ({1*cos(\x)}, {1*sin(\x)});
\draw [phil, ultra thick, domain=180:270] plot ({1*cos(\x)}, {1*sin(\x)});
\draw [phir, ultra thick, domain=270:360] plot ({1*cos(\x)}, {1*sin(\x)}); 
\node at (-1,0) {$\times $};	
\node at (1,0) {$\times $};	
\node at (0,-2) {\Large $(-i\sigma_4)^2B_{MR}(k_s)$};	

\node at (-2.25,1.5) {$\bf 1$};	
\node at (-2.25,-1.5) {$\bf 2$};	
\node at (2.25,-1.5) {$\bf 3$};	
\node at (2.25,1.5) {$\bf 4$};	
	
\end{scope}

\begin{scope}[shift={(3,-15)}]
\draw [phil, ultra thick] (1,0) -- (2,1.5);
\draw [phir, ultra thick] (1,0) -- (2,-1.5);
\draw [phir, ultra thick] (-1,0) -- (-2,1.5);
\draw [phir, ultra thick] (-1,0) -- (-2,-1.5);
\draw [phil, ultra thick, domain=0:90] plot ({1*cos(\x)}, {1*sin(\x)});
\draw [phil, ultra thick, domain=90:180] plot ({1*cos(\x)}, {1*sin(\x)});
\draw [phir, ultra thick, domain=180:270] plot ({1*cos(\x)}, {1*sin(\x)});
\draw [phil, ultra thick, domain=270:360] plot ({1*cos(\x)}, {1*sin(\x)}); 
\node at (-1,0) {$\times $};	
\node at (1,0) {$\times $};	
\node at (0,-2) {\Large $(-i\sigma_4)(i\sigma_4^\star)  B_{PL}(k_s)$};	

\node at (-2.25,1.5) {$\bf 1$};	
\node at (-2.25,-1.5) {$\bf 2$};	
\node at (2.25,-1.5) {$\bf 3$};	 
\node at (2.25,1.5) {$\bf 4$};	
	
\end{scope}

\begin{scope}[shift={(10,-15)}]
\draw [phil, ultra thick] (1,0) -- (2,1.5);
\draw [phir, ultra thick] (1,0) -- (2,-1.5);
\draw [phir, ultra thick] (-1,0) -- (-2,1.5);
\draw [phir, ultra thick] (-1,0) -- (-2,-1.5);
\draw [phil, ultra thick, domain=0:90] plot ({1*cos(\x)}, {1*sin(\x)});
\draw [phil, ultra thick, domain=90:180] plot ({1*cos(\x)}, {1*sin(\x)});
\draw [phil, ultra thick, domain=180:270] plot ({1*cos(\x)}, {1*sin(\x)});
\draw [phir, ultra thick, domain=270:360] plot ({1*cos(\x)}, {1*sin(\x)});  
\node at (-1,0) {$\times $};	
\node at (1,0) {$\times $};	
\node at (0,-2) {\Large $(-\lambda_\Delta)^2B_{ML}(k_s)$};	

\node at (-2.25,1.5) {$\bf 1$};	
\node at (-2.25,-1.5) {$\bf 2$};	
\node at (2.25,-1.5) {$\bf 3$};	
\node at (2.25,1.5) {$\bf 4$};	
	
\end{scope}	

\end{tikzpicture}
\end{center}
\caption{Diagrammatic representation of the Ten  $1$-Loop Integrals $ \phir\phir  \rightarrow \phir \phil  $}
\label{diag:sigmaoneloop}
\end{figure}
Again, only the Passarino-Veltman $B$ type diagrams contribute to the one loop beta function for $\sigma_4$. All the $B$ type diagrams are depicted in figure \ref{diag:sigmaoneloop}. The sum over all of them is given by
\begin{equation}
\begin{split}
&-i\sigma_4 \\
&+\frac{(-i\lambda_4)(-i\sigma_4)}{2}B_{RR}(k_s)
+\frac{(i\sigma_4^\star)(-\lambda_\Delta)}{2}B_{LL}(k_s)
+(-i\sigma_4)(-\lambda_\Delta)B_{LR}(k_s)\\
&+ \frac{(-\lambda_4)(i\sigma_4^\star)}{2}B_{PM}(k_s)
+\frac{(-\lambda_\Delta)(-i\sigma_4)}{2}B_{MP}(k_s)
+(-i\sigma_4)(-\lambda_\Delta)B_{PP}(k_s) \\
&+(-i\lambda_4)(-\lambda_\Delta)B_{PR}(k_s) 
+(-i\sigma_4)^2B_{MR}(k_s)\\
&+(-i\sigma_4)(i\sigma_4^\star)B_{PL}(k_s)
+(-\lambda_\Delta)^2B_{ML}(k_s)\\
&+(k_s\longleftrightarrow k_t)
%\nonumber\\&&
+(k_s\longleftrightarrow k_u) 
\label{sigmafour1}
\end{split}
\end{equation}
The one loop counter-term for $\sigma_4$ is given by
\begin{eqnarray}
\delta \sigma_4\Big|_\msbar	= -\frac{3}{(4\pi)^2} \Bigl[ \sigma_4^2+(\lambda_4+\sigma_4^\ast)(\sigma_4 - i \lambda_\Delta)+\lambda_\Delta^2\ \Bigr] \Bigl[ \frac{1}{d-4} +\frac{1}{2}(\gamma_E-\ln\, 4\pi) \Bigr]
\label{sigmafour2}
\end{eqnarray}
and the one loop beta function is found to be 
\begin{eqnarray}
\beta_{\sigma_4}
&=&\frac{3}{(4\pi)^2} \Bigl[ \sigma_4^2+(\lambda_4+\sigma_4^\ast)(\sigma_4 - i \lambda_\Delta)+\lambda_\Delta^2\ \Bigr]
\label{sigmafour3}
\end{eqnarray}

\subsection{One loop beta function for \texorpdfstring{$\lambda_\Delta$}{lambdadelta}}
\label{subsec:oneloopdeltafour}

\begin{figure}
\begin{center}
\begin{tikzpicture}[line width=1 pt, scale=0.8]

\begin{scope}[shift={(0,0)}]
\draw [phil, ultra thick] (1,0) -- (2,1.5);
\draw [phil, ultra thick] (1,0) -- (2,-1.5);
\draw [phir, ultra thick] (-1,0) -- (-2,1.5);
\draw [phir, ultra thick] (-1,0) -- (-2,-1.5);
\draw [phir, ultra thick, domain=0:180] plot ({1*cos(\x)}, {1*sin(\x)});
\draw [phir, ultra thick, domain=180:360] plot ({1*cos(\x)}, {1*sin(\x)}); 
\node at (-1,0) {$\times $};	
\node at (1,0) {$\times $};	
\node at (0,-2) {\Large $\frac{(-i\lambda_4)(-\lambda_\Delta)}{2}B_{RR}(k_s)$};	

\node at (-2.25,1.5) {$\bf 1$};	
\node at (-2.25,-1.5) {$\bf 2$};	
\node at (2.25,-1.5) {$\bf 3$};	
\node at (2.25,1.5) {$\bf 4$};	
	
\end{scope}

\begin{scope}[shift={(6,0)}]
\draw [phil, ultra thick] (1,0) -- (2,1.5);
\draw [phil, ultra thick] (1,0) -- (2,-1.5);
\draw [phir, ultra thick] (-1,0) -- (-2,1.5);
\draw [phir, ultra thick] (-1,0) -- (-2,-1.5);
\draw [phil, ultra thick, domain=0:90] plot ({1*cos(\x)}, {1*sin(\x)});
\draw [phil, ultra thick, domain=90:180] plot ({1*cos(\x)}, {1*sin(\x)});
\draw [phil, ultra thick, domain=180:270] plot ({1*cos(\x)}, {1*sin(\x)});
\draw [phil, ultra thick, domain=270:360] plot ({1*cos(\x)}, {1*sin(\x)}); 
\node at (-1,0) {$\times $};	
\node at (1,0) {$\times $};	
\node at (0,-2) {\Large $\frac{(i\lambda_4^\star)(-\lambda_\Delta)}{2}B_{LL}(k_s)$};	

\node at (-2.25,1.5) {$\bf 1$};	
\node at (-2.25,-1.5) {$\bf 2$};	
\node at (2.25,-1.5) {$\bf 3$};	
\node at (2.25,1.5) {$\bf 4$};	
	
\end{scope}

\begin{scope}[shift={(12,0)}]
\draw [phil, ultra thick] (1,0) -- (2,1.5);
\draw [phil, ultra thick] (1,0) -- (2,-1.5);
\draw [phir, ultra thick] (-1,0) -- (-2,1.5);
\draw [phir, ultra thick] (-1,0) -- (-2,-1.5);
\draw [phir, ultra thick, domain=0:90] plot ({1*cos(\x)}, {1*sin(\x)});
\draw [phir, ultra thick, domain=90:180] plot ({1*cos(\x)}, {1*sin(\x)});
\draw [phil, ultra thick, domain=180:270] plot ({1*cos(\x)}, {1*sin(\x)});
\draw [phil, ultra thick, domain=270:360] plot ({1*cos(\x)}, {1*sin(\x)}); 
\node at (-1,0) {$\times $};	
\node at (1,0) {$\times $};	
\node at (0,-2) {\Large $(-i\sigma_4)(i\sigma_4^\star)B_{LR}(k_s)$};	

\node at (-2.25,1.5) {$\bf 1$};	
\node at (-2.25,-1.5) {$\bf 2$};	
\node at (2.25,-1.5) {$\bf 3$};	
\node at (2.25,1.5) {$\bf 4$};	
	
\end{scope}

\begin{scope}[shift={(0,-5)}]
\draw [phil, ultra thick] (1,0) -- (2,1.5);
\draw [phil, ultra thick] (1,0) -- (2,-1.5);
\draw [phir, ultra thick] (-1,0) -- (-2,1.5);
\draw [phir, ultra thick] (-1,0) -- (-2,-1.5);
\draw [phil, ultra thick, domain=0:90] plot ({1*cos(\x)}, {1*sin(\x)});
\draw [phir, ultra thick, domain=90:180] plot ({1*cos(\x)}, {1*sin(\x)});
\draw [phir, ultra thick, domain=180:270] plot ({1*cos(\x)}, {1*sin(\x)});
\draw [phil, ultra thick, domain=270:360] plot ({1*cos(\x)}, {1*sin(\x)});  
\node at (-1,0) {$\times $};	
\node at (1,0) {$\times $};	
\node at (0,-2) {\Large $\frac{(-i\lambda_4)(i\lambda_4^\star)}{2}B_{PM}(k_s)$};	

\node at (-2.25,1.5) {$\bf 1$};	
\node at (-2.25,-1.5) {$\bf 2$};	
\node at (2.25,-1.5) {$\bf 3$};	
\node at (2.25,1.5) {$\bf 4$};	
	
\end{scope}

\begin{scope}[shift={(6,-5)}]
\draw [phil, ultra thick] (1,0) -- (2,1.5);
\draw [phil, ultra thick] (1,0) -- (2,-1.5);
\draw [phir, ultra thick] (-1,0) -- (-2,1.5);
\draw [phir, ultra thick] (-1,0) -- (-2,-1.5);
\draw [phir, ultra thick, domain=0:90] plot ({1*cos(\x)}, {1*sin(\x)});
\draw [phil, ultra thick, domain=90:180] plot ({1*cos(\x)}, {1*sin(\x)});
\draw [phil, ultra thick, domain=180:270] plot ({1*cos(\x)}, {1*sin(\x)});
\draw [phir, ultra thick, domain=270:360] plot ({1*cos(\x)}, {1*sin(\x)}); 
\node at (-1,0) {$\times $};	
\node at (1,0) {$\times $};	
\node at (0,-2) {\Large $\frac{(-\lambda_\Delta)^2}{2}B_{MP}(k_s)$};	

\node at (-2.25,1.5) {$\bf 1$};	
\node at (-2.25,-1.5) {$\bf 2$};	
\node at (2.25,-1.5) {$\bf 3$};	
\node at (2.25,1.5) {$\bf 4$};	
	
\end{scope}

\begin{scope}[shift={(12,-5)}]
\draw [phil, ultra thick] (1,0) -- (2,1.5);
\draw [phil, ultra thick] (1,0) -- (2,-1.5);
\draw [phir, ultra thick] (-1,0) -- (-2,1.5);
\draw [phir, ultra thick] (-1,0) -- (-2,-1.5);
\draw [phir, ultra thick, domain=0:90] plot ({1*cos(\x)}, {1*sin(\x)});
\draw [phil, ultra thick, domain=90:180] plot ({1*cos(\x)}, {1*sin(\x)});
\draw [phir, ultra thick, domain=180:270] plot ({1*cos(\x)}, {1*sin(\x)});
\draw [phil, ultra thick, domain=270:360] plot ({1*cos(\x)}, {1*sin(\x)}); 
\node at (-1,0) {$\times $};	
\node at (1,0) {$\times $};	
\node at (0,-2) {\Large $(-i\sigma_4)(i\sigma_4^\star)B_{PP}(k_s)$};	

\node at (-2.25,1.5) {$\bf 1$};	
\node at (-2.25,-1.5) {$\bf 2$};	
\node at (2.25,-1.5) {$\bf 3$};	
\node at (2.25,1.5) {$\bf 4$};	
	
\end{scope}

\begin{scope}[shift={(3,-10)}]
\draw [phil, ultra thick] (1,0) -- (2,1.5);
\draw [phil, ultra thick] (1,0) -- (2,-1.5);
\draw [phir, ultra thick] (-1,0) -- (-2,1.5);
\draw [phir, ultra thick] (-1,0) -- (-2,-1.5);
\draw [phir, ultra thick, domain=0:90] plot ({1*cos(\x)}, {1*sin(\x)});
\draw [phir, ultra thick, domain=90:180] plot ({1*cos(\x)}, {1*sin(\x)});
\draw [phir, ultra thick, domain=180:270] plot ({1*cos(\x)}, {1*sin(\x)});
\draw [phil, ultra thick, domain=270:360] plot ({1*cos(\x)}, {1*sin(\x)}); 
\node at (-1,0) {$\times $};	
\node at (1,0) {$\times $};	
\node at (0,-2) {\Large $(i\sigma_4^\star)(-i\lambda_4)B_{PR}(k_s)$};	

\node at (-2.25,1.5) {$\bf 1$};	
\node at (-2.25,-1.5) {$\bf 2$};	
\node at (2.25,-1.5) {$\bf 3$};	
\node at (2.25,1.5) {$\bf 4$};	
	
\end{scope}

\begin{scope}[shift={(10,-10)}]
\draw [phil, ultra thick] (1,0) -- (2,1.5);
\draw [phil, ultra thick] (1,0) -- (2,-1.5);
\draw [phir, ultra thick] (-1,0) -- (-2,1.5);
\draw [phir, ultra thick] (-1,0) -- (-2,-1.5);
\draw [phir, ultra thick, domain=0:90] plot ({1*cos(\x)}, {1*sin(\x)});
\draw [phir, ultra thick, domain=90:180] plot ({1*cos(\x)}, {1*sin(\x)});
\draw [phil, ultra thick, domain=180:270] plot ({1*cos(\x)}, {1*sin(\x)});
\draw [phir, ultra thick, domain=270:360] plot ({1*cos(\x)}, {1*sin(\x)}); 
\node at (-1,0) {$\times $};	
\node at (1,0) {$\times $};	
\node at (0,-2) {\Large $(-i\sigma_4)(-\lambda_\Delta)B_{MR}(k_s)$};	

\node at (-2.25,1.5) {$\bf 1$};	
\node at (-2.25,-1.5) {$\bf 2$};	
\node at (2.25,-1.5) {$\bf 3$};	
\node at (2.25,1.5) {$\bf 4$};	
	
\end{scope}

\begin{scope}[shift={(3,-15)}]
\draw [phil, ultra thick] (1,0) -- (2,1.5);
\draw [phil, ultra thick] (1,0) -- (2,-1.5);
\draw [phir, ultra thick] (-1,0) -- (-2,1.5);
\draw [phir, ultra thick] (-1,0) -- (-2,-1.5);
\draw [phil, ultra thick, domain=0:90] plot ({1*cos(\x)}, {1*sin(\x)});
\draw [phil, ultra thick, domain=90:180] plot ({1*cos(\x)}, {1*sin(\x)});
\draw [phir, ultra thick, domain=180:270] plot ({1*cos(\x)}, {1*sin(\x)});
\draw [phil, ultra thick, domain=270:360] plot ({1*cos(\x)}, {1*sin(\x)}); 
\node at (-1,0) {$\times $};	
\node at (1,0) {$\times $};	
\node at (0,-2) {\Large $(-i\sigma_4)(i\lambda_4^\star)B_{PL}(k_s)$};	

\node at (-2.25,1.5) {$\bf 1$};	
\node at (-2.25,-1.5) {$\bf 2$};	
\node at (2.25,-1.5) {$\bf 3$};	
\node at (2.25,1.5) {$\bf 4$};	
	
\end{scope}

\begin{scope}[shift={(10,-15)}]
\draw [phil, ultra thick] (1,0) -- (2,1.5);
\draw [phil, ultra thick] (1,0) -- (2,-1.5);
\draw [phir, ultra thick] (-1,0) -- (-2,1.5);
\draw [phir, ultra thick] (-1,0) -- (-2,-1.5);
\draw [phil, ultra thick, domain=0:90] plot ({1*cos(\x)}, {1*sin(\x)});
\draw [phil, ultra thick, domain=90:180] plot ({1*cos(\x)}, {1*sin(\x)});
\draw [phil, ultra thick, domain=180:270] plot ({1*cos(\x)}, {1*sin(\x)});
\draw [phir, ultra thick, domain=270:360] plot ({1*cos(\x)}, {1*sin(\x)});  
\node at (-1,0) {$\times $};	 
\node at (1,0) {$\times $};	
\node at (0,-2) {\Large $(i\sigma_4^\star)(-\lambda_\Delta)B_{ML}(k_s)$};	

\node at (-2.25,1.5) {$\bf 1$};	
\node at (-2.25,-1.5) {$\bf 2$};	
\node at (2.25,-1.5) {$\bf 3$};	
\node at (2.25,1.5) {$\bf 4$};	
	
\end{scope}	 

\end{tikzpicture}
\end{center}
\caption{Diagrammatic representation of the Ten  $1$-Loop Integrals $ \phir\phir \longrightarrow \phil \phil  $ (Here $k_t=k_1+k_2$)}
\label{diag:deltaoneloopI}
\end{figure}

The Passarino-Veltman $B$ type contributions for $s$-channel and $t$-channel is being shown in figure \ref{diag:deltaoneloopI} and figure  \ref{diag:deltaoneloopII} respectively. $u$-channels diagrams are obtained by interchanging $1\leftrightarrow 2$ in figure  \ref{diag:deltaoneloopII}. The sum over all the contributions is given as 
\begin{eqnarray}
&&	-\lambda_\Delta 
\nonumber\\
&&	
+\frac{(-i\lambda_4)(-\lambda_\Delta)}{2}B_{RR}(k_s)
+\frac{(i\lambda_4^\star)(-\lambda_\Delta)}{2}B_{LL}(k_s)
+(-i\sigma_4)(i\sigma_4^\star)B_{LR}(k_s)
\nonumber\\
&&	
+\frac{(-i\lambda_4)(i\lambda_4^\star)}{2}B_{PM}(k_s)
+
\frac{(-\lambda_\Delta)^2}{2}B_{MP}(k_s)
+
(-i\sigma_4)(i\sigma_4^\star)B_{PP}(k_s)
\label{deltafour1}
\\ 
&&	
+
(i\sigma_4^\star)(-i\lambda_4)B_{PR}(k_s)
+
(-i\sigma_4)(-\lambda_\Delta)B_{MR}(k_s)
\nonumber\\
&&
+
(-i\sigma_4)(i\lambda_4^\star)B_{PL}(k_s)
+
(i\sigma_4^\star)(-\lambda_\Delta) B_{ML}(k_s)
\nonumber\\ 
&&+\frac{(-i\sigma_4)(-i\sigma_4)}{2}B_{RR}(k_t) 
	+
	\frac{(i\sigma_4^\star)(i\sigma_4^\star)}{2}B_{LL}(k_t)
	+(-\lambda_\Delta)^2 B_{LR}(k_t)
\nonumber	\\
&&
+\frac{(-i\sigma_4)(i\sigma_4^\star)}{2}B_{PM}(k_t)
+\frac{(i\sigma_4^\star)(-i\sigma_4)}{2}B_{MP}(k_t)
+(-\lambda_\Delta)^2 B_{PP}(k_t)
\nonumber\\
&&
+(-i\sigma_4)(-\lambda_\Delta)B_{PR}(k_t)
+(-i\sigma_4)(-\lambda_\Delta)B_{MR}(k_t)
\nonumber\\&&
+(i\sigma_4^\star)(-\lambda_\Delta)B_{PL}(k_t)
+(i\sigma_4^\star)(-\lambda_\Delta) B_{ML}(k_t) 
\nonumber\\&&
+(k_t\longleftrightarrow k_u)
\nonumber
\end{eqnarray} 
The one-loop divergence can removed by adding the following counter-term
\begin{eqnarray}
\delta \lambda_\Delta\Big|_\msbar	&=&-\frac{1}{(4\pi)^2i} \Bigl[ (\lambda_4+2\sigma_4^\ast)(\sigma_4^\ast+ i \lambda_\Delta)+3i \sigma_4 \lambda_\Delta -c.c. \Bigr]
\nonumber\\
&&\qquad\qquad
 \Bigl[ \frac{1}{d-4} +\frac{1}{2}(\gamma_E-\ln\, 4\pi) \Bigr]
\label{deltafour2}
\end{eqnarray}
and one loop beta function for $\lambda_\Delta$ is given by  
\begin{equation}
\begin{split} 
\beta_{\lambda_\Delta}
&=\frac{1}{(4\pi)^2i} \Bigl[ \lambda_4(\sigma_4^\ast+ i \lambda_\Delta)-2\sigma_4^2+5i \sigma_4 \lambda_\Delta -\lambda_4^\ast(\sigma_4- i \lambda_\Delta)+2(\sigma_4^\ast)^2+5i \sigma_4^\ast \lambda_\Delta  \Bigr]\\
&=\frac{1}{(4\pi)^2i} \Bigl[ (\lambda_4+2\sigma_4^\ast)(\sigma_4^\ast+ i \lambda_\Delta)+3i \sigma_4 \lambda_\Delta -c.c. \Bigr]\\ 
\label{deltafour3}
\end{split}
\end{equation}

\subsection{Checking Lindblad condition for quartic couplings}
\label{subsec:oneloopLindbladquartic}

From equation \eqref{eq:lambdafour3}, equation \eqref{sigmafour3} and equation \eqref{deltafour3}, we can compute the one loop beta function for the Lindblad combination. We have
\begin{equation}
\begin{split}
\beta_{(\im  \ \lambda_4+ 4\ \im  \ \sigma_4 -3 \lambda_\Delta)}
&=\frac{6}{(4\pi)^2} (\im  \ \lambda_4+ 4\ \im  \ \sigma_4 -3 \lambda_\Delta)  (\re \ \lambda_4+ 2 \re \ \sigma_4) 
\label{Lindbladbeta3}
\end{split}
\end{equation} 
This equation, along with \eqref{Lindbladbeta1} and \eqref{Lindbladbeta2}, implies that if one starts with a Lindblad theory then one loop renormalization 
preserves the Lindblad condition. 

\begin{figure}
\begin{center}
\begin{tikzpicture}[line width=1 pt, scale=0.8]
%\begin{scope}[rotate=-80]

\begin{scope}[shift={(0,0)}]
\draw [phil, ultra thick] (0,1) -- (2,1.5);
\draw [phil, ultra thick] (0,-1) -- (2,-1.5);
\draw [phir, ultra thick] (0,-1) -- (-2,-1.5);
\draw [phir, ultra thick] (0,1) -- (-2,1.5);
\draw [phir, ultra thick, domain=-90:90] plot ({1*cos(\x)}, {1*sin(\x)});
\draw [phir, ultra thick, domain=90:270] plot ({1*cos(\x)}, {1*sin(\x)}); 
\node at (0,1) {$\times $};	
\node at (0,-1) {$\times $};	
\node at (0,-2.5) {\Large $\frac{(-i\sigma_4)(-i\sigma_4)}{2}B_{RR}(k_t)$};	

\node at (-2.25,1.5) {$\bf 1$};	
\node at (-2.25,-1.5) {$\bf 2$};	
\node at (2.25,-1.5) {$\bf 3$};	
\node at (2.25,1.5) {$\bf 4$};

\end{scope}

\begin{scope}[shift={(6,0)}]
\draw [phil, ultra thick] (0,1) -- (2,1.5);
\draw [phil, ultra thick] (0,-1) -- (2,-1.5);
\draw [phir, ultra thick] (0,-1) -- (-2,-1.5);
\draw [phir, ultra thick] (0,1) -- (-2,1.5);
\draw [phil, ultra thick, domain=-90:90] plot ({1*cos(\x)}, {1*sin(\x)});
\draw [phil, ultra thick, domain=90:270] plot ({1*cos(\x)}, {1*sin(\x)}); 
\node at (0,1) {$\times $};	
\node at (0,-1) {$\times $};		
\node at (0,-2.5) {\Large $\frac{(i\sigma_4^\star)(i\sigma_4^\star)}{2}B_{LL}(k_t)$};	

\node at (-2.25,1.5) {$\bf 1$};	
\node at (-2.25,-1.5) {$\bf 2$};	
\node at (2.25,-1.5) {$\bf 3$};	
\node at (2.25,1.5) {$\bf 4$};
	
\end{scope}

\begin{scope}[shift={(12,0)}]
\draw [phil, ultra thick] (0,1) -- (2,1.5);
\draw [phil, ultra thick] (0,-1) -- (2,-1.5);
\draw [phir, ultra thick] (0,-1) -- (-2,-1.5);
\draw [phir, ultra thick] (0,1) -- (-2,1.5);
\draw [phil, ultra thick, domain=-90:90] plot ({1*cos(\x)}, {1*sin(\x)});
\draw [phir, ultra thick, domain=90:270] plot ({1*cos(\x)}, {1*sin(\x)}); 
\node at (0,1) {$\times $};	
\node at (0,-1) {$\times $};	
\node at (0,-2.5) {\Large $(-\lambda_\Delta)^2B_{LR}(k_t)$};	

\node at (-2.25,1.5) {$\bf 1$};	
\node at (-2.25,-1.5) {$\bf 2$};	
\node at (2.25,-1.5) {$\bf 3$};	
\node at (2.25,1.5) {$\bf 4$};
	
\end{scope}

\begin{scope}[shift={(0,-5)}]
\draw [phil, ultra thick] (0,1) -- (2,1.5);
\draw [phil, ultra thick] (0,-1) -- (2,-1.5);
\draw [phir, ultra thick] (0,-1) -- (-2,-1.5);
\draw [phir, ultra thick] (0,1) -- (-2,1.5);
\draw [phir, ultra thick, domain=0:90] plot ({1*cos(\x)}, {1*sin(\x)});
\draw [phir, ultra thick, domain=90:180] plot ({1*cos(\x)}, {1*sin(\x)});
\draw [phil, ultra thick, domain=180:270] plot ({1*cos(\x)}, {1*sin(\x)});
\draw [phil, ultra thick, domain=270:360] plot ({1*cos(\x)}, {1*sin(\x)});  
\node at (0,1) {$\times $};	
\node at (0,-1) {$\times $};
\node at (0,-2.5) {\Large $\frac{(-i\sigma_4)(i\sigma_4^\star)}{2}B_{PM}(k_t)$};	 

\node at (-2.25,1.5) {$\bf 1$};	
\node at (-2.25,-1.5) {$\bf 2$};	
\node at (2.25,-1.5) {$\bf 3$};	
\node at (2.25,1.5) {$\bf 4$};
	
\end{scope}

\begin{scope}[shift={(6,-5)}]
\draw [phil, ultra thick] (0,1) -- (2,1.5);
\draw [phil, ultra thick] (0,-1) -- (2,-1.5);
\draw [phir, ultra thick] (0,-1) -- (-2,-1.5);
\draw [phir, ultra thick] (0,1) -- (-2,1.5);
\draw [phil, ultra thick, domain=0:90] plot ({1*cos(\x)}, {1*sin(\x)});
\draw [phil, ultra thick, domain=90:180] plot ({1*cos(\x)}, {1*sin(\x)});
\draw [phir, ultra thick, domain=180:270] plot ({1*cos(\x)}, {1*sin(\x)});
\draw [phir, ultra thick, domain=270:360] plot ({1*cos(\x)}, {1*sin(\x)}); 
\node at (0,1) {$\times $};	
\node at (0,-1) {$\times $};
\node at (0,-2.5) {\Large $\frac{(i\sigma_4^\star)(-i\sigma_4)}{2}B_{MP}(k_t)$};	

\node at (-2.25,1.5) {$\bf 1$};	
\node at (-2.25,-1.5) {$\bf 2$};	
\node at (2.25,-1.5) {$\bf 3$};	
\node at (2.25,1.5) {$\bf 4$};
	
\end{scope}

\begin{scope}[shift={(12,-5)}]
\draw [phil, ultra thick] (0,1) -- (2,1.5);
\draw [phil, ultra thick] (0,-1) -- (2,-1.5);
\draw [phir, ultra thick] (0,-1) -- (-2,-1.5);
\draw [phir, ultra thick] (0,1) -- (-2,1.5);
\draw [phir, ultra thick, domain=0:90] plot ({1*cos(\x)}, {1*sin(\x)});
\draw [phil, ultra thick, domain=90:180] plot ({1*cos(\x)}, {1*sin(\x)});
\draw [phir, ultra thick, domain=180:270] plot ({1*cos(\x)}, {1*sin(\x)});
\draw [phil, ultra thick, domain=270:360] plot ({1*cos(\x)}, {1*sin(\x)}); 
\node at (0,1) {$\times $};	
\node at (0,-1) {$\times $};
\node at (0,-2.5) {\Large $(-\lambda_\Delta)^2B_{PP}(k_t)$};	

\node at (-2.25,1.5) {$\bf 1$};	
\node at (-2.25,-1.5) {$\bf 2$};	
\node at (2.25,-1.5) {$\bf 3$};	
\node at (2.25,1.5) {$\bf 4$};
	
\end{scope}

\begin{scope}[shift={(3,-10)}]
\draw [phil, ultra thick] (0,1) -- (2,1.5);
\draw [phil, ultra thick] (0,-1) -- (2,-1.5);
\draw [phir, ultra thick] (0,-1) -- (-2,-1.5);
\draw [phir, ultra thick] (0,1) -- (-2,1.5);
\draw [phir, ultra thick, domain=0:90] plot ({1*cos(\x)}, {1*sin(\x)});
\draw [phir, ultra thick, domain=90:180] plot ({1*cos(\x)}, {1*sin(\x)});
\draw [phir, ultra thick, domain=180:270] plot ({1*cos(\x)}, {1*sin(\x)});
\draw [phil, ultra thick, domain=270:360] plot ({1*cos(\x)}, {1*sin(\x)}); 
\node at (0,1) {$\times $};	
\node at (0,-1) {$\times $};
\node at (0,-2.5) {\Large $(-i\sigma_4)(-\lambda_\Delta)B_{PR}(k_t)$};	

\node at (-2.25,1.5) {$\bf 1$};	
\node at (-2.25,-1.5) {$\bf 2$};	
\node at (2.25,-1.5) {$\bf 3$};	 
\node at (2.25,1.5) {$\bf 4$};
	
\end{scope}

\begin{scope}[shift={(10,-10)}]
\draw [phil, ultra thick] (0,1) -- (2,1.5);
\draw [phil, ultra thick] (0,-1) -- (2,-1.5);
\draw [phir, ultra thick] (0,-1) -- (-2,-1.5);
\draw [phir, ultra thick] (0,1) -- (-2,1.5);
\draw [phil, ultra thick, domain=0:90] plot ({1*cos(\x)}, {1*sin(\x)});
\draw [phir, ultra thick, domain=90:180] plot ({1*cos(\x)}, {1*sin(\x)});
\draw [phir, ultra thick, domain=180:270] plot ({1*cos(\x)}, {1*sin(\x)});
\draw [phir, ultra thick, domain=270:360] plot ({1*cos(\x)}, {1*sin(\x)}); 
\node at (0,1) {$\times $};	
\node at (0,-1) {$\times $};  
\node at (0,-2.5) {\Large $(-i\sigma_4)(-\lambda_\Delta)B_{MR}(k_t)$};	

\node at (-2.25,1.5) {$\bf 1$};	
\node at (-2.25,-1.5) {$\bf 2$};	
\node at (2.25,-1.5) {$\bf 3$};	
\node at (2.25,1.5) {$\bf 4$};	
	
\end{scope} 

\begin{scope}[shift={(3,-15)}]
\draw [phil, ultra thick] (0,1) -- (2,1.5);
\draw [phil, ultra thick] (0,-1) -- (2,-1.5);
\draw [phir, ultra thick] (0,-1) -- (-2,-1.5);
\draw [phir, ultra thick] (0,1) -- (-2,1.5);
\draw [phil, ultra thick, domain=0:90] plot ({1*cos(\x)}, {1*sin(\x)});
\draw [phil, ultra thick, domain=90:180] plot ({1*cos(\x)}, {1*sin(\x)});
\draw [phir, ultra thick, domain=180:270] plot ({1*cos(\x)}, {1*sin(\x)});
\draw [phil, ultra thick, domain=270:360] plot ({1*cos(\x)}, {1*sin(\x)}); 
\node at (0,1) {$\times $};	
\node at (0,-1) {$\times $};
\node at (0,-2.5) {\Large $(i\sigma_4^\star)(-\lambda_\Delta)B_{ML}(k_t)$};	
	
\node at (-2.25,1.5) {$\bf 1$};	
\node at (-2.25,-1.5) {$\bf 2$};	
\node at (2.25,-1.5) {$\bf 3$};	
\node at (2.25,1.5) {$\bf 4$};	
	
\end{scope}

\begin{scope}[shift={(10,-15)}]
\draw [phil, ultra thick] (0,1) -- (2,1.5);
\draw [phil, ultra thick] (0,-1) -- (2,-1.5);
\draw [phir, ultra thick] (0,-1) -- (-2,-1.5);
\draw [phir, ultra thick] (0,1) -- (-2,1.5);
\draw [phil, ultra thick, domain=0:90] plot ({1*cos(\x)}, {1*sin(\x)});
\draw [phir, ultra thick, domain=90:180] plot ({1*cos(\x)}, {1*sin(\x)});
\draw [phil, ultra thick, domain=180:270] plot ({1*cos(\x)}, {1*sin(\x)});
\draw [phil, ultra thick, domain=270:360] plot ({1*cos(\x)}, {1*sin(\x)});   
\node at (0,1) {$\times $};	
\node at (0,-1) {$\times $};
\node at (0,-2.5) {\Large $(i\sigma_4^\star)(-\lambda_\Delta)B_{PL}(k_t)$};	

\node at (-2.25,1.5) {$\bf 1$};	
\node at (-2.25,-1.5) {$\bf 2$};	
\node at (2.25,-1.5) {$\bf 3$};	
\node at (2.25,1.5) {$\bf 4$};
	
\end{scope}	  

\end{tikzpicture}
\end{center}
\caption{Diagrammatic representation of the Ten  $1$-Loop Integrals $ \phir\phir \rightarrow \phil \phil  $ (Here $k_t=k_1+k_4$)}
\label{diag:deltaoneloopII}
\end{figure}

\newpage
\subsection{Summary of the results}
\label{subsec:oneloopsummary}

%\begin{equation}
%\begin{split}
%\frac{d m^2}{d\ln\, \mu}
%&=\frac{1}{(4\pi)^2} \Bigl[ (\lambda_3)^2 -(\sigma_3^\star)^2+
%2\left\{\lambda_3\sigma_3+|\sigma_3|^2 \right\}\ 
%+(\lambda_4-i\lambda_\Delta+2\sigma_4)\left(\re\,m^2\right) 
%\Bigr]
%\\
%\frac{d m_\Delta^2}{d\ln\, \mu} 
%&= \frac{1}{(4\pi)^2} \left[-4(\re\,  \lambda_3\,+\re\,   \sigma_3\,)\, \im\,  \sigma_3 
% +(2\lambda_\Delta - 2 \im\,\sigma_4)\re\,m^2
% \right] 
%\\
%\frac{d \lambda_3}{d\ln\, \mu}
%&=\frac{3}{(4\pi)^2} \Bigl[\lambda_4\lambda_3 -2\lambda_\Delta\im\, \sigma_3
%+\lambda_4\sigma_3+\sigma_4\lambda_3+\sigma_4\sigma_3^\ast \Bigr]
%\\
%\frac{d \sigma_3}{d\ln\, \mu}
%&=\frac{1}{(4\pi)^2}
%\Bigg[2i\lambda_4\im\,  \sigma_3 -i\lambda_\Delta(\sigma_3^\star+\lambda_3^\star)
% +\sigma_4\lambda^\star_3+\sigma_4\sigma_3
%\\ 
%&\qquad\qquad 
%\,  
%+2\sigma_4\lambda_3
%+4i\im[\sigma_4\sigma_3]-2i\lambda_\Delta(\lambda_3+\sigma_3^\ast)+ 2\sigma^\ast_4\sigma_3 \Bigg]  
%\\
%\frac{d \lambda_4}{d\ln\, \mu}
%&=\frac{3}{(4\pi)^2} \Bigl[ \lambda^2+2\  \sigma(\lambda+i \Delta)+\Delta^2\ \Bigr]= \frac{3}{(4\pi)^2}  (\lambda+2\ \sigma-i\Delta)(\lambda+i \Delta)\\
%\frac{d \sigma_4}{d\ln\, \mu}
%&=\frac{3}{(4\pi)^2} \Bigl[ \sigma^2+(\lambda+\sigma^\ast)(\sigma - i \Delta)+\Delta^2\ \Bigr]=\frac{3}{(4\pi)^2} (\lambda+\sigma+\sigma^\ast+i\Delta)(\sigma - i \Delta)\\
%\frac{d \lambda_\Delta}{d\ln\, \mu}
%&=\frac{1}{(4\pi)^2i} \Bigl[ \lambda_4(\sigma^\ast_4+ i \lambda_\Delta)-2\sigma^2_4+5i \sigma_4 \lambda_\Delta -\lambda^\ast_4(\sigma_4- i \lambda_\Delta)+2(\sigma^\ast_4)^2-5i \sigma^\ast_4 \lambda_\Delta  \Bigr]
%\end{split}
%\end{equation} 
We started with the most general Lagrangian of a mixed system described by a scalar field with cubic and quartic coupling in \eqref{Smacro}. Using this action, we have demonstrated that  the standard counter-term technique of unitary QFTs can be extended  to deal with the one-loop UV divergences of the open EFT. We have then computed the beta functions of this open EFT, summarized in equation \eqref{eq:betamassintro}, \eqref{eq:betacubicintro} and \eqref{eq:betaquarticintro} of the introduction. One can then use these beta functions to determine the running of the Lindblad violating combinations   $(\im\,m^2 - m_\Delta^2)$, $(\im\,  \lambda_3+3\im\,  \sigma_3)$ and $(\im\, \lambda_4 +4\im\, \sigma_4 -3\lambda_\Delta)$ giving Eq. \eqref{eq:betalindlbladintro}. The equation Eq.\eqref{eq:betalindlbladintro} shows that the beta function of the Lindblad violating couplings are 
proportional to the Lindblad violating couplings. In other words, if we set the Lindblad violating coupling to zero at tree level then the Lindblad violating coupling will not be generated under one loop renormalization.

% \begin{equation}
% \begin{split}
% \beta_{\lambda_\Delta}
% &=\frac{1}{(4\pi)^2i} \Bigl[ \lambda_4(\sigma_4^\ast+ i \lambda_\Delta)-2\sigma_4^2+5i \sigma_4 \lambda_\Delta -\lambda_4^\ast(\sigma_4- i \lambda_\Delta)+2(\sigma_4^\ast)^2+5i \sigma_4^\ast \lambda_\Delta  \Bigr]\\
% &=\frac{1}{(4\pi)^2i} \Bigl[ (\lambda_4+2\sigma_4^\ast)(\sigma_4^\ast+ i \lambda_\Delta)+3i \sigma_4 \lambda_\Delta -c.c. \Bigr]\\ 
% \label{deltafour3}
% \end{split}
% \end{equation}
% 
% 
% \begin{eqnarray}
%  \beta_{(\im\,m^2 - m_\Delta^2)}&=&\frac{2}{(4\pi)^2}\Bigg[(\im\, \lambda_3+3\, \im\,  \sigma_3)(\re\, \lambda_3+\re\,  \sigma_3)
% \nonumber\\
% &&\qquad\quad
% +  (\im\lambda_4\ +4\im\,\sigma_4  -3\lambda_\Delta )(\re\,m^2)\Bigg]
%    \nonumber\\
% \beta_{(\im\,  \lambda_3+3\im\,  \sigma_3)}
% &=&\frac{3}{(4\pi)^2}\Bigg[(\re\, \lambda_4+2\re\, \sigma_4) (\im\, \lambda_3+3\im\, \sigma_3)
% \\
% &&\qquad\quad
% +(\re\, \lambda_3+\re\, \sigma_3)(\im\, \lambda_4 +4\im\, \sigma_4 -3\lambda_\Delta)\Bigg]
%    \nonumber\\
% \beta_{(\im\,  \lambda_4+4\im\,  \sigma_4-3\lambda_\Delta)}
% &=&
% \frac{6}{(4\pi)^2} (\im  \ \lambda_4+ 4\ \im  \ \sigma_4 -3 \lambda_\Delta)  (\re \ \lambda_4+ 2 \re \ \sigma_4) 
% %\nonumber\\
% \nonumber 
% \end{eqnarray}

%\chapter{Computation in the average-difference basis}

\section{Computation in the average-difference basis}
\label{sec:avgdifbasis}

In section \ref{sec:oneloopmass}, we had computed the one loop beta functions for various couplings of an open $\phi^3+\phi^4$ theory. In particular, by looking at 
the Lindblad violating couplings, we found that the Lindblad condition is preserved under one loop renormalization. In this section, 
we will rewrite the perturbation theory in a different basis where this fact is manifest. We would also like to prove that the 
preservation of Lindblad conditions hold to arbitrary perturbative order. The proof that we present here is very much inspired by 
a correponding argument in the context of cutting rules in a unitary theory and uses a version of Feynman tree theorem.

The basis we shift to is often termed the Keldysh basis. It is made of the average and difference of bra and ket fields. This 
basis has an advantage that the difference operator decouplings are more manifest in this basis while it obscures
the cutting  rule interpretation of various diagrams involved. While the unitary vertices are mixed up with the Feynman-Vernon
couplings in this basis, the computations do greatly simplify owing to lesser number of divergent
diagrams and vanishing of difference-difference propagator. Our discussion here would necessarily be brief, since the details 
are straightforward and similar to the computation in the previous section. For a more detailed presentation, we refer the reader to
appendix \ref{sec:appendixavgdiff}

\subsection{Action in the average-difference basis}
\label{subsec:avgdifaction}

We define $\phidif$ and $\phiav$ such that
\begin{eqnarray}
\phidif=\phir-\phil
\qquad\qquad
\phiav=\frac{1}{2}(\phir+\phil)
\label{avdif1}
\end{eqnarray}
where the subscripts $d$ and $a$ denote `difference' and `average' respectively.

The Lagrangian in this basis is given by  
\begin{equation} 
\begin{split}
i\mathcal{L}&= +\frac{1}{2\times 2!}  (\im  \ z +z_\Delta)(\partial\phidif)^2 + (-i)(\re \ z)(\partial\phiav).(\partial \phidif) 
\\
& +\frac{1}{2\times 2!}  (\im  \ m^2 +m_\Delta^2)\phidif^2 + (-i)(\re \ m^2)\phiav \phidif 
\\
&+(-i)(\re \ \lambda_3+ \ \re \ \sigma_3) \frac{\phiav^2\phidif}{2!}
+  \frac{1}{2} (\im  \ \lambda_3- \ \im  \ \sigma_3)  \frac{\phiav\phidif^2}{2!}
%\\&
+
\frac{(-i)}{4}(\re \ \lambda_3-3\ \re \ \sigma_3) \frac{\phidif^3}{3!}
\\
&+(-i) (\re \ \lambda_4+ 2\ \re \ \sigma_4) \frac{\phiav^3\phidif}{3!}
+\frac{1}{2}(\im  \ \lambda_4+  \lambda_\Delta)   \frac{\phiav^2\phidif^2}{2!2!}+  \frac{(-i)}{4} (\re \ \lambda_4- 2\ \re \ \sigma_4)  \frac{\phiav\phidif^3}{3!}\\
 &+
\frac{1}{8}(\im  \ \lambda_4-4\ \im  \ \sigma_4 -3 \lambda_\Delta) \frac{\phidif^4}{4!}\\ 
 &+\frac{1}{2!}  2(\im  \ z - z_\Delta)(\partial\phiav)^2+\frac{1}{2!}  2(\im  \ m^2 - m_\Delta^2)\phiav^2 \\
 &+ 2(\im  \ \lambda_3+ 3\ \im  \ \sigma_3) \frac{\phiav^3}{3!}
 +2(\im  \ \lambda_4+ 4\ \im  \ \sigma_4 -3 \lambda_\Delta) \frac{\phiav^4}{4!}
 \label{avdif2}
\end{split}
\end{equation}
The Feynman rules in this basis are given in figure \ref{fig:avgdifffeynrules}. Note that the terms in the last two lines of the Lagrangian involves only the average fields $\phiav$. The coefficients of  the purely average couplings are exactly the Lindblad violating couplings. This is expected for the following reason :  since $\phidif$ vanishes when $\phir = \phil$,  the terms that can contribute to the imaginary part of the action, in that limit, are the pure $\phiav$ vertices. Since all Linblad terms vanish in this limit, it follows that pure $\phiav$ vertices should be Lindblad violating. In addition, we observe that in the open $\phi^3+\phi^4$ theory, all  Lindblad violating couplings are of pure average type. This clear separation of the Lindblad violating couplings is the most salient aspect of this basis, making it easy to trace their renormalisation. 

The propagators in this basis are given by \cite{Chou:1984es,Kamenev,Haehl:2016pec}  
\begin{equation}
\begin{split}
\text{a} : \langle \mathcal{T}_{SK} \phiav \phiav \rangle&=\frac{1}{2}\langle \mathcal{T}_{SK} \phir \phir \rangle +\frac{1}{2} \langle \mathcal{T}_{SK} \phil \phil \rangle\\
\text{f} :\langle \mathcal{T}_{SK} \phiav \phidif \rangle&=\langle \mathcal{T}_{SK} \phir \phir \rangle - \langle \mathcal{T}_{SK} \phir \phil \rangle\\
\text{b} :\langle \mathcal{T}_{SK} \phidif \phiav \rangle&=\langle \mathcal{T}_{SK} \phir \phir \rangle - \langle \mathcal{T}_{SK} \phil \phir \rangle\\
\text{d} :\langle \mathcal{T}_{SK} \phidif \phidif \rangle&=0
\end{split}
\label{avdif3}
\end{equation}
Please note that we will use a different color for propagators in the average-difference basis. 
% Average and difference fields will be denoted by bold and dashed lines respectively. 
Also, we shall be using results presented in appendix \ref{sec:onelooppvavgdiff}.
\begin{figure}[ht]
\begin{center}

\begin{tikzpicture} 

\phipropagatora{0}{0}{0}{1}{}
\node at (-3,0) {Propagator $a$};
%\node at (5,0) {phipropagatora\{0\}\{0\}\{2\}\{0\}};
\node at (-.5,0) {$\phiav$}; 
\node at (2.5,0) {$\phiav$};

\phipropagatorf{0}{-1}{0}{1}{}
\node at (-3,-1) {Propagator $f$};
%\node at (5,-2) {phipropagatorf\{0\}\{0\}\{2\}\{0\}};
\node at (-.5,-1) {$\phiav$}; 
\node at (2.5,-1) {$\phidif$};

\phipropagatorb{0}{-2}{0}{1}{}
\node at (-3,-2) {Propagator $b$};
%\node at (5,-4) {phipropagatorb\{0\}\{0\}\{2\}\{0\}};
\node at (-.5,-2) {$\phidif$};
\node at (2.5,-2) {$\phiav$};

\phipropagatord{0}{-3}{0}{1}{}
\node at (-3,-3) {Propagator $d$};
%\node at (5,-6) {phipropagatorb\{0\}\{0\}\{2\}\{0\}};
\node at (-.5,-3) {$\phidif$};
\node at (2.5,-3) {$\phidif$};
	   
\end{tikzpicture}	

\end{center}
\caption{Propagators in the average-difference basis}
%\label{fig3pt} 
\end{figure}
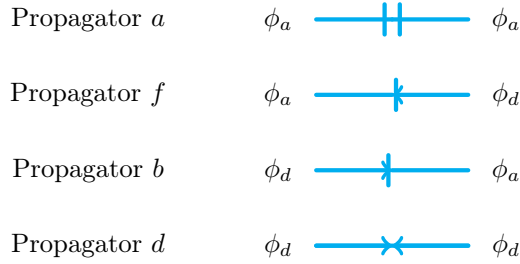 
In this basis, only the tadpole $A_a$ diverges \eqref{avgdifpvint1}  and its divergence is same as the divergence of usual PV $A$ diagram. All other $A$ integrals, $A_f, A_b$ and $A_d$, vanish. Similarly, only the bubbles $B_{af}$, $B_{ab}$ diverge and their divergence is half the divergence of the usual PV $B$ diagram \eqref{avgdifpvint3}.

\subsection{One loop computations}
\label{subsec:avgdifoneloop}

As mentioned before, the computation greatly simplifies in this basis. All the computations in average-difference basis can be found in appendix 
\ref{sec:appendixavgdiff}. Here we shall demonstrate only a few examples.  For instance, let us compute the beta function of one of the Lindblad violating terms,
$(\im  \ m^2- m_\Delta^2)$. In figure \ref{fig:avgdiffmass10},  we have considered all the divergent diagrams (i.e., the diagrams involving $A_a$,  $B_{af}$ and $B_{ab}$)
that contribute to the process $\phidif \rightarrow \phidif $. The total contribution to the process is given by

\begin{figure}[ht] 
\begin{center}
\begin{tikzpicture}[ scale=0.65]

%\drawphiavdifdiagaa{0}{0}{90}

\phipropagatorb{-3}{0}{0}{1}{}
\phipropagatorb{3}{0}{180}{1}{} 
\drawphiavdifdiagbaf{0}{0}{0}

\begin{scope}[shift={(7,0)}]
\phipropagatorb{-3}{0}{0}{1}{}
\phipropagatorb{3}{0}{180}{1}{}
\drawphiavdifdiagbab{0}{0}{0}	
\end{scope}

\begin{scope}[shift={(14,1)}]
\drawphiavdifdiagaa{0}{0}{90}

\phipropagatorb{-2}{-1}{0}{1}{}
\phipropagatorb{2}{-1}{0}{-1}{}
	
\end{scope}

\end{tikzpicture}
\end{center}
\caption{Renormalization of the Lindblad violating mass term in the average-difference basis}
\label{fig:avgdiffmass10}  
\end{figure}
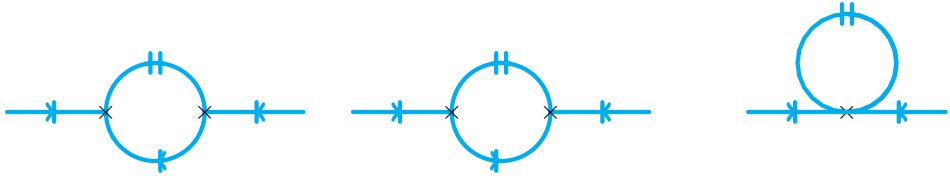
\begin{equation}
\begin{split}
2(\im  \ m^2- m_\Delta^2) & + 2(\im  \ \lambda_3+ 3\ \im  \ \sigma_3)\times (-i)  (\re \ \lambda_3+ \ \re \ \sigma_3)(B_{af}+B_{ab})\\
&+ 2(\im  \ \lambda_4+ 4\ \im  \ \sigma_4 -3 \lambda_\Delta) (\re \ m^2) A_a 
\end{split}
\end{equation}  
Hence, the one loop beta function for the Lindblad violating mass term is given by 
\begin{equation} 
\begin{split}
\frac{d}{d\ln\,  \mu}(\im  \ m^2- m_\Delta^2)
&= \frac{1}{(4\pi)^2}\Big[ 2(\im  \ \lambda_3+ 3\ \im  \ \sigma_3) (\re \ \lambda_3+ \ \re \ \sigma_3)
\\&
\qquad\qquad
+
(\re \ m^2)(\im  \ \lambda_4+ 4\ \im  \ \sigma_4 -3 \lambda_\Delta) \Big]
\end{split} 
\end{equation}
We had obtained the same result (equation \eqref{Lindbladbeta2}) in the other basis.  Notice that the beta function of the Lindblad violating term can easily be computed 
just by computing one process in this basis. 

\begin{figure}[ht]
\begin{center}
\begin{tikzpicture}[ scale=0.65]

\phipropagatorb{-3}{0}{0}{1}{}
\phipropagatorf{1}{0}{30}{1.2}{}
\phipropagatorf{1}{0}{-30}{1.2}{}

\node at (-3.5,0) {$\bf 1$};	
\node at (3.5,-1) {$\bf 2$};	
\node at (3.5,1) {$\bf 3$};

\drawphiavdifdiagbaf{0}{0}{0}

\begin{scope}[shift={(10,0)}]
\phipropagatorb{-3}{0}{0}{1}{}
\phipropagatorf{1}{0}{30}{1.2}{}
\phipropagatorf{1}{0}{-30}{1.2}{}
\drawphiavdifdiagbab{0}{0}{0}	
\node at (-3.5,0) {$\bf 1$};	
\node at (3.5,-1) {$\bf 2$};	 
\node at (3.5,1) {$\bf 3$};	
\end{scope}
 
\end{tikzpicture}
\end{center}
\caption{Renormalization of the Lindblad violating cubic coupling in the average-difference basis}
\label{fig:avgdiffcubic10} 
\end{figure}
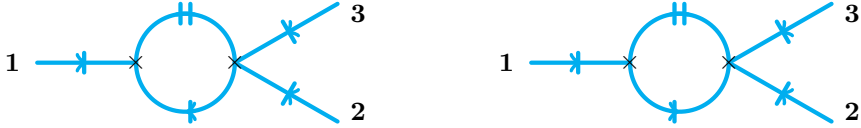

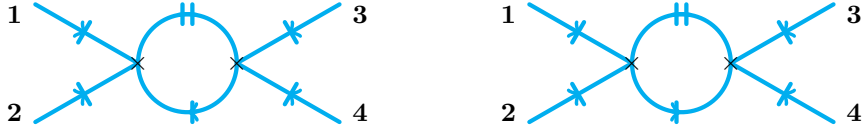
\begin{figure}[ht]
\begin{center}
\begin{tikzpicture}[ scale=0.65]

\phipropagatorf{-1}{0}{150}{1.2}{}
\phipropagatorf{-1}{0}{-150}{1.2}{}
\phipropagatorf{1}{0}{30}{1.2}{}
\phipropagatorf{1}{0}{-30}{1.2}{}

\node at (-3.5,1) {$\bf 1$};	
\node at (-3.5,-1) {$\bf 2$};	
\node at (3.5,1) {$\bf 3$};	
\node at (3.5,-1) {$\bf 4$};

\drawphiavdifdiagbaf{0}{0}{0}

\begin{scope}[shift={(10,0)}]
\phipropagatorf{-1}{0}{150}{1.2}{}
\phipropagatorf{-1}{0}{-150}{1.2}{} 
\phipropagatorf{1}{0}{30}{1.2}{}
\phipropagatorf{1}{0}{-30}{1.2}{}
\drawphiavdifdiagbab{0}{0}{0}
	
\node at (-3.5,1) {$\bf 1$};	
\node at (-3.5,-1) {$\bf 2$};	
\node at (3.5,1) {$\bf 3$};	
\node at (3.5,-1) {$\bf 4$};	
\end{scope}
 
\end{tikzpicture}
\end{center}
\caption{Renormalization of the Lindblad violating quartic coupling in the average-difference basis}
\label{fig:avgdiffquartic10} 
\end{figure}

Similarly, one can calculate the beta function of the Lindblad violating term $(\im  \ \lambda_3+ 3\ \im  \ \sigma_3)$. Divergent diagrams for one particular channel
is depicted in figure \ref{fig:avgdiffcubic10}. There are two more channels. The total contribution is given as
 
\begin{equation}
\begin{split}
2&(\im  \ \lambda_3+ 3\ \im  \ \sigma_3)\\
&+ 2(\im  \ \lambda_3+ 3\ \im  \ \sigma_3)  \times (-i) (\re \ \lambda_4+ 2\ \re \ \sigma_4) [B_{af}(k_1)+B_{af}(k_2)+B_{af}(k_3)] \\
& + 2(\im  \ \lambda_4+ 4\ \im  \ \sigma_4 -3 \lambda_\Delta)  \times (-i) (\re \ \lambda_3+ \ \re \ \sigma_3)[B_{ab}(k_1)+B_{ab}(k_2)+B_{ab}(k_3)]
\end{split}
\end{equation}
Following the standard procedures, we can very easily compute the one loop beta function and it is given by 
\begin{equation}
\begin{split}
\frac{d}{d\ln\,  \mu}(\im  \ \lambda_3+ 3\ \im  \ \sigma_3)
&=\frac{3}{(4\pi)^2} (\im  \ \lambda_3+ 3\ \im  \ \sigma_3) (\re \ \lambda_4+ 2\ \re \ \sigma_4)\\
&+\frac{3}{(4\pi)^2} (\im  \ \lambda_4+ 4\ \im  \ \sigma_4 -3 \lambda_\Delta) (\re \ \lambda_3+ \ \re \ \sigma_3)
\end{split}
\end{equation}
Now, let us we compute the beta function of Lindblad violating term $(\im\ \lambda_4+ 4\ \im\ \sigma_4 -3 \lambda_\Delta)$ by computing the process 
$\phidif\phidif \rightarrow \phidif \phidif$ via $\phiav^4$ vertex, which is depicted in figure \ref{fig:avgdiffquartic10}. The total contribution is given by

\begin{equation} 
\begin{split}
2&(\im  \ \lambda_4+ 4\ \im  \ \sigma_4 -3 \lambda_\Delta)\\
& + (2\ \text{diagrams}) (3\ \text{channels})\times 2(\im  \ \lambda_4+ 4\ \im  \ \sigma_4 -3 \lambda_\Delta)\\
&\times (-i) (\re \ \lambda_4+ 2\ \re \ \sigma_4) \times \frac{i}{2(4\pi)^2}\left(\frac{2}{d-4} +   \ln\  \frac{1}{4\pi e^{-\gamma_E}}\right)\\
\end{split}
\end{equation}
Hence, the one loop beta function is
\begin{equation}
\begin{split}
\frac{d}{d\, \ln\, \mu}(\im  \ \lambda_4+ 4\ \im  \ \sigma_4 -3 \lambda_\Delta)
&=\frac{6}{(4\pi)^2} (\im  \ \lambda_4+ 4\ \im  \ \sigma_4 -3 \lambda_\Delta)  (\re \ \lambda_4+ 2 \re \ \sigma_4) 
\end{split}
\end{equation}
The usefulness of average-difference basis is quite evident from these three calculations. The complete computation in average-difference basis can be found in appendix \ref{sec:appendixavgdiff}.

\subsection{Lindblad condition is never violated by perturbative corrections}
\label{subsec:avgdifallorder}

In this section, we will give an all order perturbative argument for why Lindblad conditions are not violated to arbitrary order 
in perturbation theory. Consider the action in the average-difference basis given in \eqref{avdif2}. From this expression we note  that all the Lindblad violating couplings of open $\phi^3+\phi^4$ theory  appear as the coupling constants for the pure average vertices. Our argument 
below can be easily extended to any open QFT which has the property that all Lindblad violating vertices are pure average vertices.
Note that the converse is always true in an open EFT : any pure average vertex is necessarily  Lindblad violating (since it contributes to 
the action even in the $\phir=\phil$ limit).

Now we want to show that if we start from the open $\phi^3+\phi^4$ theory, then the Lindblad condition(s) are never violated under 
perturbative corrections using the fact that they are all of pure average type. In other words, by assuming that  there is
no pure average vertex at tree level, and that there is no difference-difference propagator, we would like to  show that such a vertex/propagator
can never be generated under loop corrections.  We will prove it in three steps. 

We will begin with an
\begin{itemize}
 \item {\bf Assumption:} At tree level, one has no pure-average vertex and no pure-difference propagator. All Lindblad violating couplings are assumed to 
be pure average vertices and hence are taken to vanish at tree level.
 \item {\bf  Statement 1 }: Say we assume that  there is no pure-average 1PI 2 point vertex generated at $g$ loop. Then, it implies that 
 there is no pure-difference 1 PI propagator generated at $g$ loop.\\
 {\bf Proof }: According to our starting assumption, there is no tree level pure-difference propagator. Such a propagator can then only be generated by a Feynman diagram of type depicted in figure \ref{fig:nonrenorm1}. 

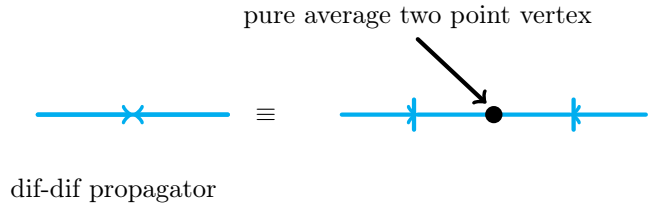
\begin{figure}[ht] 
\begin{center}

\begin{tikzpicture} 

\phipropagatord{0}{0}{0}{1.25}{}
\node at (1,-1) {dif-dif propagator};

\node at (3,0) {$\equiv$};
\phipropagatorb{4}{0}{0}{1}{}
\phipropagatorf{6}{0}{0}{1}{}

\filldraw (6,0) circle (.1cm);
 
\draw [ultra thick] [->] (5,1) -- (5.9,.15);
\node at (5,1.3) {pure average two point vertex};
 	  
\end{tikzpicture}	

\end{center}
\caption{Pure difference propagator from loop correction(s)}
\label{fig:nonrenorm1} 
\end{figure}

So, any contribution to pure-difference 1 PI propagator is of the form
\begin{center}
 (difference to average tree level propagator)$\times$ (pure-average 1PI 2 point vertex) $\times$ (average to diff tree level propagator).
\end{center}
Thus, if there are no pure-average 1PI vertices, there are no  pure-difference 1PI propagator.(QED)
\item  {\bf  Statement 2 }:  If there is no  pure-average 1PI vertex at $g$ loops, there is no such vertex at $g+1$ loops. This statement 
via induction, then implies that  pure-average 1PI vertices are never generated at any loop order. By our previous statement, this implies
then that  pure-difference 1 PI propagators are also never generated at any loop order. In order to prove this, we first prove 
\begin{itemize}
\item {\bf  Statement 2a}:  Consider a Feynman diagram contributing to a pure-average 1PI vertex. There must be at least one vertex (internal or external) such that the following is true :  there exists a closed path completely made of $b$-type propagators which begins and ends in that vertex (it may or may not pass through  external vertices). \\
{\bf  Proof :} Since we are considering a diagram contributing to a pure-average 1PI vertex, all the external propagators at every external vertex are of $a$-type. By our assumption, there is no pure-average 1PI vertex. Thus, there should be at least one $d$-type line leaving at a given external vertex. Since there is  no pure-difference propagator, this $d$-type line necessarily converts itself into an $a$-type line : thus the  propagator is of  average-difference $b$-type with the arrow leaving the external vertex. This propagator thus ends as an $a$-type line either in the vertex that one began with, or another external vertex or an internal vertex.\\
In the first case, we have obtained the desired result : there exists a closed path completely made of $b$-type propagators which begins and ends in that vertex.\\
In the second case, we note that the external vertex has an external $a$-type  leg, and the $b$-type propagator which went from the starting vertex also ends with an $a$-type leg on the second vertex. Since there are no pure-average vertices, there should necessarily be a $d$-type leg which is going out of the second vertex. The $d$-type leg can again only be a part of an $b$-type propagator since there is no pure-difference propagator. A similar argument also applies to the third case of an internal vertex.\\
We can now follow the  $b$-type propagators and repeat the argument again. This process would terminate (since we are looking at a finite graph) 
and we would return to some vertex on the path for second time, closing the loop. (QED)\\
\item {\bf  Statement 2b}: Any Feynman diagram with a closed path completely made of $b$-type propagators is zero. \\
{\bf Proof}: In position space, $b$-type propagators are just retarted propagators. The $b$-type propagator, using equation \eqref{avdif3}, in position space is given by, 
\begin{equation}
\begin{split}
& \int \frac{d^4p}{(2\pi)^4} \left( \frac{-i}{p^2+m^2-i\varepsilon} - 2\pi \delta_{-}(p^2+m^2) \right) e^{ip(x-y)} \\
&= \int \frac{d^4p}{(2\pi)^4} \frac{-i}{p^2+m^2-i\varepsilon} e^{ip(x-y)} - \int \frac{d^4p}{(2\pi)^3} \Theta(-p_0) \delta(p^2+m^2) e^{ip(x-y)} \\
&= \int \frac{d^3p}{(2\pi)^3} e^{i\vec{p} \cdot (\vec{x}-\vec{y})} \left(\Theta(x_0 - y_0) \frac{e^{-i\omega_p (x_0-y_0)}}{2 \omega_p} + \Theta(y_0 - x_0) \frac{e^{i\omega_p (x_0-y_0)}}{2 \omega_p}\right) \\
&- \int \frac{d^3p}{(2\pi)^3} e^{i\vec{p} \cdot (\vec{x}-\vec{y})} \frac{e^{i\omega_p (x_0-y_0)}}{2 \omega_p} \left(\Theta(x_0 - y_0) + \Theta(y_0 - x_0) \right) \\
&= \Theta(x_0 - y_0) \int \frac{d^3p}{(2\pi)^3} e^{i\vec{p} \cdot (\vec{x}-\vec{y})} \left(\frac{e^{-i\omega_p (x_0-y_0)}}{2 \omega_p} - \frac{e^{i\omega_p (x_0-y_0)}}{2 \omega_p} \right) \\
&= G_R(x-y) 
\end{split}
\end{equation}
where, $\omega_p = (\vec{p}^2+m^2)^{\frac{1}{2}}$ and $G_R(x-y)$ denotes the retarted propagator. \\
We will now use the result that a closed loop of retarted propagators is identically zero. This statement is a part of the Feynman tree theorem \cite{Feynman:1972mt}. A closed loop of retarted propagators can be written as  
\begin{equation}
\begin{split} 
\int \frac{d^dp}{(2\pi)^d} \prod_i G_R^i (p+k_i) = 0
\end{split}
\end{equation}
where, $k_i$ denotes the external momenta. Since all the poles in a retarded propagator are below the real $p_0$ axis. So, one can close the contour from above, picking no residues and, as a result, the integral vanishes. (QED)
\end{itemize}
\item Statement 2a and 2b imply that if there is no pure average vertex operator or a pure difference propagator at $g$ loop then there will be no such vertex/propagator at $g+1$ loop.  From this we conclude, via induction, that if there is no Lindblad violating coupling at tree level, such a coupling is never generated by perturbative corrections.(QED)
\end{itemize}

This then concludes our argument in the average-difference basis that the Lindblad violating couplings are never generated in loops. The readers familiar 
with cutting rule arguments ala Veltman in unitary theories  would recognise the style of the above argument. The proof that difference operators 
decouple at arbitrary loops in a unitary theory, or equivalently the proof that Keldysh causal structure is preserved under loop corrections for a 
unitary theory bear a close resemblance to the proof above. The surprise here is that the argument goes through even without assuming unitarity. We also note the perturbative nature of the above argument, since it invokes the fact that the graphs at any given loop order are finite. It would be interesting to try
and give a non-perturbative proof of the statement of this section.

With this formal proof in hand, in next section, we will now turn to a preliminary study of the RG running in our open EFT. The interesting question is to  map out behaviour novel to open EFTs which cannot be found in unitary QFTs.

\section{Running of the coupling constants and physical meaning}

In this section, we will perform an analysis of the running of couplings from our $1$-loop beta functions. Given the many couplings involved in the 
the RG equations in \eqref{eq:betamassintro},\eqref{eq:betacubicintro} and \eqref{eq:betaquarticintro}, we will begin with
a judicious rewriting of our equations. Once the Lindblad conditions are imposed, we obtain the following count for the couplings :  
\begin{enumerate}
	\item 5 quartic couplings + 1 lindblad condition \eqref{eq:LindbladCondn} $\implies$ 4 independent quartic couplings 
	\item 4 cubic couplings+ 1 lindblad condition \eqref{eq:LindbladCondn} $\implies$ 3 independent cubic couplings 
	\item 3 mass terms + 1 lindblad condition \eqref{eq:LindbladCondn}  $\implies$ 2 independent mass terms
\end{enumerate}

Our RG equations for these $9$ independent variables can then be recast into the following convenient form :
\begin{equation}
\label{eq:Coupleddiff}
\begin{split}
\beta_{\re \ \lambda_4+2\re \ \sigma_4} &= \frac{3}{(4\pi)^2}\left(\re \ \lambda_4+2\re \ \sigma_4\right)^2\\
\beta_{\im  \ \lambda_4+\im  \ \sigma_4} &= \frac{5}{(4\pi)^2}\left(\im  \ \lambda_4+\im  \ \sigma_4\right)\left(\re \ \lambda_4+2\re \ \sigma_4\right) \\
\beta_{\re \ \lambda_4} &= \frac{1}{3(4\pi)^2}\left(9\re \ \lambda_4\left(\re \ \lambda_4+2\re \ \sigma_4 \right) - 8\left(\im  \ \lambda_4+\im  \ \sigma_4\right)^2\right)\\
\beta_{\im  \ \lambda_4-4\im  \ \sigma_4} &= \frac{10}{(4\pi)^2}\re \ \lambda_4 \left(\im  \ \lambda_4 +\im  \ \sigma_4 \right)\\
\beta_{\re \ \lambda_3+\re \ \sigma_3} &= \frac{3}{(4\pi)^2}\left(\re \ \lambda_3+\re \ \sigma_3\right)\left(\re \ \lambda_4+2\re \ \sigma_4\right)\\
\beta_{\im  \ \lambda_3 } &=-3\beta_{\im  \ \sigma_3}= \frac{1}{(4\pi)^2}\left(3\left(\im  \ \lambda_4+\im  \ \sigma_4\right)\left(\re \ \lambda_3+\re \ \sigma_3\right) + 2\im  \ \lambda_3 \left(\re \ \lambda_4+2\re \ \sigma_4\right)\right)\\
\beta_{\re \ \lambda_3-\re \ \sigma_3} &=\frac{1}{3(4\pi)^2}\left( 9\re \ \lambda_4 \left(\re \ \lambda_3+\re \ \sigma_3\right) - 8\im  \ \lambda_3 \left(\im  \ \lambda_4+\im  \ \sigma_4\right)\right)\\
\beta_{\re \ m^2} &=\frac{1}{(4\pi)^2}\left( \left(\re \ \lambda_3+\re \ \sigma_3\right)^2 + \re \ m^2 \left(\re \ \lambda_4+2\re \ \sigma_4\right)\right)\\
\beta_{\im  \ m^2} &= \beta_{ m^2_{\Delta}}=\frac{2}{3(4\pi)^2}\left( 2\im  \ \lambda_3 \left(\re \ \lambda_3+\re \ \sigma_3\right) + \re \ m^2 \left(\im  \ \lambda_4+\im  \ \sigma_4\right)\right)
\end{split}
\end{equation}
Note the simple structure of the  $9$ coupled  differential equations given above. We have ordered them such that the $j$th equation depends only on the variables appearing in the first $j-1$ equations.  As a result, a step by step method of solution becomes viable : one can start by solving the first equation for a given initial condition and then use the solution of the first equation as an input to solve the second equation and so on, for all the subsequent equations.

The first, the second and the fifth equation imply the existence of the fixed point, given by  
\begin{eqnarray}
&&\re \ \lambda_4 = -2\re \ \sigma_4, 
\qquad\qquad
\im  \ \lambda_4 = -\im  \ \sigma_4
\\
&& \qquad\qquad
\re \ \lambda_3 = -\re \ \sigma_3. 
\end{eqnarray}
To analyse the nature of this fixed point, we turn to the first equation which drives them all. It can be written as  
\begin{eqnarray}
\beta_{|\re \ \lambda_4+2\re \ \sigma_4|} &=\textrm{sign}(\re \ \lambda_4+2\re \ \sigma_4) \frac{3}{(4\pi)^2}\left|\re \ \lambda_4+2\re \ \sigma_4\right|^2
\end{eqnarray}
This implies that depending on the sign of the initial value, $\re \ \lambda_4+2\re \ \sigma_4$ either increases or  decreases as we go to higher energy scales. As we will see, this sign controls whether the theory is UV free or IR free.  We recognise in the RG equation for $\re \ \lambda_4+2\re \ \sigma_4$
the usual $\phi^4$ coupling  RG equation with $\re \ \lambda_4+2\re \ \sigma_4$ serving as an effective $\phi^4$ coupling. The asymptotically free regime 
and the negative beta function corresponds to this  effective $\phi^4$ coupling turning negative and is hence akin to the theory studied by Symanzik \cite{Symanzik1973}.\footnote{We would like to thank Nima Arkani-Hamed for a discussion of this issue and bringing the relevant literature to our attention.} 

We will begin by performing a linearised analysis around the fixed point mentioned above and follow it up with a more detailed 
numerical analysis.

\subsection{Linearized analysis around the fixed point}

In this section, we study linearized beta functions around the fixed points and find the eigenvalues and eigenvectors of the beta function matrix. Consider small deviations around the fixed points
\begin{equation}
\begin{split}
\re \ \lambda_4+2\, \re \ \sigma_4 &= \epsilon_1 \\
\im  \ \lambda_4+\im  \ \sigma_4 &= \epsilon_2 \\
\re \ \lambda_3+\re \ \sigma_3 &= \epsilon_3
\end{split}
\end{equation}
where, we have assumed 
\begin{eqnarray}
|\epsilon_i|<< 1\qquad \qquad \ \forall \ i=1,2,3 
\end{eqnarray}
% \begin{equation}
% \begin{split}
% \beta_{\re \ \lambda_4+2\re \ \sigma_4} &= 0\\
% \beta_{\im  \ \lambda_4+\im  \ \sigma_4} &= 0 \\
% \beta_{\re \ \lambda_4} &= \frac{1}{(4\pi)^2}\left(3 \epsilon_1 \re \ \lambda_4 \right)\\
% \beta_{\im  \ \lambda_4-4\im  \ \sigma_4} &= \frac{1}{(4\pi)^2}\left(10 \epsilon_2 \re \ \lambda_4 \right) \\
% \beta_{\re \ \lambda_3+\re \ \sigma_3} &= 0 \\
% \beta_{\im  \ \lambda_3} &=\frac{1}{(4\pi)^2}\left(2 \epsilon_1 \im  \ \lambda_3 \right) \\
% \beta_{\re \ \lambda_3-\re \ \sigma_3} &=\frac{1}{3(4\pi)^2}\left( 9 \epsilon_3 \re \ \lambda_4 - 8 \epsilon_2 \im  \ \lambda_3 \right)\\
% \beta_{\re \ m^2} &=\frac{1}{(4\pi)^2}\left(3 \epsilon_1 \re \ m^2 \right)\\
% \beta_{\im  \ m^2} &=\frac{2}{3(4\pi)^2}\left( 2 \epsilon_3 \im  \ \lambda_3 + \epsilon_2 \re \ m^2 \right)
% \end{split}
% \end{equation}
The linearized beta functions for $\re \ \lambda_4+2\re \ \sigma_4$, $\im  \ \lambda_4+\im  \ \sigma_4$ and $\re \ \lambda_3+\re \ \sigma_3$ are zero. This suggests that $\epsilon_1$, $\epsilon_2$ and $\epsilon_3$ remain constant (i.e., they are marginal couplings at the fixed point).

The rest of the  linearized beta functions about the fixed point  can be written as
\begin{equation}
\begin{split}
\frac{d}{dt} \mathcal{G}(t) = \mathcal{B} \mathcal{G}(t)
\end{split}
\end{equation}
in terms of the RG time $t \equiv \frac{\ln\ \mu}{(4\pi)^2}$. Here we have defined the coupling constant matrix  $\mathcal{G}$ as 
\begin{equation}
\mathcal{G}
\equiv
\begin{pmatrix}
 \re \ \lambda_4 \\ \im  \ \lambda_4-4\im  \ \sigma_4 \\ \im  \ \lambda_3 \\ \re \ \lambda_3-\re \ \sigma_3
 \\ \re \ m^2 \\ \im  \ m^2
\end{pmatrix}
\end{equation}
and the beta function matrix $\mathcal{B}$ is given by 
\begin{equation}
\mathcal{B}
\equiv
\begin{pmatrix}
   3\epsilon_1 & 0 & 0 & 0 & 0 & 0 \\ 
   10\epsilon_2 & 0 & 0 & 0 & 0 & 0 \\
   0 & 0 & 2\epsilon_1 & 0 & 0 & 0 \\ 
   3\epsilon_3 & 0 & -\frac{8\epsilon_2}{3} & 0 & 0 & 0 \\ 
   0 & 0 & 0 & 0 & \epsilon_1 & 0 \\ 
   0 & 0 & \frac{4\epsilon_3}{3} & 0 & \frac{2\epsilon_2}{3} & 0 
\end{pmatrix}
\end{equation}

The six eigenvalues of the matrix $\mathcal{B}$ are - $0, 0, 0, \epsilon_1, 2\epsilon_1, 3\epsilon_1$.
The corresponding eigenvectors are given by:
\begin{equation}
\begin{pmatrix}
 0 \\ 1 \\ 0 \\ 0 \\ 0 \\ 0  
\end{pmatrix}
,\
\begin{pmatrix}
 0 \\ 0 \\ 0 \\ 1 \\ 0 \\ 0  
\end{pmatrix}
,\
\begin{pmatrix}
 0 \\ 0 \\ 0 \\ 0 \\ 0 \\ 1  
\end{pmatrix}
,\
\begin{pmatrix}
 0 \\ 0 \\ 0 \\ 0 \\ 3\epsilon_1 \\ 2\epsilon_2 
\end{pmatrix}
,\
\begin{pmatrix}
 0 \\ 0 \\ 3\epsilon_1 \\ -4\epsilon_2 \\ 0 \\ 2\epsilon_3  
\end{pmatrix}
,\
\begin{pmatrix}
 3\epsilon_1 \\ 10\epsilon_2 \\ 0 \\ 3\epsilon_3 \\ 0 \\ 0  
\end{pmatrix}
\end{equation}
\\
Eigenvalues of the matrix $\mathcal{B}$ suggest that three out of the six coupling combinations are marginal at the fixed point.  The asymptotic behavior of the rest of  the variables depend only on the sign of $\epsilon_1$ or $\re \ \lambda_4+2\re \ \sigma_4$. A positive $\epsilon_1$ would mean that the couplings become relevant in UV, whereas a negative $\epsilon_1$ would mean that the couplings are relevant in IR. This conforms to the intuition we presented in the beginning of this section : the coupling  $\re \ \lambda_4+2\re \ \sigma_4$  runs like the quartic coupling of an ordinary $\phi^4$ theory : the theory is IR free for positive value of this combination whereas it is UV free (asymptotically free) for negative value of this combination. This coupling then drives all other couplings to be either IR free or asymptotically free. 

Let us now extend our analysis beyond the linearised regime around tthe fixed points, given by  
\begin{eqnarray}
&&\re \ \lambda_4 = -2\re \ \sigma_4, 
\qquad\qquad
\im  \ \lambda_4 = -\im  \ \sigma_4
\\
&& \qquad\qquad
\re \ \lambda_3 = -\re \ \sigma_3. 
\end{eqnarray}
We will begin by re-examining eqn.\eqref{eq:Coupleddiff} to gain more qualitative insight on the nature of running in this theory:
\begin{enumerate}
\item We will begin with the statement that,  depending on the sign of the initial value, $\re \ \lambda_4+2\re \ \sigma_4$ either increases or  decreases as we go to higher energy scales . Thus, we can have two distinct scenarios  
\begin{enumerate}
\item $\re \ \lambda_4 + 2\re \ \sigma_4 > 0$
\item  $\re \ \lambda_4+2\re \ \sigma_4 < 0$.
\end{enumerate}
\item The second equation depends upon the sign of $(\re \ \lambda_4+2\re \ \sigma_4)$ as well as on the sign of the initial value
of $\im  \ \lambda_4+\im  \ \sigma_4$. For instance, keeping a  positive $\re \ \lambda_4+2\re \ \sigma_4$ and a negative initial value
of $\im  \ \lambda_4+\im  \ \sigma_4$ would result in a decreasing  behavior as shown in figure \ref{fig:case2} and figure \ref{fig:case4}. Thus, we have two further sub-cases, depending upon the sign of $\im  \ \lambda_4+\im  \ \sigma_4$. 

\item The third and fourth equation implies that the evolution of $\re \ \lambda_4$ and $\im  \ \lambda_4-4\, \im  \ \sigma_4$ depends only on the values of $\re \ \lambda_4+2\, \re \ \sigma_4$ and $\im  \ \lambda_4+\im  \ \sigma_4$, given the assumption that the imaginary couplings are small compared to the real ones. 
\item The fifth equation is similar to  the second equation. Hence, there will again be two sub-cases. 
\item It's easy to verify that, with the  help of similar reasonings, the remaining equations will not provide us with further sub-cases. 
\end{enumerate}
We found that the key conclusion remains unchanged for $\re \ \lambda_3+\re \ \sigma_3 \gtrless 0$. So we will always be considering the case $\re \ \lambda_3+\re \ \sigma_3 \gtrless 0$ together.  
Thus, we conclude that we can broadly have 8 cases in total and they basically correspond to the two sides of either of these three fixed points: each fixed point will provide two cases and we have $2^3$ cases altogether. 

With this insight, we will proceed to a more detailed numerical analysis.

\subsection{Numerical analysis of RG equations}

In this subsection, we continue our analysis of the various possible cases in the RG evolution equations. It is useful to have a rough criteria to check the validity of our analysis 
and as to when the analysis can be interpreted physically.  We will perform this analysis only for the Lindblad theory, where the coupling constants obey the Lindblad conditions.
We shall always work in a regime where the imaginary couplings are smaller compared to the real ones (since this is the regime where  our beta functions were derived). Moreover, 
we will demand the following bounds 
\begin{equation}
\lambda_{\Delta} < 0,\qquad\qquad m^2_{\Delta} < 0
%,\qquad\qquad \lambda_{\Delta}m^2_{\Delta} - (\im  \ \sigma_3)^2 > 0
\label{couplingbound1}
\end{equation}
which seem to be reasonable from the point of microsocopic unitarity\cite{Baidya:2017ab}. We will deem the couplings which do not satisfy this bound as unphysical
in the following.  The initial conditions are chosen keeping these physical bounds in consideration   and we shall analyse the dynamics corresponding to all the possible behaviors. 

\subsubsection{I: $\re \ \lambda_4+2\re \ \sigma_4 > 0$, $\im  \ \lambda_4+\im  \ \sigma_4 > 0$ and $\re \ \lambda_3+\re \ \sigma_3 \gtrless 0$}
\begin{figure}
\centering
\subfloat[]{\includegraphics[clip=true, totalheight=0.4\textheight, width=1\textwidth]{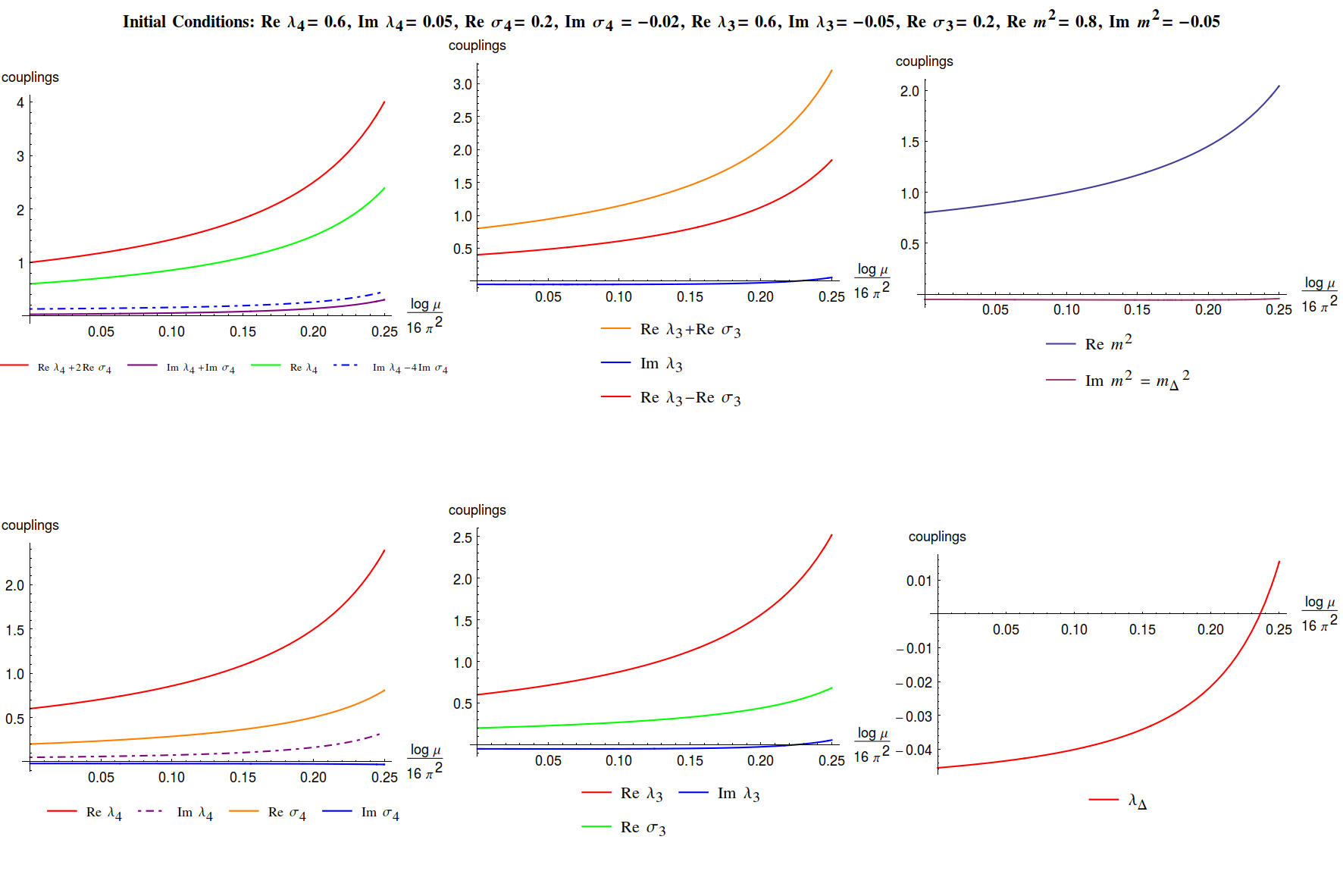}} \\
\subfloat[]{\includegraphics[clip=true, totalheight=0.4\textheight, width=1\textwidth]{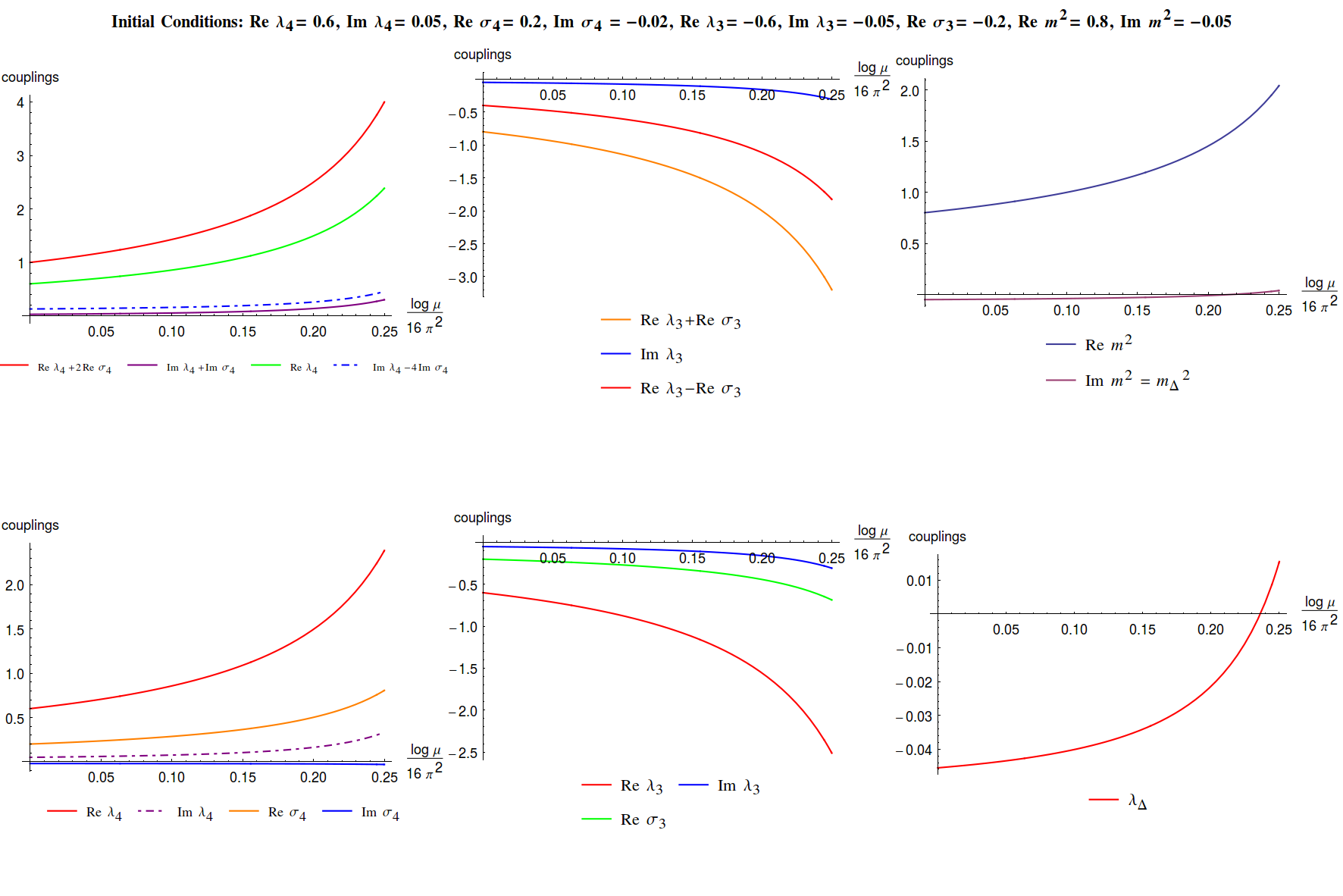}}
\caption{Figures showing the evolution of all the couplings for the case $\re \,  \lambda_4+2\re \,  \sigma_4 > 0$, 
$\im  \,  \lambda_4+\im  \,  \sigma_4 > 0$ and the two subcases - $\re \,  \lambda_3+\re \,  \sigma_3 > 0$(figure a) and
$\re \,  \lambda_3+\re \,  \sigma_3 < 0$ (figure b)}
\label{fig:case1}
\end{figure}

The first equation in \eqref{eq:Coupleddiff} tells us that the sign of $\re \ \lambda_4+2\re \ \sigma_4$ will remain positive in this regime. In particular,
$\re \ \lambda_4+2\re \ \sigma_4$ evolves in the same way as $\lambda_{\text{Unitary}}$ \footnote{$\lambda_{\text{Unitary}}$ denotes the coupling constant of an unitary $\phi^4$ theory} . Now, from the second equation, one can see that
$\im  \ \lambda_4+\im  \ \sigma_4$ will keep increasing if it starts at a positive initial value, but at a slower rate compared to $\re \ \lambda_4+2\re \ \sigma_4$. Similarly, from the third and fourth equation, one can see that, keeping in mind the
assumptions, both $\re \ \lambda_4$ and $\im  \ \lambda_4-4\im  \ \sigma_4$ will increase in the way as shown in figure \ref{fig:case1}. Note here that
$\im  \ \lambda_4-4\im  \ \sigma_4$ rises faster than $\im  \ \lambda_4+\im  \ \sigma_4$ and thus, it results in a continuously increasing 
$\im  \ \lambda_4$ and a decreasing $\im  \ \sigma_4$ as shown in the second diagram in figure  \ref{fig:case1}. Also, the increase of $\im  \ \lambda_4$ is faster than the decrease of $\im  \ \sigma_4$ and thus, under the RG flow, $\lambda_{\Delta}$ becomes positive, which is unphysical. Evolution of the remaining cubic couplings and 
mass terms variables will not affect the evolution $\im  \ \lambda_4$ and $\im  \ \sigma_4$. So, both the sub-cases due to different signs of
$\re \ \lambda_3+\re \ \sigma_3 $ would have a positive $\lambda_{\Delta}$ and thus, these two cases can be deemed as unphysical.

\subsubsection{II: $\re \ \lambda_4+2\re \ \sigma_4 > 0$, $\im  \ \lambda_4+\im  \ \sigma_4 < 0$ and $\re \ \lambda_3+\re \ \sigma_3 \gtrless 0$ }
\begin{figure}
\centering
\subfloat[]{\includegraphics[clip=true, totalheight=0.4\textheight, width=1\textwidth]{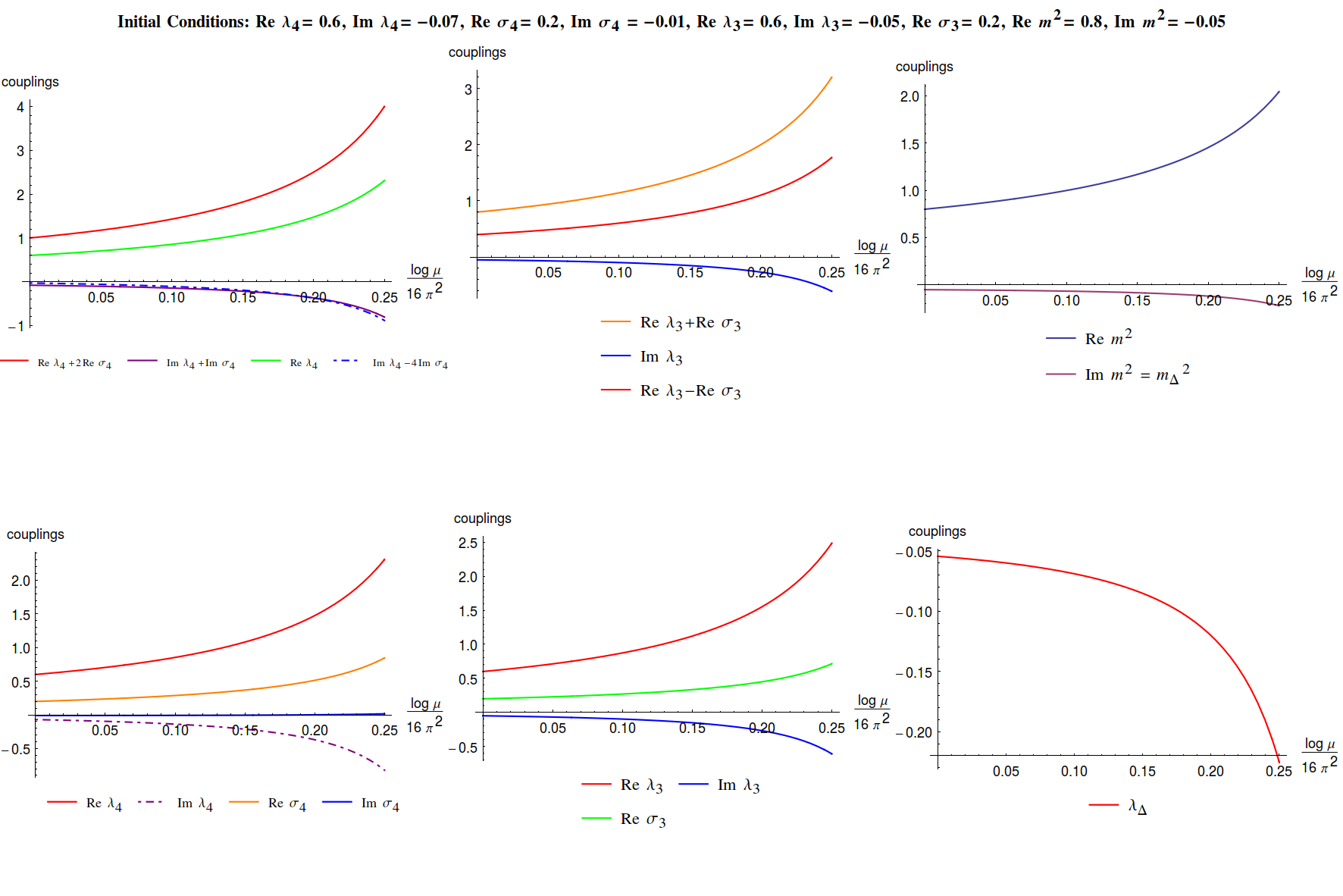}} \\
\subfloat[]{\includegraphics[clip=true, totalheight=0.4\textheight, width=1\textwidth]{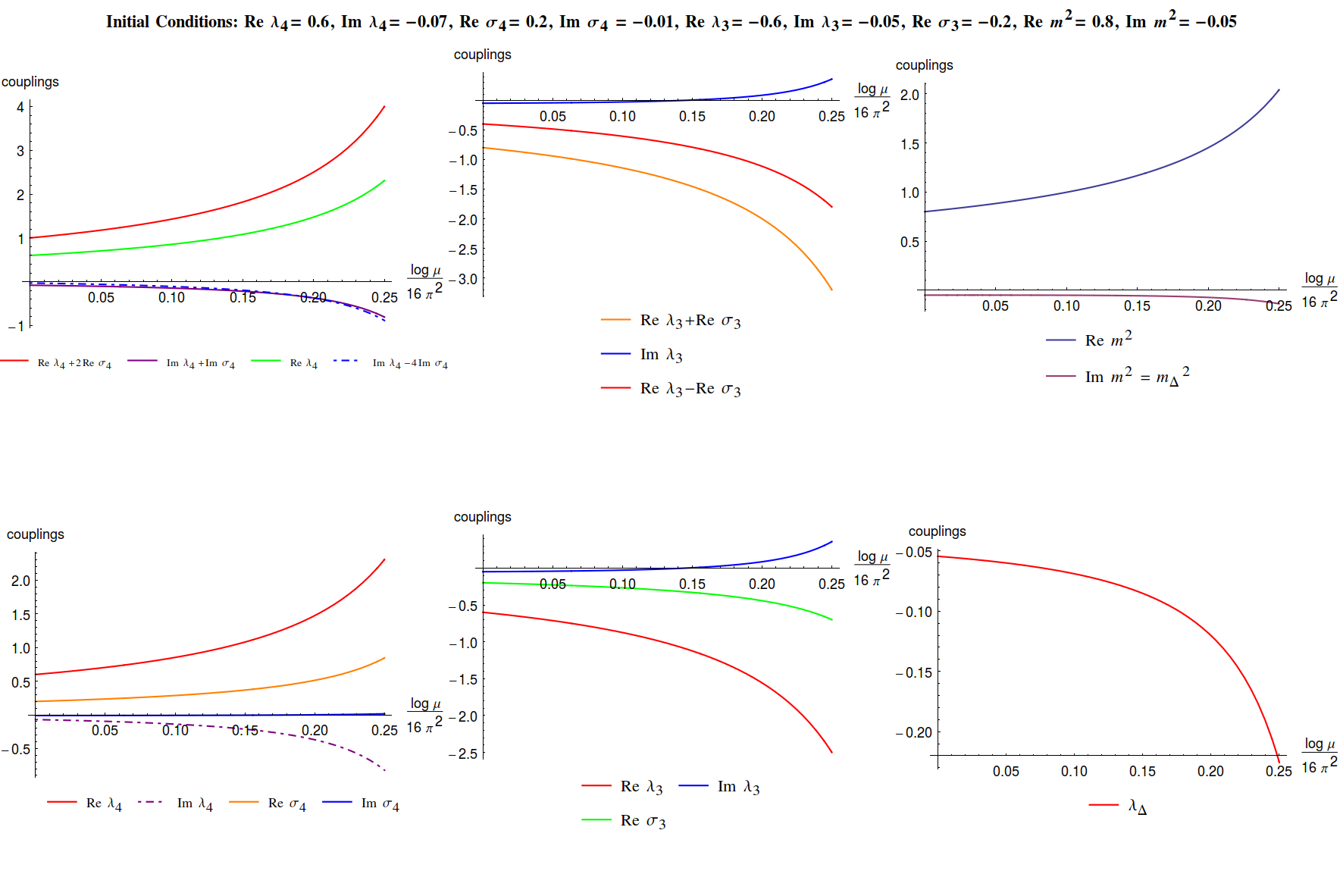}}
\caption{Figures showing the evolution of all the couplings for the case $\re \,  \lambda_4+2\re \,  \sigma_4 > 0$, 
$\im  \,  \lambda_4+\im  \,  \sigma_4 < 0$ and the two subcases - $\re \,  \lambda_3+\re \,  \sigma_3 > 0$(figure a) and
$\re \,  \lambda_3+\re \,  \sigma_3 < 0$(figure b)}
\label{fig:case2} 
\end{figure}

The evolution of each variable for this case is depicted in figure \ref{fig:case2}. We observe that the couplings do not violate the physical conditions throughout. 
One can see that the couplings become stronger in the UV and attain a Landau pole.

\subsubsection{III: $\re \ \lambda_4+2\re \,  \sigma_4 < 0$, $\im  \,  \lambda_4+\im  \,  \sigma_4 > 0$ and $\re \,  \lambda_3+\re \,  \sigma_3 \gtrless 0$ }
\begin{figure}
\centering
\subfloat[]{\includegraphics[clip=true, totalheight=0.4\textheight, width=1\textwidth]{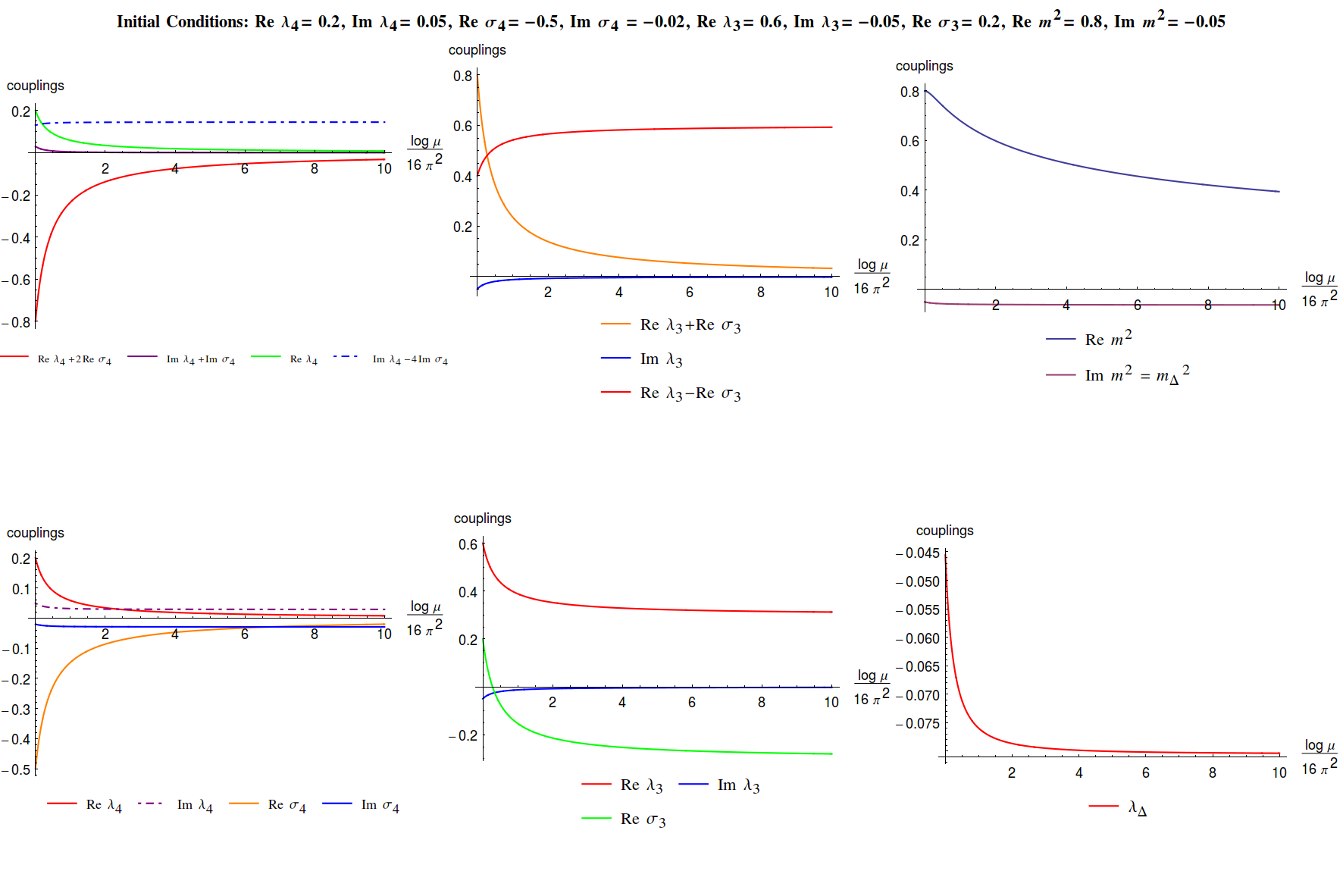}} \\
\subfloat[]{\includegraphics[clip=true, totalheight=0.4\textheight, width=1\textwidth]{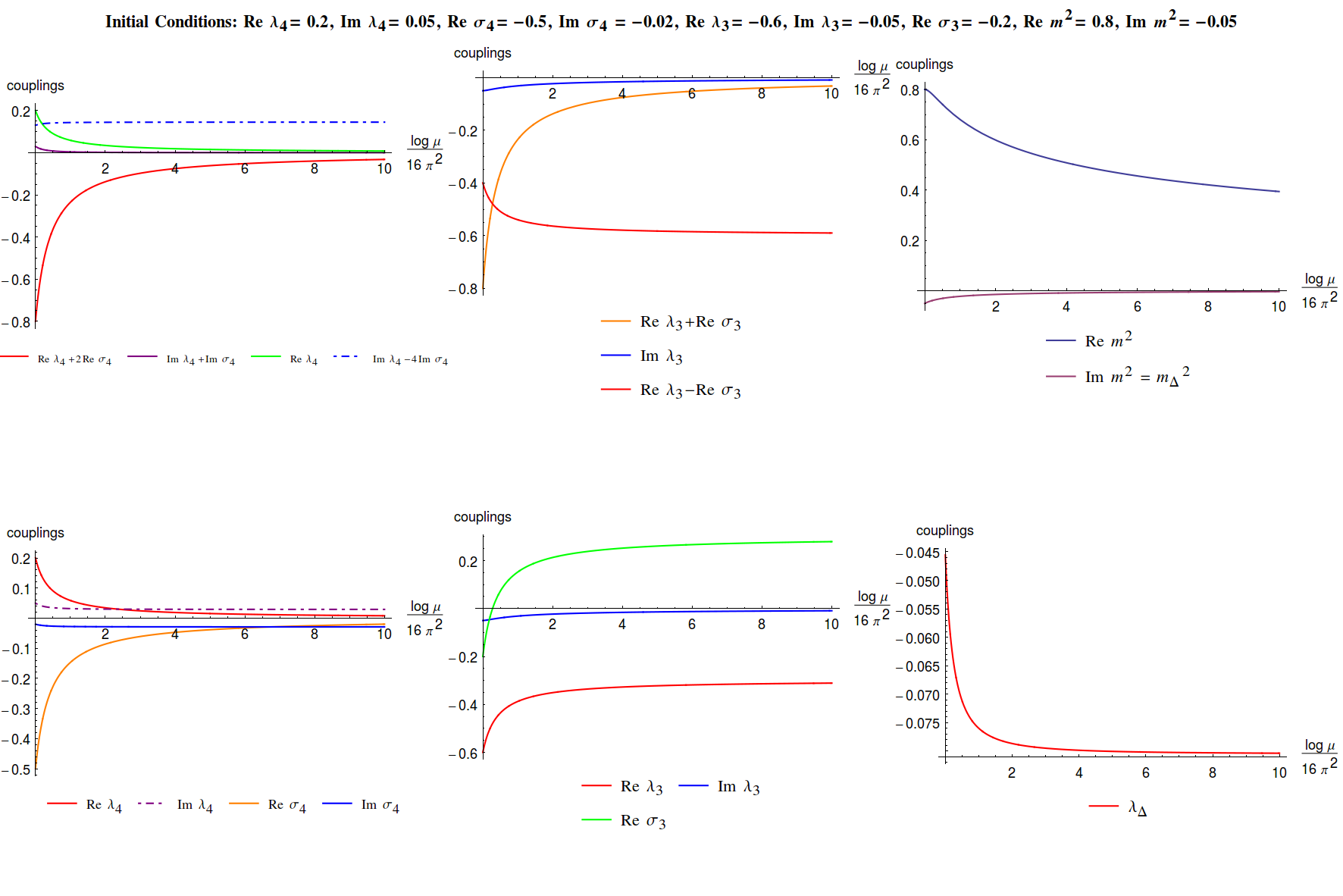}}
\caption{Figures showing the evolution of all the couplings for the case $\re \,  \lambda_4+2\re \,  \sigma_4 < 0$, 
$\im  \,  \lambda_4+\im  \,  \sigma_4 > 0$ and the two sub-cases - $\re \,  \lambda_3+\re \,  \sigma_3 > 0$(figure a) and
$\re \,  \lambda_3+\re \,  \sigma_3 < 0$(figure b)}
\label{fig:case3}
\end{figure}

This is a case where the couplings are relevant in IR and remain within the physical bounds throughout as can be observed in figure \ref{fig:case3}. In this case, 
$\re \,  \lambda_4+2\re \,  \sigma_4 $ becomes asymptotically free as can be seen from the first equation in \eqref{eq:Coupleddiff}. The second and fifth equation, meanwhile, tells us that $\im  \,  \lambda_4+\im  \,  \sigma_4 $ and $\re \,  \lambda_3 +\re \,  \sigma_3 $ would go to zero as we go to higher energies. 
This would also mean that $\im  \,  \lambda_4-4\,\im  \,  \sigma_4 $ becomes constant as $\im  \,  \lambda_4+\im  \,  \sigma_4 $ goes to zero. $\im  \,  \lambda_4$ 
and $\im  \,  \sigma_4$ become constant at higher energies and $\lambda_{\Delta}$ attains a fixed point. With similar reasonings, one can predict
the behavior of other couplings.

\subsubsection{IV: $\re \,  \lambda_4+2\, \re \,  \sigma_4 < 0$, $\im  \,  \lambda_4+\im  \,  \sigma_4 < 0$ and $\re \,  \lambda_3+\re \,  \sigma_3 \gtrless 0$}
\begin{figure}
\centering
\subfloat[]{\includegraphics[clip=true, totalheight=0.4\textheight, width=1\textwidth]{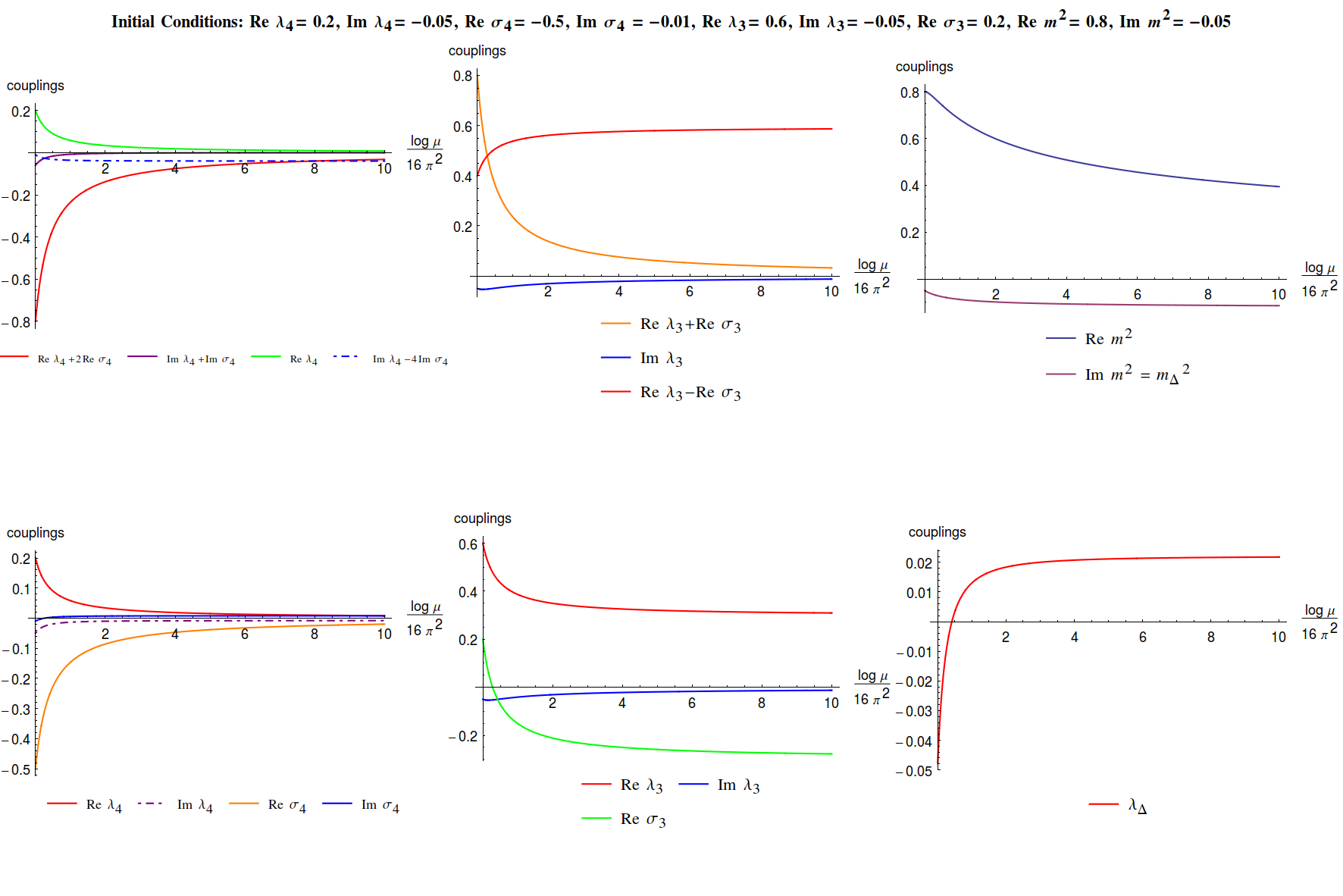}} \\
\subfloat[]{\includegraphics[clip=true, totalheight=0.4\textheight, width=1\textwidth]{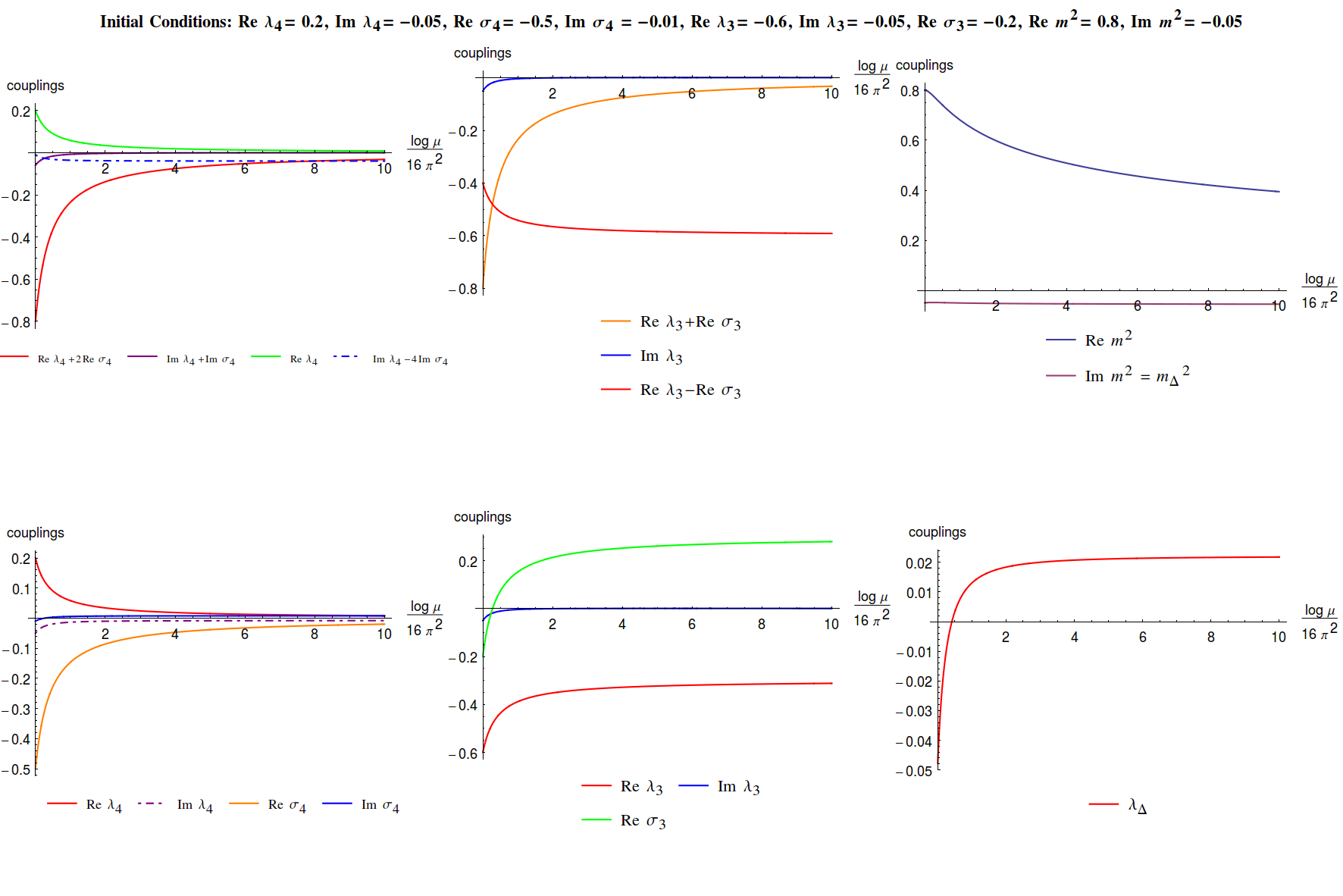}}
\caption{Figures showing the evolution of all the couplings for the case $\re \ \lambda_4+2\re \ \sigma_4 < 0$, 
$\im  \ \lambda_4+\im  \ \sigma_4 < 0$ and the two sub-cases - $\re \ \lambda_3+\re \ \sigma_3 > 0$(figure (a)) and
$\re \ \lambda_3+\re \ \sigma_3 < 0$(figure b)}
\label{fig:case4}
\end{figure}

One can observe from figure \ref{fig:case4} that this case can be deemed as unphysical as $\lambda_{\Delta}$ attains a positive value. It basically comes about due to the sign of $\im  \,  \lambda_4+\im  \,  \sigma_4$ as can be seen from the second equation in \eqref{eq:Coupleddiff}.

%\paragraph{Summary}
%Here we summarize our analysis 

\begin{table}
\begin{center}
\begin{tabular}{|c|c|c|c|}  
\hline
\hline
$\re \, \lambda_4+2\re \, \sigma_4 $ & $\im  \, \lambda_4+\im  \, \sigma_4$ &$\re \, \lambda_3+\re \, \sigma_3$ & Conclusion  \\
\hline
\hline
$>0$ & $>0$ &  $ \gtrless 0$ &  Unphysical \\ [0.1cm]
$>0$ & $<0$ &  $ \gtrless 0$ &  Landau pole\\ [0.1cm]
$<0$ & $>0$ &  $ \gtrless 0$ &  Relevant in IR\\ [0.1cm]
$<0$ & $<0$ &  $ \gtrless 0$ &  Unphysical \\ [0.1cm]
\hline
\hline
\end{tabular}
\caption{\label{tab:runningcoup} Running of coupling constants} 
\end{center}
\end{table}

%
%\arnote{High energy behavior of scalar four point function} 
%
%The running of coupling constants is very crucial in understanding the high and low energy behavior of a quantum field theory. For example, if there is a Landau pole in the running of the coupling constant then that would imply that the theory is incomplete above certain energy scale. Historically the sign of the beta function in QCD (and hence asymptotic freedom) resolved the apparent paradox that why  free quarks or gluons weren't seen in nature.  One would like to see whether the theory has landau poles or the couplings become asymptotically free at high energies. As has been shown in \avnote{reference}, one does get some physical bounds on the couplings and one need to check if these conditions are violated by the renormalization group flow. In this section, we show a plot of various coupling constants against $\ln\, \mu$ and consider  the behavior for different sets of initial conditions. The beta function describes a dynamical systems. Different initial conditions would correspond to different type of 
%evolution of the dynamical system. 
%
%
%
%Now we list some of the key features of these equations and will try to understand various distinct scenarios of the dynamical system : 

\section{Conclusion and Future directions}
\label{sec:conclusion}

In this work, we considered a simple $\phi^3+\phi^4$ toy model of an open quantum field theory in which the renormalisation and the running of couplings could be studied. By enumerating all power-counting renormalisable terms, we demonstrate that the theory is $1$-loop renormalisable whereby all UV divergences can be absorbed into appropriate counter-terms. This is in analogy with the standard result for a unitary QFT. The novelty lies in the  non-unitary Feynman-Vernon couplings and the corresponding UV divergences which result in a $\beta$ function for such non-unitary
couplings. One of the main results of our paper is that these beta functions surprisingly protect a particular fine-tuning of couplings which is associated with demanding that the non-unitary evolution be that of Lindblad form. We end with an all loop argument on why this protection should extend to any order on perturbation theory.

The work described in this article has various natural extensions - to large $N$ models, to theories with fermions and theories with gauge fields. Given our 
experience with supersymmetric field theories, open versions of supersymmetric theories may well provide an exactly solvable model of an open QFT 
where one can study non-perturbative physics as well as dualities. We hope to return to these issues in the immediate future.

With this work, we hope to have convinced the reader of the charms of hitherto unexplored world of open quantum field theories. In many aspects, they closely
mimic the familiar paradigm of unitary quantum field theories but yet deviate from them in interesting ways. Very basic conceptual issues like renormalisation or anomalies or non-pertubative physics (as that of instantons) are yet ill-understood.

\paragraph{Acknowledgements} 
We would like to thank  David Poulin and John Preskill for their unpublished notes \cite{Preskill:2000cy} and John Preskill for discussions. We thank Chi-Ming Chang, Michael Geracie, Felix Haehl, William R. Kelly, Manas Kulkarni,  Shiraz Minwalla,  David M. Ramirez, Mukund Rangamani and Krishnendu Ray for useful discussions.  We thank Amin A. Nizami for pointing out various typos in the first version of this draft. CJ and RL gratefully acknowledge support from International Centre for Theoretical Sciences, Tata Institute of Fundamental Research (ICTS-TIFR), Bengaluru. A.R. would like to thank ICTS-TIFR, Bengaluru for hospitality during the initial stages of this work. A. would like to thank support from Kishore Vaigyanik Protsahana Yojana (KVPY)  funded by the Department of Science and Technology, Government of India. A. would also like to thank Indian Institute of Science (IISc) and ICTS-TIFR for the hospitality provided during this work. RL would also like to acknowledge his debt to all those who have generously supported and encouraged the pursuit of science in India. 

\appendix

\section{Notations and Conventions}
\label{sec:conventions}

\subsection{Most commonly used acronyms}

\begin{itemize}
\item SK - Schwinger-Keldysh 
\item FV - Feynman-Vernon 
\item PV - Passarino-Veltman 
\item The loop integrals are named as the following. We start from the left(bottom) vertex and move in the counter-clockwise direction for s-channel(t, u channel) diagrams(s).  
\end{itemize}

\subsection{Conventions for Feynman integrals }

Since more general diagrams can appear in this context we will introduce a suitable notation. Following the standard notation \cite{tHooft:1978jhc, Passarino:1978jh}, we will use $A$ for tadpole diagrams and $B$ for bubble diagrams. In addition, we will use $R,L,P,M$ as subscripts to denote the corresponding propagators as present in the diagrams. \ref{fig:microskphipropagator}. 

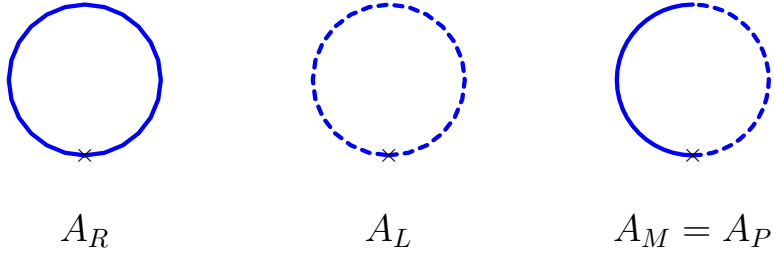
\begin{figure}[ht]
\begin{center}
\begin{tikzpicture}[line width=1 pt, scale=1]
%\begin{scope}[rotate=90]

\begin{scope}[shift={(0,0)}, rotate=-90]
\draw [phir, ultra thick, domain=0:360] plot ({1*cos(\x)}, {1*sin(\x)});
\node at (1,0) {$\times $};
\node at (2,0) {\Large $  A_{R}$};
\end{scope}

\begin{scope}[shift={(4,0)}, rotate=-90]
\draw [phil, ultra thick, domain=0:360] plot ({1*cos(\x)}, {1*sin(\x)});
\node at (1,0) {$\times $};
\node at (2,0) {\Large $A_{L}$};
\end{scope}

\begin{scope}[shift={(8,0)}, rotate=-90]
\draw [phil, ultra thick, domain=0:180] plot ({1*cos(\x)}, {1*sin(\x)});
\draw [phir, ultra thick, domain=180:360] plot ({1*cos(\x)}, {1*sin(\x)});
\node at (1,0) {$\times $};
\node at (2,0) {\Large $A_{M}=A_{P}$};
\end{scope}

\end{tikzpicture} 
\end{center}
\caption{PV One loop $A$ type integrals in SK theory }
\label{fig:pvaing1}
\end{figure}

We are using slightly different normalization from \cite{Passarino:1978jh} for Passarino-Veltman integrals. The relation between our integrals and the integrals in \cite{Passarino:1978jh} is given below  
\begin{eqnarray}
A_0^{PV}&=&	-(4\pi)^2
A_{_{R}}
\\
B_0^{PV}	
&=&i(4\pi)^2B_{_{RR}}
\\
C_0^{PV}	
&=&(4\pi)^2C_{_{RRR}}  
\\
D_0^{PV}	
&=&-i(4\pi)^2D_{_{RRRR}} 
\end{eqnarray} 
We also note that Passarino-Veltman definitions use mostly negative  metric $\eta_{\mu \nu} = \textrm{diag}(1, −1, −1, −1)$ , while in this work we use mostly positive metric $\eta_{\mu \nu} = \textrm{diag}(−1, 1, 1, 1)$. This fact has to be  taken  into account while comparing our expressions in terms of  momentum-square against the standard expressions in discussions of PV integrals.

In SK theory there are four $A$ type integrals. They are given by
\begin{equation}
\begin{split}
A_R(k)&\equiv \  \mu^{4-d}\int\frac{d^d p}{(2\pi)^d i}\ \frac{1}{p^2+m^2-i\varepsilon} \\
A_L(k)&\equiv \  \mu^{4-d}\int\frac{d^d p}{(2\pi)^d i}\ \frac{(-1)}{p^2+m^2+i\varepsilon} \\
A_P(k)&\equiv  \mu^{4-d}\int\frac{d^d p}{(2\pi)^d i}\ 2\pi i\delta_+(p^2+m^2)\\
A_M(k)&\equiv  \mu^{4-d}\int\frac{d^d p}{(2\pi)^d i}\ 2\pi i\delta_-(p^2+m^2)
\end{split}
\end{equation}
Here 
\begin{eqnarray}
\delta_+(p^2+m^2)\equiv	 \Theta(p^0)\delta(p^2+m^2)
\\
\delta_-(p^2+m^2)\equiv	 \Theta(-p^0)\delta(p^2+m^2)
\end{eqnarray}

%and 
%\begin{eqnarray} 
%	
%\frac{\re\, m^2}{16\pi^2}\left[ \frac{2}{(d-4)} -1+\gamma_E+\ln \,\left(\re\,m^2/\mu^2\right)-\ln 4\pi 
%	\right] 
%\end{eqnarray}

% \subsection{Passarino-Veltman $B$ type integral}

In SK theory of one-scalar, there are ten $B$ type integrals (compared to one $B$ type integral in ordinary QFT of a single scalar). For the sake of generality, we will evaluate the most general scalar B-type integrals with unequal masses, $m$ and $\bar{m}$, that  can occur in an open EFT perturbation theory. 
These are 16 in number and they are defined as:

\begin{figure}[ht]
\begin{center}
\begin{tikzpicture}[line width=1 pt, scale=1]
%\begin{scope}[rotate=-80]

\draw [phir, ultra thick, domain=0:180] plot ({1*cos(\x)}, {1*sin(\x)});
\draw [phir, ultra thick, domain=180:360] plot ({1*cos(\x)}, {1*sin(\x)}); 
\node at (-1,0) {$\times $};	
\node at (1,0) {$\times $};	
\node at (0,-2) {\Large $B_{RR}(k)$};	
	
\begin{scope}[shift={(3,0)}]
\draw [phil, ultra thick, domain=0:180] plot ({1*cos(\x)}, {1*sin(\x)});
\draw [phil, ultra thick, domain=180:360] plot ({1*cos(\x)}, {1*sin(\x)}); 
\node at (-1,0) {$\times $};	
\node at (1,0) {$\times $};	
\node at (0,-2) {\Large $B_{LL}(k)$};	
\end{scope}

\begin{scope}[shift={(6,0)}]
\draw [phir, ultra thick, domain=0:180] plot ({1*cos(\x)}, {1*sin(\x)});
\draw [phil, ultra thick, domain=180:360] plot ({1*cos(\x)}, {1*sin(\x)}); 
\node at (-1,0) {$\times $};	
\node at (1,0) {$\times $};	
\node at (0,-2) {\Large $B_{LR}(k)$};
\end{scope}

\begin{scope}[shift={(9,0)}]
\draw [phil, ultra thick, domain=0:180] plot ({1*cos(\x)}, {1*sin(\x)});
\draw [phir, ultra thick, domain=180:360] plot ({1*cos(\x)}, {1*sin(\x)}); 
\node at (-1,0) {$\times $};	
\node at (1,0) {$\times $};	
\node at (0,-2) {\Large $B_{RL}(k)$};
\end{scope}

\begin{scope}[shift={(0,-4)}]
\draw [phil, ultra thick, domain=0:90] plot ({1*cos(\x)}, {1*sin(\x)});
\draw [phir, ultra thick, domain=90:180] plot ({1*cos(\x)}, {1*sin(\x)});
\draw [phir, ultra thick, domain=180:270] plot ({1*cos(\x)}, {1*sin(\x)}); 
\draw [phil, ultra thick, domain=270:360] plot ({1*cos(\x)}, {1*sin(\x)}); 
\node at (-1,0) {$\times $};	
\node at (1,0) {$\times $};	
\node at (0,-2) {\Large $B_{PM}(k)$};
\end{scope}

\begin{scope}[shift={(3,-4)}]
\draw [phir, ultra thick, domain=0:90] plot ({1*cos(\x)}, {1*sin(\x)});
\draw [phil, ultra thick, domain=90:180] plot ({1*cos(\x)}, {1*sin(\x)});
\draw [phil, ultra thick, domain=180:270] plot ({1*cos(\x)}, {1*sin(\x)}); 
\draw [phir, ultra thick, domain=270:360] plot ({1*cos(\x)}, {1*sin(\x)}); 
\node at (-1,0) {$\times $};	
\node at (1,0) {$\times $};	
\node at (0,-2) {\Large $B_{MP}(k)$};	
\end{scope}

\begin{scope}[shift={(6,-4)}]
\draw [phir, ultra thick, domain=0:90] plot ({1*cos(\x)}, {1*sin(\x)});
\draw [phil, ultra thick, domain=90:180] plot ({1*cos(\x)}, {1*sin(\x)});
\draw [phir, ultra thick, domain=180:270] plot ({1*cos(\x)}, {1*sin(\x)}); 
\draw [phil, ultra thick, domain=270:360] plot ({1*cos(\x)}, {1*sin(\x)}); 
\node at (-1,0) {$\times $};	
\node at (1,0) {$\times $};	
\node at (0,-2) {\Large $B_{PP}(k)$};	
\end{scope}

\begin{scope}[shift={(9,-4)}]
\draw [phil, ultra thick, domain=0:90] plot ({1*cos(\x)}, {1*sin(\x)});
\draw [phir, ultra thick, domain=90:180] plot ({1*cos(\x)}, {1*sin(\x)});
\draw [phil, ultra thick, domain=180:270] plot ({1*cos(\x)}, {1*sin(\x)}); 
\draw [phir, ultra thick, domain=270:360] plot ({1*cos(\x)}, {1*sin(\x)}); 
\node at (-1,0) {$\times $};	
\node at (1,0) {$\times $};	
\node at (0,-2) {\Large $B_{MM}(k)$};	
\end{scope}

\begin{scope}[shift={(0,-8)}]
\draw [phir, ultra thick, domain=0:180] plot ({1*cos(\x)}, {1*sin(\x)});
\draw [phir, ultra thick, domain=180:270] plot ({1*cos(\x)}, {1*sin(\x)}); 
\draw [phil, ultra thick, domain=270:360] plot ({1*cos(\x)}, {1*sin(\x)}); 
\node at (-1,0) {$\times $};	
\node at (1,0) {$\times $};	
\node at (0,-2) {\Large $B_{PR}(k)$};
\end{scope}

\begin{scope}[shift={(3,-8)}]
\draw [phir, ultra thick, domain=0:180] plot ({1*cos(\x)}, {1*sin(\x)});
\draw [phil, ultra thick, domain=180:270] plot ({1*cos(\x)}, {1*sin(\x)}); 
\draw [phir, ultra thick, domain=270:360] plot ({1*cos(\x)}, {1*sin(\x)}); 
\node at (-1,0) {$\times $};	
\node at (1,0) {$\times $};	
\node at (0,-2) {\Large $B_{MR}(k)$};	
\end{scope}

\begin{scope}[shift={(6,-8)}]
\draw [phil, ultra thick, domain=0:180] plot ({1*cos(\x)}, {1*sin(\x)});
\draw [phil, ultra thick, domain=180:270] plot ({1*cos(\x)}, {1*sin(\x)}); 
\draw [phir, ultra thick, domain=270:360] plot ({1*cos(\x)}, {1*sin(\x)}); 
\node at (-1,0) {$\times $};	
\node at (1,0) {$\times $};	
\node at (0,-2) {\Large $B_{ML}(k)$};
\end{scope}

\begin{scope}[shift={(9,-8)}]
\draw [phil, ultra thick, domain=0:180] plot ({1*cos(\x)}, {1*sin(\x)});
\draw [phir, ultra thick, domain=180:270] plot ({1*cos(\x)}, {1*sin(\x)}); 
\draw [phil, ultra thick, domain=270:360] plot ({1*cos(\x)}, {1*sin(\x)}); 
\node at (-1,0) {$\times $};	
\node at (1,0) {$\times $};	
\node at (0,-2) {\Large $B_{PL}(k)$};	
\end{scope}

\begin{scope}[shift={(0,-12)}]
\draw [phir, ultra thick, domain=0:90] plot ({1*cos(\x)}, {1*sin(\x)}); 
\draw [phil, ultra thick, domain=90:180] plot ({1*cos(\x)}, {1*sin(\x)});
\draw [phir, ultra thick, domain=180:360] plot ({1*cos(\x)}, {1*sin(\x)});
\node at (-1,0) {$\times $};	
\node at (1,0) {$\times $};	
\node at (0,-2) {\Large $B_{RP}(k)$};
\end{scope}

\begin{scope}[shift={(3,-12)}]
\draw [phir, ultra thick, domain=180:360] plot ({1*cos(\x)}, {1*sin(\x)});
\draw [phil, ultra thick, domain=0:90] plot ({1*cos(\x)}, {1*sin(\x)}); 
\draw [phir, ultra thick, domain=90:180] plot ({1*cos(\x)}, {1*sin(\x)}); 
\node at (-1,0) {$\times $};	
\node at (1,0) {$\times $};	
\node at (0,-2) {\Large $B_{RM}(k)$};	
\end{scope}

\begin{scope}[shift={(6,-12)}]
\draw [phil, ultra thick, domain=180:360] plot ({1*cos(\x)}, {1*sin(\x)});
\draw [phil, ultra thick, domain=0:90] plot ({1*cos(\x)}, {1*sin(\x)}); 
\draw [phir, ultra thick, domain=90:180] plot ({1*cos(\x)}, {1*sin(\x)}); 
\node at (-1,0) {$\times $};	
\node at (1,0) {$\times $};	
\node at (0,-2) {\Large $B_{LM}(k)$};
\end{scope}

\begin{scope}[shift={(9,-12)}]
\draw [phil, ultra thick, domain=180:360] plot ({1*cos(\x)}, {1*sin(\x)});
\draw [phir, ultra thick, domain=0:90] plot ({1*cos(\x)}, {1*sin(\x)}); 
\draw [phil, ultra thick, domain=90:180] plot ({1*cos(\x)}, {1*sin(\x)}); 
\node at (-1,0) {$\times $};	
\node at (1,0) {$\times $};	
\node at (0,-2) {\Large $B_{LP}(k)$};	
\end{scope}

\end{tikzpicture}
\end{center}
\caption{PV one loop $B$ type integrals in SK theory. The momentum and mass corresponding to the lower propagator is denoted by $p^\mu$ and $m$ respectively, whereas the momentum and mass corresponding to the upper propagator are $q^\mu$ and $\bar{m}$ respectively. The momenta p and q are taken to flow anti-clockwise in the loop.}
\label{diag:pvoneloopI}
\end{figure}
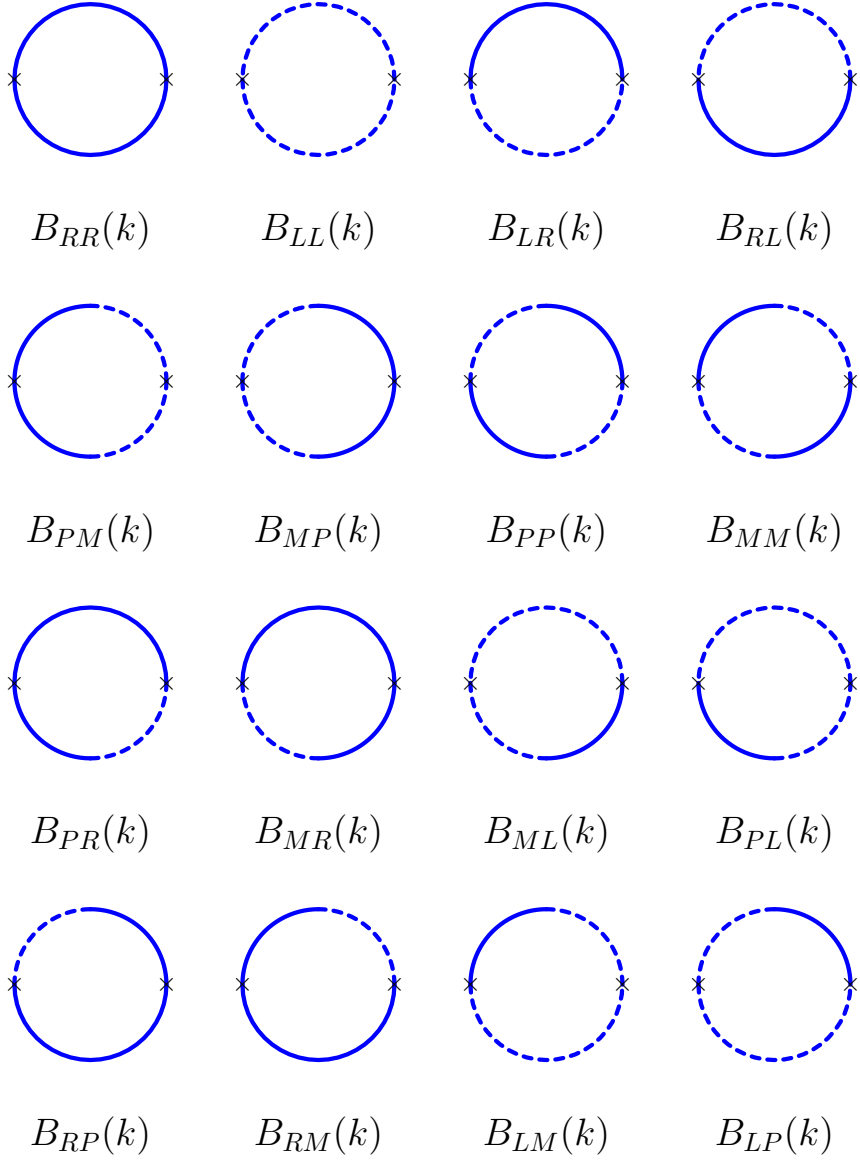

\begin{equation}
\begin{split}
B_{RR}(k)&\equiv \mu^{4-d}\int\frac{d^d p}{(2\pi)^d i}\frac{d^d q}{(2\pi)^d i}\ \frac{1}{p^2+m^2-i\varepsilon} \times\frac{1}{q^2+\bar{m}^2-i\varepsilon} \times (2\pi)^d \ \delta^d(p-q-k) \\
B_{LL}(k)&\equiv \mu^{4-d}\int\frac{d^d p}{(2\pi)^d i}\frac{d^d q}{(2\pi)^d i}\ \frac{(-1)}{p^2+m^2+i\varepsilon} \times\frac{(-1)}{q^2+\bar{m}^2+i\varepsilon} \times (2\pi)^d \ \delta^d(p-q-k) \\
B_{RL}(k)&\equiv \mu^{4-d}\int\frac{d^d p}{(2\pi)^d i}\frac{d^d q}{(2\pi)^d i}\ \frac{1}{p^2+m^2-i\varepsilon} \times\frac{(-1)}{q^2+\bar{m}^2+i\varepsilon} \times (2\pi)^d \ \delta^d(p-q-k) \\
B_{LR}(k)&\equiv \mu^{4-d}\int\frac{d^d p}{(2\pi)^d i}\frac{d^d q}{(2\pi)^d i}\ \frac{(-1)}{p^2+m^2+i\varepsilon} \times\frac{1}{q^2+\bar{m}^2-i\varepsilon} \times (2\pi)^d \ \delta^d(p-q-k) \\
\end{split}
\end{equation}
 
\begin{equation}
\begin{split}
B_{PM}(k)&\equiv \mu^{4-d}\int\frac{d^d p}{(2\pi)^d i}\frac{d^d q}{(2\pi)^d i}\ 2\pi i\delta_+(p^2+m^2)\times 2\pi i\delta_-(q^2+\bar{m}^2) \times (2\pi)^d \ \delta^d(p-q-k) \\
B_{MP}(k)&\equiv \mu^{4-d}\int\frac{d^d p}{(2\pi)^d i}\frac{d^d q}{(2\pi)^d i}\ 2\pi i\delta_-(p^2+m^2)\times 2\pi i\delta_+(q^2+\bar{m}^2) \times (2\pi)^d \ \delta^d(p-q-k) \\
B_{PP}(k)&\equiv \mu^{4-d}\int\frac{d^d p}{(2\pi)^d i}\frac{d^d q}{(2\pi)^d i}\ 2\pi i\delta_+(p^2+m^2)\times 2\pi i\delta_+(q^2+\bar{m}^2) \times (2\pi)^d \ \delta^d(p-q-k) \\
B_{MM}(k)&\equiv \mu^{4-d}\int\frac{d^d p}{(2\pi)^d i}\frac{d^d q}{(2\pi)^d i}\ 2\pi i\delta_-(p^2+m^2)\times 2\pi i\delta_-(q^2+\bar{m}^2) \times (2\pi)^d \ \delta^d(p-q-k) \\
\end{split} 
\end{equation}

\begin{equation}
\begin{split}
B_{PR}(k)&\equiv \mu^{4-d}\int\frac{d^d p}{(2\pi)^d i}\frac{d^d q}{(2\pi)^d i}\  2\pi i\delta_+(p^2+m^2) \frac{1}{q^2+\bar{m}^2-i\varepsilon} \times (2\pi)^d \ \delta^d(p-q-k) \\
B_{MR}(k)&\equiv \mu^{4-d}\int\frac{d^d p}{(2\pi)^d i}\frac{d^d q}{(2\pi)^d i}\ 2\pi i\delta_-(p^2+m^2) \frac{1}{q^2+\bar{m}^2-i\varepsilon}  \times (2\pi)^d \ \delta^d(p-q-k) \\
B_{PL}(k)&\equiv \mu^{4-d}\int\frac{d^d p}{(2\pi)^d i}\frac{d^d q}{(2\pi)^d i}\ 2\pi i\delta_+(p^2+m^2) \frac{(-1)}{q^2+\bar{m}^2+i\varepsilon} \times (2\pi)^d \ \delta^d(p-q-k) \\
B_{ML}(k)&\equiv \mu^{4-d}\int\frac{d^d p}{(2\pi)^d i}\frac{d^d q}{(2\pi)^d i}\  2\pi i\delta_-(p^2+m^2) \frac{(-1)}{q^2+\bar{m}^2+i\varepsilon} \times (2\pi)^d \ \delta^d(p-q-k) \\
\end{split}
\end{equation}

\begin{equation}
\begin{split}
B_{RP}(k)&\equiv \mu^{4-d}\int\frac{d^d p}{(2\pi)^d i}\frac{d^d q}{(2\pi)^d i}\ \frac{1}{p^2+m^2-i\varepsilon}\ 2\pi i\delta_+(q^2+\bar{m}^2) \times (2\pi)^d \ \delta^d(p-q-k) \\
B_{RM}(k)&\equiv \mu^{4-d}\int\frac{d^d p}{(2\pi)^d i}\frac{d^d q}{(2\pi)^d i}\ \frac{1}{p^2+m^2-i\varepsilon} \ 2\pi i\delta_-(q^2+\bar{m}^2) \times (2\pi)^d \ \delta^d(p-q-k) \\
B_{LP}(k)&\equiv \mu^{4-d}\int\frac{d^d p}{(2\pi)^d i}\frac{d^d q}{(2\pi)^d i}\  \frac{(-1)}{p^2+m^2+i\varepsilon}\ 2\pi i\delta_+(q^2+\bar{m}^2) \times (2\pi)^d \ \delta^d(p-q-k) \\
B_{LM}(k)&\equiv \mu^{4-d}\int\frac{d^d p}{(2\pi)^d i}\frac{d^d q}{(2\pi)^d i}\   \frac{(-1)}{p^2+m^2+i\varepsilon}\ 2\pi i\delta_-(q^2+\bar{m}^2) \times (2\pi)^d \ \delta^d(p-q-k) \\
\end{split}
\end{equation}
% Now some of them are related by complex conjugation 
% \begin{eqnarray}
% 	&& B_{RR}(k)=[B_{LL}(-k)]^\ast 
% \qquad \qquad
% B_{RL}(k)=[B_{RL}(k)]^\ast 
% \qquad \qquad
% B_{PM}(k)=[B_{PM}(-k)]^\ast 
% \\
% && B_{PL}(k)=[B_{MR}(-k)]^\ast 
% \qquad \qquad
% B_{PR}(k)=[B_{ML}(-k)]^\ast 
% \qquad \qquad
% B_{PP}(k)=[B_{PP}(-k)]^\ast 
% \end{eqnarray}
% These relations imply that we need to evaluate only $6$ integrals.

In the following appendix, we will evaluate these integrals and their divergences.
\section{Evaluating Passarino-Veltman Loop Integrals for open $\phi^3+\phi^4$ theory}
\label{sec:onelooppvintegrals}
In this section, we describe in some detail the loop integrals that appear in the perturbation theory of  open $\phi^3+\phi^4$ theory. While some of the integrals are familiar
from usual QFT textbooks and a few other integrals occur in discussions of cutting rules, as far as the authors are aware, the majority of the integrals described in this section
are not analyzed elsewhere. Hence, these integrals are described in some detail with a special focus on the new kind of features that occur when we try to do integrals 
in the real time (most of the integrals in this section do not admit Wick rotation because of their unusual $i\epsilon$ prescriptions) .

\subsection{Passarino-Veltman $A$ type integrals}
% \label{sec:onelooppvatypeintegrals}

There are four $A$ type PV integrals : $A_{R}$, $A_{L}$, $A_{P}, A_{M}$. They satisfy the following relations 
\begin{eqnarray}
	&A_{R}=A_{L}^\ast \qquad A_{P}=A_{P}^\ast 
\label{pvaintg1}	
\\
	&A_{R}+A_{L}=A_{P}+A_{M} \equiv 2 A_{P}
\label{pvaintg2}	
\end{eqnarray}
First we compute the integral $A_{R}$ 
\begin{equation}
\begin{split}
A_{R} &= \mu^{4-d}\int \frac{d^dp}{(2\pi)^d}\frac{-i}{p^2+m^2-i\varepsilon}=\int \frac{d^dp}{(2\pi)^di}\frac{1}{\omega_p^2-(p^0)^2-i\varepsilon}\\ 
&=\mu^{4-d} \int \frac{d^{d-1}p}{(2\pi)^{d-1}}\frac{1}{2\omega_p}
=	\mu^{4-d}\frac{\textrm{Vol}(\mathbb{S}^{d-2})}{2(2\pi)^{d-1}}\int_m\left(\sqrt{\omega_p^2-m^2}\right)^{d-3} d\omega_p\\
&=\Gamma\Bigl[ \frac{1}{2}(2-d) \Bigr]  \frac{m^2}{(4\pi)^2} \Bigl( \frac{m^2}{4\pi \mu^2}\Bigr)^{\frac{d-4}{2}}\\
&=  \frac{m^2}{(4\pi)^2} \frac{2}{d-4} +  \frac{m^2}{(4\pi)^2}\Bigl\{ \ln\  \frac{m^2}{4\pi\mu^2e^{-\gamma_E}} -1\Bigr\} +O(d-4)\\
%&=\frac{1}{16\pi^2}\left[ \frac{2}{(d-4)} -1+\gamma_E+\ln \,\left(m^2/\mu^2\right)-\ln 4\pi 
%	\right]m^2
\label{pvaintg3}	
\end{split}
\end{equation}
Using this result and the relations \eqref{pvaintg1}-\eqref{pvaintg2}, we can determine $A_{L}$ and $A_{P}$. $A_{L}  $, $A_{P}  $ and $A_{M}  $ are given by 
\begin{eqnarray} 
	A_{R}= A_{L}  =A_{P} =A_{M}  =	\frac{m^2}{(4\pi)^2}\left(\frac{2}{d-4} +   \ln\  \frac{m^2}{4\pi\mu^2e^{-\gamma_E}} -1\right)
\label{pvaintg4}	
\end{eqnarray}

\subsection{Integrals \texorpdfstring{$B_{PM}(k)$}{BPM} and \texorpdfstring{$B_{MP}(k)$}{BMP}}
 We will now consider the following two integrals 
\begin{equation}
\begin{split}
B_{PM}(k)&\equiv \mu^{4-d}\int\frac{d^d p}{(2\pi)^d i}\frac{d^d q}{(2\pi)^d i}\ 2\pi i\delta_+(p^2+m^2)\times 2\pi i\delta_-(q^2+\bar{m}^2) \times (2\pi)^d \ \delta^d(p-q-k) \\
B_{MP}(k)&\equiv \mu^{4-d}\int\frac{d^d p}{(2\pi)^d i}\frac{d^d q}{(2\pi)^d i}\ 2\pi i\delta_-(p^2+m^2)\times 2\pi i\delta_+(q^2+\bar{m}^2) \times (2\pi)^d \ \delta^d(p-q-k) \\
\end{split}
\end{equation}
These integrals  are well-known in discussions of cutting rules and we mainly discuss them here for completeness. They 
represent the amplitudes for the two body decay of a particle with mass $\sqrt{-k^2}$ to decay into two particles of mass $m$ and $\bar{m}$.
It follows that the integral  $B_{PM}(k)$ vanishes unless $k^\mu$ is future time-like, whereas $B_{MP}(k)$ is zero unless $k^\mu$ is past time-like. 
Since $B_{MP}(k)=B_{PM}(-k)$, it suffices to argue this for $B_{PM}(k)$. The mathematical reasoning is as follows : 
if $k^\mu$ were to be space-like, we can go to a frame where $k^0=0$ and the energy delta function then gives $\delta(\omega_p+\bar{\omega}_q) =0$, thus 
reducing the amplitude to zero. If $k^\mu$ is time-like and we move to its rest frame by setting $k^\mu=\{M, \vec{0}\}$ where $M=\text{sgn}(k^0)\sqrt{-k^2}$.
The energy delta function for $B_{PM}(k)$ then gives $\delta(\omega_p+\bar{\omega}_q-M) =0$, thus forcing $M>0$, i.e., $k^\mu$ should be future time-like.

By performing $p^0$ and $q^0$ integrals in the rest frame of $k^\mu$, we can reduce both integrals to the two body phase space integral. Let us begin by 
defining the basic kinematics of a particle of mass $M=\sqrt{-k^2}$ decaying into two particles of mass $m$ and $\bar{m}$. Let $p_*$ be the momentum with 
which these two particles fly away in the rest frame of $M$. Energy-momentum conservation then fixes 
\begin{equation}
\begin{split}
p_\star &= \frac{1}{2M} \Bigl[ M^2- (m+\bar{m})^2\Bigr]^{1/2}\Bigl[ M^2- (m-\bar{m})^2\Bigr]^{1/2}\\
&= \frac{M}{2} \Bigl[ 1 - 2\left(\frac{m^2 + \bar{m}^2}{M^2}\right) + \left(\frac{m^2 - \bar{m}^2}{M^2}\right)^2\ \Bigr]^{1/2}\\
\omega_p^\star&\equiv  \sqrt{p_\star^2+m^2}= \frac{1}{2M} (M^2+m^2-\bar{m}^2)\\
\bar{\omega}_p^\star&\equiv  \sqrt{p_\star^2+\bar{m}^2}= \frac{1}{2M} (M^2+\bar{m}^2-m^2)
\end{split}
\end{equation}
These expressions make sense only when the argument of the square root is positive, i.e., when $M\geq m+\bar{m}$ which is the condition for 
the two body decay to be kinematically possible. In this regime, $\omega_p^\star$ and $\bar{\omega}_p^\star$ are both positive as we would expect. Another 
useful identity is 
\begin{equation}
 \frac{1}{(2\omega_p) (2\bar{\omega}_p)}\delta(\omega_p+\bar{\omega}_p-M) = \frac{1}{4Mp_\star}\delta( |p|-p_\star)
\end{equation}

Let us now compute this integral in terms  of these kinematic variables. We have
\begin{equation}
\begin{split}
B_{PM}(k)&= \mu^{4-d}\int\frac{d^{d-1} p}{(2\pi)^{d-1}(2\omega_p) }\frac{d^{d-1} q}{(2\pi)^{d-1} (2\bar{\omega}_q) }\ \ \times (2\pi)^d \ \delta^{d-1}(\vec{p}-\vec{q}) \ \delta(\omega_p+\omega_q-M) \\
&= \mu^{4-d}\int\frac{d^{d-1} p}{(2\pi)^{d-1}(2\omega_p) (2\bar{\omega}_p)}\ 2\pi \ \delta(\omega_p+\bar{\omega}_p-M) \\
&= \mu^{4-d}\frac{\text{Vol}(\mathbb{S}^{d-2})}{(2\pi)^{d-2} } \int^\infty_{0}
\frac{p^{d-2}\ d^dp}{(2\omega_p) (2\bar{\omega_p})}\delta(\omega_p+\bar{\omega_p}-M)
=  \mu^{4-d}\frac{\text{Vol}(\mathbb{S}^{d-2})p_\star^{d-3}}{4M(2\pi)^{d-2} }\\
&= \frac{\text{Vol}(\mathbb{S}^{d-2})}{32\pi^2}\left(\frac{M}{4\pi\mu}\right)^{d-4} \left(\frac{2p_\star}{M}\right)^{d-3}
\end{split}
\end{equation}
which is the two-body phase space as advertised.

Restoring the kinematic constraints, we get
\begin{equation}
\begin{split}
B_{PM}(k)
%&= \mu^{4-d}\frac{\text{Vol}(\mathbb{S}^{d-2}) M^{d-4}}{2(4\pi)^{d-2} }\Bigl[1 - 2\left(\frac{m^2 + \bar{m}^2}{M^2}\right)
%+ \left(\frac{m^2 - \bar{m}^2}{M^2}\right)^2 \Bigr]^{\frac{d-3}{2}} \\
%&\qquad \times \Theta\left(M^2-(m+\bar{m})^2\right)\Theta(M)\\
&= \frac{\text{Vol}(\mathbb{S}^{d-2})}{32 \pi^2 } \left(\frac{-k^2}{(4\pi \mu)^2}\right)^{\frac{d-4}{2}} \Bigl[1 
+ 2\left(\frac{m^2 + \bar{m}^2}{k^2}\right) + \left(\frac{m^2 - \bar{m}^2}{k^2}\right)^2 \Bigr]^{\frac{d-3}{2}} \\
&\qquad \times \Theta\left(-k^2-(m+\bar{m})^2\right)\Theta(k^0) 
\end{split}
\end{equation}
The integral can then be expanded near $d=4$ to get  
\begin{equation}
B_{PM}(k)= \frac{\Theta\left(-k^2-(m+\bar{m})^2\right)\Theta(k^0)}{8\pi} \Bigl[1 + 2\left(\frac{m^2 + \bar{m}^2}{k^2}\right) 
+ \left(\frac{m^2 - \bar{m}^2}{k^2}\right)^2 \Bigr]^{\frac{1}{2}}
\end{equation}
Replacing $k^\mu$ by $-k^\mu$ we get  
\begin{equation}
B_{MP}(k)= \frac{\Theta\left(-k^2-(m+\bar{m})^2\right)\Theta(-k^0)}{8\pi} \Bigl[1 + 2\left(\frac{m^2 + \bar{m}^2}{k^2}\right) 
+ \left(\frac{m^2 - \bar{m}^2}{k^2}\right)^2 \Bigr]^{\frac{1}{2}}\ .
\end{equation}
These two expressions can be added to get
\begin{equation}
B_{PM}(k)+B_{MP}(k)=   \frac{\Theta\left(-k^2-(m+\bar{m})^2\right)}{8\pi} \Bigl[
1 + 2\left(\frac{m^2 + \bar{m}^2}{k^2}\right) + \left(\frac{m^2 - \bar{m}^2}{k^2}\right)^2 \Bigr]^{\frac{1}{2}}
\end{equation}

When $m = \bar{m}$, we can write
\begin{equation}
\boxed{
B_{PM}(k)= \frac{\Theta(k^0)\Theta(-k^2-4m^2)}{8\pi}\sqrt{1+\frac{4m^2}{k^2}}
}
\end{equation}
and
\begin{equation}
\boxed{
B_{MP}(k)= \frac{\Theta(-k^0)\Theta(-k^2-4m^2)}{8\pi}\sqrt{1+\frac{4m^2}{k^2}}
}
\end{equation}

\subsection{Integrals \texorpdfstring{$B_{PP}(k)$}{BPP} and \texorpdfstring{$B_{MM}(k)$}{BMM}}
We now turn to the `cross-cut' integrals $B_{PP}(k)$ and $B_{MM}(k)$ which do not occur in the usual discussions of cutting rules 
in a unitary theory. They are loop integrals peculiar to open QFTs with their own characteristic kinematic behavior. 

\subsubsection{Time-like $k^\mu$}
We will begin by examining  $B_{PP}(k)$ for time-like $k^\mu$. In the rest frame of $k^\mu$ i.e., $k^\mu=(M,\vec{0})$, we can do 
similar manipulations as in the previous subsection to get
\begin{equation}
\begin{split} 
B_{PP}(k) &\equiv \mu^{4-d} \int \frac{d^dp}{(2\pi)^di}\frac{d^dq}{(2\pi)^di} 2\pi i \delta_+ (p^2+m^2)\ 2\pi i \delta_+ (q^2+\bar{m}^2)\ (2\pi)^d\ \delta^d (p-q-k)\\
&= \mu^{4-d} \int \frac{d^{d-1}p}{(2\pi)^{d-1}2\omega_p}\frac{d^{d-1}q}{(2\pi)^{d-1}2\bar \omega_q} \delta(\omega_p-\bar{\omega}_q-M) (2\pi)^d\ \delta^{d-1} (\vec{p}-\vec{q})\\ 
&= \mu^{4-d} \int \frac{d^{d-1}p}{(2\pi)^{d-2}} \frac{1}{2\omega_p} \frac{1}{2\bar{\omega}_p} \delta(\omega_p-\bar{\omega}_p-M)\\
&= \frac{\mu^{4-d} \text{Vol}(\mathbb{S}^{d-2})}{(2\pi)^{d-2}}\int_0^\infty \frac{p^{d-2}dp}{(2\omega_p)(2 \bar{\omega}_p)}\delta(\omega_p-\bar{\omega}_p-M)
= \frac{\text{Vol}(\mathbb{S}^{d-2})}{32\pi^2}\left(\frac{|M|}{4\pi\mu}\right)^{d-4} \left(\frac{2p_\star}{|M|}\right)^{d-3}
\end{split}
\end{equation}
where we have used 
\begin{equation}
 \frac{1}{(2\omega_p) (2\bar{\omega}_p)}\delta(\omega_p-\bar{\omega}_p-M) = \frac{1}{4|M|p_\star}\delta( |p|-p_\star)
\end{equation}
with $p_\star$ being the appropriate momentum which solves the kinematics  (see below).

For $M>0$, i.e., $k^\mu$ being future time-like, we recognize the integral as the one describing the phase space for a deep in-elastic scattering process :
 $\bar{m}$ with momentum $p$ strikes against the target $M$ at rest converting it into the particle $m$ traveling with momentum $p$. The kinematics is solved
by 
\begin{equation}
\begin{split}
p_\star&= \frac{1}{2M} \Bigl[ (m+\bar{m})^2-M^2\Bigr]^{1/2}\Bigl[ (m-\bar{m})^2-M^2\Bigr]^{1/2}\\
 &= \frac{M}{2} \Bigl[ 1 - 2\left(\frac{m^2 + \bar{m}^2}{M^2}\right) + \left(\frac{m^2 - \bar{m}^2}{M^2}\right)^2\ \Bigr]^{1/2}\\
\omega_p^\star&\equiv  \sqrt{p_\star^2+m^2}= \frac{1}{2M} (m^2+M^2-\bar{m}^2)\\
\bar{\omega}_p^\star&\equiv  \sqrt{p_\star^2+\bar{m}^2}= \frac{1}{2M} (m^2-M^2-\bar{m}^2)
\end{split}
\end{equation}
which is sensible for $M<m-\bar{m}$.Thus, in this kinematic regime we get 
\begin{equation}
\begin{split}
B_{PP}(k)
&\ni \frac{\text{Vol}(\mathbb{S}^{d-2})}{32 \pi^2 } \left(\frac{-k^2}{(4\pi \mu)^2}\right)^{\frac{d-4}{2}} \Bigl[1 
+ 2\left(\frac{m^2 + \bar{m}^2}{k^2}\right) + \left(\frac{m^2 - \bar{m}^2}{k^2}\right)^2 \Bigr]^{\frac{d-3}{2}} \\
&\qquad \times \Theta\left((m-\bar{m})^2+k^2\right)\Theta(k^0) 
\end{split}
\end{equation}

For $M<0$, i.e., $k^\mu$ being past time-like, we recognize the integral as the one describing the phase space for the two body decay of $\bar{m}$ into a particle of mass $|M|$ and $m$.
The kinematics is solved by 
\begin{equation}
\begin{split}
p_\star&= \frac{1}{2|M|} \Bigl[ (m+\bar{m})^2-M^2\Bigr]^{1/2}\Bigl[ (m-\bar{m})^2-M^2\Bigr]^{1/2}\\
 &= \frac{|M|}{2} \Bigl[ 1 - 2\left(\frac{m^2 + \bar{m}^2}{M^2}\right) + \left(\frac{m^2 - \bar{m}^2}{M^2}\right)^2\ \Bigr]^{1/2}\\
\omega_p^\star&\equiv  \sqrt{p_\star^2+m^2}= \frac{1}{2|M|} (\bar{m}^2-m^2-M^2)\\
\bar{\omega}_p^\star&\equiv  \sqrt{p_\star^2+\bar{m}^2}= \frac{1}{2|M|} (\bar{m}^2-m^2+M^2)
\end{split}
\end{equation}
which is sensible for $\bar{m}>|M|+m$.Thus, in this kinematic regime we get 
\begin{equation}
\begin{split}
B_{PP}(k)
&\ni \frac{\text{Vol}(\mathbb{S}^{d-2})}{32 \pi^2 } \left(\frac{-k^2}{(4\pi \mu)^2}\right)^{\frac{d-4}{2}} \Bigl[1 
+ 2\left(\frac{m^2 + \bar{m}^2}{k^2}\right) + \left(\frac{m^2 - \bar{m}^2}{k^2}\right)^2 \Bigr]^{\frac{d-3}{2}} \\
&\qquad \times \Theta\left((\bar{m}-m)^2+k^2\right)\Theta(-k^0) 
\end{split}
\end{equation}

Thus, we conclude that for time-like $k^\mu$, 
\begin{equation}
\begin{split}
B_{PP}(k)
&\ni \frac{\text{Vol}(\mathbb{S}^{d-2})}{32 \pi^2 } \left(\frac{-k^2}{(4\pi \mu)^2}\right)^{\frac{d-4}{2}} \Bigl[1 
+ 2\left(\frac{m^2 + \bar{m}^2}{k^2}\right) + \left(\frac{m^2 - \bar{m}^2}{k^2}\right)^2 \Bigr]^{\frac{d-3}{2}} \\
&\qquad \times \Theta\left((\bar{m}-m)^2+k^2\right)\Theta(-k^2) 
\end{split}
\end{equation}
Note that $B_{PP}(k)= B_{PP}(-k)$ could have been directly deduced from the integral form.

\

\subsubsection{Space-like $k^\mu$}
We will next study $B_{PP}(k)$ when $k^\mu$ is space-like. We set  $k^\mu=\{0,Q=\sqrt{k^2}, \vec{0}_{d-2}\}$ where we take $Q>0$ without loss of generality.
We have
\begin{equation}\label{eq:BPP}
\begin{split}
B_{PP}(k)
&\equiv \mu^{4-d} \int \frac{d^dp}{(2\pi)^di}\frac{d^dq}{(2\pi)^di} 2\pi i \delta_+ (p^2+m^2)\ 2\pi i \delta_+ (q^2+\bar{m}^2)\ (2\pi)^d\ \delta^d (p-q-k)\\
&= \mu^{4-d}\int\frac{d^{d-1} p}{(2\pi)^{d-1}(2\omega_p) }\frac{d^{d-1} q}{(2\pi)^{d-1} (2\bar{\omega}_q) }\ \ \times (2\pi)^d \ \delta^{d-2}(\vec{p}_\perp-\vec{q}_\perp) \ \delta(p_{||}-q_{||}-Q)  \delta(\omega_p-\bar{\omega}_q) \\
&= \mu^{4-d}\int\frac{d^{d-1} p}{(2\pi)^{d-1}(2\omega_p)(2\bar{\omega}_{p\perp}) }\ 2\pi \ \delta(\omega_{p}-\bar{\omega}_{p\perp}) 
= \frac{\mu^{4-d}}{(2Q)}\int\frac{d^{d-2} p_\perp}{(2\pi)^{d-2}(2\omega_{p\perp}) }
\end{split}
\end{equation}
where, in the penultimate step  we have defined  $\bar{\omega}_{p\perp}\equiv p_\perp^2+ (Q-p_{||})^2+  \bar{m}^2$. In the last step, we have 
used 
\begin{equation}
\begin{split}
 \frac{1}{(2\omega_p) (2\bar{\omega}_{p\perp})}\delta(\omega_{p}-\bar{\omega}_{p\perp})&= \frac{1}{4Q \omega_{p\perp}}\delta( p_{||}-p_{||}^\star)\\
\end{split}
\end{equation}
with the definitions 
\begin{equation}
\begin{split}
p_{||}^\star&\equiv \frac{1}{2Q}(\bar{m}^2 -m^2 +Q^2)\\
(\omega_{p\perp})^2&\equiv p_{\perp}^2 +(p_{||}^\star)^2+ m^2=p_{\perp}^2 +\frac{1}{4Q^2} \Bigl[ (m+\bar{m})^2+Q^2\Bigr]\Bigl[ (m-\bar{m})^2+Q^2\Bigr]\ .
\end{split}
\end{equation}
The rest of the integral is $(d-2)$ dimensional transverse phase space with the transverse mass given by 
\begin{equation}
\begin{split}
m_{\perp}&\equiv \frac{1}{2Q} \Bigl[ (m+\bar{m})^2+Q^2\Bigr]^{1/2}\Bigl[ (m-\bar{m})^2+Q^2\Bigr]^{1/2}\\
&= \frac{Q}{2} \Bigl[ 1 + 2\left(\frac{m^2 + \bar{m}^2}{Q^2}\right) + \left(\frac{m^2 - \bar{m}^2}{Q^2}\right)^2\ \Bigr]^{1/2}\\ .
\end{split}
\end{equation}
We thus get
\begin{equation}
\begin{split}
B_{PP}(k)
&= \mu^{4-d}\frac{\text{Vol}(\mathbb{S}^{d-3})}{4Q(2\pi)^{d-2} } \int^\infty_{m_\perp}  \Bigl(\sqrt{\omega_{p_\perp}^2-m_\perp^2}\ \Bigr)^{d-4}  d\omega_{p_\perp} \\
&=  \mu^{4-d}\frac{\text{Vol}(\mathbb{S}^{d-2})}{(2\pi)^{d-1} } \frac{m_\perp^{d-3}}{4Q} \Gamma\Bigl(\frac{3}{2}-\frac{d}{2}\Bigr)\Gamma\Bigl(\frac{d}{2}-\frac{1}{2}\Bigr)\\
&= \frac{\text{Vol}(\mathbb{S}^{d-2})}{32\pi^2}\left(\frac{Q}{4\pi\mu}\right)^{d-4} \left(\frac{2m_{\perp}}{Q}\right)^{d-3}\frac{1}{2\pi}\Gamma\Bigl(\frac{3}{2}-\frac{d}{2}\Bigr)\Gamma\Bigl(\frac{d}{2}-\frac{1}{2}\Bigr)\
\end{split}
\end{equation}
Restoring the kinematical constraints, we get 
\begin{equation}
\begin{split}
B_{PP}(k)
&\ni \frac{\text{Vol}(\mathbb{S}^{d-2})}{32\pi^2}\left(\frac{k^2}{(4\pi\mu)^2}\right)^{\frac{d-4}{2}}
\Bigl[1 
+ 2\left(\frac{m^2 + \bar{m}^2}{k^2}\right) + \left(\frac{m^2 - \bar{m}^2}{k^2}\right)^2 \Bigr]^{\frac{d-3}{2}} \\
&\qquad \times \Theta(k) \frac{1}{2\pi}\Gamma\Bigl(\frac{3}{2}-\frac{d}{2}\Bigr)\Gamma\Bigl(\frac{d}{2}-\frac{1}{2}\Bigr)\
\end{split}
\end{equation}

Putting together the various kinematical regimes, we obtain 
\begin{equation}
\begin{split}
B_{PP}(k)
&= \frac{\text{Vol}(\mathbb{S}^{d-2})}{32\pi^2}\Bigl[1 
+ 2\left(\frac{m^2 + \bar{m}^2}{k^2}\right) + \left(\frac{m^2 - \bar{m}^2}{k^2}\right)^2 \Bigr]^{\frac{d-3}{2}} \\
&\quad \times \Bigl\{ \Theta(k^2) \frac{1}{2\pi}\Gamma\Bigl(\frac{3}{2}-\frac{d}{2}\Bigr)\Gamma\Bigl(\frac{d}{2}-\frac{1}{2}\Bigr)\left(\frac{k^2}{(4\pi\mu)^2}\right)^{\frac{d-4}{2}}\Bigr.\\
&\qquad \Bigl.+ \Theta\left((\bar{m}-m)^2+k^2\right)\Theta(-k^2)\left(\frac{-k^2}{(4\pi\mu)^2}\right)^{\frac{d-4}{2}} \Bigr\}\
\end{split}
\end{equation}
Expanding around $d=4$ we get 
\begin{equation}
\begin{split}
B_{PP}(k)= B_{MM}(k)
&= \frac{1}{8\pi}\Bigl[1 
+ 2\left(\frac{m^2 + \bar{m}^2}{k^2}\right) + \left(\frac{m^2 - \bar{m}^2}{k^2}\right)^2 \Bigr]^{\frac{1}{2}} \\
&\quad \times \Bigl\{- \frac{1}{2}\ \Theta(k^2)+ \Theta\left((\bar{m}-m)^2+k^2\right)\Theta(-k^2) \Bigr\}\
\end{split}
\end{equation}
where we have used $B_{MM}(k)= B_{PP}(-k)= B_{PP}(k)$. Taking $m=\bar{m}$, the second factor vanishes and we obtain 
\begin{equation}
\boxed{
B_{PP}(k)= B_{MM}(k)
= \frac{[-  \frac{1}{2}\ \Theta(k^2)]}{8\pi}\sqrt{1 
+ \frac{4m^2}{k^2} }\
}
\end{equation}

\subsection{Reduction of divergent integrals to  \texorpdfstring{$B_{RP}(k)$}{BRP}}
We now turn to the `quarter-cut' integrals $B_{RP}(k),B_{RM}(k),B_{LP}(k)$ and $B_{LM}(k)$ which also do not occur in the usual discussions of cutting rules 
in a unitary theory. They are also  loop integrals peculiar to open QFTs. However, unlike the integrals considered in the last section, they do not evaluate to 
on-shell phase space for various processes. This off-shell nature means that  they  exhibit UV divergences and hence are crucial to the issue of
renormalizability of open QFTs. When the open QFT is renormalisable,  these diagrams contribute to $\beta$ functions of an open QFT.  As before, we will evaluate this integrals
in various kinematic regimes and then put together the answers at the end. 

We will consider the integral 
\begin{equation}
\begin{split}
B_{RP}(k) &= \mu^{4-d} \int \frac{d^dp}{(2\pi)^d i} \int \frac{d^dq}{(2\pi)^d i}\ \frac{1}{p^2+m^2-i\varepsilon}\ 2\pi i\ \delta_{+}(q^2+\bar{m}^2)\ (2\pi)^d\ \delta^d(p-q-k)\\
\end{split}
\end{equation}
This is the characteristic integral which leads to UV divergences in open QFT.  Before analyzing this integral further, we will show that the other divergent integrals can  be 
reduced to this integral. We start with
\begin{equation}
\begin{split}
B_{RM}(k) &\equiv \mu^{4-d} \int \frac{d^dp}{(2\pi)^d i} \int \frac{d^dq}{(2\pi)^d i}\ \frac{1}{p^2+m^2-i\varepsilon}\ 2\pi i\ \delta_{-}(q^2+\bar{m}^2)\ (2\pi)^d\ \delta^d(p-q-k)\\
&= B_{RP}(-k) \\
\end{split}
\end{equation}
Similarly
\begin{equation}
\begin{split}
B_{LP}(k) &\equiv \mu^{4-d} \int \frac{d^dp}{(2\pi)^d i} \int \frac{d^dq}{(2\pi)^d i}\ \frac{-1}{p^2+m^2+i\varepsilon}\ 2\pi i\ \delta_{+}(q^2+\bar{m}^2)\ (2\pi)^d\ \delta^d(p-q-k)\\
&=  [B_{RP}(k)]^*\\
B_{LM}(k)  &\equiv \mu^{4-d} \int \frac{d^dp}{(2\pi)^d i} \int \frac{d^dq}{(2\pi)^d i}\ \frac{-1}{p^2+m^2+i\varepsilon}\ 2\pi i\ \delta_{-}(q^2+\bar{m}^2)\ (2\pi)^d\ \delta^d(p-q-k)\\
&= [B_{RM}(k)]^*= [B_{RP}(-k)]^*
\end{split}
\end{equation}
The integrals with the subscripts exchanged can be obtained by exchanging $m$ and $\bar{m}$ (thus exchanging $p^\mu$ and $q^\mu$) and reversing $k^\mu$. 
For example, $B_{PR}(k) = B_{RP}(-k)|_{m\leftrightarrow\bar{m}}$ and similarly for other integrals : 
\begin{equation}
\begin{split}
B_{PR}(k) &= B_{RP}(-k)|_{m\leftrightarrow\bar{m}}\ ,\qquad
B_{MR}(k)  = B_{RP}(k)|_{m\leftrightarrow\bar{m}}\ ,\qquad\\
B_{PL}(k)  &= B_{RP}(-k)^*|_{m\leftrightarrow\bar{m}}\ ,\qquad
B_{ML}(k)  = B_{RP}(k)^*|_{m\leftrightarrow\bar{m}}\ .
\end{split}
\end{equation}

It is convenient to define the following combination of integrals :
\begin{equation}\label{eq:BpDef}
\begin{split}
B_{RL}^+(k) &\equiv B_{RP}(k)+B_{ML}(k)= B_{RP}(k) +[B_{RP}(k)]^*_{m\leftrightarrow\bar{m}} \\
 B_{RR}^+(k) &\equiv \Theta(m>\bar{m}) (B_{RM}(k)-B_{ML}(k)) + \Theta(m<\bar{m}) (B_{PR}(k)-B_{LP}(k)) \\
&= \Theta(m>\bar{m}) (B_{RP}(-k)-B_{RP}(k)^*_{m\leftrightarrow\bar{m}}) 
+ \Theta(m<\bar{m}) (B_{RP}(-k)_{m\leftrightarrow\bar{m}}-B_{RP}(k)^*) \
 \end{split}
\end{equation}
As we will see in next subsection, using these combinations, the rest of the divergent integrals can also be reduced to $B_{RP}(k)$.

\subsubsection{Time-like $k^\mu$ : reduction of divergent integrals}
We will begin with the case of time-like $k^\mu$ and move to  the rest frame of $k^\mu$ i.e., set $k^\mu=(M,\vec{0})$. Let us begin by evaluating 
\begin{equation}
\begin{split}
B_{RP}(k) &= \mu^{4-d} \int \frac{d^dp}{(2\pi)^d i} \int \frac{d^dq}{(2\pi)^d i}\ \frac{1}{p^2+m^2-i\varepsilon}\ 2\pi i\ \delta_{+}(q^2+\bar{m}^2)\ (2\pi)^d\ \delta^d(p-q-k)\\
&= \mu^{4-d} \int \frac{d^dp}{(2\pi)^d i} \int \frac{d^{d-1}q}{(2\pi)^{d-1} (2\bar{\omega}_q)}\ \frac{1}{p^2+m^2-i\varepsilon}\  \ (2\pi)^d\ \delta^{d-1}(\vec{p}-\vec{q}) \ \delta(p^0-\bar{\omega}_q-M)\\
&= \mu^{4-d} \int \frac{d^dp}{(2\pi)^d i}  \frac{1}{2\bar{\omega}_p}\ \frac{1}{p^2+m^2-i\varepsilon}\  \ 2\pi\ \delta(p^0-\bar{\omega}_p-M)\\
&= \mu^{4-d} \int \frac{d^{d-1}p}{(2\pi)^{d-1} i}  \frac{1}{2\bar{\omega}_p}\ \frac{1}{\omega_p^2-(M+\bar{\omega}_p)^2-i\varepsilon}\ \\
&= \frac{\mu^{4-d}\ \text{Vol}(\mathbb{S}^{d-2})}{2(2\pi)^{d-1}\ i} \int_{\bar{m}}^\infty d\bar{\omega}_p\ \frac{(\bar{\omega}_p^2-\bar{m}^2)^{\frac{d-3}{2}}}{\omega_p^2-(M+\bar{\omega}_p)^2-i\varepsilon}
\end{split}
\end{equation}
 
We will now show how the rest of the one-loop integrals can be reduced  to  $B_{PR}(k)$ for time-like $k^\mu$. We have
\begin{equation}
\begin{split}
B_{RL}(k) &\equiv \mu^{4-d} \int \frac{d^dp}{(2\pi)^d i} \int \frac{d^dq}{(2\pi)^d i}\ \frac{1}{p^2+m^2-i\varepsilon}\ \frac{-1}{q^2+\bar{m}^2+i\varepsilon} (2\pi)^d\ \delta^d(p-q-k)\\
&= -\mu^{4-d} \int\ \frac{d^dp}{(2\pi)^di^2}\ \frac{1}{p^2+m^2-i\varepsilon}\ \frac{1}{(p-k)^2+\bar{m}^2+ i\varepsilon}\\
&=- \mu^{4-d} \int\ \frac{d^dp}{(2\pi)^d i^2}\ \frac{1}{(\omega_p-i\varepsilon)^2-(p^0)^2}\ \frac{1}{(\bar{\omega}_p+i\varepsilon)^2-(p^0-M)^2} \times e^{i\delta \frac{p^0}{M}}\\
\end{split}
\end{equation}
where in the last step, we have put $\epsilon,\delta$ in different places to help in contour integration. Now we perform the contour integral by closing the contour 
in the upper half plane for $M>0$ and in the lower half plane for $M<0$. This gives
\begin{equation}
\begin{split}
B_{RL}(k) 
&= \mu^{4-d} \int\ \frac{d^{d-1}p}{(2\pi)^{d-1}i} \Bigl( \frac{1}{\omega_p^2-(|M|+\bar{\omega}_p)^2-i \varepsilon\ }\ \frac{1}{2\bar{\omega}_p}-\frac{1}{2\omega_p}\ \frac{1}{\bar{\omega}_p^2-(|M|+\omega_p)^2+i \varepsilon\ } \Bigr)\\
&=\frac{\mu^{4-d}\text{Vol}(\mathbb{S}^{d-2})}{2(2\pi)^{d-1}i}\left(\ \int_{\bar{m}}^\infty d\bar{\omega}_p\ \frac{(\bar{\omega}_p^2-\bar{m}^2)^{\frac{d-3}{2}}}{\omega_p^2-(|M|+\bar{\omega}_p)^2-i \varepsilon} -\int_m^\infty d\omega_p\ \frac{(\omega_p^2-m^2)^{\frac{d-3}{2}}}{\bar{\omega}_p^2-(|M|+\omega_p)^2+i \varepsilon} \ \right)\\
&=\Theta(k^0) [B_{RP}(k)+B_{ML}(k)] +\Theta(-k^0) [B_{RM}(k)+B_{PL}(k)] \\
&=\Theta(k^0) \Bigl( B_{RP}(k) +[B_{RP}(k)]^*_{m\leftrightarrow\bar{m}} \Bigr)+\Theta(-k^0)  \Bigl( B_{RP}(-k)+ [B_{RP}(-k)]^*_{m\leftrightarrow\bar{m}} \Bigr)\\
&=\Theta(k^0)B_{RL}^+(k) +\Theta(-k^0)  B_{RL}^+(-k)
\end{split}
\end{equation}
where we have used the definition given in equation\eqref{eq:BpDef}. Next, we turn to 
\begin{equation}
\begin{split}
B_{LR}(k) 
&\equiv \mu^{4-d} \int \frac{d^dp}{(2\pi)^d i} \int \frac{d^dq}{(2\pi)^d i}\ \frac{-1}{p^2+m^2+i\varepsilon}\ \frac{1}{q^2+\bar{m}^2-i\varepsilon} (2\pi)^d\ \delta^d(p-q-k)\\
&=B_{RL}(k)_{m\leftrightarrow\bar{m}}=\Theta(k^0) [B_{MR}(k)+B_{LP}(k)] +\Theta(-k^0) [B_{PR}(k)+B_{LM}(k)] \\
 &=\Theta(k^0) \Bigl( B_{RP}(k)_{m\leftrightarrow\bar{m}}  +[B_{RP}(k)]^*\Bigr)+\Theta(-k^0)  \Bigl( B_{RP}(-k)_{m\leftrightarrow\bar{m}}+ [B_{RP}(-k)]^* \Bigr)\\
 &=\Theta(k^0) [B_{RL}^+(k)]^*+\Theta(-k^0)  [B_{RL}^+(-k)]^*
 \end{split}
\end{equation}

We then look at 
\begin{equation}
\begin{split}
B_{RR}(k) &= \mu^{4-d} \int \frac{d^dp}{(2\pi)^d i} \int \frac{d^dq}{(2\pi)^d i}\ \frac{1}{p^2+m^2-i\varepsilon}\ \frac{1}{q^2+\bar{m}^2-i\varepsilon} 
(2\pi)^d\ \delta^d(p+q-k)\\
&= \mu^{4-d} \int\ \frac{d^dp}{(2\pi)^d\ i^2}\ \frac{1}{p^2+m^2-i\varepsilon}\ \frac{1}{(k-p)^2+\bar{m}^2-i\varepsilon}\\
&= \mu^{4-d} \int\ \frac{d^dp}{(2\pi)^d\ i^2}\ \frac{1}{(\omega_p-i\varepsilon)^2-(p^0)^2}\ \frac{1}{(\bar{\omega}_p-i\varepsilon)^2-(p^0-M)^2}\\
\end{split}
\end{equation}
We want to now  write the  answer of the contour integral with a definite $\varepsilon$ prescription. An examination of the sign of resulting $\varepsilon$'s shows that 
the form depends now on the sign of $M$ as well as $m-\bar{m}$.  A careful examination of $\varepsilon$'s give
\begin{equation}
\begin{split}
&B_{RR}(k)\\
 &= \Theta(m>\bar{m}) \mu^{4-d} \int\ \frac{d^{d-1}p}{(2\pi)^{d-1}\ i} \left(\frac{1}{2\omega_p}\ \frac{1}{\bar{\omega}_p^2-(|M|+\omega_p)^2+i \varepsilon} + \frac{1}{\omega_p^2-
(|M|-\bar{\omega}_p)^2- i\varepsilon}\ \frac{1}{2\bar{\omega}_p}\right)\\
&+\Theta(m<\bar{m}) \mu^{4-d} \int\ \frac{d^{d-1}p}{(2\pi)^{d-1}\ i} \left(\frac{1}{2\omega_p}\ \frac{1}{\bar{\omega}_p^2-(|M|-\omega_p)^2-i \varepsilon} + \frac{1}{\omega_p^2-
(|M|+\bar{\omega}_p)^2+ i\varepsilon}\ \frac{1}{2\bar{\omega}_p}\right)\\
\end{split}
\end{equation}
Transcribing it into $B_{RP}$ integrals, we obtain
\begin{equation}
\begin{split}
&B_{RR}(k)\\
&=\Theta(k^0) \Bigl\{ \Theta(m>\bar{m}) (B_{RP}(-k)-B_{RP}(k)^*_{m\leftrightarrow\bar{m}}) 
+ \Theta(m<\bar{m}) (B_{RP}(-k)_{m\leftrightarrow\bar{m}}-B_{RP}(k)^*) \Bigr\}\\
&\quad +\Theta(-k^0) \Bigl\{ \Theta(m>\bar{m}) (B_{RP}(k)-B_{RP}(-k)^*_{m\leftrightarrow\bar{m}}) + \Theta(m<\bar{m}) (B_{RP}(k)_{m\leftrightarrow\bar{m}}-B_{RP}(-k)^*) \Bigr\}\\
&=\Theta(k^0) B_{RR}^+(k) +\Theta(-k^0) B_{RR}^+(-k) 
\end{split}
\end{equation}
It follows that 
\begin{equation}
\begin{split}
&B_{LL}(k)= B_{RR}(k)^*\\
&=\Theta(k^0) \Bigl\{ \Theta(m>\bar{m}) (B_{RP}(-k)^*-B_{RP}(k)_{m\leftrightarrow\bar{m}}) 
+ \Theta(m<\bar{m}) (B_{RP}(-k)^*_{m\leftrightarrow\bar{m}}-B_{RP}(k)) \Bigr\}\\
&\quad +\Theta(-k^0) \Bigl\{ \Theta(m>\bar{m}) (B_{RP}(k)^*-B_{RP}(-k)_{m\leftrightarrow\bar{m}}) + \Theta(m<\bar{m}) (B_{RP}(k)^*_{m\leftrightarrow\bar{m}}-B_{RP}(-k)) \Bigr\}\\
&=\Theta(k^0) [B_{RR}^+(k)]^*+\Theta(-k^0)  [B_{RR}^+(-k)]^*
\end{split}
\end{equation}
%To summarize, we can write the answers in this subsection in the form 
%\begin{equation}
%\begin{split}
%B_{RL}(k) &=\Theta(k^0)B_{RL}^+(k) +\Theta(-k^0)  B_{RL}^+(-k)\\
%B_{LR}(k)  &=\Theta(k^0) [B_{RL}^+(k)]^*+\Theta(-k^0)  [B_{RL}^+(-k)]^*\\
%B_{RR}(k)&=\Theta(k^0) B_{RR}^+(k) +\Theta(-k^0) B_{RR}^+(-k) \\
%B_{LL}(k) &=\Theta(k^0) [B_{RR}^+(k)]^*+\Theta(-k^0)  [B_{RR}^+(-k)]^*\\
%\end{split}
%\end{equation}
We will now turn to the case of space-like $k^\mu$ to prove similar relations in that case.

\subsubsection{Space-like $k^\mu$ : reduction of divergent integrals}
We will study $B_{RP}(k)$ when $k^\mu$ is space-like. We set  $k^\mu=\{0,Q=\sqrt{k^2}, \vec{0}_{d-2}\}$ where we can take $Q>0$ without loss of generality.
\begin{equation}
\begin{split}
B_{RP}(k) &= \mu^{4-d} \int \frac{d^dp}{(2\pi)^d i} \int \frac{d^dq}{(2\pi)^d i} \frac{1}{p^2+m^2-i\varepsilon}\ \ 2\pi i\ \delta_{+}(q^2+\bar{m}^2)\ (2\pi)^d\ \delta^d(p-q-k)\\
&= \mu^{4-d} \int \frac{d^dp}{(2\pi)^d i} \int \frac{d^dq}{(2\pi)^d i} \frac{1}{p^2+m^2-i\varepsilon}\ \ 2\pi i\ \delta_{+}(q^2+\bar{m}^2)\\
&\qquad \times(2\pi)^d\ \delta(p^0-q^0)\delta(p_{||}-q_{||}-Q)\delta(p_\perp-q_\perp)\\
%&= \mu^{4-d} \int\ \frac{d^dq}{(2\pi)^{d}i}\ 2\pi \delta_{+}(q^2+\bar{m}^2)\ \frac{1}{\omega_{q_\perp}^2+ (Q+q_{||})^2-q^2_0-i\varepsilon} \\
&= \mu^{4-d} \int\ \frac{d^{d-1}q}{(2\pi)^{d-1}i}\ \frac{1}{2\bar{\omega}_q}\ \frac{1}{q_\perp^2+ (Q+q_{||})^2+m^2-\bar{\omega}_{q}^2-i\varepsilon}\\
&= \mu^{4-d} \int\ \frac{d^{d-1}q}{(2\pi)^{d-1}i} \frac{1}{2\bar{\omega}_q}\ \frac{1}{m^2-\bar{m}^2  + Q^2 +2Q q_{||} -i\varepsilon}\\
&= \frac{\mu^{4-d}}{4Q} \int\ \frac{d^{d-1}q}{(2\pi)^{d-1}i} \frac{1}{\bar{\omega}_q}\ \frac{1}{q_{||}^* +q_{||} -i\varepsilon}
\end{split}
\end{equation}
where, we have defined 
\[ q_{||}^* \equiv \frac{m^2-\bar{m}^2 + Q^2}{2Q}\ .\]

Now, we move on to calculating $B_{RL}(k)$, given by
\begin{equation}
\begin{split}
B_{RL}(k) &= \mu^{4-d} \int \frac{d^dp}{(2\pi)^d i} \int \frac{d^dq}{(2\pi)^d i}\ \frac{1}{p^2+m^2-i\varepsilon}\ \frac{-1}{q^2+\bar{m}^2+i\varepsilon} (2\pi)^d\ \delta^d(p-q-k)\\
&= \mu^{4-d} \int\ \frac{d^dp}{(2\pi)^d}\ \frac{1}{p^2+m^2-i\varepsilon}\ \frac{1}{(p-k)^2+\bar{m}^2+i\varepsilon}\\
&= \mu^{4-d} \int\ \frac{d^dp}{(2\pi)^d}\ \frac{1}{\omega_p^2-(p^0)^2-i\varepsilon}\ \frac{1}{\bar{\omega}_{p_\perp}^2+(Q-p_{||})^2-(p^0)^2+i\varepsilon}\\
&= \mu^{4-d} \int\ \frac{d^{d-1}p}{(2\pi)^{d-1}i} \Bigg(-\frac{1}{2\omega_p}\ \frac{1}{\bar{m}^2 - m^2 + Q^2 - 2Q p_{||} +i\varepsilon } \\
&\qquad \qquad \qquad+ \frac{1}{m^2 -\bar{m}^2  - Q^2 + 2Q p_{||} -i\varepsilon}\ \frac{1}{2\bar{\omega}_{p-k}}\Bigg)
\end{split}
\end{equation}
Now, we take $p=-q$ in the first integral and $p = k + q$ for the second integral to write
\begin{equation}
\begin{split}
B_{RL}(k) &=\mu^{4-d}  \int\ \frac{d^{d-1}q}{(2\pi)^{d-1}i}  \Bigg(-\frac{1}{2\omega_q}\ \frac{1}{\bar{m}^2 - m^2 + Q^2 +2Q q_{||} +i\varepsilon } \\
&\qquad \qquad \qquad+ \frac{1}{m^2 -\bar{m}^2  + Q^2 + 2Q q_{||} -i\varepsilon}\ \frac{1}{2\bar{\omega}_{q}}\Bigg)\\
&= B_{RP}(k)^*|_{m\leftrightarrow\bar{m}} + B_{RP}(k) = B^+_{RL}(k)
\end{split}
\end{equation}
Similarly, 
\begin{equation}
\begin{split}
B_{LR}(k) &=B_{RL}(k)_{m\leftrightarrow\bar{m}} =  B_{RP}(k)|_{m\leftrightarrow\bar{m}} + B_{RP}(k)^* = [B^+_{RL}(k)]^*
\end{split}
\end{equation}

Let us now do the $B_{RR}(k)$ integral for space-like $k^\mu$  :
\begin{equation}
\begin{split}
B_{RR}(k) &= \mu^{4-d} \int \frac{d^dp}{(2\pi)^d i} \int \frac{d^dq}{(2\pi)^d i}\ \frac{1}{p^2+m^2-i\varepsilon}\ \frac{1}{q^2+\bar{m}^2-i\varepsilon} (2\pi)^d\ \delta^d(p+q-k)\\
&= \mu^{4-d} \int\ \frac{d^dp}{(2\pi)^d\ i^2}\ \frac{1}{p^2+m^2-i\varepsilon}\ \frac{1}{(k-p)^2+\bar{m}^2-i\varepsilon}\\
&= \mu^{4-d} \int\ \frac{d^dp}{(2\pi)^d\ i^2}\ \frac{1}{\omega_p^2-(p^0)^2-i\varepsilon}\ \frac{1}{\bar{\omega}_{p_\perp}^2+(Q-p_{||})^2-(p^0)^2-i\varepsilon}\\
&= \mu^{4-d} \int\ \frac{d^{d-1}p}{(2\pi)^{d-1} i} \Bigg(\frac{1}{2\sqrt{\omega_{p_\perp}^2 + p_{||}^2}}\ \frac{1}{\bar{m}^2 - m^2 + Q^2 - 2Q p_{||} -i\varepsilon\ sgn(\bar{\omega}_{p_\perp} - \omega_{p_\perp})} \\
&\qquad \qquad \qquad+ \frac{1}{m^2 -\bar{m}^2  - Q^2 + 2Q p_{||} -i\varepsilon\ sgn( \omega_{p_\perp} - \bar{\omega}_{p_\perp})}\ \frac{1}{2\sqrt{\bar{\omega}_{p_\perp}^2+(Q-p_{||})^2}}\Bigg)
\end{split}
\end{equation}
Now, we take $p_{||} = -q_{||}$ in the first integral and  $p_{||} = Q + q_{||}$ in the second, to get
\begin{equation}
\begin{split}
B_{RR}(k) &= \mu^{4-d}  \int\ \frac{d^{d-1}q}{(2\pi)^{d-1}i} \Bigg(\frac{1}{2\sqrt{\omega_{q_\perp}^2 + q_{||}^2}}\ \frac{1}{\bar{m}^2 - m^2 + Q^2 +2Q q_{||} -i\varepsilon\ sgn(\bar{\omega}_{p_\perp} - \omega_{p_\perp})} \\
&\qquad \qquad+  \frac{1}{m^2 -\bar{m}^2  + Q^2 +2Q q_{||} -i\varepsilon\ sgn( \omega_{q_\perp} - \bar{\omega}_{q_\perp})}\ \frac{1}{2\sqrt{\bar{\omega}_{q_\perp}^2+q_{||}^2}}\Bigg) \\
&= \Theta(m > \bar{m})\mu^{4-d} \int\ \frac{d^{d-1}q}{(2\pi)^{d-1} i} \Bigg( \frac{1}{m^2 -\bar{m}^2  + Q^2 - 2Q q_{||} -i\varepsilon}\ \frac{1}{2\bar{\omega}_q} \\
&\qquad \qquad+ \frac{1}{2\omega_q}\ \frac{1}{\bar{m}^2 - m^2 + Q^2 + 2Q q_{||} +i\varepsilon}
\Bigg)\\
&+ \Theta(m < \bar{m})\mu^{4-d} \int\ \frac{d^{d-1}q}{(2\pi)^{d-1} i} \Bigg(\frac{1}{2\omega_q}\frac{1}{\bar{m}^2 - m^2 + Q^2 - 2Q q_{||} -i\varepsilon} \\
&\qquad \qquad+ \frac{1}{m^2 -\bar{m}^2  + Q^2 +2Q q_{||} +i\varepsilon}\ \frac{1}{2\bar{\omega}_q}\Bigg)\\
&= \Theta(m>\bar{m}) (B_{RP}(-k)-B_{RP}(k)^*_{m\leftrightarrow\bar{m}}) 
+ \Theta(m<\bar{m}) (B_{RP}(-k)_{m\leftrightarrow\bar{m}}-B_{RP}(k)^*) \\
&= B^+_{RR}(k)
\end{split}
\end{equation}
Here, in the penultimate step, we have done some variable redefinitions to obtain an answer similar to the time-like case. We can finally compute 
\begin{equation}
\begin{split}
B_{LL}(k)&= B_{RR}(k)^*= [B^+_{RR}(k)]^*\ .
\end{split}
\end{equation}

\subsubsection{Summary of divergent integrals}
We can now put together various cases and write 
\begin{equation}
\begin{split}
B_{RL}(k) &=\Theta(k^0)\Theta(-k^2)B_{RL}^+(k) +\Theta(-k^0) \Theta(-k^2) B_{RL}^+(-k)+\Theta(k^2)B_{RL}^+(k)\\
B_{LR}(k)  &=\Theta(k^0)\Theta(-k^2) [B_{RL}^+(k)]^*+\Theta(-k^0) \Theta(-k^2) [B_{RL}^+(-k)]^*+\Theta(k^2) [B_{RL}^+(k)]^*\\
B_{RR}(k)&=\Theta(k^0)\Theta(-k^2) B_{RR}^+(k) +\Theta(-k^0) \Theta(-k^2) B_{RR}^+(-k) +\Theta(k^2)B_{RR}^+(k) \\
B_{LL}(k) &=\Theta(k^0) \Theta(-k^2) [B_{RR}^+(k)]^*+\Theta(-k^0)\Theta(-k^2)  [B_{RR}^+(-k)]^*+\Theta(k^2)  [B_{RR}^+(k)]^*\ .
\end{split}
\end{equation}
where
\begin{equation}\label{eq:BpDef2}
\begin{split}
B_{RL}^+(k) &\equiv B_{RP}(k)+B_{ML}(k)= B_{RP}(k) +[B_{RP}(k)]^*_{m\leftrightarrow\bar{m}} \\
 B_{RR}^+(k) &\equiv \Theta(m>\bar{m}) (B_{RM}(k)-B_{ML}(k)) + \Theta(m<\bar{m}) (B_{PR}(k)-B_{LP}(k)) \\
&= \Theta(m>\bar{m}) (B_{RP}(-k)-B_{RP}(k)^*_{m\leftrightarrow\bar{m}}) 
+ \Theta(m<\bar{m}) (B_{RP}(-k)_{m\leftrightarrow\bar{m}}-B_{RP}(k)^*) \
 \end{split}
\end{equation}
This is apart from the other divergent integrals :
\begin{equation}
\begin{split}
B_{RM}(k) &= B_{RP}(-k)\ ,\qquad \\
B_{LP}(k) &=  [B_{RP}(k)]^* \ ,\qquad
B_{LM}(k)  = [B_{RP}(-k)]^*\ ,\\
B_{PR}(k) &= B_{RP}(-k)|_{m\leftrightarrow\bar{m}}\ ,\qquad
B_{MR}(k)  = B_{RP}(k)|_{m\leftrightarrow\bar{m}}\ ,\qquad\\
B_{PL}(k)  &= B_{RP}(-k)^*|_{m\leftrightarrow\bar{m}}\ ,\qquad
B_{ML}(k)  = B_{RP}(k)^*|_{m\leftrightarrow\bar{m}}\ .
\end{split}
\end{equation}
We note that all these integrals can be written in terms of $B_{RP}(k)$ as advertised.

\subsubsection{Reduction and identities due to largest time equations}
A further reduction is possible using largest time equations and their concomitant cutting rules :
\begin{equation}
\begin{split}
B_{RR}(k) +B_{RL} (k) &= B_{RP}(k) +B_{RM}(k)\\
B_{LR}(k) +B_{LL} (k) &= B_{LP}(k) +B_{LM}(k)\\
B_{PR}(k) +B_{PL} (k) &= B_{PP}(k) +B_{MM}(k)\\
B_{MR}(k) +B_{ML} (k) &= B_{MP}(k) +B_{RM}(k)\\
\end{split}
\end{equation}

From applying these identities, we can conclude that $\text{Re}[B_{RP}(k) + B_{RP}(-k)]$ and $\text{Im}[B_{RP}(k)- B_{RP}(-k)]$ is symmetric 
under $m\leftrightarrow \bar{m}$ exchange. Further, the real part(viz., the cut) of $B_{RP}(k) $ integral is given by 
 \begin{equation}
\begin{split}
B_{RP}(k) + B_{RP}(k)^*&=\text{Re}\ B^+_{RL}(k)\\
&= \frac{\text{Vol}(\mathbb{S}^{d-2})}{32\pi^2}\Bigl[1 
+ 2\left(\frac{m^2 + \bar{m}^2}{k^2}\right) + \left(\frac{m^2 - \bar{m}^2}{k^2}\right)^2 \Bigr]^{\frac{d-3}{2}} \\
&\quad \times \Bigl\{ \Theta(k^2) \frac{1}{2\pi}\Gamma\Bigl(\frac{3}{2}-\frac{d}{2}\Bigr)\Gamma\Bigl(\frac{d}{2}-\frac{1}{2}\Bigr)\left(\frac{k^2}{(4\pi\mu)^2}\right)^{\frac{d-4}{2}}\Bigr.\\
&\qquad \Bigl.+ [\Theta(k^0)\Theta\left((\bar{m}-m)^2+k^2\right)+\Theta(-k^0)]\Theta(-k^2)\left(\frac{-k^2}{(4\pi\mu)^2}\right)^{\frac{d-4}{2}} \Bigr\}\
\end{split}
\end{equation}

These conditions  in turn  lead to the identities : 
\begin{equation}
\begin{split}
B_{RP}(k) +B_{LP} (k) &= B_{MR}(k) +B_{ML}(k)\\
B_{RP}(k) +B_{LM}(k) &= B_{MR}(k) +B_{PL}(k) \\
B_{RP}(k) +B_{PR}(k) &= B_{RM}(k) +B_{MR}(k) \\
B_{PR}(k) +B_{PL} (k) &= B_{RM}(k) +B_{LM}(k)\\
B_{PR}(k) +B_{ML}(k) &= B_{RM}(k) +B_{LP}(k) \\
B_{LP}(k) +B_{PL}(k) &= B_{LM}(k) +B_{ML}(k)\\
\end{split}
\end{equation}
 From,  these identities we get 
\begin{equation}\label{eq:BpDef3}
\begin{split}
B_{RL}^+(k) &\equiv B_{RP}(k)+B_{ML}(k)= B_{RP}(k) +[B_{RP}(k)]^*_{m\leftrightarrow\bar{m}} \\
 B_{RR}^+(k) &= B_{RP}(-k)-B_{RP}(k)^*_{m\leftrightarrow\bar{m}} = B_{RP}(-k)_{m\leftrightarrow\bar{m}}-B_{RP}(k)^* 
 \end{split}
\end{equation}
which in turn obey
\begin{equation}
\begin{split}
 B_{RR}^+(k)+B_{RL}^+(k) &= B_{RP}(k) +B_{RP}(-k) = B_{RP}(k)+B_{RM}(k)
 \end{split}
\end{equation}

Another implication is 
\begin{equation}
\begin{split}
\text{Re}\ B^+_{RR}(k)
&= \frac{\text{Vol}(\mathbb{S}^{d-2})}{32\pi^2}\Bigl[1 
+ 2\left(\frac{m^2 + \bar{m}^2}{k^2}\right) + \left(\frac{m^2 - \bar{m}^2}{k^2}\right)^2 \Bigr]^{\frac{d-3}{2}} \\
&\quad \times \frac{1}{2}  [\Theta(k^0)-\Theta(-k^0)]\Theta\left(-(\bar{m}-m)^2-k^2\right)\Theta(-k^2)\left(\frac{-k^2}{(4\pi\mu)^2}\right)^{\frac{d-4}{2}} 
\end{split}
\end{equation}
The following combination is symmetric under $m\leftrightarrow \bar{m}$ as well as $k\leftrightarrow -k$ exchange.
\begin{equation}
\begin{split}
B_{RP}(k) &+B_{RP}(k)^* +B_{PM}(k)\\
&= \frac{\text{Vol}(\mathbb{S}^{d-2})}{32\pi^2}\Bigl[1 
+ 2\left(\frac{m^2 + \bar{m}^2}{k^2}\right) + \left(\frac{m^2 - \bar{m}^2}{k^2}\right)^2 \Bigr]^{\frac{d-3}{2}} \\
&\quad \times \Bigl\{ \Theta(k^2) \frac{1}{2\pi}\Gamma\Bigl(\frac{3}{2}-\frac{d}{2}\Bigr)\Gamma\Bigl(\frac{d}{2}-\frac{1}{2}\Bigr)\left(\frac{k^2}{(4\pi\mu)^2}\right)^{\frac{d-4}{2}}\Bigr.\\
&\qquad \Bigl.+ \Theta(-k^2)\left(\frac{-k^2}{(4\pi\mu)^2}\right)^{\frac{d-4}{2}} \Bigr\}\
\end{split}
\end{equation}

\subsection{Evaluation of  \texorpdfstring{$B_{RP}(k)$}{BRP}}
The basic integral $B_{RP}(k)$ has the following form on the time-like case : 
\begin{equation}
\begin{split}
B_{RP}(k) 
&= \frac{\mu^{4-d}\ \text{Vol}(\mathbb{S}^{d-2})}{2(2\pi)^{d-1}\ i} \int_{\bar{m}}^\infty d\bar{\omega}_p\ \frac{(\bar{\omega}_p^2-\bar{m}^2)^{\frac{d-3}{2}}}{\omega_p^2-(M+\bar{\omega}_p)^2-i\varepsilon}\\
&= \frac{\mu^{4-d}\ \text{Vol}(\mathbb{S}^{d-2})}{2(2\pi)^{d-1}\ i} \int_{\bar{m}}^\infty d\bar{\omega}_p\ \frac{(\bar{\omega}_p^2-\bar{m}^2)^{\frac{d-3}{2}}}{2M(\bar{\omega}_p^*-\bar{\omega}_p)-i\varepsilon}\\
\end{split}
\end{equation}
with 
\[ \bar{\omega}_p^* \equiv \frac{m^2-\bar{m}^2 - M^2}{2M}\ .\]
The same integral in the space-like case takes the form
\begin{equation}
\begin{split}
B_{RP}(k) 
&= \frac{\mu^{4-d}}{2Q} \int\ \frac{d^{d-1}q}{(2\pi)^{d-1}i} \frac{1}{2\bar{\omega}_q}\ \frac{1}{q_{||}^* +q_{||} -i\varepsilon}
\end{split}
\end{equation}
where, we have defined 
\[ q_{||}^* \equiv \frac{m^2-\bar{m}^2 + Q^2}{2Q}\ .\]
Our aim in this subsection is to evaluate these integrals and extract out the appropriate divergences.

\subsubsection{Time-like $k^\mu$ : computation of divergences}
We begin by setting $\bar{\omega}_p= \bar{m} \cosh\eta$ in the time-like case to get 
\begin{equation}
\begin{split}
B_{RP}(k) 
&= \frac{\mu^{4-d}\ \text{Vol}(\mathbb{S}^{d-2})\bar{m}^{d-3}}{4M(2\pi)^{d-1}\ i} \int_0^\infty d\eta\ \frac{\sinh^{d-2}\eta}{\bar{\gamma}-\cosh \eta}\\
&= \frac{\text{Vol}(\mathbb{S}^{d-2})}{32\pi^2\ } \frac{\bar{m}}{M} \Bigl(\frac{\bar{m}^2}{4\pi\mu^2}\Bigr)^{\frac{d-4}{2}}\ \int_0^\infty \frac{d\eta}{i\pi^{d/2-1}}\ \frac{\sinh^{d-2}\eta}{\bar{\gamma}-\cosh \eta}
\end{split}
\end{equation}
where we have defined
\begin{equation}
\begin{split}
\bar{\gamma}  &= \frac{\bar{\omega}_p^*}{\bar{m}} \equiv \frac{m^2-\bar{m}^2 - M^2 -i\varepsilon}{2M\bar{m}}\ .
\end{split}
\end{equation}
Thus, we have reduced our analysis to the integral 
\begin{equation}
\begin{split}
F(\bar{\gamma}) \equiv \int_0^\infty \frac{d\eta}{i\pi^{d/2-1}}\ \frac{\sinh^{d-2}\eta}{\bar{\gamma}-\cosh \eta}
\end{split}
\end{equation}
This integral can then analyzed in detail to study the analytic structure of this integral. But, for our purposes, it is sufficient to extract the divergences.
%For small $\bar{\gamma}$, this integral can be performed to give
%\begin{equation}
%\begin{split}
%F(\bar{\gamma})&= \frac{i}{4}\frac{  \Gamma\Bigl(d-1\Bigr) }{\Gamma\Bigl(\frac{d}{2}\Bigr)} \frac{  \Gamma\Bigl(\frac{3}{2} - \frac{d}{2}\Bigr)  }{(4\pi)^{\frac{d}{2} - \frac{3}{2}} }(1 - \bar{\gamma}^2)^{\frac{1}{2} (d-3)} \\
%&+i \frac{ \Gamma\Bigl(d-1\Bigr)  }{  \Gamma\Bigl(\frac{d}{2}\Bigr)}\
%\frac{\Gamma\Bigl(1 - \frac{d}{2}\Bigr)}{(4\pi)^{ \frac{d}{2}-1} } \frac{ 
%    {}_2F_1[1, 1 - \frac{d}{2}, \frac{1}{2}, \bar{\gamma}^2]-1}{
%2 \bar{\gamma} }
%\end{split}
%\end{equation}
%Substituting this back, we obtain
%\begin{equation}
%\begin{split}
%B_{RP}(k) 
%&= -\frac{i}{(4\pi)^2}\frac{ \bar{\gamma}\bar{m}}{M} \Bigl\{ \frac{2}{d-4}+ \ln\Bigl(\frac{\bar{m}^2}{4\pi\mu^2e^{-\gamma_E}}\Bigr)-1+\frac{1}{2 \bar{\gamma}^2 } {}_2F_1^{0,1,0,0}[1, -1, \frac{1}{2}, \bar{\gamma}^2]\Bigr\}\\
%& -\frac{i}{16\pi}\frac{\bar{m}}{M}\sqrt{1- \bar{\gamma}^2}
%\end{split}
%\end{equation}

For our computation of $\beta$ function, we need to extract out the divergent part of these integrals. Focusing on the large $\bar{\omega}_p$ contribution, we can approximate  $B_{RP}(k)$ by 
\begin{equation}
\begin{split}
B_{RP}(k)&=\frac{\mu^{4-d}\ \text{Vol}(\mathbb{S}^{d-2})}{2(2\pi)^{d-1}\ i} \int_{\bar{m}}^\infty d\bar{\omega}_p\  (\bar{\omega}_p^2-\bar{m}^2)^{\frac{d}{2}} 
\Bigl[ -\frac{1}{2M\bar{\omega}_p^4} +\frac{M^2+\bar{m}^2-m^2}{4M^2\bar{\omega}_p^5}+ O(\bar{\omega}_p^{-6}) \Bigr]  \\
&= \frac{i}{128\pi^2} \frac{\Gamma\Bigl(\frac{d+2}{2}\Bigr)}{\Gamma\Bigl(\frac{d-1}{2}\Bigr)} \Bigl(\frac{\bar{m}^2}{4\pi\mu^2}\Bigr)^{\frac{d-4}{2}}\
\Bigl\{ \frac{16\bar{m}}{3M} \Gamma\Bigl(\frac{3-d}{2}\Bigr) -\sqrt{\pi}  \Gamma\Bigl(\frac{4-d}{2}\Bigr) \frac{M^2+\bar{m}^2-m^2}{M^2}+ \ldots\Bigr\}
\end{split}
\end{equation}
Near $d=4$, this gives  
\begin{equation}
\begin{split}
B_{RP}(k)&= \frac{i}{(4\pi)^2}\
\Bigl\{- \frac{16\bar{m}}{3M}+ \frac{M^2+\bar{m}^2-m^2}{2M^2}  \Bigl[\frac{2}{d-4} + \ln\Bigl(\frac{4\bar{m}^2}{4\pi\mu^2e^{-\gamma_E}} \Bigr)-\frac{1}{2}\Bigr]+ \ldots\Bigr\}
\end{split}
\end{equation}
so that

\subsubsection{Space-like $k^\mu$ : computation of divergences}
In this subsection, we will consider the space-like case and confirm that the  divergence structure is same in the space-like case.
Let us first get the real part of $B_{RP}(k)$, which is given by
\begin{equation}
\begin{split}
B_{RP}(k)&= \frac{\mu^{4-d}\ \text{Vol}(\mathbb{S}^{d-2})}{2(2\pi)^{d-1}\ i} \int_{\bar{m}}^\infty d\bar{\omega}_p\ \frac{(\bar{\omega}_p^2-\bar{m}^2)^{\frac{d-3}{2}}}{2M(\bar{\omega}_p^*-\bar{\omega}_p)-i\varepsilon}\\
\end{split}
\end{equation}

\begin{equation}\label{eq:spaceBPR}
\begin{split}
B_{RP}(k) 
&= \frac{\mu^{4-d}}{2Q} \int\ \frac{d^{d-1}q}{(2\pi)^{d-1}i} \frac{1}{2\bar{\omega}_q}\ \frac{1}{q_{||}^* +q_{||} -i\varepsilon} \\
&= \frac{\mu^{4-d}}{4Q(2\pi)^{d-1}\ i} \int_{\bar{m}}^{\infty}\ d\bar{\omega}_{q_\perp}\ \bar{\omega}_{q_\perp}(\bar{\omega}_{q_\perp}^2 - \bar{m}^2)^{\frac{d-4}{2}} \int_{-\infty}^{\infty}\ dq_{||} \frac{1}{{\sqrt{\bar{\omega}_{q_\perp}^2+q_{||}^2}}}\ \frac{1}{q_{||}^* +q_{||} -i\varepsilon}
\end{split}
\end{equation}
where, we have defined 
\[ q_{||}^* \equiv \frac{m^2-\bar{m}^2 + Q^2}{2Q}\ .\]

Let us first get the real part of $B_{RP}(k)$, which is given by
\begin{equation}
\begin{split}
&B_{RP}(k)\bigg|_{\text{real part}}\\
&= \frac{\mu^{4-d}}{4Q(2\pi)^{d-1}\ i} \int_{\bar{m}}^{\infty}\ d\bar{\omega}_{q_\perp}\ \bar{\omega}_{q_\perp}(\bar{\omega}_{q_\perp}^2 - \bar{m}^2)^{\frac{d-4}{2}} \\
&\int_{-\infty}^{\infty}\ dq_{||} \frac{1}{{\sqrt{\bar{\omega}_{q_\perp}^2+q_{||}^2}}}\ \left(\frac{1}{q_{||}^* +q_{||} -i\varepsilon} 
-\frac{1}{q_{||}^* +q_{||} +i\varepsilon} \right) \\
&= \frac{\mu^{4-d}}{4Q(2\pi)^{d-1}\ i} \int_{\bar{m}}^{\infty}\ d\bar{\omega}_{q_\perp}\ \bar{\omega}_{q_\perp}(\bar{\omega}_{q_\perp}^2 - \bar{m}^2)^{\frac{d-4}{2}} \int_{-\infty}^{\infty}\ dq_{||} \frac{1}{{\sqrt{\bar{\omega}_{q_\perp}^2+q_{||}^2}}}\ 2\pi i \ \delta(q_{||}^* +q_{||}) 
\end{split}
\end{equation}
The above equation can easily be seen to be equal to $B_{PP}(k)$ for the space-like case in \eqref{eq:BPP}. So, we have in the space-like case
\begin{equation}
 B_{RP}(k)\bigg|_{\text{real part}} = B_{PP}(k)
\end{equation}

Let us now get the imaginary part of $B_{RP}(k)$, which is also the divergent part. It is given by

\begin{equation}
\begin{split}
B_{RP}(k)\bigg|_{\text{imaginary part}}
&=\frac{i}{2(4\pi)^2}\ \frac{Q^2+m^2-\bar{m}^2}{Q^2} \left(\frac{2}{d-4} + \ln \frac{\bar{m}^2}{4\pi \mu^2e^{-\gamma_E}}-1 \right) + ....
\end{split} 
\end{equation}
when, $m= \bar{m}$, we get

\begin{equation}
\begin{split}
B_{RP}(k)\bigg|_{\text{imaginary part}}
&=\frac{i}{2(4\pi)^2}\ \left(\frac{2}{d-4}   + \ln \frac{m^2}{4\pi \mu^2e^{-\gamma_E}} -1  \right) + ....
\end{split} 
\end{equation}
We see that these divergences are same as the time-like case.

\subsubsection{Summary of divergences}
We now summarize the divergences in various integrals. We start with  
\begin{equation}
\begin{split}
\diver[B_{RP}]\Big|_{\msbar} &= \frac{i}{(4\pi)^2}\
 \frac{k^2-\bar{m}^2+m^2}{2k^2}  \Bigl[\frac{2}{d-4} + \ln\Bigl(\frac{1}{4\pi e^{-\gamma_E}} \Bigr)\Bigr]
\end{split}
\end{equation}
and
\begin{equation}
\begin{split}
\diver[B^+_{RL}]\Big|_{\msbar} &= \frac{i}{(4\pi)^2}\
 \frac{m^2-\bar{m}^2}{k^2}  \Bigl[\frac{2}{d-4} + \ln\Bigl(\frac{1}{4\pi e^{-\gamma_E}} \Bigr)\Bigr] \\
 \diver[B^+_{RR}]\Big|_{\msbar}&= \frac{i}{(4\pi)^2}\
  \Bigl[\frac{2}{d-4} + \ln\Bigl(\frac{1}{4\pi e^{-\gamma_E}} \Bigr)\Bigr] \\
\end{split}
\end{equation}
Thus, 
\begin{equation}
\begin{split}
\diver[B_{RP}]\Big|_{\msbar} &=\diver[B_{RM}]\Big|_{\msbar}= \frac{i}{(4\pi)^2}\
 \frac{k^2-\bar{m}^2+m^2}{2k^2}  \Bigl[\frac{2}{d-4} + \ln\Bigl(\frac{1}{4\pi e^{-\gamma_E}} \Bigr)\Bigr]\\
\diver[B_{PR}]\Big|_{\msbar} &= \diver[B_{MR}]\Big|_{\msbar}=\frac{i}{(4\pi)^2}\
 \frac{k^2+\bar{m}^2-m^2}{2k^2}  \Bigl[\frac{2}{d-4} + \ln\Bigl(\frac{1}{4\pi e^{-\gamma_E}} \Bigr)\Bigr]\\
 \diver[B_{RL}]\Big|_{\msbar} &= -\diver[B_{LR}]\Big|_{\msbar}= \frac{i}{(4\pi)^2}\
 \frac{m^2-\bar{m}^2}{k^2}  \Bigl[\frac{2}{d-4} + \ln\Bigl(\frac{1}{4\pi e^{-\gamma_E}} \Bigr)\Bigr] \\
 \diver[B_{RR}]\Big|_{\msbar} & =- \diver[B_{LL}]\Big|_{\msbar} = \frac{i}{(4\pi)^2}\  \Bigl[\frac{2}{d-4} + \ln\Bigl(\frac{1}{4\pi e^{-\gamma_E}} \Bigr)\Bigr]
\end{split}
\end{equation}

When $m=\bar{m}$, we can thus summarize the divergence of  `quarter-cut' integrals  as
\begin{equation}
\begin{split}
\diver[B_{RP}]\Big|_{\msbar} &=\diver[B_{PR}]\Big|_{\msbar}= \diver[B_{RM}]\Big|_{\msbar}= \diver[B_{MR}]\Big|_{\msbar}= \frac{i}{2(4\pi)^2}\
 \Bigl[\frac{2}{d-4} + \ln\Bigl(\frac{1}{4\pi e^{-\gamma_E}} \Bigr)\Bigr]\\
 \diver[B_{LP}]\Big|_{\msbar} &=\diver[B_{PL}]\Big|_{\msbar}= \diver[B_{LM}]\Big|_{\msbar}= \diver[B_{ML}]\Big|_{\msbar}= -\frac{i}{2(4\pi)^2}\
 \Bigl[\frac{2}{d-4} + \ln\Bigl(\frac{1}{4\pi e^{-\gamma_E}} \Bigr)\Bigr]\\
\end{split}
\end{equation}
This along with 
\begin{equation}
\begin{split}
\diver[B_{RR}]\Big|_{\msbar}&= \frac{i}{(4\pi)^2}\
 \Bigl[\frac{2}{d-4} + \ln\Bigl(\frac{1}{4\pi e^{-\gamma_E}} \Bigr)\Bigr]\\
 \diver[B_{LL}]\Big|_{\msbar}&= -\frac{i}{(4\pi)^2}\
 \Bigl[\frac{2}{d-4} + \ln\Bigl(\frac{1}{4\pi e^{-\gamma_E}} \Bigr)\Bigr]\\
\end{split}
\end{equation}
summarizes all the divergences needed in this work.

\subsection{UV divergences and symmetry factors}
\label{sec:onelooppvudivergences}

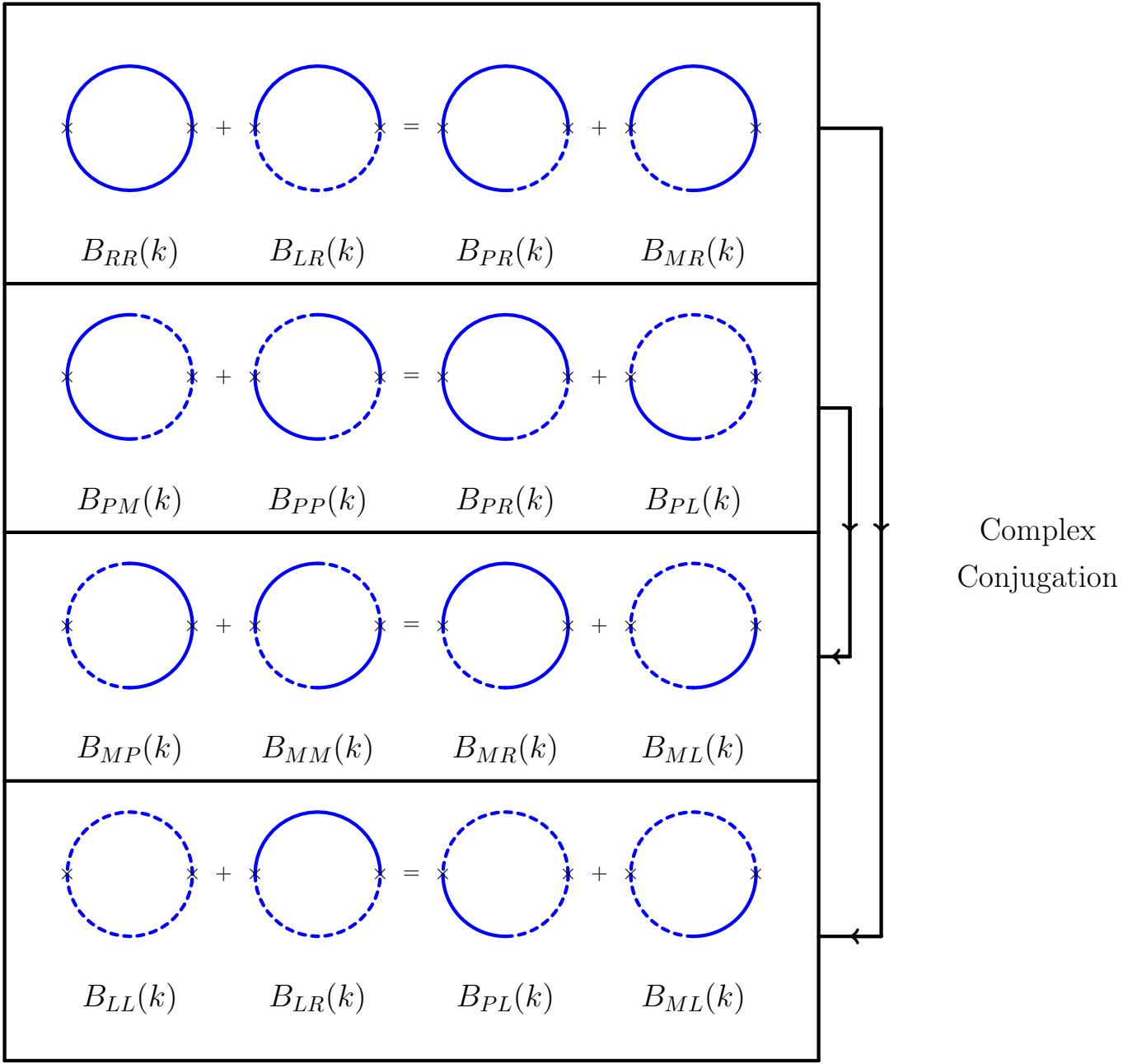
\begin{figure}[ht] 
\begin{center}
\begin{tikzpicture}[line width=1 pt, scale=1]
%\begin{scope}[rotate=-80]

\draw [ultra thick] (-2,2) -- (11,2);	
\draw [ultra thick] (-2,-2.5) -- (11,-2.5);	
\draw [ultra thick] (-2,-6.5 ) -- (11,-6.5);	 
 \draw [ultra thick] (-2,-10.5) -- (11,-10.5);	
\draw [ultra thick] (-2,-15) -- (11,-15);	

\draw [ultra thick] (-2,-6.5) -- (-2,2);	 
\draw [ultra thick] (-2,-6.5) -- (-2,-15);	 
\draw [ultra thick] (11,-6.5) -- (11, 2);	 
\draw [ultra thick] (11,-6.5) -- (11,-15);	

\draw [ultra thick] (11,-4.5) -- (11.5,-4.5);	 
\draw [ultra thick][<-] (11.2,-8.5) -- (11.5,-8.5);	 
\draw [ultra thick] (11,-8.5) -- (11.5,-8.5);	 
\draw [ultra thick] (11.5,-4.5) -- (11.5,-8.5);	 
\draw [ultra thick][->] (11.5,-4.5) -- (11.5,-6.5);

\draw [ultra thick] (11,0) -- (12,0);	 
\draw [ultra thick][<-] (11.5,-13) -- (12,-13);	 
\draw [ultra thick] (11,-13) -- (12,-13);	 
\draw [ultra thick] (12,0) -- (12,-13);	 
\draw [ultra thick][->] (12,-4.5) -- (12,-6.5);

\node at (14.5,-6.5) {\Large Complex};
\node at (14.5,-7.25) {\Large Conjugation};

\draw [phir, ultra thick, domain=0:180] plot ({1*cos(\x)}, {1*sin(\x)});
\draw [phir, ultra thick, domain=180:360] plot ({1*cos(\x)}, {1*sin(\x)}); 
\node at (-1,0) {$\times $};	
\node at (1,0) {$\times $};	
\node at (0,-2) {\Large $B_{RR}(k)$};	
\begin{scope}[shift={(0.5,0)}]
\node at (1,0) {$+ $};	
\end{scope}
\begin{scope}[shift={(3,0)}]
\draw [phir, ultra thick, domain=0:180] plot ({1*cos(\x)}, {1*sin(\x)});
\draw [phil, ultra thick, domain=180:360] plot ({1*cos(\x)}, {1*sin(\x)}); 
\node at (-1,0) {$\times $};	
\node at (1,0) {$\times $};	
\node at (0,-2) {\Large $B_{LR}(k)$};
\end{scope}
\begin{scope}[shift={(3.5,0)}]
\node at (1,0) {$=$};	
\end{scope}
\begin{scope}[shift={(6,0)}]
\draw [phir, ultra thick, domain=0:180] plot ({1*cos(\x)}, {1*sin(\x)});
\draw [phir, ultra thick, domain=180:270] plot ({1*cos(\x)}, {1*sin(\x)}); 
\draw [phil, ultra thick, domain=270:360] plot ({1*cos(\x)}, {1*sin(\x)}); 
\node at (-1,0) {$\times $};	
\node at (1,0) {$\times $};	
\node at (0,-2) {\Large $B_{PR}(k)$};
\end{scope}
\begin{scope}[shift={(6.5,0)}]
\node at (1,0) {$+$};	
\end{scope}
\begin{scope}[shift={(9,0)}]
\draw [phir, ultra thick, domain=0:180] plot ({1*cos(\x)}, {1*sin(\x)});
\draw [phil, ultra thick, domain=180:270] plot ({1*cos(\x)}, {1*sin(\x)}); 
\draw [phir, ultra thick, domain=270:360] plot ({1*cos(\x)}, {1*sin(\x)}); 
\node at (-1,0) {$\times $};	
\node at (1,0) {$\times $};	
\node at (0,-2) {\Large $B_{MR}(k)$}; 
\end{scope}

\begin{scope}[shift={(0,-4)}]
\draw [phil, ultra thick, domain=0:90] plot ({1*cos(\x)}, {1*sin(\x)});
\draw [phir, ultra thick, domain=90:180] plot ({1*cos(\x)}, {1*sin(\x)});
\draw [phir, ultra thick, domain=180:270] plot ({1*cos(\x)}, {1*sin(\x)}); 
\draw [phil, ultra thick, domain=270:360] plot ({1*cos(\x)}, {1*sin(\x)}); 
\node at (-1,0) {$\times $};	
\node at (1,0) {$\times $};	
\node at (0,-2) {\Large $B_{PM}(k)$};
\end{scope}
\begin{scope}[shift={(0.5,-4)}]
\node at (1,0) {$+ $};	
\end{scope}
\begin{scope}[shift={(3,-4)}]
\draw [phir, ultra thick, domain=0:90] plot ({1*cos(\x)}, {1*sin(\x)});
\draw [phil, ultra thick, domain=90:180] plot ({1*cos(\x)}, {1*sin(\x)});
\draw [phir, ultra thick, domain=180:270] plot ({1*cos(\x)}, {1*sin(\x)}); 
\draw [phil, ultra thick, domain=270:360] plot ({1*cos(\x)}, {1*sin(\x)}); 
\node at (-1,0) {$\times $};	
\node at (1,0) {$\times $};	
\node at (0,-2) {\Large $B_{PP}(k)$};	
\end{scope}
\begin{scope}[shift={(3.5,-4)}]
\node at (1,0) {$=$};	
\end{scope}
\begin{scope}[shift={(6,-4)}]
\draw [phir, ultra thick, domain=0:180] plot ({1*cos(\x)}, {1*sin(\x)});
\draw [phir, ultra thick, domain=180:270] plot ({1*cos(\x)}, {1*sin(\x)}); 
\draw [phil, ultra thick, domain=270:360] plot ({1*cos(\x)}, {1*sin(\x)}); 
\node at (-1,0) {$\times $};	
\node at (1,0) {$\times $};	
\node at (0,-2) {\Large $B_{PR}(k)$};
\end{scope}
\begin{scope}[shift={(6.5,-4)}]
\node at (1,0) {$+$};	
\end{scope}
\begin{scope}[shift={(9,-4)}]
\draw [phil, ultra thick, domain=0:180] plot ({1*cos(\x)}, {1*sin(\x)});
\draw [phir, ultra thick, domain=180:270] plot ({1*cos(\x)}, {1*sin(\x)}); 
\draw [phil, ultra thick, domain=270:360] plot ({1*cos(\x)}, {1*sin(\x)}); 
\node at (-1,0) {$\times $};	
\node at (1,0) {$\times $};	
\node at (0,-2) {\Large $B_{PL}(k)$};	
\end{scope}

\begin{scope}[shift={(0,-8)}]
\draw [phir, ultra thick, domain=0:90] plot ({1*cos(\x)}, {1*sin(\x)});
\draw [phil, ultra thick, domain=90:180] plot ({1*cos(\x)}, {1*sin(\x)});
\draw [phil, ultra thick, domain=180:270] plot ({1*cos(\x)}, {1*sin(\x)}); 
\draw [phir, ultra thick, domain=270:360] plot ({1*cos(\x)}, {1*sin(\x)}); 
\node at (-1,0) {$\times $};	
\node at (1,0) {$\times $};	
\node at (0,-2) {\Large $B_{MP}(k)$}; 
\end{scope}
\begin{scope}[shift={(0.5,-8)}]
\node at (1,0) {$+ $};	
\end{scope}
\begin{scope}[shift={(3,-8)}]
\draw [phil, ultra thick, domain=0:90] plot ({1*cos(\x)}, {1*sin(\x)});
\draw [phir, ultra thick, domain=90:180] plot ({1*cos(\x)}, {1*sin(\x)});
\draw [phil, ultra thick, domain=180:270] plot ({1*cos(\x)}, {1*sin(\x)}); 
\draw [phir, ultra thick, domain=270:360] plot ({1*cos(\x)}, {1*sin(\x)}); 
\node at (-1,0) {$\times $};	
\node at (1,0) {$\times $};	
\node at (0,-2) {\Large $B_{MM}(k)$};	
\end{scope}
\begin{scope}[shift={(3.5,-8)}]
\node at (1,0) {$=$};	
\end{scope}
\begin{scope}[shift={(6,-8)}]
\draw [phir, ultra thick, domain=0:180] plot ({1*cos(\x)}, {1*sin(\x)});
\draw [phil, ultra thick, domain=180:270] plot ({1*cos(\x)}, {1*sin(\x)}); 
\draw [phir, ultra thick, domain=270:360] plot ({1*cos(\x)}, {1*sin(\x)}); 
\node at (-1,0) {$\times $};	
\node at (1,0) {$\times $};	
\node at (0,-2) {\Large $B_{MR}(k)$};
\end{scope}
\begin{scope}[shift={(6.5,-8)}]
\node at (1,0) {$+$};	
\end{scope}
\begin{scope}[shift={(9,-8)}]
\draw [phil, ultra thick, domain=0:180] plot ({1*cos(\x)}, {1*sin(\x)});
\draw [phil, ultra thick, domain=180:270] plot ({1*cos(\x)}, {1*sin(\x)}); 
\draw [phir, ultra thick, domain=270:360] plot ({1*cos(\x)}, {1*sin(\x)}); 
\node at (-1,0) {$\times $};	
\node at (1,0) {$\times $};	
\node at (0,-2) {\Large $B_{ML}(k)$};		
\end{scope}

\begin{scope}[shift={(0,-12)}]
\draw [phil, ultra thick, domain=0:180] plot ({1*cos(\x)}, {1*sin(\x)});
\draw [phil, ultra thick, domain=180:360] plot ({1*cos(\x)}, {1*sin(\x)}); 
\node at (-1,0) {$\times $};	
\node at (1,0) {$\times $};	
\node at (0,-2) {\Large $B_{LL}(k)$};	
\end{scope}
\begin{scope}[shift={(0.5,-12)}]
\node at (1,0) {$+ $};	
\end{scope}
\begin{scope}[shift={(3,-12)}]
\draw [phir, ultra thick, domain=0:180] plot ({1*cos(\x)}, {1*sin(\x)});
\draw [phil, ultra thick, domain=180:360] plot ({1*cos(\x)}, {1*sin(\x)}); 
\node at (-1,0) {$\times $};	
\node at (1,0) {$\times $};	
\node at (0,-2) {\Large $B_{LR}(k)$};
\end{scope}
\begin{scope}[shift={(3.5,-12)}]
\node at (1,0) {$=$};	
\end{scope}
\begin{scope}[shift={(6,-12)}]
\draw [phil, ultra thick, domain=0:180] plot ({1*cos(\x)}, {1*sin(\x)});
\draw [phir, ultra thick, domain=180:270] plot ({1*cos(\x)}, {1*sin(\x)}); 
\draw [phil, ultra thick, domain=270:360] plot ({1*cos(\x)}, {1*sin(\x)}); 
\node at (-1,0) {$\times $};	
\node at (1,0) {$\times $};	
\node at (0,-2) {\Large $B_{PL}(k)$};	
\end{scope}
\begin{scope}[shift={(6.5,-12)}]
\node at (1,0) {$+ $};	
\end{scope}
\begin{scope}[shift={(9,-12)}]
\draw [phil, ultra thick, domain=0:180] plot ({1*cos(\x)}, {1*sin(\x)});
\draw [phil, ultra thick, domain=180:270] plot ({1*cos(\x)}, {1*sin(\x)}); 
\draw [phir, ultra thick, domain=270:360] plot ({1*cos(\x)}, {1*sin(\x)}); 
\node at (-1,0) {$\times $};	
\node at (1,0) {$\times $};	
\node at (0,-2) {\Large $B_{ML}(k)$};	
\end{scope}

\end{tikzpicture}
\end{center}
\caption{Identities between the ten  one loop $B$ type integrals}
\label{diag:pvonelooprelnI}
\end{figure}

In this subsection, we will collect the UV divergences of various $B$ type diagrams for the convenience of the reader.
\begin{subequations}	
\begin{eqnarray}
 \diver[B_{RR}(k)]\Big|_{\msbar}
&=&  \frac{i}{(4\pi)^2}\left[ \frac{2}{d-4} + \ln\Bigl(\frac{1}{4\pi e^{-\gamma_E}} \Bigr) \right]
\label{skpvsummary1a}
\\
\diver[B_{LR}(k)]\Big|_{\msbar}&=& 0
\label{skpvsummary1b} 
\\
\diver[B_{PP}(k)]\Big|_{\msbar}&=& 0
\label{skpvsummary1c}
\\
\diver[B_{PR}(k)]\Big|_{\msbar}
&=&
%\frac{i}{2(4\pi)^2}\Bigl[ \frac{2}{d-4} +\gamma_E-1-\ln\, 4\pi +\ln \frac{m^2}{\mu^2}  \Bigr]
%=
\frac{1}{2}\diver[B_{RR}(k)]\Big|_{\msbar}
\label{skpvsummary1d}
\end{eqnarray}
\end{subequations}
Further, we have 
\begin{subequations}	
\begin{eqnarray}
&&\diver[B_{RR}]\Big|_{\msbar}=	\diver[B_{LL}]^\star \Big|_{\msbar}
\label{skpvsummary2a}
\\
&&\diver[B_{LR}]\Big|_{\msbar}=	\diver[B_{PM}]^\ast \Big|_{\msbar}  =	\diver[B_{MP}]\Big|_{\msbar} =	\diver[B_{PP}]\Big|_{\msbar} =0
\label{skpvsummary2b}
\\
&&	\diver[B_{PR}]\Big|_{\msbar} 
	=	\diver[B_{MR}]\Big|_{\msbar} =	\diver[B_{PL}]^\ast \Big|_{\msbar}  =	\diver[B_{ML}]^\ast \Big|_{\msbar}=\frac{1}{2}\diver[B_{RR}]\Big|_{\msbar}
\label{skpvsummary2c} 
\end{eqnarray}
\end{subequations} 
We also give below the symmetry factors of the corresponding diagrams in figure \ref{fig:pvsymmetryfactors}. The divergences given above along with the symmetry factors 
provides a quick way to write down appropriate $\beta$ functions for the open QFT. In the ensuing figure \ref{diag:pvoneloopcut} and figure \ref{diag:pvonelooprelnI}, we tabulate a set of useful diagrammatic identities which relate the various SK loop integrals.
 
\begin{figure}[ht]
\begin{center}
\begin{tikzpicture}[line width=1 pt, scale=0.8]
%\begin{scope}[rotate=-80]

\draw [phir, ultra thick, domain=0:180] plot ({1*cos(\x)}, {1*sin(\x)});
\draw [phir, ultra thick, domain=180:360] plot ({1*cos(\x)}, {1*sin(\x)}); 
\node at (-1,0) {$\times $};	
\node at (1,0) {$\times $};	
\node at (0,-1.5) { $\frac{1}{2}$};	

%\node at (0,-3) {\Large $B_{RR}(k)$};	

\begin{scope}[shift={(3,0)}]
\draw [phil, ultra thick, domain=0:180] plot ({1*cos(\x)}, {1*sin(\x)});
\draw [phil, ultra thick, domain=180:360] plot ({1*cos(\x)}, {1*sin(\x)}); 
\node at (-1,0) {$\times $};	
\node at (1,0) {$\times $};	
\node at (0,-1.5) { $\frac{1}{2}$};	

%\node at (0,-2) {\Large $B_{LL}(k)$};	
\end{scope}

\begin{scope}[shift={(6,0)}]
\draw [phir, ultra thick, domain=0:180] plot ({1*cos(\x)}, {1*sin(\x)});
\draw [phil, ultra thick, domain=180:360] plot ({1*cos(\x)}, {1*sin(\x)}); 
\node at (-1,0) {$\times $};	
\node at (1,0) {$\times $};	

\node at (0,-1.5) { $1$};	

%\node at (0,-3) {\Large $B_{LR}(k)$};
\end{scope}

\begin{scope}[shift={(0,-4)}]
\draw [phil, ultra thick, domain=0:90] plot ({1*cos(\x)}, {1*sin(\x)});
\draw [phir, ultra thick, domain=90:180] plot ({1*cos(\x)}, {1*sin(\x)});
\draw [phir, ultra thick, domain=180:270] plot ({1*cos(\x)}, {1*sin(\x)}); 
\draw [phil, ultra thick, domain=270:360] plot ({1*cos(\x)}, {1*sin(\x)}); 
\node at (-1,0) {$\times $};	
\node at (1,0) {$\times $};	
\node at (0,-1.5) { $\frac{1}{2}$};	

%\node at (0,-2) {\Large $B_{MP}(k)$};
\end{scope}

\begin{scope}[shift={(3,-4)}]
\draw [phir, ultra thick, domain=0:90] plot ({1*cos(\x)}, {1*sin(\x)});
\draw [phil, ultra thick, domain=90:180] plot ({1*cos(\x)}, {1*sin(\x)});
\draw [phil, ultra thick, domain=180:270] plot ({1*cos(\x)}, {1*sin(\x)}); 
\draw [phir, ultra thick, domain=270:360] plot ({1*cos(\x)}, {1*sin(\x)}); 
\node at (-1,0) {$\times $};	
\node at (1,0) {$\times $};	
\node at (0,-1.5) { $\frac{1}{2}$};	

%\node at (0,-2) {\Large $B_{PR}(k)$};	
\end{scope}

\begin{scope}[shift={(6,-4)}]
\draw [phil, ultra thick, domain=0:90] plot ({1*cos(\x)}, {1*sin(\x)});
\draw [phir, ultra thick, domain=90:180] plot ({1*cos(\x)}, {1*sin(\x)});
\draw [phil, ultra thick, domain=180:270] plot ({1*cos(\x)}, {1*sin(\x)}); 
\draw [phir, ultra thick, domain=270:360] plot ({1*cos(\x)}, {1*sin(\x)}); 
\node at (-1,0) {$\times $};	
\node at (1,0) {$\times $};	
\node at (0,-1.5) { $1$};	

%\node at (0,-2) {\Large $B_{MM}(k)$};	
\end{scope}

\begin{scope}[shift={(0,-8)}]
\draw [phir, ultra thick, domain=0:180] plot ({1*cos(\x)}, {1*sin(\x)});
\draw [phir, ultra thick, domain=180:270] plot ({1*cos(\x)}, {1*sin(\x)}); 
\draw [phil, ultra thick, domain=270:360] plot ({1*cos(\x)}, {1*sin(\x)}); 
\node at (-1,0) {$\times $};	
\node at (1,0) {$\times $};	
\node at (0,-1.5) { $1$};	

%\node at (0,-2) {\Large $B_{MR}(k)$};
\end{scope}

\begin{scope}[shift={(3,-8)}]
\draw [phil, ultra thick, domain=0:180] plot ({1*cos(\x)}, {1*sin(\x)});
\draw [phil, ultra thick, domain=180:270] plot ({1*cos(\x)}, {1*sin(\x)}); 
\draw [phir, ultra thick, domain=270:360] plot ({1*cos(\x)}, {1*sin(\x)}); 
\node at (-1,0) {$\times $};	
\node at (1,0) {$\times $};	
\node at (0,-1.5) { $1$};	

%\node at (0,-2) {\Large $B_{PR}(k)$};	
\end{scope}

\begin{scope}[shift={(6,-8)}]
\draw [phir, ultra thick, domain=0:180] plot ({1*cos(\x)}, {1*sin(\x)});
\draw [phil, ultra thick, domain=180:270] plot ({1*cos(\x)}, {1*sin(\x)}); 
\draw [phir, ultra thick, domain=270:360] plot ({1*cos(\x)}, {1*sin(\x)}); 
\node at (-1,0) {$\times $};	
\node at (1,0) {$\times $};	
\node at (0,-1.5) { $1$};	

%\node at (0,-2) {\Large $B_{ML}(k)$};
\end{scope}

\begin{scope}[shift={(9,-8)}]
\draw [phil, ultra thick, domain=0:180] plot ({1*cos(\x)}, {1*sin(\x)});
\draw [phir, ultra thick, domain=180:270] plot ({1*cos(\x)}, {1*sin(\x)}); 
\draw [phil, ultra thick, domain=270:360] plot ({1*cos(\x)}, {1*sin(\x)}); 
\node at (-1,0) {$\times $};	
\node at (1,0) {$\times $};	
\node at (0,-1.5) { $1$};	

%\node at (0,-2) {\Large $B_{PL}(k)$};	
\end{scope}

\end{tikzpicture}
\end{center}
\caption{Symmetry factors for all the ten  one loop integrals}
\label{fig:pvsymmetryfactors}
\end{figure}
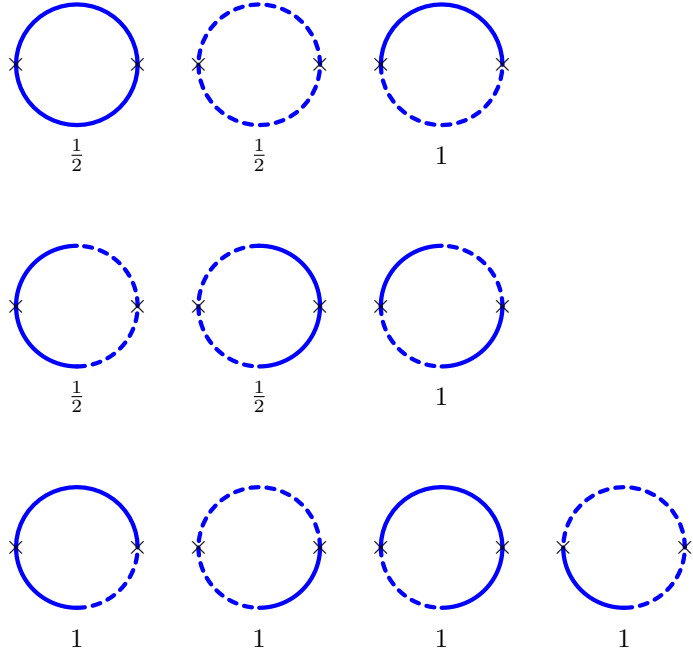

\begin{figure}[ht]
\begin{center}
\begin{tikzpicture}[line width=1 pt, scale=0.8]
%\begin{scope}[rotate=-80]

\draw [phir, ultra thick, domain=0:180] plot ({1*cos(\x)}, {1*sin(\x)});
\draw [phir, ultra thick, domain=180:360] plot ({1*cos(\x)}, {1*sin(\x)}); 
\node at (-1,0) {$\times $};	
\node at (1,0) {$\times $};	
\node at (0,-2) {\Large $B_{RR}(k)$};	
\begin{scope}[shift={(0.5,0)}]
\node at (1,0) {$+ $};	
\end{scope}
\begin{scope}[shift={(3,0)}]
\draw [phil, ultra thick, domain=0:180] plot ({1*cos(\x)}, {1*sin(\x)});
\draw [phil, ultra thick, domain=180:360] plot ({1*cos(\x)}, {1*sin(\x)}); 
\node at (-1,0) {$\times $};	
\node at (1,0) {$\times $};	
\node at (0,-2) {\Large $B_{LL}(k)$};	
\end{scope}
\begin{scope}[shift={(3.5,0)}]
\node at (1,0) {$+ $};	
\end{scope}
\begin{scope}[shift={(6.5,0)}]
\draw [phir, ultra thick, domain=0:180] plot ({1*cos(\x)}, {1*sin(\x)});
\draw [phil, ultra thick, domain=180:360] plot ({1*cos(\x)}, {1*sin(\x)}); 
\node at (-1.5,0) {\Large $2 $};	
\node at (-1,0) {$\times $};	
\node at (1,0) {$\times $};	
\node at (0,-2) {\Large $2B_{LR}(k)$};
\end{scope}
\begin{scope}[shift={(0,-4)}]
\node at (1,0) {\Large $=$};	
\end{scope}

\begin{scope}[shift={(3,-4)}]
\draw [phil, ultra thick, domain=0:90] plot ({1*cos(\x)}, {1*sin(\x)});
\draw [phir, ultra thick, domain=90:180] plot ({1*cos(\x)}, {1*sin(\x)});
\draw [phir, ultra thick, domain=180:270] plot ({1*cos(\x)}, {1*sin(\x)}); 
\draw [phil, ultra thick, domain=270:360] plot ({1*cos(\x)}, {1*sin(\x)}); 
\node at (-1,0) {$\times $};	
\node at (1,0) {$\times $};	
\node at (0,-2) {\Large $B_{PM}(k)$};
\end{scope}
\begin{scope}[shift={(3.5,-4)}]
\node at (1,0) {$+ $};	
\end{scope}
\begin{scope}[shift={(6,-4)}]
\draw [phir, ultra thick, domain=0:90] plot ({1*cos(\x)}, {1*sin(\x)});
\draw [phil, ultra thick, domain=90:180] plot ({1*cos(\x)}, {1*sin(\x)});
\draw [phil, ultra thick, domain=180:270] plot ({1*cos(\x)}, {1*sin(\x)}); 
\draw [phir, ultra thick, domain=270:360] plot ({1*cos(\x)}, {1*sin(\x)}); 
\node at (-1,0) {$\times $};	
\node at (1,0) {$\times $};	
\node at (0,-2) {\Large $B_{MP}(k)$};
\end{scope}
\begin{scope}[shift={(6.5,-4)}]
\node at (1,0) {$+ $};	
\end{scope}
\begin{scope}[shift={(9.5,-4)}]
\draw [phil, ultra thick, domain=0:90] plot ({1*cos(\x)}, {1*sin(\x)});
\draw [phir, ultra thick, domain=90:180] plot ({1*cos(\x)}, {1*sin(\x)});
\draw [phil, ultra thick, domain=180:270] plot ({1*cos(\x)}, {1*sin(\x)}); 
\draw [phir, ultra thick, domain=270:360] plot ({1*cos(\x)}, {1*sin(\x)}); 
\node at (-1.5,0) {\Large$2 $};
\node at (-1,0) {$\times $};	
\node at (1,0) {$\times $};	
\node at (0,-2) {\Large $2 B_{MM}(k)$};	
\end{scope}

\end{tikzpicture}
\end{center}
\caption{Cutting identity of  one loop integrals}
\label{diag:pvoneloopcut}
\end{figure}

\section{Passarino-Veltman diagrams in the average-difference basis} 
\label{sec:onelooppvavgdiff}
 
Let us now take a look at the Passarino-Veltman diagrams in average-difference basis. It's worth remembering here that only three out of the four propagators,
in this basis, are non-vanishing: the `$d$' propagator vanishes. This means that we have lesser number of non-vanishing diagrams in this basis. As a matter
of fact, some of the non-vanishing diagrams (in average-difference basis) do not diverge. All these facts add up to give only a 
few divergent one loop diagrams - only one $A$ type and two $B$ type integrals. Thus, computations for the beta functions greatly simplifies in this 
basis. We will not try to evaluate the PV integrals from scratch. We will express the integrals in the average-difference basis in terms of the integrals in 
the $\phir$-$\phil$ basis and then, use the results from the previous sections to determine the former. 

\subsection{Passarino-Veltman $A$ type integral in the average-difference basis}
\label{sec:avg-diff_onelooppvatypeintegrals}

\begin{figure}[ht]
\begin{center}
\begin{tikzpicture}[line width=1 pt, scale=1]
%\begin{scope}[rotate=90]

\begin{scope}[shift={(0,0)}] 
\drawphiavdifdiagaa{0}{0}{90}
\node at (0,-2) {\Large $A_{a}$};	
\end{scope}

\begin{scope}[shift={(8,0)}]
\drawphiavdifdiagaf{0}{0}{90}
\node at (0,-2) {\Large $A_{b}=A_{f}=0$};
\end{scope}

\end{tikzpicture} 
\end{center}
\caption{PV one loop $A$ type integrals in the average-difference basis}
\label{fig:pvaing2}
\end{figure}

There are two $A$ type PV integrals in this basis: $A_{a}$ and $A_{b}=A_{f}$. Using the relations in equation \eqref{avdif3} and equation \eqref{pvaintg3}-\eqref{pvaintg4}, it's easy to check that we get 
\begin{equation}
\begin{split}
A_{a} &= \frac{1}{2}\left(A_R + A_L \right) = \frac{1}{(4\pi)^2}\left[ \frac{2}{(d-4)} +\ln\Bigl(\frac{m^2}{4\pi\mu^2e^{-\gamma_E}} \Bigr) -1 \right]m^2 \\
A_{b} &= A_R - A_M = 0\\ 
A_{f} &= A_R - A_P = 0
\end{split}
\label{avgdifpvint1}
\end{equation}

\subsection{Passarino-Veltman $B$ type integral in the average-difference basis}
\label{sec:avg-diff_onelooppvbtypeintegrals}

\begin{figure}[ht]
\begin{center}
\begin{tikzpicture}[line width=1 pt, scale=1]
%\begin{scope}[rotate=90]

\begin{scope}[shift={(0,0)}]
\drawphiavdifdiagbaa{0}{0}{0}
\node at (0,-2) {\Large $B_{aa}$};	
\end{scope}

\begin{scope}[shift={(4,0)}]
\drawphiavdifdiagbaf{0}{0}{0}
\node at (0,-2) {\Large $B_{ab}$};
\end{scope}

\begin{scope}[shift={(8,0)}]
\drawphiavdifdiagbab{0}{0}{0}
\node at (0,-2) {\Large $B_{af}$};
\end{scope}

\begin{scope}[shift={(0,-4)}]
\drawphiavdifdiagbfb{0}{0}{0}  
\node at (0,-2) {\Large $B_{bf}$};
\end{scope}

\begin{scope}[shift={(4,-4)}] 
\drawphiavdifdiagbfb{0}{0}{180}
\node at (0,-2) {\Large $B_{fb}$};
\end{scope}

\begin{scope}[shift={(8,-4)}]
\drawphiavdifdiagbff{0}{0}{0}
\node at (0,-2) {\Large $B_{ff} = B_{bb}=0$};
\end{scope}

\end{tikzpicture} 
\end{center}
\caption{PV one loop $B$ type integrals in the average-difference basis}
\label{fig:pvaing3}
\end{figure}

There are six PV $B$ type integrals in this basis: $B_{aa}$, $B_{af}$, $B_{ab}$, $B_{bf}$, $B_{fb}$ and $B_{ff}= B_{bb}$. Using the relations given in equation \eqref{avdif3}, it is easy to check that 
\begin{equation} 
\begin{split}
B_{aa} &= \frac{1}{4}\left(B_{RR} + B_{RL} + B_{LR} + B_{LL} \right)\\
B_{af} &= \frac{1}{2}\left(B_{RR} - B_{RP} + B_{LR} - B_{LP} \right)\\
B_{ab} &= \frac{1}{2}\left(B_{RR} - B_{RM} + B_{LR} - B_{LM} \right)\\
B_{fb} &= B_{RR} - B_{RM} - B_{PR} + B_{PM} \\
B_{bf} &= B_{RR} - B_{RP} - B_{MR} + B_{MP} \\
B_{ff} &= B_{bb} = B_{RR} - B_{RP} - B_{PR} + B_{PP}
\end{split}
\label{avgdifpvint2}
\end{equation}
To compute the divergences for the above-mentioned integrals, we use the results given in equations \eqref{skpvsummary1a}-\eqref{skpvsummary2c} in section \ref{sec:onelooppvudivergences}.   So, we have
\begin{equation}
\begin{split}
\diver[B_{aa}]\Big|_{\msbar} &= 0 \\
\diver[B_{af}]\Big|_{\msbar} &= \frac{1}{2}\diver[B_{RR}]\Big|_{\msbar} = \frac{i}{2(4\pi)^2}\left[ \frac{2}{d-4} + \ln\Bigl(\frac{1}{4\pi e^{-\gamma_E}} \Bigr) \right]\\
\diver[B_{ab}]\Big|_{\msbar} &= \frac{1}{2}\diver[B_{RR}]\Big|_{\msbar} = \frac{i}{2(4\pi)^2}\left[ \frac{2}{d-4} + \ln\Bigl(\frac{1}{4\pi e^{-\gamma_E}} \Bigr) \right] \\
\diver[B_{fb}]\Big|_{\msbar} &=  0\\
\diver[B_{bf}]\Big|_{\msbar} &= 0 \\
\diver[B_{ff}]\Big|_{\msbar} &= \diver[B_{bb}]\Big|_{\msbar} = 0
\end{split}
\label{avgdifpvint3}
\end{equation}
We shall use these results for the computations in section \ref{sec:avgdifbasis} and in the next section.

\section{Computations in the average-difference basis}
\label{sec:appendixavgdiff}

In section \ref{sec:avgdifbasis}, we have already computed the beta functions for the Lindblad violating combinations in the average-difference basis and we found that it matches with our computations in the $\phir$-$\phil$ basis. For the sake of completion, we calculate the beta function for rest of the mass terms and the rest of the coupling constants in this basis. This computation enables one to verify the beta functions computed in $\phir$-$\phil$ basis. We shall start off by providing the set of Feynman rules in this basis.

The propagators in this basis are given in equation \eqref{avdif3}. The vertex factors in this basis are given by 

\begin{center}
\begin{tabular}{ | m{5em} | m{8 cm}|  } 
\hline
Vertex & Factor  \\ 
\hline
\hline
$\phidif^3$ & $\frac{(-i)}{4}(\re \ \lambda_3-3\ \re \ \sigma_3) (2\pi)^d \delta\left(\sum p\right)$ \\ 
\hline
$\phiav^3$ & $ 2(\im  \ \lambda_3+ 3\ \im  \ \sigma_3) (2\pi)^d \delta\left(\sum p\right)$  \\ 
\hline
$\phidif^2\phiav$  & $ \frac{1}{2} (\im  \ \lambda_3- \ \im  \ \sigma_3) (2\pi)^d \delta\left(\sum p\right)$ \\
\hline
$\phidif\phiav^2$  & $(-i)(\re \ \lambda_3+ \ \re \ \sigma_3) (2\pi)^d \delta\left(\sum p\right)$ \\
\hline
$\phidif^4$  & $\frac{1}{8}(\im  \ \lambda_4-4\ \im  \ \sigma_4 -3 \lambda_\Delta) (2\pi)^d \delta\left(\sum p\right)$ \\
\hline
$\phiav^4$  & $ 2(\im  \ \lambda_4+ 4\ \im  \ \sigma_4 -3 \lambda_\Delta) (2\pi)^d \delta\left(\sum p\right)$ \\
\hline
$\phidif^3\phiav$  & $\frac{(-i)}{4} (\re \ \lambda_4- 2\ \re \ \sigma_4) (2\pi)^d \delta\left(\sum p\right)$ \\
\hline
$\phidif\phiav^3$  & $(-i) (\re \ \lambda_4+ 2\ \re \ \sigma_4) (2\pi)^d \delta\left(\sum p\right)$ \\
\hline
$\phidif^2\phiav^2$  & $\frac{1}{2}(\im  \ \lambda_4+  \lambda_\Delta) (2\pi)^d \delta\left(\sum p\right)$ \\
\hline
\end{tabular} 
\end{center}

\begin{figure}[ht]
\begin{center}
\begin{tikzpicture}[ scale=0.5]

%%%%%%%%%%%%%%%%%%%%%%%%%%%%%%%%%%%%%%%%%%%%%%%%%%%%%%%%%%

\begin{scope}[shift={(0,0)}] 
\node at (0,-2) {$2(\im\,  m^2 -m^2_\Delta)$};
%\node at (0,-3) {vertex};
 
\phihalfpropagatora{0}{0}{0}{-1.2}{}
\phihalfpropagatora{0}{0}{0}{1.2}{}
%\counterterm{0}{0}{0.2}
\node at (0,0) {$\times $};	
\end{scope}

\begin{scope}[shift={(10,0)}]
\node at (0,-2) {$(-i)\re \ m^2$};
%\node at (0,-3) {vertex};

\phihalfpropagatora{0}{0}{0}{-1.2}{}
\phihalfpropagatord{0}{0}{0}{1.2}{}
%\counterterm{0}{0}{0.2}
\node at (0,0) {$\times $};	 
\end{scope}

\begin{scope}[shift={(20,0)}]
\node at (0,-2) {$\frac{1}{2}  (\im  \ m^2 +m_\Delta^2)$};
%\node at (0,-3) {vertex};

\phihalfpropagatord{0}{0}{0}{-1.2}{}
\phihalfpropagatord{0}{0}{0}{1.2}{}
%\counterterm{0}{0}{0.2}
\node at (0,0) {$\times $};	

\end{scope}

%%%%%%%%%%%%%%%%%%%%%%%%%%%%%%%%%%%%%%%%%%%%%%%%%%%%%%%%%%

\begin{scope}[shift={(0,-7)}]

\begin{scope}[shift={(0,0)}] 

\node at (-1,-3) {$2(\im  \ \lambda_3+ 3\ \im  \ \sigma_3) $};
%\node at (-1,-3) {vertex};

\phihalfpropagatora{-1}{0}{180}{1.2}{}
\phihalfpropagatora{-1}{0}{60}{1.2}{}
\phihalfpropagatora{-1}{0}{-60}{1.2}{} 
\node at (-1,0) {$\times $};	

\node at (-3.5,0) {$\bf 1$};	
\node at (1.5,-1) {$\bf 2$};	 
\node at (1.5,1) {$\bf 3$};
\end{scope}

\begin{scope}[shift={(10,0)}]
\node at (-1,-3) {$(-i)(\re \ \lambda_3+ \ \re \ \sigma_3) $};
%\node at (-1,-3) {vertex};

\phihalfpropagatora{-1}{0}{180}{1.2}{}
\phihalfpropagatora{-1}{0}{60}{1.2}{}
\phihalfpropagatord{-1}{0}{-60}{1.2}{} 
\node at (-1,0) {$\times $};

\node at (-3.5,0) {$\bf 1$};	
\node at (1.5,-1) {$\bf 2$};	 
\node at (1.5,1) {$\bf 3$};
\end{scope}

\begin{scope}[shift={(20,0)}]
\node at (-1,-3) {$ \frac{1}{2} (\im  \ \lambda_3- \ \im  \ \sigma_3)$};
%\node at (-1,-3) {vertex};

\phihalfpropagatora{-1}{0}{180}{1.2}{}
\phihalfpropagatord{-1}{0}{60}{1.2}{}
\phihalfpropagatord{-1}{0}{-60}{1.2}{} 
\node at (-1,0) {$\times $};	

\node at (-3.5,0) {$\bf 1$};	
\node at (1.5,-1) {$\bf 2$};	 
\node at (1.5,1) {$\bf 3$};
\end{scope}

\begin{scope}[shift={(10,-7)}]
\node at (-1,-3) {$\frac{-i}{4} (\re   \ \lambda_3- 3\ \re   \ \sigma_3)$};
%\node at (-1,-3) {vertex};

\phihalfpropagatord{-1}{0}{180}{1.2}{}
\phihalfpropagatord{-1}{0}{60}{1.2}{}
\phihalfpropagatord{-1}{0}{-60}{1.2}{} 
\node at (-1,0) {$\times $};	

\node at (-3.5,0) {$\bf 1$};	
\node at (1.5,-1) {$\bf 2$};	 
\node at (1.5,1) {$\bf 3$};
\end{scope}
	
\end{scope}

\begin{scope}[shift={(0,-21)}] 

\begin{scope}[shift={(0,0)}]
\node at (-1,-3) {$  2(\im  \ \lambda_4+ 4\ \im  \ \sigma_4 -3 \lambda_\Delta) $};
%\node at (-1,-3) {vertex};

\phihalfpropagatora{-1}{0}{135}{1.2}{} 
\phihalfpropagatora{-1}{0}{-135}{1.2}{} 
\phihalfpropagatora{-1}{0}{45}{1.2}{}
\phihalfpropagatora{-1}{0}{-45}{1.2}{} 
\node at (-1,0) {$\times $};	

\node at (-3.5,1) {$\bf 1$};	
\node at (-3.5,-1) {$\bf 2$}; 
\node at (1.5,1) {$\bf 3$};	 
\node at (1.5,-1) {$\bf 4$};
\end{scope}

\begin{scope}[shift={(10,0)}]
\node at (-1,-3) {$ (-i) (\re \ \lambda_4+ 2\ \re \ \sigma_4)$};
%\node at (-1,-3) {vertex};

\phihalfpropagatora{-1}{0}{135}{1.2}{} 
\phihalfpropagatora{-1}{0}{-135}{1.2}{} 
\phihalfpropagatora{-1}{0}{45}{1.2}{}
\phihalfpropagatord{-1}{0}{-45}{1.2}{} 
%\counterterm{-1}{0}{0.2} 
\node at (-1,0) {$\times $};	

\node at (-3.5,1) {$\bf 1$};	
\node at (-3.5,-1) {$\bf 2$};
\node at (1.5,1) {$\bf 3$};	 
\node at (1.5,-1) {$\bf 4$};
\end{scope}

%%%%%%%%%%%%%%%%%%%%%%%%%%%%%%%%%%%%%%%%%%%%%%%%%%%%%%%%%%%%%

\begin{scope}[shift={(20,0)}]
\node at (-1,-3) {$ \frac{1}{2}(\im  \ \lambda_4+  \lambda_\Delta)$};
%\node at (-1,-3) {vertex};

\phihalfpropagatora{-1}{0}{135}{1.2}{} 
\phihalfpropagatora{-1}{0}{-135}{1.2}{} 
\phihalfpropagatord{-1}{0}{45}{1.2}{}
\phihalfpropagatord{-1}{0}{-45}{1.2}{} 
\node at (-1,0) {$\times $};	

\node at (-3.5,1) {$\bf 1$};	
\node at (-3.5,-1) {$\bf 2$};
\node at (1.5,1) {$\bf 3$};	 
\node at (1.5,-1) {$\bf 4$};
\end{scope}

\begin{scope}[shift={(4,-7)}]
\node at (-1,-3) {$ \frac{(-i)}{4} (\re \ \lambda_4- 2\ \re \ \sigma_4)$};
%\node at (-1,-3) {vertex};

\phihalfpropagatora{-1}{0}{135}{1.2}{}  
\phihalfpropagatord{-1}{0}{-135}{1.2}{} 
\phihalfpropagatord{-1}{0}{45}{1.2}{}
\phihalfpropagatord{-1}{0}{-45}{1.2}{} 
\node at (-1,0) {$\times $};	

\node at (-3.5,1) {$\bf 1$};	
\node at (-3.5,-1) {$\bf 2$};
\node at (1.5,1) {$\bf 3$};	 
\node at (1.5,-1) {$\bf 4$};
\end{scope}

\begin{scope}[shift={(16,-7)}]
\node at (-1,-3) {$\frac{1}{8}(\im  \ \lambda_4- 4\ \im  \ \sigma_4 -3 \lambda_\Delta)$};
%\node at (-1,-3) {vertex};

\phihalfpropagatord{-1}{0}{135}{1.2}{} 
\phihalfpropagatord{-1}{0}{-135}{1.2}{} 
\phihalfpropagatord{-1}{0}{45}{1.2}{}
\phihalfpropagatord{-1}{0}{-45}{1.2}{} 
\node at (-1,0) {$\times $};	

\node at (-3.5,1) {$\bf 1$};	
\node at (-3.5,-1) {$\bf 2$};
\node at (1.5,1) {$\bf 3$};	 
\node at (1.5,-1) {$\bf 4$};
\end{scope} 
	
\end{scope}

\end{tikzpicture}
\end{center}
\caption{Feynman rules in the average-difference basis}
\label{fig:avgdifffeynrules}  
\end{figure}

\subsection{Beta functions for the mass terms}
In order to compute the beta functions of the masses $\re\ m^2,\im\ m^2$ and  $m_\Delta^2$, we need to compute  three different correlators. As usual, 
we omit all the finite terms which  are irrelevant for  beta function computation.

We have chosen the following three correlators:

\subsubsection{$\phiav^2$ vertex}
First we consider $\phiav \rightarrow \phiav $ via $\phiav^2$ vertex. There are three divergent one loop contribution, as depicted in the first row of the figure \ref{fig:avgdiffmass1}. The total contribution is given by 
\begin{equation}
\begin{split}
2(\im\ m^2- m_\Delta^2) & + (2\ \text{diagrams})\times 2(\im\ \lambda_3+ 3\ \im\ \sigma_3)\\
&\times (-i)  (\re\ \lambda_3+ \ \re\ \sigma_3)\frac{i}{2(4\pi)^2}\left(\frac{2}{d-4} +   \ln\  \frac{1}{4\pi e^{-\gamma_E}}\right)\\
&+ 2(\im\ \lambda_4+ 4\ \im\ \sigma_4 -3 \lambda_\Delta)   \frac{\re\ m^2}{2(4\pi)^2} \left(\frac{2}{d-4} +   \ln\  \frac{1}{4\pi e^{-\gamma_E}}\right)
\end{split}
\label{eq:avdiff111}
\end{equation} 
where, the first term is from tree level contribution and the rest are from loop level.  Thus, from equation \eqref{eq:avdiff111}, the beta function for ($\im\ m^2- m_\Delta^2$) is given by 
\begin{equation}
\begin{split}
\frac{d}{d\ln\, \mu}(\im\ m^2- m_\Delta^2)
&= \frac{2}{(4\pi)^2} (\im\ \lambda_3+ 3\ \im\ \sigma_3) (\re\ \lambda_3+ \ \re\ \sigma_3)+
\frac{\re\ m^2}{(4\pi)^2} (\im\ \lambda_4+ 4\ \im\ \sigma_4 -3 \lambda_\Delta) 
\end{split}
\label{eq:avdiff112}
\end{equation}

% 
% If we keep all terms we get
% \begin{equation}
% \begin{split}
% 2(\im\ m^2- m_\Delta^2) & + (B_{af}+B_{ab})\times 2(\im\ \lambda_3+ 3\ \im\ \sigma_3)\times (-i)  (\re\ \lambda_3+ \ \re\ \sigma_3)\\
% & + \frac{1}{2}B_{aa}\times 2(\im\ \lambda_3+ 3\ \im\ \sigma_3)\times 2(\im\ \lambda_3+ 3\ \im\ \sigma_3)\\
% & +\frac{1}{2} (B_{bb}+B_{ff})\times 2(\im\ \lambda_3+ 3\ \im\ \sigma_3)\times \frac{1}{2}(\im\ \lambda_3-\ \im\ \sigma_3)\\
% & + B_{fb} \times (-i)(\re\ \lambda_3+ \ \re\ \sigma_3)\times  (-i)(\re\ \lambda_3+ \ \re\ \sigma_3)\\
% &+\frac{1}{2} \times 2(\im\ \lambda+ 4\ \im\ \sigma -3 \Delta)   \frac{\re\ m^2}{(4\pi)^2} \frac{2}{d-4}
% \end{split}
% \end{equation} 

\subsubsection{$\phiav\phidif$ vertex}
Next we consider $\phiav \rightarrow \phiav $ via $\phiav\phidif$ vertex. It has same divergent diagrams as that of \eqref{eq:avdiff111}, but with different vertex factors. The corresponding Feynman diagrams are depicted in the second row of the figure \ref{fig:avgdiffmass1}. The total contribution is given by 
\begin{equation}
\begin{split}
(-i)\re\ m^2& +(-i)  (\re\ \lambda_3+ \ \re\ \sigma_3)\times (-i)  (\re\ \lambda_3+ \ \re\ \sigma_3)\frac{i}{2(4\pi)^2}\left(\frac{2}{d-4} +   \ln\  \frac{1}{4\pi e^{-\gamma_E}}\right)\\
& +2  (\im\ \lambda_3+ 3 \ \im\ \sigma_3)\times \frac{1}{2}  (\im\ \lambda_3- \ \im\ \sigma_3)\frac{i}{2(4\pi)^2}\left(\frac{2}{d-4} +   \ln\  \frac{1}{4\pi e^{-\gamma_E}}\right)\\
&+ (-i) (\re\ \lambda_4+ 2\ \re\ \sigma_4)   \frac{\re\ m^2}{2(4\pi)^2} \left(\frac{2}{d-4} +   \ln\  \frac{1}{4\pi e^{-\gamma_E}}\right)\\
\end{split}
\label{eq:avdiff121}
\end{equation}
Again, the first term is the tree level and the rest are the one loop contributions. Thus, from \eqref{eq:avdiff121} the beta function for ($\re\ m^2$) is as follows 
\begin{equation}
\begin{split}
\frac{d}{d\ln\, \mu}\re\ m^2
&=\frac{1}{(4\pi)^2} (\re\ \lambda_3+ \ \re\ \sigma_3)^2-\frac{1}{(4\pi)^2} (\im\ \lambda_3+ 3 \ \im\ \sigma_3) (\im\ \lambda_3- \ \im\ \sigma_3)\\
&+ \frac{\re\ m^2}{(4\pi)^2} (\re\ \lambda_4+ 2\ \re\ \sigma_4)
\end{split}
\label{eq:avdiff122}
\end{equation}
 
\subsubsection{$\phidif^2$ vertex}
\begin{figure}[ht]
\begin{center}
\begin{tikzpicture}[ scale=0.5]

%\draw [ultra thick] (-3.5,1) -- (-3.5,-12); 
 
%%%%%%%%%%%%%%%%%%%%%%%%%%%%%%%%%%%%%%%%%%%%%%%%%%%%%%%%%%

\node at (-5,.5) {$(\phiav^2)$};
\node at (-5,-.5) {vertex};

\begin{scope}[shift={(0,0)}]
\phipropagatora{0}{0}{0}{-1}{}
\phipropagatora{0}{0}{0}{1}{}
\counterterm{0}{0}{0.2}

\end{scope}

\node at (3,0) {\large $+$};

\begin{scope}[shift={(7,0)}]
\phipropagatora{-1}{0}{0}{-1}{}
\phipropagatora{1}{0}{0}{1}{}
\drawphiavdifdiagbaf{0}{0}{0}
\end{scope}
\node at (11,0) {\large $+$};

\begin{scope}[shift={(15,0)}]
\phipropagatora{-1}{0}{0}{-1}{}
\phipropagatora{1}{0}{0}{1}{}
\drawphiavdifdiagbab{0}{0}{0}	
\end{scope}

\node at (19,0) {\large $+$};
  
\begin{scope}[shift={(22,1)}]
\drawphiavdifdiagaa{0}{0}{90}
 
\phipropagatora{-2}{-1}{0}{1}{}
\phipropagatora{2}{-1}{0}{-1}{}
\end{scope}

%%%%%%%%%%%%%%%%%%%%%%%%%%%%%%%%%%%%%%%%%%%%%%%%%%%%%%%%%%

\node at (-5,-5.5) {$ (\phiav \phidif )$};
\node at (-5,-6.5) {vertex};

\begin{scope}[shift={(0,-6)}]
\phipropagatora{0}{0}{0}{-1}{}
\phipropagatorb{0}{0}{0}{1}{}
\counterterm{0}{0}{0.2}

\end{scope}

\node at (3,-6) {\large $+$};

\begin{scope}[shift={(7,-6)}]
\phipropagatora{-1}{0}{0}{-1}{}
\phipropagatorb{1}{0}{0}{1}{}
\drawphiavdifdiagbaf{0}{0}{0}
\end{scope}
\node at (11,-6) {\large $+$};

\begin{scope}[shift={(15,-6)}]
\phipropagatora{-1}{0}{0}{-1}{}
\phipropagatorb{1}{0}{0}{1}{}
\drawphiavdifdiagbab{0}{0}{0}	
\end{scope}

\node at (19,-6) {\large $+$};
  
\begin{scope}[shift={(22,-5)}]
\drawphiavdifdiagaa{0}{0}{90}

\phipropagatora{0}{-1}{0}{-1}{}
\phipropagatorb{0}{-1}{0}{1}{}
\end{scope}

%%%%%%%%%%%%%%%%%%%%%%%%%%%%%%%%%%%%%%%%%%%%%%%%%%%%%%%%%%

\node at (-5,-11.5) {$(\phidif^2)$}; 
\node at (-5,-12.5) {vertex};

\begin{scope}[shift={(0,-12)}]
\phipropagatorb{0}{0}{0}{-1}{}
\phipropagatorb{0}{0}{0}{1}{}
\counterterm{0}{0}{0.2}

\end{scope}

\node at (3,-12) {\large $+$};

\begin{scope}[shift={(7,-12)}]
\phipropagatorb{-1}{0}{0}{-1}{}
\phipropagatorb{1}{0}{0}{1}{}
\drawphiavdifdiagbaf{0}{0}{0}
\end{scope}
\node at (11,-12) {\large $+$};

\begin{scope}[shift={(15,-12)}]
\phipropagatorb{-1}{0}{0}{-1}{}
\phipropagatorb{1}{0}{0}{1}{}
\drawphiavdifdiagbab{0}{0}{0}	
\end{scope}

\node at (19,-12) {\large $+$};
  
\begin{scope}[shift={(22,-11)}]
\drawphiavdifdiagaa{0}{0}{90}

\phipropagatorb{0}{-1}{0}{-1.25}{}
\phipropagatorb{0}{-1}{0}{1.25}{} 
\end{scope}

\node at (11,-15) {\LARGE $\bf \phiav \rightarrow \phiav$};

\end{tikzpicture}
\end{center}
\caption{Mass renormalization in the average-difference basis}
\label{fig:avgdiffmass1}  
\end{figure}
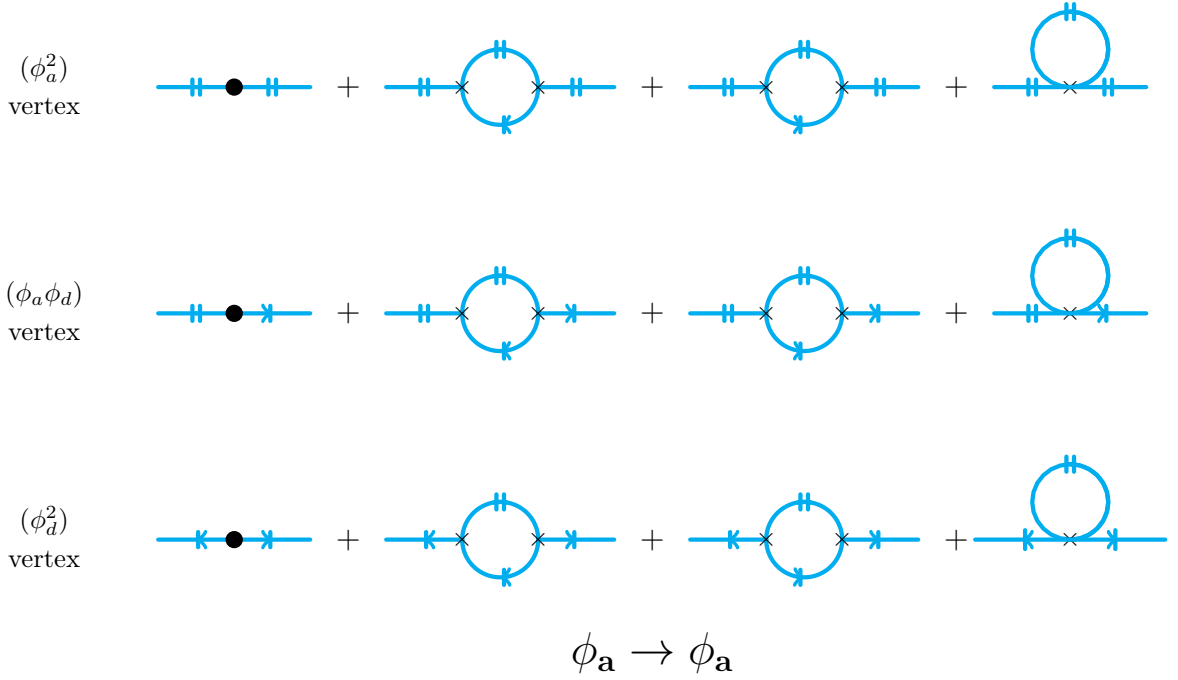

The Feynman diagrams for $\phiav \rightarrow \phiav $ via $\phidif^2$ vertex are depicted in the third row of the figure \ref{fig:avgdiffmass1}. The tree level and one loop contributions are given by 
\begin{equation}
\begin{split}
\frac{1}{2}(\im\ m^2+ m_\Delta^2)& + (2\ \text{diagrams})\times \frac{1}{2}  (\im\ \lambda_3- \ \im\ \sigma_3)\\
&\times (-i)  (\re\ \lambda_3+ \ \re\ \sigma_3)\frac{i}{2(4\pi)^2}\left(\frac{2}{d-4} +   \ln\  \frac{1}{4\pi e^{-\gamma_E}}\right)\\
&+ \frac{1}{2} (\im\ \lambda_4+  \lambda_\Delta)   \frac{\re\ m^2}{2(4\pi)^2} \left(\frac{2}{d-4} +   \ln\  \frac{1}{4\pi e^{-\gamma_E}}\right)\\
\end{split}
\label{eq:avdiff131}
\end{equation}
From this equation we obtain the beta function for ($\im\ m^2+ m_\Delta^2$)
\begin{equation}
\begin{split}
\frac{d}{d\ln\, \mu}(\im\ m^2+ m_\Delta^2)
&=\frac{2}{(4\pi)^2}  (\im\ \lambda_3- \ \im\ \sigma_3) (\re\ \lambda_3+ \ \re\ \sigma_3)+\frac{\re\ m^2}{(4\pi)^2} (\im\ \lambda_4+  \lambda_\Delta) 
\end{split}
\label{eq:avdiff132}
\end{equation}

\subsubsection{Final $\beta$ function}

The beta functions of $m^2$ ($=\re\ m^2+ i\ \im\ m^2$) and $m_\Delta^2$  can be obtained from the equations \eqref{eq:avdiff112}, \eqref{eq:avdiff122} and \eqref{eq:avdiff132}
\begin{equation}
\begin{split}
\frac{d m^2}{d\ln\, \mu}
&=\frac{1}{(4\pi)^2} (\lambda_3+\sigma_3^\ast)  (\lambda_3+\sigma_3+2i\ \im\ \sigma_3 )+ \frac{m^2}{(4\pi)^2} \Bigl[ \lambda_4+2\sigma_4 -i\lambda_\Delta\ \Bigr]\\
\frac{d m_\Delta^2}{d\ln\, \mu}
&=-\frac{4}{(4\pi)^2} \im\ \sigma_3\ (\re\ \lambda_3+ \ \re\ \sigma_3)+ \frac{2}{(4\pi)^2} \re\Bigl[ m^2 (\lambda_\Delta+i\sigma_4) \Bigr]
\end{split}
\label{eq:avdiff141}
\end{equation}

\subsection{Beta functions for the cubic couplings}
We have four  cubic coupling constants and the corresponding vertices are $\phiav^3$, $\phiav^2\phidif$, $\phiav\phidif^2$, $\phidif^3$ and  we need to compute four  correlators. In each case, we will keep only the divergent parts as before. 
\subsubsection{$\phiav^3$ vertex}
The tree level and one loop Feynman diagram for  $\phidif \rightarrow \phidif \phidif$ via $\phiav^3$ vertex is depicted in the first row of the figure  \ref{fig:avgdiffcubic1}. These contributions are given by 
\begin{equation}
\begin{split}
2&(\im\ \lambda_3+ 3\ \im\ \sigma_3)\\
&+ (3\ \text{channels})\times 2(\im\ \lambda_3+ 3\ \im\ \sigma_3)  \times (-i) (\re\ \lambda_4+ 2\ \re\ \sigma_4)   \times \frac{i}{2(4\pi)^2}\left(\frac{2}{d-4} +   \ln\  \frac{1}{4\pi e^{-\gamma_E}}\right) \\
& + (3\ \text{channels})\times 2(\im\ \lambda_4+ 4\ \im\ \sigma_4 -3 \lambda_\Delta)  \times (-i) (\re\ \lambda_3+ \ \re\ \sigma_3)\times \frac{i}{2(4\pi)^2}\left(\frac{2}{d-4} +   \ln\  \frac{1}{4\pi e^{-\gamma_E}}\right)\\
\end{split}
\label{eq:avdiff211}
\end{equation}
and from this equation we determine the beta function for $\im\ \lambda_3+ 3\ \im\ \sigma_3$
\begin{equation}
\begin{split}
\frac{d}{d\ln\, \mu}(\im\ \lambda_3+ 3\ \im\ \sigma_3)
=&\frac{3}{(4\pi)^2} (\im\ \lambda_3+ 3\ \im\ \sigma_3) (\re\ \lambda_4+ 2\ \re\ \sigma_4)\\
&+\frac{3}{(4\pi)^2} (\im\ \lambda_4+ 4\ \im\ \sigma_4 -3 \lambda_\Delta) (\re\ \lambda_3+ \ \re\ \sigma_3)
\end{split}
\label{eq:avdiff212}
\end{equation}

\subsubsection{$\phiav^2 \phidif$ vertex}
Now we compute $\phidif \rightarrow \phiav \phiav$ via $\phiav^2 \phidif$ vertex. The relevant tree level and one loop Feynman diagrams are shown in the second row of the figure \ref{fig:avgdiffcubic1} and these contributions are given by 
\begin{equation}
\begin{split}
(-i)&(\re\ \lambda_3+\ \re\ \sigma_3)\\
&+ (3\ \text{diagrams})\times (-i)(\re\ \lambda_3+ \ \re\ \sigma_3)  \times (-i) (\re\ \lambda_4+ 2\ \re\ \sigma_4)  \times \frac{i}{2(4\pi)^2}\left(\frac{2}{d-4} +   \ln\  \frac{1}{4\pi e^{-\gamma_E}}\right) \\
& + (1\ \text{diagram})\times 2(\im\ \lambda_4+ 4\ \im\ \sigma_4 -3 \lambda_\Delta)  \times \frac{1}{2} (\im\ \lambda_3- \ \im\ \sigma_3)\times \frac{i}{2(4\pi)^2}\left(\frac{2}{d-4} +   \ln\  \frac{1}{4\pi e^{-\gamma_E}}\right)\\
& + (2\ \text{diagram})\times \frac{1}{2} (\im\ \lambda_4+  \lambda_\Delta)  \times 2 (\im\ \lambda_3+3\  \im\ \sigma_3)\times \frac{i}{2(4\pi)^2}\left(\frac{2}{d-4} +   \ln\  \frac{1}{4\pi e^{-\gamma_E}}\right)\\
\end{split} 
\label{eq:avdiff221}
\end{equation}
This implies that the beta function for $(\re\ \lambda_3+\ \re\ \sigma_3)$ is given by 
\begin{equation}
\begin{split}
\frac{d}{d\ln\, \mu}(\re\ \lambda_3+\ \re\ \sigma_3)
=&\frac{3}{(4\pi)^2} (\re\ \lambda_3+ \ \re\ \sigma_3) (\re\ \lambda_4+2\ \re\ \sigma_4) \\
&-\frac{1}{(4\pi)^2}(\im\ \lambda_4+ 4\ \im\ \sigma_4 -3 \lambda_\Delta) (\im\ \lambda_3- \ \im\ \sigma_3)\\
&-\frac{2}{(4\pi)^2} (\im\ \lambda_4+  \lambda_\Delta)  (\im\ \lambda_3+ 3\ \im\ \sigma_3)
\end{split} 
\label{eq:avdiff222}
\end{equation}

\subsubsection{$\phiav \phidif^2$ vertex}
The contribution (upto one loop) to $\phiav \rightarrow \phiav \phiav$ via $\phiav \phidif^2$ vertex is given by 
\begin{equation}
\begin{split}
\frac{1}{2}&(\im\ \lambda_3-\ \im\ \sigma_3)\\
&+ (3\ \text{diagrams})\times (-i)(\re\ \lambda_3+ \ \re\ \sigma_3)  \times\frac{1}{2}(\im\ \lambda_4+\lambda_\Delta)  \times \frac{i}{2(4\pi)^2}\left(\frac{2}{d-4} +   \ln\  \frac{1}{4\pi e^{-\gamma_E}}\right) \\
& + (2\ \text{diagrams})\times (-i)(\re\ \lambda_4+ 2\ \re\ \sigma_4)  \times \frac{1}{2} (\im\ \lambda_3- \ \im\ \sigma_3)\times \frac{i}{2(4\pi)^2}\left(\frac{2}{d-4} +   \ln\  \frac{1}{4\pi e^{-\gamma_E}}\right)\\
& + \frac{(-i)}{4}(\re\ \lambda_4-2\ \re\ \sigma_4)  \times 2 (\im\ \lambda_3+3\ \im\ \sigma_3)\times \frac{i}{2(4\pi)^2}\left(\frac{2}{d-4} +   \ln\  \frac{1}{4\pi e^{-\gamma_E}}\right)\\
\end{split}
\label{eq:avdiff231}
\end{equation}
The corresponding Feynman diagram can be found in the third row of the figure \ref{fig:avgdiffcubic1}. Hence the beta function for $(\im\ \lambda_3-\ \im\ \sigma_3)$ is as follows 
\begin{equation}
\begin{split}
\frac{d}{d\ln\, \mu}(\im\ \lambda_3-\ \im\ \sigma_3)
=&\frac{3}{(4\pi)^2} (\re\ \lambda_3+ \ \re\ \sigma_3) (\im\ \lambda_4+\lambda_\Delta) \\
&+\frac{2}{(4\pi)^2}(\re\ \lambda_4+ 2\ \re\ \sigma_4)  (\im\ \lambda_3- \ \im\ \sigma_3)\\
&+\frac{1}{(4\pi)^2}(\re\ \lambda_4- 2\ \re\ \sigma_4)  (\im\ \lambda_3+3\ \im\ \sigma_3)
\end{split}
\label{eq:avdiff232}
\end{equation}
 
\subsubsection{$\phidif^3$ vertex}
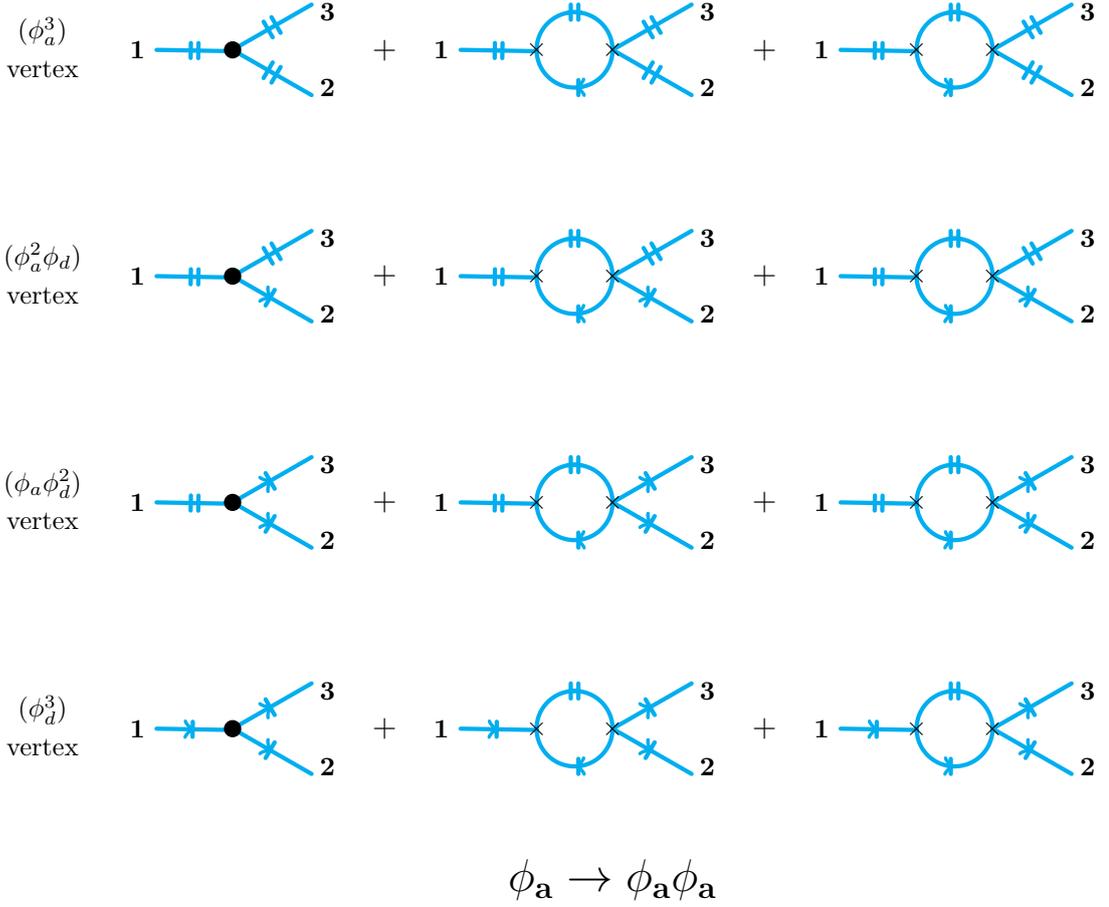
\begin{figure}[ht]
\begin{center}
\begin{tikzpicture}[ scale=.5]

%%%%%%%%%%%%%%%%%%%%%%%%%%%%%%%%%%%%%%%%%%%%%%%%%%%%%%%%%%%%%%

\node at (-6,.5) {$ (\phiav^3 )$};
\node at (-6,-.5) {vertex};

\begin{scope}[shift={(0,0)}]
\phipropagatora{-3}{0}{-1}{1}{}
\phipropagatora{-1}{0}{30}{1.2}{}
\phipropagatora{-1}{0}{-30}{1.2}{} 
\counterterm{-1}{0}{0.2}

\node at (-3.5,0) {$\bf 1$};	
\node at (1.5,-1) {$\bf 2$};	 
\node at (1.5,1) {$\bf 3$};
\end{scope}

\node at (3,0) {\large $+$};

\begin{scope}[shift={(8,0)}]
\phipropagatora{-3}{0}{-1}{1}{}
\phipropagatora{1}{0}{30}{1.2}{}
\phipropagatora{1}{0}{-30}{1.2}{} 

\node at (-3.5,0) {$\bf 1$};	
\node at (3.5,-1) {$\bf 2$};	
\node at (3.5,1) {$\bf 3$};

\drawphiavdifdiagbaf{0}{0}{0}
\end{scope}

\node at (13,0) {\large $+$};

\begin{scope}[shift={(18,0)}]
\phipropagatora{-3}{0}{-1}{1}{}
\phipropagatora{1}{0}{30}{1.2}{}
\phipropagatora{1}{0}{-30}{1.2}{} 
\drawphiavdifdiagbab{0}{0}{0}	 
\node at (-3.5,0) {$\bf 1$};	
\node at (3.5,-1) {$\bf 2$};	 
\node at (3.5,1) {$\bf 3$};	
\end{scope}

%%%%%%%%%%%%%%%%%%%%%%%%%%%%%%%%%%%%%%%%%%%%%%%%%%%%%%%%%%%%%%

\node at (-6,-5.5) {$ (\phiav^2 \phidif )$}; 
\node at (-6,-6.5) {vertex};

\begin{scope}[shift={(0,-6)}]
\phipropagatora{-3}{0}{-1}{1}{}
\phipropagatora{-1}{0}{30}{1.2}{}
\phipropagatorb{-1}{0}{-30}{1.2}{} 
\counterterm{-1}{0}{0.2}

\node at (-3.5,0) {$\bf 1$};	
\node at (1.5,-1) {$\bf 2$};	 
\node at (1.5,1) {$\bf 3$};
\end{scope}

\node at (3,-6) {\large $+$};

\begin{scope}[shift={(8,-6)}]
\phipropagatora{-3}{0}{-1}{1}{}
\phipropagatora{1}{0}{30}{1.2}{}
\phipropagatorb{1}{0}{-30}{1.2}{} 

\node at (-3.5,0) {$\bf 1$};	
\node at (3.5,-1) {$\bf 2$};	
\node at (3.5,1) {$\bf 3$};

\drawphiavdifdiagbaf{0}{0}{0}
\end{scope}

\node at (13,-6) {\large $+$};

\begin{scope}[shift={(18,-6)}]
\phipropagatora{-3}{0}{-1}{1}{}
\phipropagatora{1}{0}{30}{1.2}{}
\phipropagatorb{1}{0}{-30}{1.2}{} 
\drawphiavdifdiagbab{0}{0}{0}	
\node at (-3.5,0) {$\bf 1$};	
\node at (3.5,-1) {$\bf 2$};	 
\node at (3.5,1) {$\bf 3$};	
\end{scope}

%%%%%%%%%%%%%%%%%%%%%%%%%%%%%%%%%%%%%%%%%%%%%%%%%%%%%%%%%%%%%%

\node at (-6,-11.5) {$ (\phiav \phidif^2 )$};
\node at (-6,-12.5) {vertex};

\begin{scope}[shift={(0,-12)}]
\phipropagatora{-3}{0}{-1}{1}{}
\phipropagatorb{-1}{0}{30}{1.2}{}
\phipropagatorb{-1}{0}{-30}{1.2}{} 
\counterterm{-1}{0}{0.2}

\node at (-3.5,0) {$\bf 1$};	
\node at (1.5,-1) {$\bf 2$};	 
\node at (1.5,1) {$\bf 3$};
\end{scope}

\node at (3,-12) {\large $+$};

\begin{scope}[shift={(8,-12)}]
\phipropagatora{-3}{0}{-1}{1}{}
\phipropagatorb{1}{0}{30}{1.2}{}
\phipropagatorb{1}{0}{-30}{1.2}{} 

\node at (-3.5,0) {$\bf 1$};	
\node at (3.5,-1) {$\bf 2$};	
\node at (3.5,1) {$\bf 3$};

\drawphiavdifdiagbaf{0}{0}{0}
\end{scope}

\node at (13,-12) {\large $+$};

\begin{scope}[shift={(18,-12)}]
\phipropagatora{-3}{0}{-1}{1}{}
\phipropagatorb{1}{0}{30}{1.2}{}
\phipropagatorb{1}{0}{-30}{1.2}{} 
\drawphiavdifdiagbab{0}{0}{0}	
\node at (-3.5,0) {$\bf 1$};	
\node at (3.5,-1) {$\bf 2$};	 
\node at (3.5,1) {$\bf 3$};	
\end{scope}

%%%%%%%%%%%%%%%%%%%%%%%%%%%%%%%%%%%%%%%%%%%%%%%%%%%%%%%%%%%%%%

\node at (-6,-17.5) {$ (\phidif^3 )$};
\node at (-6,-18.5) {vertex};

\begin{scope}[shift={(0,-18)}]
\phipropagatorb{-3}{0}{-1}{1}{}
\phipropagatorb{-1}{0}{30}{1.2}{}
\phipropagatorb{-1}{0}{-30}{1.2}{} 
\counterterm{-1}{0}{0.2}

\node at (-3.5,0) {$\bf 1$};	
\node at (1.5,-1) {$\bf 2$};	 
\node at (1.5,1) {$\bf 3$};
\end{scope}

\node at (3,-18) {\large $+$};

\begin{scope}[shift={(8,-18)}]
\phipropagatorb{-3}{0}{-1}{1}{}
\phipropagatorb{1}{0}{30}{1.2}{}
\phipropagatorb{1}{0}{-30}{1.2}{} 

\node at (-3.5,0) {$\bf 1$};	
\node at (3.5,-1) {$\bf 2$};	
\node at (3.5,1) {$\bf 3$};

\drawphiavdifdiagbaf{0}{0}{0}
\end{scope}

\node at (13,-18) {\large $+$};

\begin{scope}[shift={(18,-18)}]
\phipropagatorb{-3}{0}{-1}{1}{}
\phipropagatorb{1}{0}{30}{1.2}{}
\phipropagatorb{1}{0}{-30}{1.2}{} 
\drawphiavdifdiagbab{0}{0}{0}	
\node at (-3.5,0) {$\bf 1$};	
\node at (3.5,-1) {$\bf 2$};	 
\node at (3.5,1) {$\bf 3$};	
\end{scope}

\node at (9,-22) {\LARGE $\bf \phiav \rightarrow \phiav \phiav$}; 

\end{tikzpicture}
\end{center}
\caption{Renormalization of the cubic couplings in the average-difference basis}
\label{fig:avgdiffcubic1} 
\end{figure}

The tree level and the one loop Feynman diagrams for $\phiav \rightarrow \phiav \phiav$ via $\phidif^3$ vertex is depicted in the fourth row of the figure \ref{fig:avgdiffcubic1} and the contribution from these diagrams are given by 
\begin{equation}
\begin{split}
\frac{-i}{4}&(\re\ \lambda_3-3\ \re\ \sigma_3)\\
&+ (3\ \text{channels})\times (-i)(\re\ \lambda_3+ \ \re\ \sigma_3)  \times \frac{(-i)}{4} (\re\ \lambda_4- 2\ \re\ \sigma_4)   \times \frac{i}{2(4\pi)^2}\left(\frac{2}{d-4} +   \ln\  \frac{1}{4\pi e^{-\gamma_E}}\right) \\
& + (3\ \text{channels})\times \frac{1}{2}(\im\ \lambda_4+ \lambda_\Delta)  \times \frac{1}{2} (\im\ \lambda_3- \ \im\ \sigma_3)\times \frac{i}{2(4\pi)^2}\left(\frac{2}{d-4} +   \ln\  \frac{1}{4\pi e^{-\gamma_E}}\right)\\
\end{split}
\label{eq:avdiff241}
\end{equation}
which leads to the following beta function for $(\re\ \lambda_3-3\ \re\ \sigma_3)$
\begin{equation}
\begin{split}
\frac{d}{d\ln\, \mu}(\re\ \lambda_3-3\ \re\ \sigma_3)
=&\frac{3}{(4\pi)^2} (\re\ \lambda_3+ \ \re\ \sigma_3) (\re\ \lambda_4- 2\ \re\ \sigma_4) \\
&-\frac{3}{(4\pi)^2} (\im\ \lambda_4+ \lambda_\Delta)(\im\ \lambda_3- \ \im\ \sigma_3)
\end{split}
\label{eq:avdiff242}
\end{equation}

\subsubsection{Final $\beta$ function} 

From the equations \eqref{eq:avdiff212}, \eqref{eq:avdiff222}, \eqref{eq:avdiff232} and \eqref{eq:avdiff242} we can compute the beta functions of $\lambda_3$, $\sigma_3$ 
\begin{equation}
\begin{split}
\frac{d \lambda_3}{d\ln\, \mu}
&=\frac{3}{(4\pi)^2} \Bigl[ \lambda_4(\lambda_3+\sigma_3)+ \sigma_4(\lambda_3+\sigma_3^\ast)+i \lambda_\Delta(\sigma_3-\sigma_3^\ast) \Bigr] \\
\frac{d \sigma_3}{d\ln\, \mu}
&=\frac{1}{(4\pi)^2} \Bigl[ (\lambda_4+2\sigma^\ast_4)(\sigma_3-\sigma_3^\ast) +\sigma_4 (\lambda_3^\ast+2\lambda_3+3\sigma_3)-i\lambda_\Delta(\lambda_3^\ast+2\lambda_3+3\sigma_3^\ast)
 \Bigr] \\
\end{split}
\label{eq:avdiff251}
\end{equation}
 
\subsection{Beta functions for the quartic couplings}
In this subsection we compute the beta functions for the quartic couplings in the average-difference basis. There are five different quartic coupling constants: these are the coupling constants multiplying the operators $\phiav^4$, $\phiav^3\phidif$, $\phiav^2\phidif^2$, $\phiav\phidif^3$, $\phidif^4$. For all these couplings there are two distinct divergent diagrams (similar to the cubic coupling constants). As in the last subsection, we keep  only the divergent terms from the one loop contributions.

\subsubsection{\texorpdfstring{$\phiav^4$}{phia} vertex} 
We start by computing $\phidif\phidif \rightarrow \phidif \phidif$ via $\phiav^4$ vertex. The Feynman diagrams are depicted in the first row of the figure \ref{fig:avgdiffquartic1} and the corresponding contributions are given by 
\begin{equation}
\begin{split}
2&(\im\ \lambda_4+ 4\ \im\ \sigma_4 -3 \lambda_\Delta)\\
& + (2\ \text{diagrams}) (3\ \text{channels})\times 2(\im\ \lambda_4+ 4\ \im\ \sigma_4 -3 \lambda_\Delta)  \\
&\times (-i) (\re\ \lambda_4+ 2\ \re\ \sigma_4) \times \frac{i}{2(4\pi)^2}\left(\frac{2}{d-4} +   \ln\  \frac{1}{4\pi e^{-\gamma_E}}\right)\\
% &= 2(\im\ \lambda_4+ 4\ \im\ \sigma_4 -3\ \lambda_\Delta) + 2(\im\ \lambda_4+ 4\ \im\ \sigma_4 -3\ \lambda_\Delta)  \times (\re\ \lambda_4+ 2\ \re\ \sigma_4) \times \frac{6}{(4\pi)^2} \frac{1}{d-4}
\end{split}
\label{eq:avdiff311}
\end{equation}
Thus the beta function for $(\im\ \lambda_4+ 4\ \im\ \sigma_4 -3 \lambda_\Delta)$ is given by 
\begin{equation}
\begin{split}
\frac{d}{d\ln\, \mu}(\im\ \lambda_4+ 4\ \im\ \sigma_4 -3 \lambda_\Delta)
&=\frac{6}{(4\pi)^2} (\im\ \lambda_4+ 4\ \im\ \sigma_4 -3 \lambda_\Delta)  (\re\ \lambda_4+ 2 \re\ \sigma_4) 
\end{split}
\label{eq:avdiff312}
\end{equation}

\subsubsection{$\phiav^3\phidif$ vertex}
Next we compute $\phidif\phidif \rightarrow \phiav \phidif$ via $\phiav^3\phidif$ vertex, which is depicted in the second row of the figure \ref{fig:avgdiffquartic1}. The tree level and one loop contributions from these Feynman diagrams are given by
\begin{equation}
\begin{split}
 \frac{d}{d\ln\, \mu}(\re\ \lambda_4+2\ \re\ \sigma_4)
& +(3\ \text{channels})\times  2(\im\ \lambda_4+ 4\ \im\ \sigma_4 -3 \lambda_\Delta)  \\
&\times \frac{1}{2}(\im\ \lambda_4+  \lambda_\Delta)  \times \frac{i}{2(4\pi)^2}\left(\frac{2}{d-4} +   \ln\  \frac{1}{4\pi e^{-\gamma_E}}\right)\\
& +(3\ \text{channels})\times  (-i) (\re\ \lambda_4+2\ \re\ \sigma_4) \\
&\times  (-i) (\re\ \lambda_4+2\ \re\ \sigma_4)  \times \frac{i}{2(4\pi)^2}\left(\frac{2}{d-4} +   \ln\  \frac{1}{4\pi e^{-\gamma_E}}\right)\\
% &=(-i) (\re\ \lambda_4+2\ \re\ \sigma_4)+ i (\im\ \lambda_4+ 4\ \im\ \sigma_4 -3 \lambda_\Delta) (\im\ \lambda+  \lambda_\Delta)   \frac{3}{(4\pi)^2} \frac{1}{d-4}\\
% &\qquad - i   (\re\ \lambda_4+ 2\ \re\ \sigma_4)^2 \times \frac{3}{(4\pi)^2} \frac{1}{d-4}
\end{split}
\label{eq:avdiff321}
\end{equation}
we can determine the beta function for $(\re\ \lambda_4+2\ \re\ \sigma_4)$ - 
\begin{equation}
\begin{split}
\frac{d}{d\ln\, \mu}(\re\ \lambda_4+2\ \re\ \sigma_4)
&=\frac{3}{(4\pi)^2} \Bigl[ (\re\ \lambda_4+2\ \re\ \sigma_4)^2  -(\im\ \lambda_4+ 4\ \im\ \sigma_4 -3 \lambda_\Delta) (\im\ \lambda_4+  \lambda_\Delta) \Bigr]
\end{split}
\label{eq:avdiff322}
\end{equation} 

\subsubsection{$\phiav^2\phidif^2$ vertex}
The tree level and one loop Feynman diagrams for $\phiav\phiav \rightarrow \phiav \phiav$ via $\phiav^2\phidif^2$ vertex, which is depicted in the third row of the figure  \ref{fig:avgdiffquartic1}, contributes as follows 
\begin{equation}
\begin{split}
 \frac{1}{2}&(\im\ \lambda_4+  \lambda_\Delta)  \\
& +(3\ \text{channels})\times   \frac{1}{2}(\im\ \lambda_4+  \lambda_\Delta) \times (-i) (\re\ \lambda+2\ \re\ \sigma_4)   \times \frac{i}{2(4\pi)^2}\left(\frac{2}{d-4} +   \ln\  \frac{1}{4\pi e^{-\gamma_E}}\right)\\
& +(2\ \text{channels})\times   \frac{1}{2}(\im\ \lambda_4+  \lambda_\Delta) \times (-i) (\re\ \lambda_4+2\ \re\ \sigma_4)   \times \frac{i}{2(4\pi)^2}\left(\frac{2}{d-4} +   \ln\  \frac{1}{4\pi e^{-\gamma_E}}\right)\\
& + 2(\im\ \lambda_4+ 4\ \im\ \sigma_4 -3 \lambda_\Delta) \times \frac{(-i)}{4} (\re\ \lambda_4-2\ \re\ \sigma_4)  \times \frac{i}{2(4\pi)^2}\left(\frac{2}{d-4} +   \ln\  \frac{1}{4\pi e^{-\gamma_E}}\right)\\
% &= \frac{1}{2}(\im\ \lambda_4+  \lambda_\Delta)+  (\im\ \lambda_4+  \lambda_\Delta)(\re\ \lambda_4+2\ \re\ \sigma_4)   \frac{5}{2(4\pi)^2} \frac{1}{d-4}\\
% &\qquad +  (\im\ \lambda_4+ 4\ \im\ \sigma_4 -3 \lambda_\Delta)(\re\ \lambda_4-2\ \re\ \sigma_4)\frac{1}{2(4\pi)^2} \frac{1}{d-4}
\end{split}
\label{eq:avdiff331}
\end{equation} 
From  this equation we evaluate the beta function for $(\im\ \lambda_4+  \lambda_\Delta)$,
\begin{equation}
\begin{split}
\frac{d}{d\ln\, \mu}(\im\ \lambda_4+  \lambda_\Delta) 
=&\frac{1}{(4\pi)^2} \Bigl[ 5 (\im\ \lambda_4+  \lambda_\Delta)(\re\ \lambda_4+2\ \re\ \sigma_4) \\
&+ (\im\ \lambda_4+ 4\ \im\ \sigma_4 -3 \lambda_\Delta)(\re\ \lambda_4-2\ \re\ \sigma_4)\ \Bigr]
\end{split}
\label{eq:avdiff332}
\end{equation}

\subsubsection{$\phiav\phidif^3$ vertex}
Here we determine the tree level and one loop contribution to $\phiav\phiav \rightarrow \phiav \phiav$ via $\phiav\phidif^3$ vertex. The Feynman diagrams can be found  in the fourth row in figure \ref{fig:avgdiffquartic1} and the corresponding contributions are 
\begin{equation}
\begin{split}
\frac{(-i)}{4}& (\re\ \lambda_4-2\ \re\ \sigma_4)  \\
& +(3\ \text{channels})\times   \frac{1}{2}(\im\ \lambda_4+  \lambda_\Delta) \times \frac{1}{2}(\im\ \lambda_4+  \lambda_\Delta)   \times \frac{i}{2(4\pi)^2}\left(\frac{2}{d-4} +   \ln\  \frac{1}{4\pi e^{-\gamma_E}}\right)\\
& +(3\ \text{channels})\times   \frac{(-i)}{4} (\re\ \lambda_4-2\ \re\ \sigma_4)  \\
&\times (-i) (\re\ \lambda_4+2\ \re\ \sigma_4)   \times \frac{i}{2(4\pi)^2}\left(\frac{2}{d-4} +   \ln\  \frac{1}{4\pi e^{-\gamma_E}}\right)\\
% =&\frac{(-i)}{4} (\re\ \lambda_4-2\ \re\ \sigma_4) + i  (\im\ \lambda_4+  \lambda_\Delta)^2(\re\ \lambda_4+2\ \re\ \sigma_4)   \frac{3}{4(4\pi)^2} \frac{1}{d-4}\\
% &\qquad -i  (\re\ \lambda_4-2\ \re\ \sigma_4)(\re\ \lambda_4+2\ \re\ \sigma_4)\frac{3}{4(4\pi)^2} \frac{1}{d-4}
\end{split}
\label{eq:avdiff341}
\end{equation}
The beta function for $(\re\ \lambda_4-2\ \re\ \sigma_4)$ is given by 
\begin{equation}
\begin{split}
\frac{d}{d\ln\, \mu}(\re\ \lambda_4-2\ \re\ \sigma_4)
&=\frac{3}{(4\pi)^2} \Bigl[ (\re\ \lambda_4-2\ \re\ \sigma_4)(\re\ \lambda_4+2\ \re\ \sigma_4)-  (\im\ \lambda_4+  \lambda_\Delta)^2\ \Bigr]
\end{split}
\label{eq:avdiff342}
\end{equation}

\subsubsection{$\phidif^4$ vertex}
\begin{figure}[ht]
\begin{center}
\begin{tikzpicture}[ scale=0.5]

%%%%%%%%%%%%%%%%%%%%%%%%%%%%%%%%%%%%%%%%%%%%%%%%%%%%%%%%%%%%%
\node at (-6,.5) {$ (\phiav^4 )$}; 
\node at (-6,-.5) {vertex};

\begin{scope}[shift={(0,0)}]
\phipropagatora{-1}{0}{150}{1.2}{} 
\phipropagatora{-1}{0}{-150}{1.2}{} 
\phipropagatora{-1}{0}{30}{1.2}{}
\phipropagatora{-1}{0}{-30}{1.2}{} 
\counterterm{-1}{0}{0.2}

\node at (-3.5,1) {$\bf 1$};	
\node at (-3.5,-1) {$\bf 2$};
\node at (1.5,1) {$\bf 3$};	 
\node at (1.5,-1) {$\bf 4$};
\end{scope}

\node at (3,0) {\large $+$};

\begin{scope}[shift={(8,0)}]
\phipropagatora{-1}{0}{150}{1.2}{} 
\phipropagatora{-1}{0}{-150}{1.2}{} 
\phipropagatora{1}{0}{30}{1.2}{}
\phipropagatora{1}{0}{-30}{1.2}{} 

\node at (-3.5,1) {$\bf 1$};	
\node at (-3.5,-1) {$\bf 2$};	
\node at (3.5,1) {$\bf 3$};	
\node at (3.5,-1) {$\bf 4$};

\drawphiavdifdiagbaf{0}{0}{0}
\end{scope}

\node at (13,0) {\large $+$};

\begin{scope}[shift={(18,0)}]
\phipropagatora{-1}{0}{150}{1.2}{} 
\phipropagatora{-1}{0}{-150}{1.2}{} 
\phipropagatora{1}{0}{30}{1.2}{}
\phipropagatora{1}{0}{-30}{1.2}{} 
\drawphiavdifdiagbab{0}{0}{0}
	
\node at (-3.5,1) {$\bf 1$};	
\node at (-3.5,-1) {$\bf 2$};	
\node at (3.5,1) {$\bf 3$};	
\node at (3.5,-1) {$\bf 4$};	
\end{scope}

%%%%%%%%%%%%%%%%%%%%%%%%%%%%%%%%%%%%%%%%%%%%%%%%%%%%%%%%%%%%%
\node at (-6,-5.5) {$ (\phiav^3 \phidif )$}; 
\node at (-6,-6.5) {vertex};

\begin{scope}[shift={(0,-6)}]
\phipropagatora{-1}{0}{150}{1.2}{} 
\phipropagatora{-1}{0}{-150}{1.2}{} 
\phipropagatora{-1}{0}{30}{1.2}{}
\phipropagatorb{-1}{0}{-30}{1.2}{} 
\counterterm{-1}{0}{0.2} 

\node at (-3.5,1) {$\bf 1$};	
\node at (-3.5,-1) {$\bf 2$};
\node at (1.5,1) {$\bf 3$};	 
\node at (1.5,-1) {$\bf 4$};
\end{scope}

\node at (3,-6) {\large $+$};

\begin{scope}[shift={(8,-6)}]
\phipropagatora{-1}{0}{150}{1.2}{} 
\phipropagatora{-1}{0}{-150}{1.2}{} 
\phipropagatora{1}{0}{30}{1.2}{}
\phipropagatorb{1}{0}{-30}{1.2}{} 

\node at (-3.5,1) {$\bf 1$};	
\node at (-3.5,-1) {$\bf 2$};	
\node at (3.5,1) {$\bf 3$};	
\node at (3.5,-1) {$\bf 4$};

\drawphiavdifdiagbaf{0}{0}{0}
\end{scope}

\node at (13,-6) {\large $+$};

\begin{scope}[shift={(18,-6)}]
\phipropagatora{-1}{0}{150}{1.2}{} 
\phipropagatora{-1}{0}{-150}{1.2}{} 
\phipropagatora{1}{0}{30}{1.2}{}
\phipropagatorb{1}{0}{-30}{1.2}{} 
\drawphiavdifdiagbab{0}{0}{0}
	
\node at (-3.5,1) {$\bf 1$};	
\node at (-3.5,-1) {$\bf 2$};	
\node at (3.5,1) {$\bf 3$};	
\node at (3.5,-1) {$\bf 4$};	
\end{scope}

%%%%%%%%%%%%%%%%%%%%%%%%%%%%%%%%%%%%%%%%%%%%%%%%%%%%%%%%%%%%%
\node at (-6,-11.5) {$ (\phiav^2 \phidif^2 )$}; 
\node at (-6,-12.5) {vertex};

\begin{scope}[shift={(0,-12)}]
\phipropagatora{-1}{0}{150}{1.2}{} 
\phipropagatora{-1}{0}{-150}{1.2}{} 
\phipropagatorb{-1}{0}{30}{1.2}{}
\phipropagatorb{-1}{0}{-30}{1.2}{} 
\counterterm{-1}{0}{0.2}

\node at (-3.5,1) {$\bf 1$};	
\node at (-3.5,-1) {$\bf 2$};
\node at (1.5,1) {$\bf 3$};	 
\node at (1.5,-1) {$\bf 4$};
\end{scope}

\node at (3,-12) {\large $+$};

\begin{scope}[shift={(8,-12)}]
\phipropagatora{-1}{0}{150}{1.2}{} 
\phipropagatora{-1}{0}{-150}{1.2}{} 
\phipropagatorb{1}{0}{30}{1.2}{}
\phipropagatorb{1}{0}{-30}{1.2}{} 

\node at (-3.5,1) {$\bf 1$};	
\node at (-3.5,-1) {$\bf 2$};	
\node at (3.5,1) {$\bf 3$};	
\node at (3.5,-1) {$\bf 4$};

\drawphiavdifdiagbaf{0}{0}{0}
\end{scope}

\node at (13,-12) {\large $+$};

\begin{scope}[shift={(18,-12)}]
\phipropagatora{-1}{0}{150}{1.2}{} 
\phipropagatora{-1}{0}{-150}{1.2}{} 
\phipropagatorb{1}{0}{30}{1.2}{}
\phipropagatorb{1}{0}{-30}{1.2}{} 
\drawphiavdifdiagbab{0}{0}{0}
	
\node at (-3.5,1) {$\bf 1$};	
\node at (-3.5,-1) {$\bf 2$};	
\node at (3.5,1) {$\bf 3$};	 
\node at (3.5,-1) {$\bf 4$};	
\end{scope}

%%%%%%%%%%%%%%%%%%%%%%%%%%%%%%%%%%%%%%%%%%%%%%%%%%%%%%%%%%%%%
\node at (-6,-17.5) {$ (\phiav \phidif^3 )$}; 
\node at (-6,-18.5) {vertex};

\begin{scope}[shift={(0,-18)}]
\phipropagatora{-1}{0}{150}{1.2}{} 
\phipropagatorb{-1}{0}{-150}{1.2}{} 
\phipropagatorb{-1}{0}{30}{1.2}{}
\phipropagatorb{-1}{0}{-30}{1.2}{} 
\counterterm{-1}{0}{0.2}

\node at (-3.5,1) {$\bf 1$};	
\node at (-3.5,-1) {$\bf 2$};
\node at (1.5,1) {$\bf 3$};	 
\node at (1.5,-1) {$\bf 4$};
\end{scope}

\node at (3,-18) {\large $+$};

\begin{scope}[shift={(8,-18)}]
\phipropagatora{-1}{0}{150}{1.2}{} 
\phipropagatorb{-1}{0}{-150}{1.2}{} 
\phipropagatorb{1}{0}{30}{1.2}{}
\phipropagatorb{1}{0}{-30}{1.2}{} 

\node at (-3.5,1) {$\bf 1$};	
\node at (-3.5,-1) {$\bf 2$};	
\node at (3.5,1) {$\bf 3$};	
\node at (3.5,-1) {$\bf 4$};

\drawphiavdifdiagbaf{0}{0}{0}
\end{scope}

\node at (13,-18) {\large $+$};

\begin{scope}[shift={(18,-18)}]
\phipropagatora{-1}{0}{150}{1.2}{} 
\phipropagatorb{-1}{0}{-150}{1.2}{} 
\phipropagatorb{1}{0}{30}{1.2}{}
\phipropagatorb{1}{0}{-30}{1.2}{} 
\drawphiavdifdiagbab{0}{0}{0}
	
\node at (-3.5,1) {$\bf 1$};	
\node at (-3.5,-1) {$\bf 2$};	
\node at (3.5,1) {$\bf 3$};	
\node at (3.5,-1) {$\bf 4$};	
\end{scope}

%%%%%%%%%%%%%%%%%%%%%%%%%%%%%%%%%%%%%%%%%%%%%%%%%%%%%%%%%%%%%
\node at (-6,-23.5) {$ (\phidif^4 )$}; 
\node at (-6,-24.5) {vertex};

\begin{scope}[shift={(0,-24)}]
\phipropagatorb{-1}{0}{150}{1.2}{} 
\phipropagatorb{-1}{0}{-150}{1.2}{} 
\phipropagatorb{-1}{0}{30}{1.2}{}
\phipropagatorb{-1}{0}{-30}{1.2}{} 
\counterterm{-1}{0}{0.2}

\node at (-3.5,1) {$\bf 1$};	
\node at (-3.5,-1) {$\bf 2$};
\node at (1.5,1) {$\bf 3$};	 
\node at (1.5,-1) {$\bf 4$};
\end{scope}

\node at (3,-24) {\large $+$};

\begin{scope}[shift={(8,-24)}]
\phipropagatorb{-1}{0}{150}{1.2}{} 
\phipropagatorb{-1}{0}{-150}{1.2}{} 
\phipropagatorb{1}{0}{30}{1.2}{}
\phipropagatorb{1}{0}{-30}{1.2}{} 

\node at (-3.5,1) {$\bf 1$};	
\node at (-3.5,-1) {$\bf 2$};	
\node at (3.5,1) {$\bf 3$};	
\node at (3.5,-1) {$\bf 4$};

\drawphiavdifdiagbaf{0}{0}{0}
\end{scope}

\node at (13,-24) {\large $+$};

\begin{scope}[shift={(18,-24)}]
\phipropagatorb{-1}{0}{150}{1.2}{} 
\phipropagatorb{-1}{0}{-150}{1.2}{} 
\phipropagatorb{1}{0}{30}{1.2}{}
\phipropagatorb{1}{0}{-30}{1.2}{} 
\drawphiavdifdiagbab{0}{0}{0}
	
\node at (-3.5,1) {$\bf 1$};	
\node at (-3.5,-1) {$\bf 2$};	
\node at (3.5,1) {$\bf 3$};	
\node at (3.5,-1) {$\bf 4$};	
\end{scope} 
\node at (9,-27.5) {\LARGE $\bf \phiav \phiav \rightarrow \phiav \phiav$};	
\end{tikzpicture}
\end{center}
\caption{Renormalization of the quartic couplings in the average-difference basis}
\label{fig:avgdiffquartic1} 
\end{figure}
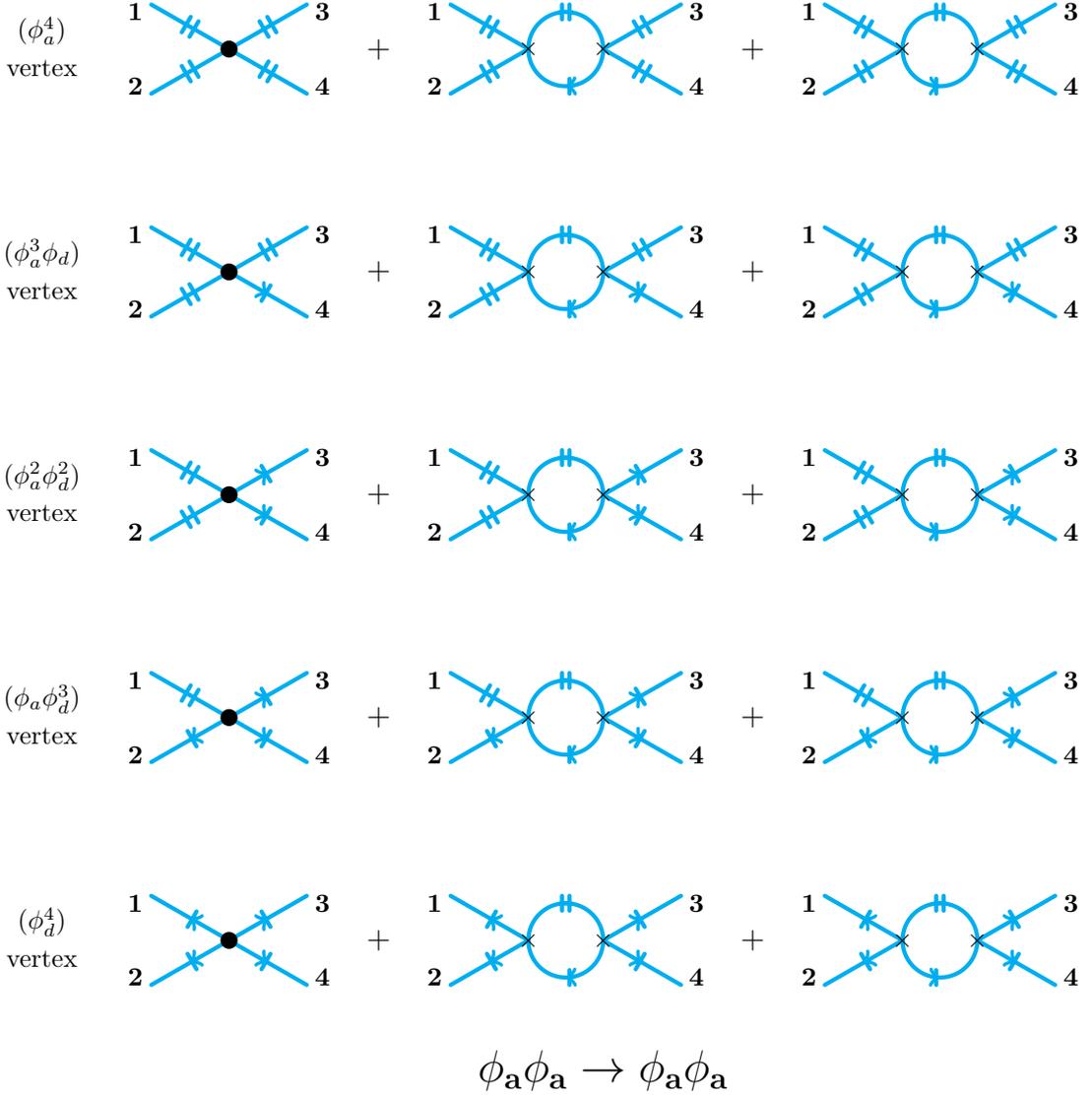

The last computation of this section is $\phiav\phiav \rightarrow \phiav \phiav$ via $\phidif^4$ vertex, which is depicted in the fifth row in figure \ref{fig:avgdiffquartic1}. The tree level and one loop contribution to this process is given by 
\begin{equation}
\begin{split}
\frac{1}{8}&(\im\ \lambda_4-4\ \im\ \sigma_4 -3 \lambda_\Delta)\\
& + (2\ \text{diagrams}) (3\ \text{channels})\times \frac{1}{2}(\im\ \lambda_4+  \lambda_\Delta)  \\
&\times \frac{(-i)}{4} (\re\ \lambda_4- 2\ \re\ \sigma_4) \times \frac{i}{2(4\pi)^2}\left(\frac{2}{d-4} +   \ln\  \frac{1}{4\pi e^{-\gamma_E}}\right)\\
% &= \frac{1}{8}(\im\ \lambda_4-4\ \im\ \sigma_4 -3 \lambda_\Delta) + (\im\ \lambda_4+  \lambda_\Delta)  \times (\re\ \lambda_4- 2\ \re\ \sigma_4) \times \frac{3}{4(4\pi)^2} \frac{1}{d-4}
\end{split} 
\label{eq:avdiff351}
\end{equation}
this equation determines the the beta function for $(\im\ \lambda_4- 4\ \im\ \sigma_4 -3 \lambda_\Delta)$ and it is given by 
\begin{equation}
\begin{split}
\frac{d}{d\ln\, \mu}(\im\ \lambda_4- 4\ \im\ \sigma_4 -3 \lambda_\Delta)
&=\frac{6}{(4\pi)^2} (\im\ \lambda_4+  \lambda_\Delta)  \times (\re\ \lambda_4- 2\ \re\ \sigma_4)
\end{split}
\label{eq:avdiff352}
\end{equation}

\subsubsection{Final $\beta$ functions}

From the five equations - \eqref{eq:avdiff312}, \eqref{eq:avdiff322}, \eqref{eq:avdiff332}, \eqref{eq:avdiff342} and \eqref{eq:avdiff352} we can determine the beta functions of $\lambda_4$ ($=\re\ \lambda_4+i\ \im\ \lambda_4$), $\sigma_4$ ($=\re\ \sigma_4+i\ \im\ \sigma_4$) and $\lambda_\Delta$ and they are given by 
\begin{equation}
\begin{split}
\frac{d \lambda_4}{d\ln\, \mu}
&=\frac{3}{(4\pi)^2} \Bigl[ \lambda_4^2+2\  \sigma_4(\lambda_4+i \lambda_\Delta)+\lambda_\Delta^2\ \Bigr]= \frac{3}{(4\pi)^2}  (\lambda_4+2\ \sigma_4-i\lambda_\Delta)(\lambda_4+i \lambda_\Delta)\\
\frac{d \sigma_4}{d\ln\, \mu}
&=\frac{3}{(4\pi)^2} \Bigl[ \sigma_4^2+(\lambda_4+\sigma_4^\ast)(\sigma_4 - i \lambda_\Delta)+\lambda_\Delta^2\ \Bigr]=\frac{3}{(4\pi)^2} (\lambda_4+\sigma_4+\sigma_4^\ast+i\lambda_\Delta)(\sigma - i \lambda_\Delta)\\
\frac{d \lambda_\Delta}{d\ln\, \mu}
&=\frac{1}{(4\pi)^2i} \Bigl[ \lambda(\sigma_4^\ast+ i \lambda_\Delta)-2\sigma_4^2+5i \sigma_4 \lambda_\Delta -\lambda_4^\ast(\sigma_4- i \lambda_\Delta)+2(\sigma_4^\ast)^2+5i \sigma_4^\ast \lambda_\Delta  \Bigr]\\
&=\frac{1}{(4\pi)^2i} \Bigl[ (\lambda_4+2\sigma_4^\ast)(\sigma_4^\ast+ i \lambda_\Delta)+3i \sigma_4 \lambda_\Delta -c.c. \Bigr]\\
\label{eq:avdiff3.6.1}
\end{split}
\end{equation}

%\tableofcontents{}

\section{Tadpoles}
\label{appendix:tadpole}

In this appendix we compute various one loop tadpoles in this theory. 

\paragraph{$\phir$ 1 loop tadpole}

\begin{figure}[ht]
\begin{center}
\begin{tikzpicture}[line width=1 pt, scale=1]
%\begin{scope}[rotate=90]

\begin{scope}[shift={(0,0)}, rotate=90]
\draw [phir, ultra thick, domain=0:360] plot ({1*cos(\x)}, {1*sin(\x)});
\draw [phir, ultra thick] (1,0) -- (2.5,0);
\node at (1,0) {$\times $};
\node at (-2,0) {\Large $ \frac{(-i\lambda_3)}{2} A_{R}$};
\end{scope}

\begin{scope}[shift={(5,0)}, rotate=90]
\draw [phil, ultra thick, domain=0:360] plot ({1*cos(\x)}, {1*sin(\x)});
\draw [phir, ultra thick] (1,0) -- (2.5,0);
\node at (1,0) {$\times $};
\node at (-2,0) {\Large $\frac{(i\sigma_3^\star)}{2}A_{L}$};
\end{scope}

\begin{scope}[shift={(10,0)}, rotate=90] 
%\begin{scope}[shift={(2.5,-4)}, rotate=-90]
\draw [phil, ultra thick, domain=0:180] plot ({1*cos(\x)}, {1*sin(\x)});
\draw [phir, ultra thick, domain=180:360] plot ({1*cos(\x)}, {1*sin(\x)});
\draw [phir, ultra thick] (1,0) -- (2.5,0);
\node at (1,0) {$\times $};
\node at (-2,0) {\Large $(-i\sigma_3) A_{M}$};
\end{scope}

\end{tikzpicture}
\end{center}
\caption{1 loop tadpole to $\phir$}
\label{fig:phirtadpole1}
\end{figure}
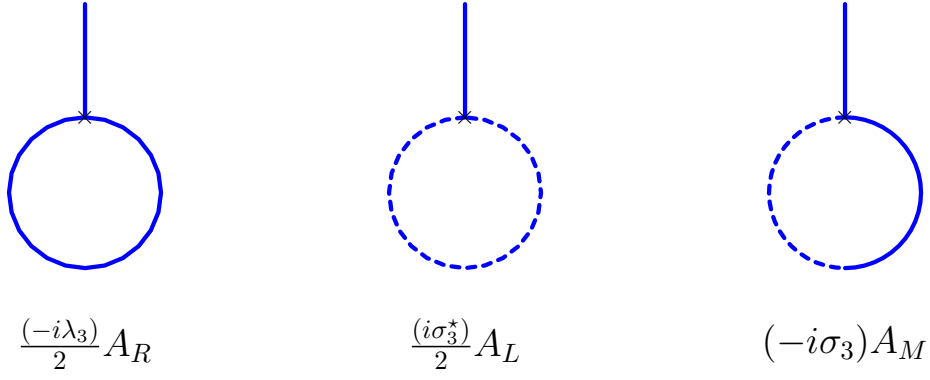
The  Feynman diagrams contributing to one loop tadpole of $\phir$ is being drawn in figure \ref{fig:phirtadpole1}. The sum of the contribution from all the Feynman diagrams is given by 
\begin{eqnarray}
i\mathcal{M}&=&
+\frac{(-i\lambda_3)}{2}A_R+\frac{(i\sigma_3^\star)}{2}A_L
+(-i\sigma_3)A_M 
\label{tadpole1}  
\end{eqnarray}
Now using result from appendix \ref{sec:onelooppvintegrals} we get the following divergent contribution  
\begin{eqnarray}
&=&
\frac{-i\re\ m^2}{(4\pi)^2} \left[\frac{\lambda_3-\sigma_3^\star +2\sigma_3}{2}  
	\right]\left(\frac{2}{d-4} +   \ln\  \frac{1}{4\pi\mu^2e^{-\gamma_E}}\right)
%\nonumber\\
%&=&\frac{-i\re\ [m^2]}{(4\pi)^2} \left[\frac{\re\, \lambda_3+\re\, \sigma_3+i(\im\, \lambda_3+3\, \im\,\sigma_3)}{2}  
%	\right]\left[ \frac{2}{(d-4)} -1+\gamma_E-\ln \,4\pi  
%	\right]
%	\nonumber\\
\label{tadpole12}  
\end{eqnarray}
This can be  removed a counter-term of the form 
\begin{eqnarray}
\kappa  &=&\frac{-i\re\ m^2}{(4\pi)^2} \left[\frac{\re\, \lambda_3+\re\, \sigma_3+i(\im\, \lambda_3+3\, \im\,\sigma_3)}{2}  
	\right]\left(\frac{2}{d-4} +   \ln\  \frac{1}{4\pi\mu^2e^{-\gamma_E}}\right)
	\nonumber\\
\label{tadpole13}  
\end{eqnarray}

\paragraph{Checking lindblad condition}
Now we want to check whether the counter-terms that were added to remove the tadpoles satisfy the Lindblad condition or not.
\begin{eqnarray}
i \Big[\kappa-
\kappa^\star \Big]	=
\frac{-1}{(4\pi)^2} \left[\im\, \lambda_3  +3\, \im\, \sigma_3
	\right]\left(\frac{2}{d-4} +   \ln\  \frac{1}{4\pi\mu^2e^{-\gamma_E}}\right) \re\ m^2
\label{tadpole31}  
\end{eqnarray}
So, the counter-terms obey Lindblad condition if there is no Lindblad violating cubic couplings.

\newpage 
 
%\begin{thebibliography}{99} 

%\providecommand{\href}[2]{#2}
\addcontentsline{toc}{section}{References}
\bibliographystyle{utphys}
\bibliography{lindblad}   
   
%\end{thebibliography} 

\end{document}